\documentclass[11pt]{article}         
\usepackage{epsfig}                   
\usepackage{graphicx}                   
\usepackage{a4p}                      
\usepackage{cite}                     
\usepackage{rotating}                 
\usepackage{psfrag}                 
\begin{document}
%
\newcommand {\dravsn}     {DRAFT 0.33}
\newcommand {\dradate}    {\today}
\newcommand {\commdate}   {never}
\newcommand {\commtime}   {24:00}
\newcommand {\commentto}  {Tatsuo.Kawamoto@cern.ch, Guenter.Duckeck@cern.ch and Richard.Kellogg@cern.ch}
%
%
\newcommand {\etal} {{\em et al.}}
\newcommand {\Zzero}      {\mathrm{Z}}
\newcommand {\MZ}         {m_{\mathrm{Z}}}
\newcommand {\GZ}         {\Gamma_{\mathrm{Z}}}
\newcommand {\shadpol}    {\sigma^0_{\mathrm{h}}}
\newcommand {\Rl}         {R_{\ell}}
\newcommand {\Afbpol}     {A_{\mathrm{FB}}^0}
\newcommand {\Afblpol}    {A_{\mathrm{FB}}^{0,\ell}}
\newcommand {\MW}         {m_{\mathrm{W}}}
\newcommand {\GF}         {G_{\mathrm{F}}}
\newcommand {\Gmu}        {G_{\mu}}
\newcommand {\alfem}      {\alpha} 
\newcommand {\alfemmz}    {\alpha(\MZ^2)} 
\newcommand {\dalh}       {\Delta \alpha^{(5)}_{\mathrm{had}}}
\newcommand {\alfas}      {\alpha_{\mathrm{s}}}
\newcommand {\alfasmz}    {\alfas(\MZ^2)}
\newcommand {\MH}         {m_{\mathrm{H }}}
\newcommand {\Mt}         {m_{\mathrm{t}}}
\newcommand {\ebeam}      {E_{\mathrm {beam}}}
\newcommand {\ecm}        {E_{\mathrm {cm}}}
\newcommand {\roots}      {\sqrt{s}}
\newcommand {\sprime}     {s^\prime}
\newcommand {\rootsprime} {\sqrt{\sprime}}
\newcommand {\rootsprimes} {\sqrt{\sprime/s}}
\newcommand {\costhe}     {\cos\theta}
\newcommand {\abscosthe}  {|\cos\theta\,|}
\newcommand {\abscosthlm} {|\cos\theta_{\ell^-}\,|}
\newcommand {\x}          {{\mathrm X_0}}
\newcommand {\LEPI}       {\mbox{LEP$\,$1}}
\newcommand {\LEPII}      {\mbox{LEP$\,$2}}
\newcommand {\pk}         {peak}
\newcommand {\pkm}        {\mbox{peak$-2$}}
\newcommand {\pkp}        {\mbox{peak$+2$}}

\newcommand {\df}         {\Delta f/f}
\newcommand {\gev}        {\left( \mathrm{GeV} \right) }
\newcommand {\nb}         {\left( \mathrm{nb} \right)  }
\newcommand {\pkmu}       {\mbox{peak$-1$}}
\newcommand {\pkmd}       {\mbox{peak$-2$}}
\newcommand {\pkmt}       {\mbox{peak$-3$}}
\newcommand {\pkpu}       {\mbox{peak$+1$}}
\newcommand {\pkpd}       {\mbox{peak$+2$}}
\newcommand {\pkpt}       {\mbox{peak$+3$}}
\newcommand {\Alr}        {A_{\mathrm{LR}}}
\newcommand {\Afb}        {A_{\mathrm{FB}}}
\newcommand {\Afbee}      {A_{\mathrm{FB}}^{\mathrm{ee}}}
\newcommand {\Afbmumu}    {A_{\mathrm{FB}}^{\mu\mu}}
\newcommand {\Afbtautau}  {A_{\mathrm{FB}}^{\tau\tau}}
\newcommand {\Afbll}      {A_{\mathrm{FB}}^{\ell\ell}}
\newcommand {\Afbbkg}     {A_{\mathrm{FB}}^{\mathrm{bkg}}}
\newcommand {\Afbpolee}   {A_{\mathrm{FB}}^{0,\mathrm{e}}}
\newcommand {\Afbpolmumu} {A_{\mathrm{FB}}^{0,\mu}}
\newcommand{\Afbpoltautau}{A_{\mathrm{FB}}^{0,\tau}}
\newcommand {\Afbpolll}   {A_{\mathrm{FB}}^{0,\ell}}
\newcommand {\NF}         {N_{\mathrm{F}}}
\newcommand {\NB}         {N_{\mathrm{B}}}
\newcommand {\sigF}       {\sigma_{\mathrm{F}}}
\newcommand {\sigB}       {\sigma_{\mathrm{B}}}
\newcommand {\Nnu}        {N_\nu}
\newcommand {\Ree}        {R_{\mathrm{e}}}
\newcommand {\Rmu}        {R_{\mu}}
\newcommand {\Rtau}       {R_{\tau}}
\newcommand {\Rinv}       {R_{\mathrm{inv}}}
\newcommand {\Rinvsm}     {R_{\mathrm{inv}}^{\mathrm{SM}}}

\newcommand {\shad}       {\sigma_{\mathrm{had}}}
\newcommand {\sff}        {\sigma_{\mathrm{f\overline f}}}
\newcommand {\see}        {\sigma_{\mathrm{ee}}}
\newcommand {\seepol}     {\sigma_{\mathrm{ee}}^0}
\newcommand {\smu}        {\sigma_{\mu\mu}}
\newcommand {\stau}       {\sigma_{\tau\tau}}
\newcommand {\sll}        {\sigma_{\ell\ell}}

\newcommand {\thw}        {\theta_{\mathrm{W}}}
\newcommand {\thweff}     {\theta_{\mathrm{W}}^{eff}}
\newcommand {\thweffl}    {\theta_{\mathrm{W}}^{eff,\,\ell}}
\newcommand {\theffl}     {\theta_{\mathrm{eff}}^{\mathrm{lept}}}
\newcommand {\swsq}       {\sin^2\!\thw}
\newcommand {\swsqweff}   {\sin^2\!\thweff}
\newcommand {\swsqweffl}  {\sin^2\!\thweffl}
\newcommand {\swsqeffl}   {\sin^2\!\theffl}
\newcommand {\cwsq}       {\cos^2\!\thw}

\newcommand {\en}         {{\mathrm{e}}^-}
\newcommand {\ep}         {{\mathrm{e}}^+}
\newcommand {\ee}         {\ep\en}
\newcommand {\mumu}       {\mu^+\mu^-}
\newcommand {\tautau}     {\tau^+\tau^-}
\newcommand {\elel}       {\ell^+\ell^-}
\newcommand {\uu}         {{\mathrm u\overline{\mathrm u}}}
\newcommand {\dd}         {{\mathrm d\overline{\mathrm d}}}
\newcommand {\cc}         {{\mathrm c\overline{\mathrm c}}}
\newcommand {\bb}         {{\mathrm b\overline{\mathrm b}}}
\newcommand {\qq}         {{\mathrm q\overline{\mathrm q}}}
\newcommand {\ff}         {{\mathrm f\overline{\mathrm f}}}
\newcommand {\FF}         {{\mathrm F\overline{\mathrm F}}}
\newcommand {\WW}         {\mathrm{W}^+\mathrm{W}^-}
\newcommand {\gaga}       {\gamma\gamma}

\newcommand {\eeee}       {\ee\rightarrow\ee}
\newcommand {\eemumu}     {\ee\rightarrow\mumu}
\newcommand {\eetautau}   {\ee\rightarrow\tautau}
\newcommand {\eett}       {\ee\rightarrow\tautau}
\newcommand {\eehad}      {\ee\rightarrow\qq}
\newcommand {\eeqq}       {\ee\rightarrow\qq}
\newcommand {\eell}       {\ee\rightarrow\elel}
\newcommand {\eegg}       {\ee\rightarrow\gaga}
\newcommand {\eeeell}     {\ee\rightarrow\ee\elel}
\newcommand {\eeeeee}     {\ee\rightarrow\ee\ee}
\newcommand {\eeeemumu}   {\ee\rightarrow\ee\mumu}
\newcommand {\eeeetautau} {\ee\rightarrow\ee\tautau}
\newcommand {\eeeeff}     {\ee\rightarrow\ee\ff}
\newcommand {\eeeeqq}     {\ee\rightarrow\ee\qq}
\newcommand {\eemumumumu} {\ee\rightarrow\mumu\mumu}
\newcommand {\eetautaumumu} {\ee\rightarrow\tautau\mumu}
\newcommand {\eeqqmumu}   {\ee\rightarrow\qq\,\mumu}
\newcommand {\eeffmumu}   {\ee\rightarrow\ff\,\mumu}
\newcommand {\eegammaZ}   {\ee\rightarrow\gamma^*/\Zzero}

\newcommand {\eetoZ}      {\ee\rightarrow\Zzero}
\newcommand {\Ztoee}      {\Zzero\rightarrow\ee}
\newcommand {\Ztomumu}    {\Zzero\rightarrow\mumu}
\newcommand {\Ztotautau}  {\Zzero\rightarrow\tautau}
\newcommand {\Ztoll}      {\Zzero\rightarrow\elel}
\newcommand {\Ztoqq}      {\Zzero\rightarrow\qq}

\newcommand {\Gee}        {\Gamma_{\mathrm{ee}}}
\newcommand {\Gmumu}      {\Gamma_{\mu\mu}}
\newcommand {\Gtautau}    {\Gamma_{\tau\tau}}
\newcommand {\Gll}        {\Gamma_{\ell\ell}}
\newcommand {\Ghad}       {\Gamma_{\mathrm{had}}}
\newcommand {\Gqq}        {\Gamma_{\mathrm{qq}}}
\newcommand {\Gff}        {\Gamma_{\mathrm{ff}}}
\newcommand {\Ginv}       {\Gamma_{\mathrm{inv}}}
\newcommand {\Gnu}        {\Gamma_{\nu\nu}}
\newcommand {\Gtot}       {\Gamma_{\mathrm{tot}}}
\newcommand {\Guu}        {\Gamma_{\uu}}
\newcommand {\Gdd}        {\Gamma_{\dd}}
\newcommand {\Gcc}        {\Gamma_{\cc}}
\newcommand {\Gss}        {\Gamma_{\mathrm{ss}}}
\newcommand {\Gbb}        {\Gamma_{\bb}}

\newcommand {\Cagz}       {C^{\mathrm{a}}_{\gamma \mathrm{Z}}}
\newcommand {\Csgz}       {C^{\mathrm{s}}_{\gamma \mathrm{Z}}}
\newcommand {\Cazz}       {C^{\mathrm{a}}_{\mathrm{ZZ}}}
\newcommand {\Cszz}       {C^{\mathrm{s}}_{\mathrm{ZZ}}}
 
\newcommand {\gaf}        {g_{\mathrm{Af}}}
\newcommand {\gvf}        {g_{\mathrm{Vf}}}
\newcommand {\gal}        {g_{\mathrm{A\ell}}}
\newcommand {\gvl}        {g_{\mathrm{V\ell}}}
\newcommand {\gae}        {g_{\mathrm{Ae}}}
\newcommand {\gve}        {g_{\mathrm{Ve}}}
\newcommand {\gam}        {g_{\mathrm{A\mu}}}
\newcommand {\gvm}        {g_{\mathrm{V\mu}}}
\newcommand {\gatau}      {g_{\mathrm{A\tau}}}
\newcommand {\gvtau}      {g_{\mathrm{V\tau}}}
\newcommand {\gaq}        {g_{\mathrm{Aq}}}
\newcommand {\gvq}        {g_{\mathrm{Vq}}}
\newcommand {\gavsq}      {(\gal)^2(\gvl)^2}
\newcommand {\Itf}        {I^3_{\mathrm f}}
\newcommand {\Qf}         {Q_{\mathrm f}}
\newcommand {\Qq}         {Q_{\mathrm q}}
\newcommand {\Qb}         {Q_{\mathrm b}}
\newcommand {\rhof}       {\rho_{\mathrm f}}
\newcommand {\gahat}      {\hat{g}_{\mathrm{A}}}
\newcommand {\gvhat}      {\hat{g}_{\mathrm{V}}}
\newcommand {\gahatf}     {\hat{g}_{\mathrm{Af}}}   
\newcommand {\gvhatf}     {\hat{g}_{\mathrm{Vf}}}   
\newcommand {\gahatl}     {\hat{g}_{\mathrm{A\ell}}}
\newcommand {\gvhatl}     {\hat{g}_{\mathrm{V\ell}}}
\newcommand {\gahate}     {\hat{g}_{\mathrm{Ae}}}   
\newcommand {\gvhate}     {\hat{g}_{\mathrm{Ve}}}   
\newcommand {\gahatmu}    {\hat{g}_{\mathrm{A\mu}}} 
\newcommand {\gvhatmu}    {\hat{g}_{\mathrm{V\mu}}} 
\newcommand {\gahattau}   {\hat{g}_{\mathrm{A\tau}}}
\newcommand {\gvhattau}   {\hat{g}_{\mathrm{V\tau}}}
\newcommand {\gahatq}     {\hat{g}_{\mathrm{Aq}}}
\newcommand {\gvhatq}     {\hat{g}_{\mathrm{Vq}}}
\newcommand {\gahatnu}    {\hat{g}_{\mathrm{A\nu}}}
\newcommand {\gvhatnu}    {\hat{g}_{\mathrm{V\nu}}}

\newcommand {\gagvhatsq}  {\gahat^2\gvhat^2}
\newcommand {\gagvhatlsq} {\gahatl^2\gvhatl^2}
\newcommand {\gagvhatesq} {\gahate^2\gvhate^2}
\newcommand {\gagvhatmusq}{\gahatmu^2\gvhatmu^2}
\newcommand {\so}         {s_0^2}
\newcommand {\co}         {c_0^2}
\newcommand {\lept}       {\ell^+\ell^-}
\newcommand {\cms}        {\mbox{centre-of-mass}}
\newcommand {\delQED}        {\delta_{\mathrm{QED}}}
\newcommand {\delQCD}        {\delta_{\mathrm{QCD}}}
\newcommand {\gagve}         {\gae^2+\gve^2}
\newcommand {\gagvm}         {\gam^2+\gvm^2}
\newcommand {\gagvtau}       {\gatau^2+\gvtau^2}
\newcommand {\gagvl}         {\gal^2+\gvl^2}
\newcommand {\AAe}           {{\cal A}_{\mathrm{e}}}
\newcommand {\AAm}           {{\cal A}_{\mu}} 
\newcommand {\AAtau}         {{\cal A}_{\tau}} 
\newcommand {\AAl}           {{\cal A}_{\ell}} 
\newcommand {\AAf}           {{\cal A}_{\mathrm{f}}}
\newcommand {\seeF}          {\sigma_{\mathrm{F}}^{\mathrm{ee}}}
\newcommand {\seeB}          {\sigma_{\mathrm{B}}^{\mathrm{ee}}}
\newcommand {\sZFs}          {\sigma_{\mathrm{F}}^{{\mathrm{ZF}},s}}
\newcommand {\sZBs}          {\sigma_{\mathrm{B}}^{{\mathrm{ZF}},s}}
\newcommand {\sAFs}          {\sigma_{\mathrm{F}}^{{\mathrm{AL}},s}}
\newcommand {\sABs}          {\sigma_{\mathrm{B}}^{{\mathrm{AL}},s}}
\newcommand {\sAFt}          {\sigma_{\mathrm{F}}^{{\mathrm{AL}},t}}
\newcommand {\sABt}          {\sigma_{\mathrm{B}}^{{\mathrm{AL}},t}}
\newcommand {\sAFst}         {\sigma_{\mathrm{F}}^{{\mathrm{AL}},s+t}}
\newcommand {\sABst}         {\sigma_{\mathrm{B}}^{{\mathrm{AL}},s+t}}
\newcommand {\sAFBt}         {\sigma_{\mathrm{F(B)}}^{{\mathrm{AL}},t}}
\newcommand {\sAFBst}        {\sigma_{\mathrm{F(B)}}^{{\mathrm{AL}},s+t}}
\newcommand {\sAFBs}         {\sigma_{\mathrm{F(B)}}^{{\mathrm{AL}},s}}

\newcommand {\mc}         {\multicolumn {2} {|c|}}
\newcommand {\mcl}        {\multicolumn {2} {|l|}}
\newcommand {\mcea}       {\multicolumn {7} {|c|}}
\newcommand {\mceb}       {\multicolumn {6} {|c|}}
\newcommand {\mcec}       {\multicolumn {8} {|c|}}
\newcommand {\mco}        {\multicolumn {1} {|c|}}
\newcommand {\mcoo}       {\multicolumn {1} {|c}}

\newcommand {\rr}         {R_{\mathrm R}}
\newcommand {\rl}         {R_{\mathrm L}}
\newcommand {\ra}         {R_{\mathrm A}}
\newcommand {\phir}       {\phi_{\mathrm R}}
\newcommand {\phil}       {\phi_{\mathrm L}}
\newcommand {\er}         {E_{\mathrm R}}
\newcommand {\el}         {E_{\mathrm L}}
\newcommand {\SwitA}      {\mbox{\sc SwitA} }
\newcommand {\SwitR}      {\mbox{\sc SwitR} }
\newcommand {\SwitL}      {\mbox{\sc SwitL} }
\newcommand {\SwitRL}     {\mbox{\sc SwitR/L} }
\newcommand {\SwitX}      {\mbox{\sc SwitX} }
\newcommand {\La}         {L_{\mbox{A}} }
\newcommand {\Lrl}        {L_{\mbox{RL}} }
\newcommand {\Drl}        {D_{\mathrm{RL}}}
\newcommand {\Dlr}        {D_{\mathrm{LR}}}
\catcode`@=11 
\def \gsim{\mathrel{\mathpalette\@versim>}}
\def \lsim{\mathrel{\mathpalette\@versim<}}
\def \@versim#1#2{\lower0.4ex\vbox{\baselineskip\z@skip\lineskip\z@skip
     \lineskiplimit\z@\ialign{$\m@th#1\hfil##\hfil$%
     \crcr#2\crcr\sim\crcr}}}
\catcode`@=12 
\newcommand {\NCT}        {N_{\mathrm{track}}}
\newcommand {\NECAL}      {N_{\mathrm{cluster}}}
\newcommand {\NFD}        {N_{\mathrm{FD}}}
\newcommand {\Nall}       {N_{\mathrm{all}}}
\newcommand {\EECAL}      {E_{\mathrm{cluster}}}
\newcommand {\EFD}        {E_{\mathrm{FD}}}
\newcommand {\EECALi}     {\EECAL^{i}}
\newcommand {\EFDj}       {\EFD^{j}}
\newcommand {\Mhemi}      {m_{\mathrm{hemi}}}
\newcommand {\Rvis}       {R_{\mathrm{vis}}}
\newcommand {\Rcal}       {E_{\mathrm{cal}}^{\mathrm{mh}}/\sqrt{s}}
\newcommand {\Rbal}       {R_{\mathrm{bal}}^{\mathrm{energy}}}
\newcommand {\sumi}       {\sum_{i}}
\newcommand {\sumj}       {\sum_{j}}
\newcommand {\thetai}     {\theta_{i}}
\newcommand {\thetaj}     {\theta_{j}}
\newcommand {\ieff}       {\bar{\varepsilon}}
\newcommand {\effx}       {{\ieff}_x}
\newcommand {\effdat}     {{\ieff}^{\mathrm{\,corr}}}
\newcommand {\effMC}      {{\ieff}^{\mathrm{\,MC}}}
\newcommand {\effxdat}    {{\ieff}_x^{\mathrm{\,data}}}
\newcommand {\effxMC}     {{\ieff}_x^{\mathrm{\,MC}}}
\newcommand {\effz}       {{\ieff}_z}
\newcommand {\effzdat}    {{\ieff}_z^{\mathrm{\,data}}}
\newcommand {\effzMC}     {{\ieff}_z^{\mathrm{\,MC}}}
\newcommand {\thetthr}    {\theta_{\mathrm{thr}}}
\newcommand {\thetthrx}   {{\tilde\theta}_{\mathrm{thr}}}
\newcommand {\abscosthetthr} {|\cos\thetthr|}
\newcommand {\abscosthetthrx}{|\cos\thetthrx|}
\newcommand {\QO}         {Q_{0}}
\newcommand {\sigmaq}     {\sigma_q}
\newcommand {\LambdaQCD}  {\Lambda_{\mathrm{QCD}}}
%
%
\newcommand {\Etotal}        {E_{\mathrm{total}}}
\newcommand {\ptotal}        {p_{\mathrm{total}}}
\newcommand {\Eshw   }     {\sum E_{\mathrm{cluster}}^i}
\newcommand {\Etrk}        {\sum p_{\mathrm{track}}^i}
\newcommand {\Ntra}       {N_{\mathrm{track}}}
\newcommand {\Nclu}       {N_{\mathrm{cluster}}}
\newcommand {\sumE}       {\sum E}
\newcommand {\esum}       {\Eshw/\sqrt{s}}
\newcommand {\esume}      {\Eshw}
\newcommand {\thetae}     {\theta_{\mathrm{e}}}
\newcommand {\thelec}     {\theta_{\en}}
\newcommand{\abscosthetae}{|\cos\thetae|}
\newcommand{\abscosthelec}{|\cos\thelec|}
\newcommand {\thacol}     {\theta_{\mathrm{acol}}}
\newcommand {\thacop}     {\phi_{\mathrm{acop}}}
\newcommand {\fbkg}       {f_{\mathrm{bkg}}}
\newcommand {\effF}       {\varepsilon_{\mathrm{F}}}
\newcommand {\effB}       {\varepsilon_{\mathrm{B}}}
\newcommand {\costhelm}   {\cos\theta_{\ell^-}}
\newcommand {\thetaelm}   {\theta_{\ell^-}}
\newcommand {\abscosthei}  {|\cos\theta^i\,|}
\newcommand{\mpmm}{\mbox{$\mu^+\mu^-$}}
\newcommand {\mcost}      {\mbox{$\abscosthe$}}
\newcommand {\dzero}      {d_0}
\newcommand {\absdo}      {|\dzero|}
\newcommand {\sumabsdo}   {\sum{\absdo}}
\newcommand {\zo}         {z_0}
\newcommand {\abszo}      {|\zo|}
\newcommand {\abssumzo}   {|\sum{\zo}|}
\newcommand {\Nhit}       {N_{\mathrm{hit}}}
\newcommand {\delphi}     {\Delta\phi}
\newcommand {\cosdelphi}  {\cos(\delphi)}
\newcommand {\absdelphi}  {|\cos(\delphi)|}
\newcommand {\delphiexp}  {(\delphi)^{\mathrm{exp}}}
\newcommand {\delphiobs}  {(\delphi)^{\mathrm{obs}}}
\newcommand {\Evis}       {E_{\mathrm{vis}}^\mu}
\newcommand {\thetam}     {\theta_{\mu^-}}
\newcommand{\abscosthmuon}{|\cos\thetam|}
\newcommand{\costhmuon}{\cos\thetam}
\newcommand{\abscosthmup}{|\cos\theta_{\mu^+}|}
\newcommand{\abscosthmum}{|\cos\theta_{\mu^-}|}
\newcommand{\costhmum}{\cos\theta_{\mu^-}}
\newcommand{\costhmup}{\cos\theta_{\mu^-}}
%

\newcommand {\epem}        {\mathrm{e}^+\mathrm{e}^-}
\newcommand {\tptm}        {\tau^+\tau^-}
\newcommand {\Rvistau}     {E_{\mathrm{vis}}^{\tau}/\sqrt{s}}
\newcommand {\Evistau}     {E_{\mathrm{vis}}^{\tau}}
\newcommand {\Ecluster }   {E_{\mathrm{cluster}}^i}
\newcommand {\Rshw   }     {\Eshw/\sqrt{s}}
\newcommand {\ptrack}      {p_{\mathrm{track}}^i}
\newcommand {\Rtrk}        {\Etrk/ \sqrt{s}}
\newcommand {\costau}      {|\cos\theta_\tau|}
\newcommand {\costaum}     {|\cos\theta_{\tau^-}|}
\newcommand {\costaums}    {\cos\theta_{\tau^-}}
\newcommand {\zvtx}        {z_{\mathrm{vtx}}}
\newcommand {\thacoltrue}  {\theta_{\mathrm{acol(\tptm)}}}
\newcommand {\cospmis}     {cospmis}
\newcommand {\Econeo}      {econeo}
\newcommand {\Econet}      {econet}
\newcommand {\pconeo}      {pconeo}
\newcommand {\pconet}      {pconet}
\newcommand {\ptsum}       {\Sigma{p_{\mathrm{T}}}}
\newcommand {\Afbmeas}  {A_{\mathrm{FB}}^{\mathrm{meas}}}
\newcommand {\Afbt}  {A_{\mathrm{FB}}^{\tau\tau}}
 

\newcommand{\gashate}{g_{{A}{\rm e}}}
\newcommand{\gvshate}{g_{{V}{\rm e}}}
\newcommand{\gashatf}{g_{{A}{\rm f}}}
\newcommand{\gvshatf}{g_{{V}{\rm f}}}
\newcommand{\gashatl}{g_{{A}{\rm \ell}}}
\newcommand{\gvshatl}{g_{{V}{\rm \ell}}}
\newcommand{\MZbar}{\overline{m}_{\mathrm{Z}}}
\newcommand{\GZbar}{\overline{\Gamma}_{\mathrm{Z}}}
\newcommand{\rf}{{r}_{\mathrm{f}}}
\newcommand{\jf}{{j}_{\mathrm{f}}}
\newcommand{\gf}{{g}_{\mathrm{f}}}
\newcommand{\rhad}{r_{\mathrm{had}}}
\newcommand{\jhad}{j_{\mathrm{had}}}
\newcommand{\ghad}{g_{\mathrm{had}}}
\newcommand{\rtotf}{r^{\mathrm{tot}}_{\mathrm{f}}}
\newcommand{\jtotf}{j\mathrm{^{tot}_{f}}}
\newcommand{\gtotf}{g\mathrm{^{tot}_{f}}}
\newcommand{\rfbf}{r\mathrm{^{fb}_{f}}}
\newcommand{\jfbf}{j\mathrm{^{fb}_{f}}}
\newcommand{\gfbf}{g\mathrm{^{fb}_{f}}}

\newcommand{\rtott}{r^{\mathrm{tot}}_{\tau}}
\newcommand{\jtott}{j\mathrm{^{tot}_{\tau}}}
\newcommand{\rfbt}{r\mathrm{^{fb}_{\tau}}}
\newcommand{\jfbt}{j\mathrm{^{fb}_{\tau}}}

\newcommand{\rtote}{r^{\mathrm{tot}}_{\mathrm{e}}}
\newcommand{\jtote}{j\mathrm{^{tot}_{e}}}
\newcommand{\rfbe}{r\mathrm{^{fb}_{e}}}
\newcommand{\jfbe}{j\mathrm{^{fb}_{e}}}

\newcommand{\rtotm}{r^{\mathrm{tot}}_{\mu}}
\newcommand{\jtotm}{j\mathrm{^{tot}_{\mu}}}
\newcommand{\rfbm}{r\mathrm{^{fb}_{\mu}}}
\newcommand{\jfbm}{j\mathrm{^{fb}_{\mu}}}

\newcommand{\rtoth}{r^{\mathrm{tot}}_{\mathrm{had}}}
\newcommand{\jtoth}{j\mathrm{^{tot}_{had}}}
\newcommand {\eetoff}       {\ee\rightarrow\ff}
\newcommand {\gzif}         {\gamma/\Zzero}
\newcommand {\fgz}          {f_{\gamma/\Zzero}}
\newcommand {\chisq}        {\ensuremath{\chi^2}}
\newcommand {\chidof}       {\ensuremath{\chi^2 / {\mathrm{d.o.f.}}}}
\newcommand {\als}          {\alpha_{\mathrm{s}}}
\newcommand {\sleppol}      {\sigma^0_{\mathrm{\ell}}}
\newcommand {\sfpol}        {\sigma^0_{\mathrm{f}}}
\newcommand {\Afbfpol}      {A_{\mathrm{FB}}^{0,{\mathrm{f}}}}
\newcommand {\EQ}           {Equation}
\newcommand {\EQs}          {Equations}
\newcommand {\FIG}          {Figure}
\newcommand {\FIGs}         {Figures}
\newcommand {\SECT}         {Section}
\newcommand {\SECTs}        {Sections}
\newcommand {\TAB}          {Table}
\newcommand {\TABs}         {Tables}

\newcommand {\Bff}     {Br( \Zzero \rightarrow \mathrm{f \bar{f} })}
\newcommand {\Bee}     {Br( \Zzero \rightarrow \mathrm{e^+ e^-})}

\newcommand {\SLP}     {model-independent $\Zzero$ parameters}
\newcommand {\SLPfive} {5-parameter model-independent}
\newcommand {\SLPnine} {9-parameter model-independent}

\newcommand {\SM}         {SM}

\def\PLB{Phys.\ Lett.\ {\bf B}}
\def\ZPC{Z.\ Phys.\ {\bf C}}
\def\PRL{Phys.\ Rev.\ Lett.\ }
\def\EPJ{Eur.\ Phys.\ J.\ {\bf C}}

\flushbottom
\begin{titlepage}
%
%
\begin{center}{\Large
EUROPEAN ORGANISATION FOR NUCLEAR RESEARCH
}\end{center}\bigskip
\begin{flushright}
       CERN-EP-2000-148 \\ OPAL PR 328 \\ 30 November 2000 
\end{flushright}
%
%
\bigskip
\boldmath
\begin{center}{\huge\bf
%
Precise Determination of the Z Resonance Parameters at LEP:
``Zedometry'' \\
%
}
\end{center}\unboldmath\bigskip
\vspace*{1cm}
\begin{center}{\LARGE The OPAL Collaboration
}\end{center}\bigskip\bigskip
%
%
%

%
\vspace*{2cm}
%
%
\begin{center}{\large Abstract}\end{center}
\vspace*{2ex}
This final analysis of hadronic and leptonic 
cross-sections and of leptonic forward-backward asymmetries in
$\ee$ collisions with the OPAL detector
makes use of the full {\LEPI} data sample comprising
$161~\mathrm{pb}^{-1}$ of integrated luminosity
and $4.5\times10^6$ selected $\Zzero$ decays.
An interpretation of the data in terms of
contributions  from pure $\Zzero$ exchange and from 
$\gzif$ interference allows the parameters of the 
$\Zzero$ resonance to be determined in a model-independent way.
Our results are in good agreement  with lepton 
universality and consistent with the vector and
axial-vector couplings predicted in the
Standard Model. 
A fit to the complete dataset yields the fundamental $\Zzero$
resonance parameters:
$\MZ =  (91.1852 \pm 0.0030)\,\mbox{GeV}$,
$\GZ =   (2.4948 \pm 0.0041)\,\mbox{GeV}$,
$\shadpol =  (41.501 \pm 0.055) \,\mbox{nb}$,
$\Rl =   20.823 \pm 0.044$, and
$\Afbpolll  =   0.0145 \pm 0.0017$.  
Transforming these parameters gives a measurement of the ratio between
the decay width
into invisible particles and the width to a single species of charged
lepton, \mbox{$\Ginv/\Gll = 5.942 \pm 0.027$}.  Attributing the
entire invisible width to neutrino decays and 
assuming the Standard Model couplings for neutrinos, this translates into 
a measurement of the effective number of light neutrino species,
$N_{\nu} =  2.984 \pm  0.013$.
Interpreting the data within the context of the Standard Model
allows the mass of the top quark,
$\Mt =  (162 ^{+29}_{-16})\,$GeV, to be determined through its influence on
radiative corrections.
Alternatively, utilising the direct external measurement of $\Mt$
as an additional constraint leads to a measurement of
the strong coupling constant and the mass of the Higgs boson:
$\als(\MZ) = 0.127 \pm 0.005\,$ and $\MH = (390^{+750}_{-280}) \mbox{~GeV}$.

%
%
%
%
\bigskip\bigskip\bigskip\bigskip
\bigskip\bigskip
\begin{center}{\large
To be submitted to Eur. Phys. J. C}\end{center}
\bigskip\bigskip\bigskip\bigskip
\bigskip\bigskip
\begin{center}{\large
}
\end{center}
%
%

\end{titlepage}
\begin{center}{\Large        The OPAL Collaboration
}\end{center}\bigskip
\begin{center}{
G.\thinspace Abbiendi$^{  2}$,
C.\thinspace Ainsley$^{  5}$,
P.F.\thinspace {\AA}kesson$^{  3}$,
G.\thinspace Alexander$^{ 22}$,
J.\thinspace Allison$^{ 16}$,
G.\thinspace Anagnostou$^{  1}$,
K.J.\thinspace Anderson$^{  9}$,
S.\thinspace Arcelli$^{ 17}$,
S.\thinspace Asai$^{ 23}$,
S.F.\thinspace Ashby$^{  1}$
% ...
}\end{center}\bigskip
\bigskip
$^{  1}$School of Physics and Astronomy, University of Birmingham,
Birmingham B15 2TT, UK
\newline
$^{  2}$Dipartimento di Fisica dell' Universit\`a di Bologna and INFN,
I-40126 Bologna, Italy
\newline
$^{  3}$Physikalisches Institut, Universit\"at Bonn,
D-53115 Bonn, Germany
\newline
% ...
$^{  5}$Cavendish Laboratory, Cambridge CB3 0HE, UK
\newline
$^{  6}$Ottawa-Carleton Institute for Physics,
Department of Physics, Carleton University,
Ottawa, Ontario K1S 5B6, Canada
\newline
$^{  7}$Centre for Research in Particle Physics,
Carleton University, Ottawa, Ontario K1S 5B6, Canada
\newline
$^{  8}$CERN, European Organisation for Nuclear Research,
CH-1211 Geneva 23, Switzerland
\newline
$^{  9}$Enrico Fermi Institute and Department of Physics,
University of Chicago, Chicago IL 60637, USA
\newline
% ...
$^{ 16}$Department of Physics, Schuster Laboratory, The University,
Manchester M13 9PL, UK
\newline
$^{ 17}$Department of Physics, University of Maryland,
College Park, MD 20742, USA
\newline
% ...
$^{ 22}$Department of Physics and Astronomy, Tel Aviv University,
Tel Aviv 69978, Israel
\newline
$^{ 23}$International Centre for Elementary Particle Physics and
Department of Physics, University of Tokyo, Tokyo 113-0033, and
Kobe University, Kobe 657-8501, Japan
\newline
% ...
$^{  a}$ and at TRIUMF, Vancouver, Canada V6T 2A3
\newline
$^{  b}$ and Royal Society University Research Fellow
\newline
$^{  c}$ and Institute of Nuclear Research, Debrecen, Hungary
\newline
$^{  d}$ and University of Mining and Metallurgy, Cracow
\newline
$^{  e}$ and Heisenberg Fellow
\newline
$^{  f}$ and Department of Experimental Physics, Lajos Kossuth University,
 Debrecen, Hungary
\newline
$^{  g}$ and MPI M\"unchen
\newline
$^{  h}$ now at MPI f\"ur Physik, 80805 M\"unchen
\newline
% ...
$^{  m}$ and Tel Aviv University, School of Physics and Astronomy,
Tel Aviv 69978, Israel.
\newline
% ...

%
\tableofcontents
\listoftables
\listoffigures
%
%
\clearpage \newpage

%
\section{Introduction}
\label{sec-intro}

One of the principal goals of the Large Electron-Positron~(LEP) collider 
project~\cite{bib-LEP1yr} at CERN is to make precise 
measurements of the properties of the $\Zzero$ gauge boson,
which are basic parameters of nature. 
These include the $\Zzero$~mass and its total decay width,
as well as the composition and angular 
distributions of~$\Zzero$ decay products. 
In combination with other precise measurements~\cite{bib-pdg98,bib-LEPEWWG},
these physical observables provide the 
most stringent tests of the
Standard Model (SM) of electroweak interactions~\cite{bib-EWSM} yet possible, 
allowing it to be investigated at the level of higher-order electroweak 
corrections~\cite{bib-veltman}.
These tests represent a unique probe of the underlying gauge 
structure of electroweak physics.
Possible new physics or new particles beyond those of the {\SM}
might be revealed through the subtle changes which they would induce in
the precise results reported here.

From 1989 to 1995 LEP produced $\ee$ collisions with 
centre-of-mass \mbox{energies,~$\roots$}, 
close to the $\Zzero$ mass,~$\MZ$. This is referred to 
as the {\LEPI} programme. 
For much of this period $\roots$ was chosen to 
be within about 200~MeV of~$\MZ$, close to the peak of the $\Zzero$ 
resonance curve, in order to obtain the maximum number of $\Zzero$ decay 
events. 
These ``on-peak'' data provide high-statistics samples for 
measurements of production cross-sections and $\Zzero$ decay properties, 
such as partial decay widths and forward-backward asymmetries. 
Data were also recorded at several centre-of-mass energy points 
up to $ 4$ GeV above and below the peak of the $\Zzero$ resonance. 
These ``off-peak'' data samples provide sensitivity to the lineshape 
of the $\Zzero$ resonance and hence to its mass, $\MZ$, and total decay 
width,~$\GZ$.
The OPAL collaboration has previously published measurements of
$\Zzero$ properties based on {\LEPI} data recorded up to the end
of~1992~\cite{bib-opal-ls89,bib-opal-ls90,bib-opal-ls91,bib-opal-ls92}.

In this paper we report new OPAL measurements of hadronic and leptonic 
cross-sections and leptonic forward-backward asymmetries based on the 
higher luminosity {\LEPI} runs which took place during the years 
1993--1995 and resulted in a four-fold increase of our data set. 
In 1993 and 1995 LEP energy scans were performed 
with significant luminosity
collected at the off-peak points,
in order to improve substantially the determination of $\MZ$ and $\GZ$.
These off-peak points were chosen to be approximately 1.8 GeV below
and above $\MZ$. 
In 1994 all data were collected on peak.
In the following we will refer to the energy points 
below, close to and above the $\Zzero$ resonance peak
as ``peak$-2$'', ``peak'' and ``peak$+2$'', respectively.
In parallel to the large increase of the data set, 
significant progress has been made
in each of the many aspects which affect the precision of the results
and their interpretation, namely the LEP energy calibration, the luminosity
determination, the selection of the $\Zzero$ decay products and
the theoretical predictions of observable quantities.
The measurements from the 1993--1995 data
are combined with those from previous years in order to determine the
OPAL values for $\Zzero$ properties based on the full {\LEPI} data sample.

We analyse our results by first interpreting the cross-section and
forward-backward asymmetry measurements
in a model-independent fashion, in which the $\Zzero$ couplings 
to hadrons and leptons are allowed to vary freely. 
This provides a useful phenomenological description of observable $\Zzero$
properties and allows basic predictions of the {\SM}, such as
lepton universality and the vector and axial-vector structure of the couplings,
to be verified.
We then go on to make a fit within the full context of the {\SM}, 
leading
to a direct determination of the accessible {\SM} parameters.
In a forthcoming publication these OPAL measurements will be combined with 
similar results from the ALEPH~\cite{bib-ALEPH-final}, 
DELPHI ~\cite{bib-DELPHI-final} 
and L3~\cite{bib-L3-final}
collaborations, 
in order to determine the final set of LEP results for $\Zzero$ parameters. 

The structure of this paper is as follows. Section~\ref{sec-det} 
contains a brief description of the OPAL detector, simulation program
and {\LEPI} data samples.
Section~\ref{sec-sel} gives an overview of the essential
concepts we use to define our measured sample of $\Zzero$ decays.
The LEP centre-of-mass energy calibration is 
outlined in {\SECT}~\ref{sec-ene}.
The luminosity measurement is reviewed briefly in {\SECT}~\ref{sec-lum}, 
full details being available in~\cite{bib-lumi-siw}.
The details of the hadronic and leptonic event selections and 
analyses are given in \mbox{{\SECTs}~\ref{sec-had} and~\ref{sec-lept}.} 
The cross-section and leptonic forward-backward asymmetry measurements
are described in \mbox{{\SECTs}~\ref{sec-meas} and~\ref{sec-afb}.}
In {\SECT}~\ref{sec-ewp-intro} we introduce the basic formalism
for the  parametrisation of the $\Zzero$ resonance and discuss radiative 
corrections.
The determination of 
$\Zzero$ properties, their interpretation within the context of the {\SM} 
and the implications for {\SM} parameters are presented in 
{\SECT}~\ref{sec-ewp}. The results are summarised in {\SECT}~\ref{sec-sum}.

%
%
\section{The OPAL detector and its simulation}
\label{sec-det}
The OPAL detector is described in detail in~\cite{bib-det-opal}.
Therefore only those aspects which are relevant to the present analysis
are mentioned briefly here. 
In the following,
a right-handed coordinate system is used in which the origin is located 
at the geometrical centre of the tracking chambers, the \mbox{$z$-axis}
is along the electron beam direction, the \mbox{$x$-axis} points to the 
centre of the LEP ring, $r$ is the radial coordinate, normal to~$z$, and
the angles $\theta$ and~$\phi$ are respectively the polar and azimuthal
angles with respect to~$z$.
The $x$-axis defines~$\phi=0$.
We set $c = \hbar = 1$ throughout.

Charged particle trajectories are reconstructed and their momenta are 
measured using cylindrical central tracking detectors (CT). 
These consist of a silicon micro-vertex detector~\cite{bib-det-SiVtx}, 
a high precision vertex wire chamber, 
a large volume jet chamber (CJ) 
and thin $z$-chambers. 
The jet chamber is \mbox{$400\,{\mathrm{cm}}$} in length and
\mbox{$185\,{\mathrm{cm}}$} in radius.
Its 24 azimuthal sectors are formed by planes of anode and cathode
wires stretched parallel to the $z$-axis.
It provides up to 159~space 
points per track and also measures the ionisation energy loss of charged 
\mbox{particles,~${\mathrm d}E/{\mathrm d}x$~\cite{bib-det-CJ}.}
The $z$-chambers, which improve the track measurements 
in~$\theta$, are situated immediately outside and coaxial with the jet 
chamber. 
Quality cuts are made to select well-reconstructed tracks emanating from
the interaction point for use in the analysis.
In general no track is used which has a momentum component transverse to the
beam axis less than 100~MeV.
The number of hits associated with the track in the central tracking 
chambers must be at least~20.
The distance of closest approach of the track to the nominal beam 
crossing point must be less than $2\,{\mathrm{cm}}$ radially and less than
$100\,{\mathrm{cm}}$ along the \mbox{$z$-axis.}
Track finding is nearly 100\% efficient within the angular 
region \mbox{$\abscosthe<0.97$.} 
For the analyses presented here, slightly tighter tracking 
requirements are imposed which are optimised for each of the $\Zzero$ decay
channels.
The whole central detector is contained 
within a pressure vessel which maintains a constant absolute pressure of
4 bar and a solenoid which provides a uniform axial magnetic field 
\mbox{of~$0.435\,{\mathrm T}$}. 
The solenoid is surrounded by a time-of-flight 
scintillation counter array.

The electromagnetic calorimeter (ECAL),
located outside the CT pressure vessel and the solenoid,
measures the energies and positions of 
showering particles. 
The barrel consists of a presampler followed by a 
cylindrical ensemble of 9440~lead glass blocks arranged such that 
each block points towards the beam collision point, but the 
inter-block gaps point slightly away from the origin. 
Mechanically the barrel lead glass calorimeter consists of 10 C-shaped 
modules.
The two endcaps each consist of a presampler followed by 1132~lead 
glass blocks aligned parallel to the beam axis. 
The barrel covers the angular region \mbox{$\abscosthe<0.82$} 
while the endcaps cover \mbox{$0.81<\abscosthe<0.98$.}
Overall, the electromagnetic calorimeter 
 provides complete coverage
 for the entire angular range of \mbox{$\abscosthe<0.98$}.
Only in the region where the barrel and endcaps overlap, and at the
narrow boundaries between the barrel modules, is the uniformity
of response slightly degraded.
For use in the analysis electromagnetic clusters 
in the barrel are required to have
a minimum energy of 100~MeV, and clusters in the endcaps  must consist 
of energy deposits in at least 2~adjacent lead glass blocks and
must have a minimum energy of 200~MeV. 

Calorimeters close to the beam axis, and located on both sides of the 
interaction point, measure the luminosity using small-angle Bhabha 
scattering events. 
They complete the geometrical acceptance down to 25~mrad
from the beam axis. These include the forward detectors~(FD), which are 
lead-scintillator sandwich calorimeters, and at smaller angles, 
silicon-tungsten calorimeters~\cite{bib-det-siw, bib-lumi-siw}~(SiW),
which were installed in 1993, increasing the precision of the luminosity
measurement by an order of magnitude.

The iron return yoke of the magnet lies outside the electromagnetic 
calorimeter and is instrumented with streamer tubes as a hadronic
calorimeter (HCAL). 
Four layers of muon detectors~\cite{bib-det-muon} (MU)
are situated outside the hadronic calorimeter. 
Muons with momenta above 3~GeV usually penetrate to the muon detectors. 
In addition, up to nine hits may be recorded for minimum ionising 
particles traversing the hadronic calorimeter, further aiding muon 
identification.

The OPAL data-acquisition system~\cite{bib-det-daq} reads out and 
records data associated with particular events which are selected 
using a three-level system, consisting of a 
pretrigger~\cite{bib-det-pretrig}, a trigger~\cite{bib-det-trig} and 
an online event filter~\cite{bib-det-filter}. 
These make use of a large number of independent signals from a variety of 
detector components and have very high efficiency and 
redundancy for the $\Zzero$~decay events which are of interest for the 
analyses reported in this paper.
The trigger inefficiency for $\Zzero$ events decaying to charged fermions 
within the geometrical acceptance is less than~0.1\% and the trigger 
redundancy allows all efficiencies to be measured from the data.
The online event filter serves primarily for data quality monitoring,
but also allows a small number of obvious non-physics events
to be identified through software reconstruction and rejected from the
data stream.
The fact that no good events are rejected by the filter has been carefully
tested using the redundancy of selection criteria and also in numerous 
samples where rejection by the filter was temporarily disabled.
Events selected by the filter are fully reconstructed online and written 
into an offline storage facility for further analyses~\cite{bib-det-offline}.

Unless stated otherwise, all Monte Carlo event samples have been processed
using a full simulation of the OPAL detector~\cite{bib-mc-GOPAL} which 
treats in detail the detector geometry and material as well as the effects 
of detector resolution and efficiency.
The simulated events have been reconstructed using the same procedures that
were used for the OPAL data.
Monte Carlo event generators are further discussed in {\SECT}~\ref{sec-sel}.

\subsection{{\LEPI} data samples}
\label{sec-samples}

In \mbox{{\TAB}~\ref{tab-meas-data}}
the integrated luminosities are given for all of the {\LEPI} data 
samples which are used in this analysis. 
The total integrated luminosity is \mbox{$161~{\mathrm{pb}}^{-1}$,} 
which includes \mbox{$43~{\mathrm{pb}}^{-1}$} of data recorded off-peak.
The peak data are dominated by the dedicated high-statistics running in 
1992 and especially 1994,
while most of the off-peak data were collected in the precision scans in 
1993 and 1995, when the running was confined to three energies: 
on-peak and peak$\pm 2$~GeV. During the ``prescan'' periods in 1993 and 1995
running was confined to the peak while
all the necessary elements of the LEP beam energy calibration were 
commissioned. These periods also coincide with the commissioning
of the SiW luminometer (1993) or the SiW bunch tagger (1995), described below.

The additional \mbox{$117~{\mathrm{pb}}^{-1}$} of luminosity from 1993 to 1995 
has warranted a significantly improved analysis
of the hadronic decay channel.  
This has reduced the systematic error in the hadronic acceptance by about 
a factor of three, which makes it comparable with the much reduced statistical 
error.
In the analysis of the leptonic channels studies
of systematic effects have benefited from the greatly increased statistics.
Except for the measurement of the $\mumu$ asymmetry in 1992 
(see {\SECT}~\ref{sec-mumuasy})  the data
presented in our previous publications has not been reanalysed.
However, we have retrospectively applied corrections for a few small effects
(see {\SECT}~\ref{sec-meas}).

\subsubsection{LEP operation}
\label{sec-det-lep}
Many aspects of the LEP experimental programme were optimised to reduce
potential systematic effects in measuring the parameters of the $\Zzero$.
The off-peak data are essential for the measurements of~$\MZ$ and~$\GZ$.
For the 1993 and 1995 scans, the choice was made to run at only three points
to maximise the statistical precision in the measurement of these quantities:
at the peak and approximately $\pm1.8$~GeV from the peak.
The exact energies were chosen to allow a precise calibration
of the LEP beam energies by resonant depolarisation (see {\SECT}~\ref{sec-ewp}). 
In each scan the cross-sections at the two off-peak points were
typically measured in adjacent LEP fills, interspersed with fills
at the $\Zzero$~peak. 
This reduces any possible systematic biases
resulting from changes in LEP or OPAL operating conditions, and also
gives balanced data samples at the off-peak points within each year.
For the determination of~$\MZ$ the crucial experimental measurement is the
ratio of the cross-sections above and below the $\Zzero$~peak, and the
impact of inter-year systematic effects are minimised by the balance of
peak$+2$ and peak$-2$ measurements within each year.
A check of the stability of the LEP energy calibration can also be made 
by measuring $\MZ$ in each scan year (see {\SECT}~\ref{sec-ewp}).

The determination of~$\GZ$, however, chiefly depends on the measurement of
the ratio of off-peak to on-peak cross-sections.
The on-peak cross-sections are essentially determined by the 1992 and
especially the 1994 data, while the off-peak cross-sections are determined
by the 1993 and 1995 data.
The measurement of $\GZ$ therefore enjoys little
inherent protection from time-dependent systematic shifts in the
scale of the cross-section measurements.
Control of potential inter-year systematic
uncertainties from changes in LEP operation or the OPAL detector
configuration are essential. 

The operation of the LEP collider has evolved considerably over the years
to increase the luminosity delivered to the 
experiments and to improve the precision with which the centre-of-mass
energy could be calibrated (see {\SECT}~\ref{sec-ene}).
Initially LEP operated in a mode in which four bunches of electrons and
four bunches of positrons collided  at the four interaction points
every $22\,\mu\mathrm{s}$.
From 1992 to 1994 LEP ran in a mode with 8 bunches of electrons and
positrons which collide every $11\,\mu\mathrm{s}$. 
In 1995 the LEP collider was operated in a new ``bunch-train'' mode, in which
four equally spaced bunch-trains replaced the usual bunches, and crossed
at each interaction point every $22~\mu\mathrm{s}$.
Each train consisted of up to four (typically three) bunchlets, separated
from each other by~$247\,{\mathrm{ns}}$. Preparations for the bunch-train mode
required a return to 4-bunch running at the end of 1994 (periods peak(c)
and (d)). 
The impact of these changes on the OPAL set up are described below.

\subsubsection{OPAL detector stability}
\label{sec-det-opal}

The configuration of the OPAL detector has changed slightly over 
the entire period of these measurements. Of greatest significance
was the installation of the precision SiW~luminometer before the 1993 run.
This yields absolute luminosity measurements for the 1993--1995 data samples
which are an order of magnitude more precise than the 
earlier, systematics-limited measurements using the forward detectors. 

A silicon micro-vertex detector was first installed in OPAL before the 1992 
running period. The detector was augmented for the 1993 run. It was 
removed for repairs at the end of 1994 (periods peak(c) and (d)), and
replaced with an improved geometry for the 1995 run.
Data from the micro-vertex detector are not used directly in the
reconstruction of tracks for this analysis. The $\sim 0.015\x$ of 
additional material it introduces affects the conversion of photons into
electron pairs, and is adequately reproduced by the detector simulation program.

Halving the interval between bunch crossings from 22 to $11\,\mu$sec in 1992
no longer allowed sufficient time to form the full trigger information 
from the tracking chambers. OPAL adopted a 
pretrigger scheme~\cite{bib-det-pretrig} in which very loose 
trigger conditions using fast signals 
identified a small fraction ($\sim 1\%$) of 
all bunch crossings for which sensitivity to interactions in the 
next crossing would be lost while waiting for the
full track trigger information.
The pretrigger inefficiency for all relevant events 
was determined to be negligible.

The new bunch-train mode of operation in 1995
required several modifications.
In particular, the SiW~luminometer electronics were considerably modified to
operate properly under these conditions~\cite{bib-lumi-siw}.
In addition, small corrections were applied to the electromagnetic calorimeter energy
measurements on the
basis of bunchlet timing information obtained from the time-of-flight
detectors, or from the tracking chambers.
For most events, the tracking detectors are able to determine the bunchlet
crossing which produced the visible tracks, since only one potential choice
of the origin for the drift times yields good tracks passing through the 
interaction point.
The misassignment of bunchlet number could potentially cause problems in 
track reconstruction.
The influence of such bunchlet effects on the event selection efficiencies 
has been checked using redundant selections and was found to be negligible.

%
\section{\boldmath Selection and analysis of $\Zzero$ decay channels}
\label{sec-sel}
In the {\SM} the~$\Zzero$ is expected to decay into a
fermion-antifermion pair. 
With three generations of fermions, there are eleven 
possible decay channels: five quark flavours (the top quark is too 
heavy), three neutrino species and three charged leptons. 
The approximate branching ratios are \mbox{ 70:20:10 }
to hadrons, neutrinos and charged leptons, respectively.
Decays of the~$\Zzero$ to neutrinos normally go undetected and are 
referred to as invisible decays.
No attempt is made in this analysis to separate the different quark flavours,
with all hadronic $\Zzero$ decays being classified as \mbox{$\eehad$} events.
Measurements of
$\Zzero$~partial decay widths and forward-backward asymmetries using hadronic
$\Zzero$~decays in which the different quark flavours {\em are} distinguished
have been published in~\cite{bib-opal-hadew}. 

The analysis described here is focused on selecting
visible $\Zzero$ decay events 
in just four categories: \mbox{$\eehad$}, \mbox{$\ee$}, \mbox{$\mumu$} and 
\mbox{$\tautau$,} where, in each case, initial- and final-state radiation
can lead to one or more additional photons in the final state. 
Also, the classification \mbox{$\eehad$} is inclusive of all final-state
QCD interactions, including hard gluon bremsstrahlung.
In general, events in these 
categories can easily be distinguished from each other and from the 
remaining background, which is very small compared to the $\Zzero$ 
resonance signal, leading to event selections of high efficiency and purity.

Typical examples of the four event categories as observed in the
OPAL detector are shown in {\FIG}~\ref{f-opal-ev}.
Independent of the specific decay channel, the detected particles in
$\Zzero$ events generally exhibit
momentum balance along the beam direction, 
in contrast to background events from two-photon interaction processes
($\ee\rightarrow\ee\ff$).
Events produced by the passage of cosmic rays through the detector can
normally be rejected since they are rarely consistent
with the observed origin of true signal events in space and time.
The \mbox{$\eehad$} events are characterised by high-multiplicity final 
states, due to quark fragmentation and hadronisation, in contrast to 
the low-multiplicity lepton-pair events.
The \mbox{$\eeee$} events have two high-energy deposits from the 
final-state electrons in the electromagnetic calorimeter,
associated with tracks reconstructed in the central detector. 
The \mbox{$\eemumu$} events typically contain two high-momentum central 
detector tracks, 
with little energy deposited in the electromagnetic calorimeter.
The tracks are usually associated with track segments in the
muon chambers or the hadron 
calorimeter strips, which provide evidence for penetrating muons.
The \mbox{$\eetautau$} events are distinguished by two 
low-multiplicity `jets', 
each consistent with the decay of a~$\tau$, and typically a lower
measured energy than the other lepton-pair final states, due 
to the undetected neutrinos from the~$\tau$ decays.
The detailed selection criteria for events in each of these four 
categories are described in \mbox{{\SECT}s~\ref{sec-had} and \ref{sec-lept}.}
The selections have very little overlap, in particular for the three
leptonic selections it is ensured that no event is
classified in more than one category (see {\SECT}~\ref{sec-leptcor}).

\subsection{Experimental acceptance}

The limited coverage of the tracking 
chambers in the forward direction prevents us from detecting leptons
close to the beam direction, 
but the simple topologies of the lepton-pair final states allow a precise 
experimental acceptance to be defined in a restricted region of $\costhe$. 
In the case of the \mbox{$\eemumu$} and \mbox{$\tautau$} channels 
the cross-sections in these regions ($|\cos\theta|<0.95$ and $0.90$
respectively) 
can then be extrapolated to the full angular acceptance. 

In addition to the $\Zzero$ $s$-channel annihilation diagram, 
the process \mbox{$\eeee$} has 
contributions from \mbox{$t$-channel} diagrams, 
dominated by photon exchange, which lead to a divergence of the forward 
cross-section. 
This makes a similar extrapolation of the \mbox{$\eeee$} cross-sections to the 
full angular acceptance meaningless. 
We therefore restrict the \mbox{$\eeee$} measurements to 
a  region well within the detector, $|\cos\theta|<0.70$, 
which limits the $t$-channel contributions to a manageable level
(15\% at the peak), and do not extrapolate them. 

For  \mbox{$\eehad$} events, the hadronic jets are broad enough to 
ensure high efficiency even for events with decay axes close to 
the beam direction.
Hence we have almost 100\% acceptance for $\Zzero$ events decaying
hadronically, and directly measure events produced over essentially the full
angular region.
The approximately $0.5\%$ inefficiency is dominated by very narrow 2-jet
events oriented along the beam line.



\subsection{Kinematic acceptance for cross-sections and asymmetries}
\label{sec-sel-def}

In order to interpret the measured cross-section and asymmetry for each
reaction 
the limits of kinematic phase space must be specified in a 
precise manner, which is adapted to available theoretical calculations,
in order to take into account properly the effect of initial- and
final-state photon radiation.
We therefore correct our raw measurements to correspond to simple, ideal,
kinematical limits, which are chosen to correspond reasonably closely to
the experimental acceptances to reduce the resulting acceptance 
extrapolations.
Table~\ref{tab-sel-kine} summarises the limits of the ideal kinematic 
phase space within which we define our cross-sections and asymmetries 
for each species.
The experimentally observed number of events in each channel is
corrected for selection inefficiencies and background to give the total
number of events produced within the kinematic acceptance defined by these
idealised cuts.

We specify the kinematic phase space for the cross-sections for all species,
except electrons, in terms of a lower bound on
either \mbox{$\sprime / s$} or \mbox{$m_\ff^2/s$.}
Here $\sprime$ is the squared {\cms} energy after initial-state radiation
and  $m_\ff$ is the invariant mass of the final-state fermion pair.
Both definitions suffer from some degree of ambiguity.
In the case of $\eehad$, QCD effects obscure the precise meaning of
$m_\ff^2$, while $\sprime$ suffers from the impossibility of distinguishing 
initial- from final-state radiation.
There is, however, little quantitative difference in our regime
between $m_\ff^2$ and $\sprime$.
Switching between using $m_\ff^2$ and $\sprime$ in the acceptance 
definition changes the hadronic cross-section, for example, by only
a few parts in $10^{5}$.

The experimental acceptance becomes very small for events with low $\sprime$
or $m_\ff^2$, but the number of produced events in this region is also small.
The calculated number of hard radiative events which fall between
our ideal and experimental acceptance limits in proportion
to all accepted events is about $1.5\times10^{-4}$
for hadrons and $5\times10^{-3}$ for $\mumu$ and $\tautau$.
For $\eemumu$ and $\tautau$ the dominant acceptance extrapolations
(respectively $\sim 7\%$ and 15\%) are due to the limited angular region
of the experimental event selection.

For the process $\eeee$ the ideal kinematic acceptance is chosen
very close to the 
experimental acceptance, both being defined by the range of polar angles
allowed for the final-state e$^-$ and the maximum $\ee$ acollinearity.
The total acceptance correction required is less than $1\%$.

For measuring the asymmetries we treat the kinematic acceptance for
all the leptons similarly to the electrons, and limit the range of polar
angle of the final-state fermions to the nominal
experimental acceptance.
To reduce effects associated with strong initial-state radiation,
we also limit the maximum acollinearity between the pair of leptons
to $10^\circ$ or $15^\circ$, as shown in Table~\ref{tab-sel-kine}.

\subsection{Four-fermion processes and radiative photon interference}

Four-fermion diagrams, as shown in {\FIG}~\ref{f-four-fermion}, also 
contribute at a small level to the selected event samples.
Some of these, such as the conversion of initial- or
final-state photons to fermion pairs
({\FIG}~\ref{f-four-fermion}(b) and {\FIG}~\ref{f-four-fermion}(c))
are properly considered as 
radiative corrections to fermion pair production.
We therefore apply corrections to the selection efficiency where necessary
to ensure that such events are fully counted as part of the measured
signal.
Other four-fermion processes, such as multi-peripheral diagrams 
(two-photon processes, {\FIG}~\ref{f-four-fermion}(d)) are
clearly unrelated to the $\Zzero$ and we therefore subtract them as background.
A third class of four-fermion processes
({\FIG}~\ref{f-four-fermion}(e) and {\FIG}~\ref{f-four-fermion}(f))
are pair corrections to scattering in the $t$-channel.
Such events are treated as signal only in the $\eeee$ channel.
More details concerning the treatment of four-fermion final states
are given in Appendix~\ref{sec-4f}.

Our measured cross-sections and asymmetries include the effects of
interference between initial- and final-state photon radiation.
Since our primary Monte Carlo programs, JETSET and KORALZ\footnote{KORALZ
can include initial-final-state interference, but only when
radiative corrections are treated to ${\cal{O}}(\alfem)$.}
do not include such interference, we adjusted the calculated event
selection efficiency for our cross-section measurements by corrections of
${\cal{O}}(10^{-4})$ to account for this deficiency, as described in 
Appendix~\ref{sec-ifi}.
For the asymmetries there is a close correspondence between the experimental
and ideal acceptances, and no such acceptance correction is necessary.

\subsection{Monte Carlo event generators}
\label{sec-sel-mc}
The following event generator programs have been used to simulate 
signal and background processes:
$\eehad$ events have been generated using
the programs JETSET, version~7.3~\cite{bib-mc-JETSET73}, and HERWIG,
version~5.8~\cite{bib-mc-HERWIG},
with hadronisation parameters tuned using a sample of hadronic events 
selected from OPAL {\LEPI} data~\cite{bib-mc-OPALtune-j73}.
Cross checks have been made using
version~7.4 of the JETSET program~\cite{bib-mc-JETSET74} with an updated
set of parameters~\cite{bib-mc-OPALtune-j74} tuned using a larger sample
of OPAL {\LEPI} data.
The KORALZ program, version~4.02~\cite{bib-mc-KORALZ}, has been used for
\mbox{$\eemumu$} and \mbox{$\eetautau$} events and
the BHWIDE program, version~1.00~\cite{bib-mc-BHWIDE}, \mbox{for~$\eeee$}
events. Background events produced by two-photon interaction processes,
\mbox{$\eeeeff$},
have been studied using the programs PHOJET~\cite{bib-mc-PHOJET}, HERWIG
and the program of Vermaseren~\cite{bib-mc-Vermaseren}, while those from
\mbox{$\eegg$} have been calculated using the RADCOR
program~\cite{bib-mc-RADCOR}.
The four-fermion signal processes in the $s$-channel have been
studied using the FERMISV~\cite{bib-mc-FERMISV} generator.
Background from $t$-channel four-fermion events has been evaluated
using PYTHIA~\cite{bib-mc-JETSET74} for \mbox{$\eeeeqq$} and
grc4f~\cite{bib-mc-grc4f} for \mbox{$\eeeell$}.


%
%
\section{LEP energy}
\label{sec-ene}
%
The LEP energy scale is one of the crucial ingredients in the determination
of $\Zzero$~resonance parameters.
The initial energy calibration of LEP~\cite{bib-LEPENE90} 
was performed in 1989 and 1990 by circulating protons in the LEP ring
at 20~GeV and using magnetic measurements to obtain the centre-of-mass energy
in physics conditions.  
This resulted in errors on~$\roots$, and hence~$\MZ$,
of 28~MeV and 22~MeV for the 1989 and 1990 data, respectively.
Using the technique of resonant depolarisation~\cite{bib-ResDepol},
a precise calibration of the LEP energy scale
was achieved in 1991, resulting in a systematic uncertainty of 6~MeV on~$\MZ$ 
and approximately 5~MeV on~$\GZ$~\cite{bib-LEPENE91}.
The recognition of energy shifts due to the alignment of the
copper RF cavities had particular importance 
at the two interaction points equipped with cavities (OPAL and L3). 
The calibration of the LEP energy scale in 1992~\cite{bib-LEPENE92}
was performed using a similar procedure.
In 1992, however, calibrations with resonant
depolarisation were successful only late in the year and showed a 
large spread.
The central value of the energy derived from the
polarisation measurements was found to be in good agreement with
the one obtained from methods based on measuring the magnetic field in the
LEP dipoles. 
The quoted error of 18~MeV on~$\roots$
arises predominantly from the scatter of the depolarisation measurements
and the extrapolation to the beginning of the year using magnetic
measurements.  
Since the 1992 data have all been collected at the peak of the
$\Zzero$~resonance, this larger error has an insignificant impact on
the precision of the derived $\Zzero$ resonance parameters.
        
In 1993 a more complete understanding of the time-dependent parameters that
influence the LEP energy, such as tidal deformations of the Earth
and changes in magnet temperatures,
as well as more frequent polarisation measurements,
led to a further increase in the precision of the energy calibration.
A concerted effort to ensure the complete logging of
all LEP parameters relevant to the energy measurement allowed
this inherent precision to be extended to a large fraction of all fills.
This resulted in a systematic uncertainty of
1.4~MeV on the absolute centre-of-mass energy of LEP~\cite{bib-LEPENE93}, 
and the point-to-point energy errors, 
including the error on the centre-of-mass energy spread, contributed
an uncertainty of 1.5~MeV on~$\GZ$.  
The quality of measurement was
maintained throughout the 1994 running.

In 1995 vertical dispersion of the beam energy at the interaction point was
introduced by bunch-train operation, which required frequent vernier scans to
control possible shifts in the mean centre-of-mass collision
energy~\cite{bib-LEPecal-98}.
More significantly, additional instrumentation
installed in the LEP tunnel allowed the observation of an unexpected change
in the beam energy during LEP fills.  Leakage currents from
the electric railway system in the Geneva area flow through the LEP tunnel
and perturb the magnetic field of the LEP dipoles, which causes the bending
field (and hence, the beam energy) to rise steadily during a fill.
The interplay between the rise introduced by the leakage currents and that
due to temperature variations in the dipole magnets necessitated a more
detailed study of the temperature-dependence of the bending field, resulting in
a much improved model of this behaviour.
This knowledge provides the definitive description of
the LEP energy calibration for 1993--1995~\cite{bib-LEPecal-98}.  
In light of this improved understanding, the 1993 energies
and their errors, first published in~\cite{bib-LEPENE93}, 
have been revised, and the errors significantly increased.
The data needed for updating
the less critical calibrations for years earlier than 1993 are unavailable.
However, these errors are uncorrelated with
the errors for 1993--1995, so the data can easily be combined.

Taking these final calibration results into account,
the uncertainties in the LEP energy contribute
errors of 1.8~MeV and 1.3~MeV to the OPAL determination of $\MZ$ and $\GZ$
respectively, as described in Section~\ref{sec-parerr}.
At a scale a few times larger than the inherent precision
of the LEP energy calibration
an important and completely independent test of its consistency
over time can be made by studying the stability in the
value of $\MZ$ measured over the six years of data-taking
(see {\TAB}~\ref{tab-ewp-mztest}).
 
The values for the spread of the centre-of-mass energies, due to the
energy spread of the particles in the beams, for the different running
periods and energy points are summarised in {\TABs}~20 and~21 of 
reference~\cite{bib-LEPecal-98}. For the 1993--1995 data they range between 
\mbox{54.6-56.7~MeV,} increasing with~$\roots$, and with uncertainties 
of~1.1~MeV in 1993--1994 and 1.3~MeV in 1995.
The centre-of-mass energy spreads for data from years before 1993 have 
been re-evaluated using measurements made in subsequent years.
They are between \mbox{43-53~MeV}, with an error of~3~MeV.
We quote our cross-sections and asymmetries
both as measured and after correction for the energy spread,
as described in Appendix~\ref{sec-covmat}.

%
\section{The luminosity measurement}
\label{sec-lum}
All OPAL cross-section measurements rely on normalising the
observed number of events in any given final state to the
integrated luminosity measured by counting the number of Bhabha
scattering events at angles small enough that $t$-channel photon exchange
dominates the cross-section, and the influence of the
$\Zzero$ is reduced to a small correction ($\lsim 1\%$). 

The SiW~luminometer and the associated luminosity selection, which are fully
described in~\cite{bib-lumi-siw}, have been 
optimised to exploit the characteristics of Bhabha scattering, \mbox{$\eeee$.}
The small-angle cross-section of \mbox{$\eeee$} is dominated by the
\mbox{$t$-channel} exchange of a photon, leading to a~$1/\theta^3$ spectrum, 
where $\theta$~is the angle of the out-going electron and positron with 
respect to the incoming beams.  
By instrumenting a region at very small angles, the 
accepted \mbox{$\eeee$} cross-section for SiW is approximately 
\mbox{$79\,$nb,} 
sufficiently larger than the total cross-section for $\Zzero$~production to 
limit the importance of the statistical uncertainty in the luminosity
measurement.
The forward-peaked $1/\theta^3$~Bhabha spectrum
requires that the detector and luminosity analysis define the inner edge 
of the acceptance with particularly high precision in order to reduce the
systematic error.
For example, to achieve a precision of~$10^{-3}$, the inner edge of the 
acceptance must be known to \mbox{$10\,\mu$rad,} corresponding to about
\mbox{$25\,\mu$m} in the radial coordinate of the showers produced in the 
calorimeters, which are located at a distance of approximately \mbox{$2.5\,$m} 
on each side of the beam crossing point.

For the measurement of the absolute luminosity, the experimental 
error achieved by the SiW~luminometer is
\mbox{$3.4\times 10^{-4}$,} 
which includes all intrinsic and time-dependent sources of
experimental uncertainty, such as detector geometry, gain variations,
energy and positional biases in the detector response to electromagnetic
showers, variations in the LEP beam geometry, backgrounds and other
environmental influences.
In 1995 the introduction of bunch-trains in LEP led us to install a ``wagon 
tagger''~\cite{bib-lumi-siw} which allowed the luminometer to measure
Bhabha-scattering events in all bunchlet crossings in each bunch-train
with no compromise in performance.

For the first four years of the 
{\LEPI} programme OPAL used the forward detectors~\cite{bib-det-opal} for 
measuring the luminosity with a resulting systematic uncertainty of 
$4.1\times10^{-3}$.
This uncertainty dominates the total uncertainty on the hadronic pole
cross-section obtained from the 1990--1992 data. 
The larger systematic error of the FD measurement leads 
to a reduced weight from the earlier data $(\sim4\%)$ in determining
the hadronic cross-section. 
Since the SiW luminometer was installed in 
front of FD, obscuring its inner edge, no direct experimental
cross-calibration of the two detectors is possible.
A retrospective check of the hadronic peak 
cross-sections measured from the 1992 and 1994 data shows good 
agreement $(0.4\pm0.5)\%$. 

The luminosity measurement requires that the theoretical cross-section for
small-angle Bhabha scattering within the experimental acceptance be 
accurately calculated.
At the time of our last $\Zzero$ resonance
publication~\cite{bib-opal-ls92}, techniques 
based on YFS exponentiation yielded an error of
\mbox{$2.5\times10^{-3}$~\cite{bib-mc-BHLUMItwo}}.
These calculations have been extended to include second-order next-to-leading
log terms~\cite{bib-mc-BHLUMI}, and 
calculations of third-order indicate that the error is
\mbox{$6.1\times10^{-4}$~\cite{bib-bward}}.
Improved calculations of light pair corrections within the OPAL
SiW acceptance result in a final total theoretical error of
\mbox{$5.4\times10^{-4}$~\cite{bib-pavia}}.
The scale of our previously published cross-sections, normalised with the FD 
luminometer, have been corrected in this analysis to incorporate these latest 
theoretical results in small-angle Bhabha scattering.

%
\section{\boldmath Measurement of \mbox{$\eehad$ events}}
\label{sec-had}
In the {\SM} about~87\% of visible $\Zzero$~decays are expected to 
result in quark-antiquark pairs, leading to hadronic final states.
These events therefore  provide the most accurate 
determination of $\MZ$ and~$\GZ$.
In general, hadronic events can be clearly distinguished from
leptonic $\Zzero$~decays, or other background processes, due to their high 
multiplicity  and large visible energy 
of final-state particles and the balance of visible momentum
along the beam direction. 
Problems mainly arise for events consisting of two narrow jets in which most 
of the final-state particles have trajectories with only a very small angle 
relative to the beam axis. The geometrical coverage of the detector is 
incomplete in these regions, because of the beam pipe 
and associated apparatus, and therefore only a fraction of the particles 
from such an event are registered. 
The evaluation of the selection inefficiency which results from the
loss of such events depends on the modelling of 
QCD effects in the process \mbox{$\qq\rightarrow\mathrm{hadrons}$}, 
which we subsequently refer to as ``hadronisation'' modelling.
Since the treatment of hadronisation in
Monte Carlo event generators is limited by phenomenological uncertainties,
we have developed a new technique which uses the centre of the barrel region
to emulate the setup and response of the detector close to the beam axis.
This allows us to estimate the inefficiency using real data events 
and results in a factor of almost three reduction in the systematic error 
for the acceptance calculation compared to our previous 
publications~\cite{bib-opal-ls91,bib-opal-ls92}.

\subsection{Selection criteria}
\label{had-sel}

The selection of \mbox{$\eehad$} events
uses the information from tracks reconstructed in the central 
detector~(CT) and clusters of energy reconstructed in both the lead 
glass~(ECAL) and the forward detector~(FD) electromagnetic calorimeters. 
Tracks  are typically reconstructed
with good efficiency down to $\theta = 15^{\rm o}$, ECAL clusters extend 
down to $\theta = 11^{\rm o}$ and the FD covers the range
from $3^{\rm o} < \theta < 9^{\rm o}$.

The selection is based on the following event parameters.
$\NCT$~is the total number of tracks.
$\NECAL$~and $\NFD$~are the total numbers of  ECAL~clusters and  
FD~clusters, respectively.
$\EECALi$~is the energy of  ECAL cluster~$i$, \mbox{($i=1,\NECAL$).}
$\EFDj$~is the energy of  FD cluster~$j$, \mbox{($j=1,\NFD$),} where
a valid FD cluster must have at least \mbox{3 GeV.}
Each event is divided into two hemispheres with respect to the thrust axis 
of the event, which is determined using all the  tracks and  clusters 
in the event. The invariant mass of each hemisphere is calculated from the  
tracks and  clusters which lie within it, assigning the pion mass to 
tracks and zero mass to clusters.
No attempt is made to eliminate the implicit double counting of 
energy when using both track momenta and calorimeter energy
for charged particles, such as electrons.
The sum of these two hemisphere invariant masses is~$\Mhemi$.

\noindent
Five selection criteria are used to define a 
candidate \mbox{$\eehad$} event:
\begin{itemize}
\item Charged multiplicity: $\NCT \ge 2\,$.
\item Total multiplicity: $\Nall \equiv \NCT+\NECAL+\NFD \ge 11\,$.
\item Sum of the invariant masses of the two hemispheres: 
$\Mhemi > 4.5~{\mathrm{GeV}}$.
\item Visible energy in ECAL and FD:
$ \ \ \Rcal \equiv { {\left( {\sumi\EECALi} + ({\sumj\EFDj}/3) \right)} 
                \  / \, \roots} \, > \, 0.1 \, . $
\item Energy balance along the beam direction: 
$$ \Rbal \equiv \frac
  {{\left| {\sumi(\EECALi\cos\thetai)} + {\sumj(\EFDj\cos\thetaj)} \right|}}
  { ( {\sumi\EECALi} + {\sumj\EFDj} ) }  \, < \, 0.75 \, ,$$
where~$\thetai$ and $\thetaj$ are the polar angles of ECAL~cluster~$i$ and 
FD~cluster~$j$, respectively.
\end{itemize}

Most $\Zzero$ decays to leptonic final states are rejected by the 
cuts on~$\Nall$ and~$\Mhemi$. The additional loose cut on~$\NCT$
reduces the background arising from events
induced by cosmic ray muons or showers.
The requirements on~$\Rcal$ and~$\Rbal$ suppress contributions from
two-photon interaction processes.
The factor $1/3$, by which the weight of the FD energy is reduced
in the construction of $\Rcal$, is chosen to optimise the 
separation between $\eehad$ signal events and two-photon background.
A~comparison of distributions of some of the cut variables between data 
and Monte Carlo simulation is shown in {\FIG}~\ref{mhcuts}.

The Monte Carlo prediction, based on the JETSET~\cite{bib-mc-JETSET73}
event generator, for the $\eehad$ selection inefficiency on-peak
is $\effMC = (48.1 \pm 1.1) \times 10^{-4}$ within $\sprime/s > 0.01$.
The inefficiency of the selection is approximately constant as a function
of $\sprime$ down to
$\sprime/s = 0.25$. For events with still harder initial-state radiation (ISR)
it rises rapidly and approaches 100\% for $\sprime/s = 0.01$, most 
events being rejected by the requirement on $\Rbal$.
At the peak energy events with hard radiation ($\sprime/s < 0.25$)
comprise only a small fraction ($4\times 10^{-4}$) of the total cross-section
and about half of these 
fail the selection.
At {\cms} energies away from the $\Zzero$ peak the relative rate of 
hard-radiation events is higher,
which slightly reduces the efficiency compared to the peak.
Changes of 
$-4.7\times 10^{-4}$ and $-1.4\times 10^{-4}$
were calculated using dedicated Monte Carlo  samples
 at {\pkm} and {\pkp}, respectively.
In addition, a small contamination ($0.3\times 10^{-4}$ at the peak) from selected
events with $\sprime/s < 0.01$ is subtracted.
Effects from multiple ISR, which are not simulated in the JETSET 
Monte Carlo generator, have been checked and were found to be negligible for 
determining the efficiency.

The total number of \mbox{$\eehad$} events selected from the 
\mbox{1993--1995} data samples is~\mbox{$2\,908\,566$,}
giving a relative statistical 
error of~\mbox{$6\times10^{-4}$.}
In order to reduce the uncertainties on 
systematic effects to a corresponding level 
one must carefully disentangle the potential error
sources and apply new techniques which base the corrections 
largely on the observed data properties.
In the following subsections we first describe a method to determine the
inefficiency due to  events lost along the beam axis
based on well-measured data events in the barrel region of the
detector.
We then discuss corrections and systematic errors associated with the 
selection of $\eehad$ events and the background estimation.
Unless specified otherwise the quoted 
correction factors and uncertainties refer to the 1994 peak data.
Full details for all data periods are given in {\TAB}~\ref{tab-mhcor}.
Correlations between energy points and data-taking years 
are specified in {\TAB}~\ref{tab:had_matrix}.

\subsection{Acceptance hole emulation}
\label{had-frag}
The overall inefficiency of the event selection is 
\mbox{approximately~$5\times10^{-3}$}. 
Most events which are lost have narrow, two-jet final states 
pointing into the very small polar-angle region (large $\abscosthe$, 
close to the beam axis), where there is a ``hole'' in
the acceptance.
In contrast, the inefficiency in the entire barrel region is 
about \mbox{$1\times10^{-4}$.}
One of the principal systematic errors in the Monte Carlo calculation of the 
selection inefficiency arises from uncertainties in the physics modelling 
of this class of events and hence in the evaluation of
the rate of events lost in the acceptance hole. 
In order to match the goal of 
$5\times10^{-4}$ acceptance uncertainty set by the statistical precision,
the rate of such lost events  must be understood at the 10\,\% level. 

In previous OPAL analyses~\cite{bib-opal-ls91,bib-opal-ls92}
this hadronisation uncertainty was assessed by comparing the inefficiencies 
predicted by different Monte Carlo generators, JETSET and HERWIG, and
by using different parameter settings within the JETSET program,
which led to a systematic error assignment of~0.11\% in the cross-section.
To reduce substantially this source of uncertainty a new technique was developed, 
which uses data events collected in the barrel region 
of the detector as a control sample.
In this technique, which we refer to as the 
``acceptance hole emulation'', we modify the response of 
both the real detector
and its Monte Carlo simulation in the region surrounding the 
\mbox{$x$-axis}, {\em i.e.} perpendicular
to the beam axis, to correspond closely to the actual gaps and
imperfections in detector coverage around the \mbox{$z$-axis.}

The first step of the acceptance hole emulation is to identify 
a sample
of hadronic $\Zzero$ events whose thrust axes lie within a well-defined cone
around the $x$-axis, chosen to be large enough to safely cover the
corresponding problematic region around the $z$-axis 
(\mbox{$\abscosthetthr>0.90$}).
This sample is complete and unbiased due to the almost complete detection
efficiency in the barrel region.
We then emulate the detector response which would have been observed if the
defining cone of this sample had been oriented along the $z$-axis 
by degrading the actual detector response correspondingly.
Since CT plays a minor role in the selection,
the edge of the CT acceptance is emulated simply by rejecting tracks 
which point within the corresponding narrow cone about the \mbox{$x$-axis}.
For the most crucial detector component, ECAL, in addition to using such
a geometric rejection, we emulate the position-dependent
energy response and cluster separation near the edge of the hole.
Similarly, the~FD is emulated using the data from the~ECAL clusters in the 
barrel by mapping the appropriate angular range, cluster size and energy
threshold.
Small adjustments of the selection cuts are made, such that the exclusive inefficiency of
each cut\footnote{The inefficiency due to events which fail just the cut 
under discussion.}
 due to the  emulated hole   matches
the corresponding inefficiency due to the endcap hole.
We then re-apply the event selection criteria\footnote{In 
applying this event selection all axis-sensitive quantities, such as $\Rbal$, 
are transformed from the $z$-axis to the $x$-axis.}
to determine the inefficiency of the selection for the emulated, $x$-axis hole,
$\effxdat$.

This entire procedure is repeated with a sample of Monte Carlo events
to determine the corresponding inefficiency of the selection in the
$x$-axis hole, $\effxMC$.
To the extent that the emulated holes in the Monte Carlo simulation
and in the data
are identical, the differences in $\effxdat$ and $\effxMC$ will be exclusively
due to departures in the Monte Carlo model of hadronisation.
The ratio between the two emulated inefficiencies, $\effxdat/\effxMC$,
therefore 
provides a correction factor to the Monte Carlo calculation,
which accounts for the imperfections in the hadronisation modelling. 
The overall corrected \mbox{$\eehad$} event selection inefficiency,~$\effdat$, 
is calculated as
\begin{equation}
\label{eq-had-frag}
\effdat =  \frac{\effxdat}{ \effxMC}  \,\effMC \; ,
\end{equation}
where $\effMC$ is the overall selection 
inefficiency predicted by the Monte Carlo simulation.
We obtain $\effdat = (56.8 \pm 1.3) \times 10^{-4}$, which is 
$8.7 \times 10^{-4}$ larger than $\effMC$.
Accounting for detector simulation deficiencies in the barrel region,
as discussed in Section~\ref{had-frag-sys},
reduces this difference to $5.5 \times  10^{-4}$.
Note that it is sufficient to calibrate the
rate of lost events with this method at the level of several per-cent.
It is not necessary, or expected, that either~$\effxMC$ or~$\effxdat$ gives an 
accurate estimate of the true inefficiency close to the \mbox{$z$-axis.}
Small differences between the composition of the event
samples along the $x$- and $z$-axes, such as introduced by initial-state radiation, are
completely negligible for this purpose.

\subsection[Selection uncertainties for $\eehad$ events]
{\boldmath{Selection uncertainties for $\eehad$ events}}
In this subsection the various sources of  systematic uncertainties
in the selection of $\eehad$ events are discussed.
The acceptance hole emulation strongly reduces the dependence
on the hadronisation modelling.
However, it also introduces new systematic errors which are related
to the quality of the hole emulation and 
the detector simulation in the barrel for determining $\effxMC$.
The direct reliance on detector simulation for determining $\effMC$
is the other main source of uncertainty.
Here the simulation of the endcap detectors is of particular importance.
In our previous  analysis~\cite{bib-opal-ls92} 
a detector simulation uncertainty of 0.14\% was assigned,
based on the global comparison of data and Monte Carlo energy distributions.
Such broad checks cannot
distinguish between detector simulation problems 
and deficiencies in the physics modelling of the generator.
They are also insensitive to possible
local inhomogeneities in detector response.
In order to disentangle these effects
a new approach has been adopted which investigates in more 
detail the simulation of individual hadronic and electromagnetic showers 
in the calorimeters.
Finally, the performance of the detector was carefully checked.
Each aspect of triggering, data taking and data quality
which could bias the selection has been examined.

\subsubsection{Systematic errors in the acceptance hole emulation}
\label{had-frag-sys}
Three different sources contribute to systematic uncertainties of the
acceptance hole emulation:
({\em i}) the residual hadronisation dependence, ({\em ii}) limitations of the 
acceptance hole emulation program, and ({\em iii}) the quality of the detector
simulation in the barrel region for determining $\effxMC$. 
Table~\ref{tab-had-hole} summarises the corresponding uncertainties.

\noindent
{\bf Residual hadronisation dependence:}
The sensitivity of the  method  to uncertainties in the modelling of 
hadronisation has been assessed by varying the parameters of the 
fragmentation model in the JETSET~\cite{bib-mc-JETSET73} Monte Carlo 
event generator and by 
employing an alternative Monte Carlo program, 
HERWIG~\cite{bib-mc-HERWIG}, which uses a 
different fragmentation mechanism. 
As an example, we changed the JETSET parameter $\QO$,~the 
invariant mass cut-off 
below which gluon radiation stops, from its default value of 1.0~GeV to 1.8~GeV.
This change corresponds to one standard deviation as determined from 
a global tuning of the 
JETSET parameters to OPAL {\LEPI} data~\cite{bib-mc-OPALtune-j73},
and shifts $\effMC$ by $(7 \pm 1) \times 10^{-4}$.
Using the acceptance hole emulation, however, the corresponding
shift in $\effdat$ is only \mbox{$(0.8\pm 1.1) \times10^{-4}$}, 
which illustrates the effectiveness of the method.
The variation of other fragmentation parameters yielded effects of similar size.
The value of  $\effdat$ calculated using the sample of HERWIG Monte Carlo 
events differs by \mbox{$ (2.3\pm 2.0) \times10^{-4}$} compared to
the reference value obtained using JETSET Monte Carlo events.
We take this difference as the
systematic error due to the residual sensitivity to hadronisation.
A further uncertainty arises from the small inefficiency in the
barrel region which is not addressed by the hole extrapolation.
From similar Monte Carlo studies of the hadronisation dependence, {\em i.e.},
variation of fragmentation parameters and comparison of models,
the largest change in this inefficiency is found to be 
\mbox{$( 2.2\pm 0.5 )\times10^{-4}$,}
which is taken as an additional systematic error. 
The total direct hadronisation uncertainty is $3.2\times10^{-4}$.

\noindent
{\bf Limitations of the acceptance hole emulation procedure:}
The detector setup and response differs substantially between barrel and endcap,
therefore the emulation of the endcap hole in the barrel is only
approximate. 
As discussed above the selection cuts for the barrel hole are adjusted to match
the exclusive inefficiency in the endcap hole.
This scaling of the cuts changes $\effdat$ by \mbox{$1.5\times10^{-4}$} which 
we take as a systematic error.
Furthermore, the radius defining the edge of the ECAL~acceptance
in the hole emulation program was varied over the full size of an ECAL block,
which leads to an additional uncertainty of 
\mbox{$0.4\times10^{-4}$}.

\noindent
{\bf Detector simulation in the barrel region:}
The acceptance hole emulation procedure relies  on the quality of
the barrel detector simulation for determining $\effxMC$.
Figure~\ref{mhhole2} shows distributions of the emulated variables used in the 
cuts for events close to the \mbox{$x$-axis.}
There are small offsets between the data and Monte Carlo prediction
for the energy and 
multiplicity distributions, although the shapes of the distributions 
are well simulated. Rescaling the Monte Carlo energy and multiplicity 
distributions in order to correct for these differences 
changes the value of~$\effdat$ by \mbox{$3.3\times10^{-4}$.} 
Switching on and off the correction methods described in 
{\SECT}~\ref{had-detsim}, which are independent of assumptions about 
hadronisation, gives a consistent difference in~$\effdat$. 
We take this value as a correction and assign to it a 100\,\%
uncertainty.

In summary, the acceptance hole emulation leads to an efficiency correction of 
\mbox{$(5.5 \pm 4.8) \times 10^{-4}$}.

\subsubsection{\label{had-detsim}Detector simulation uncertainties}
In addition to hadronisation uncertainties, which are reduced
by the hole emulation study, the event selection 
efficiency also relies directly on the accuracy of the detector 
simulation for determining $\effMC$. Since the main source of selection inefficiency 
occurs for events with jets close to the beam axis, the
simulation of the detector response in the endcap region is
more critical than that in the barrel region. The cut on
$\Rcal$ causes the largest inefficiency and
consequently the most important issue is the simulation of the 
energy response in the ECAL and the FD.

The electromagnetic response of these calorimeters has been examined
using lepton-pair events with an identified low-energy radiated photon.
The measured photon energy is compared to the energy predicted from
the momenta of the two leptons, assuming a three-particle final state.
Checks are made for global differences between data and Monte Carlo 
distributions and for local inhomogeneities.
Most crucial for the \mbox{$\eehad$} event selection inefficiency are 
the regions just inside the edges of the acceptance,
at small angles to the beam axis. 
The stability of the energy response of the ECAL endcap inner rings has been 
found to be good and the simulation is accurate to the 5\% level. 
This corresponds to an uncertainty of \mbox{$2.0\times10^{-4}$} on the 
overall inefficiency. There are further small differences between the data and 
detector simulation for the remainder of the ECAL endcaps, leading to a 
further correction to the inefficiency of \mbox{$(2.0\pm 2.0)\times10^{-4}$.}
A similar check for the FD electromagnetic response shows good consistency
between data and Monte Carlo simulation.
The statistical precision of this check translates into an 
uncertainty of \mbox{$2.0\times10^{-4}$} on the overall inefficiency.

The studies described above provide verification of the ECAL response
to electromagnetic showers. Also important is the ECAL response to
hadrons.  The energy spectra of single ECAL clusters associated with
isolated tracks in hadronic events were studied in narrow bins of
track momentum in both data and Monte Carlo simulation.  The
simulation did not reproduce the response spectra exactly, although
the mean detector behaviour was modelled adequately.  The barrel,
endcap, and overlap regions of the ECAL were studied separately.  In
each momentum bin and detector region a correction function depending
on energy was then constructed to adjust the detailed shape of the
simulated calorimeter response to match that observed in the data.
When this correction is applied in the Monte Carlo event simulation to
all ECAL energy deposits which arise from charged or neutral hadrons
(using the measured and true particle momentum, respectively), the
overall selection efficiency shifts by \mbox{$(2.0 \pm2.0) \times
10^{-4}$}, which we take as a correction.  The effect of the
correction on the barrel ECAL response is used in deriving the hole
emulation systematic uncertainty.
Figure~\ref{clenergy} shows the distribution of ECAL 
cluster energies for all tracks summed over all momenta from the data 
and from the Monte Carlo simulation before and after correction.
The response of FD to hadrons plays a minor role for the selection since
most hadronic 
showers remain below the  threshold of 3 GeV required for a valid FD cluster.
Therefore no additional uncertainty is assigned.

Overall, general improvements in the Monte Carlo simulation
and detailed studies of the
calorimeter response to electromagnetic and hadronic showers have reduced
the direct detector simulation uncertainty, and result in an efficiency
correction of $(4.0\pm4.0)\times10^{-4}$.

\subsubsection{\label{had-detstab}Detector performance}
The OPAL trigger system~\cite{bib-det-trig}
uses a large number of independent signals from 
a variety of detector components.
This information is combined to form many different event selection criteria, 
any one of which is sufficient to trigger the OPAL detector to be read out. 
The relatively low beam-induced background conditions of {\LEPI} permit
the choice of very loose settings for these trigger conditions. 
In general, \mbox{$\eehad$} events satisfy many independent trigger criteria. 
This redundancy allows the trigger inefficiency to be determined 
directly from the data. It is found to be less than~$10^{-5}$.
Each subsequent step in the data-recording chain, from data acquisition, 
through online reconstruction, to the final writing of the data samples 
into the off{}line storage facility, was investigated.
Possible effects, such as incorrect logging of detector status, bookkeeping
discrepancies at various stages or failures of the event reconstruction 
have been examined. Only one problem has been found:
very rarely, individual events suffer from high 
electronic noise levels in the central tracking detector, which causes the
track reconstruction to fail. These events are not classified as 
\mbox{$\eehad$} candidates, which causes an inefficiency of about 
\mbox{$0.8\times10^{-4}$.}
Both the short-term and long-term stability of each of the detector 
components used for the selection of \mbox{$\eehad$} events have been examined, 
as have distributions of the cut variables themselves.
Only one significant effect has been found:
for the~FD small offsets in the energy distribution of the 1995 data
give rise to an additional inefficiency of \mbox{$(5\pm3)\times10^{-4}$.}
The stability of the selection was checked by 
using alternative selection criteria, each based on data from only a single 
detector component. No notable effect is seen.
In total an uncertainty of \mbox{$2\times10^{-4}$} is assigned 
due to irregularities in the detector performance during
the 1993 and 1994 data-taking. 

For the 1995 data further studies were performed to search for possible effects on the 
\mbox{$\eehad$} selection inefficiency due to
the bunch-train mode of operation of LEP.
As discussed in {\SECT}~\ref{sec-samples} the bunch-train mode
could potentially affect the CT reconstruction and the ECAL response.
In order to check for such problems a tracking-independent selection has 
been developed, which is designed to overlap as much
as possible with the standard selection.
Within statistics no bunchlet-dependent effects are observed. 
The \mbox{$3\times10^{-4}$} precision of the check is assigned
as a systematic uncertainty.
Together with  the FD energy offset discussed above and the general
\mbox{$2\times10^{-4}$} data-taking uncertainty,
the overall detector performance error for 1995 is 
\mbox{$4.7\times10^{-4}$}.

\subsection[Background in the \mbox{$\eehad$} channel]
{\label{had-back}\boldmath Background in the \mbox{$\eehad$} channel}
The largest backgrounds to the \mbox{$\eehad$} event selection 
arise from \mbox{$\eetautau$} events and from two-photon interaction processes,
\mbox{$\eeeeqq$.}
The background fraction from \mbox{$\eetautau$} events is estimated to be
\mbox{$14.1\times10^{-4}$} using the KORALZ Monte Carlo event 
generator~\cite{bib-mc-KORALZ}.
Detailed studies of variables sensitive to \mbox{$\eetautau$} events indicate
that the Monte Carlo simulation underestimates this background by 13\%.
Part of this underestimate is caused by deficiencies in the simulation of
the conversion of radiated photons into electron-positron pairs.
Taking this into account results in a corrected background estimate of
\mbox{$(15.9\pm2.0)\times10^{-4}$,} where the assigned systematic error
is based on the observed discrepancy.

The background from two-photon interaction processes has been 
estimated directly from the data by making use of the
characteristics of two-photon events. In general they have low visible 
energy,~$\Rcal$, and large energy imbalance along the
beam-axis,~$\Rbal$ (see {\FIG}~\ref{mhcuts}).
The cross-section for two-photon processes is proportional 
 to $\log{s}$
and therefore the cross-section
changes by less than 1\,\% in the energy range between {\pkm} and {\pkp}.
Figure~\ref{nrb} shows, for the three energy points, the 
cross-sections for low~$\Rcal$ \mbox{(0.10--0.18)}
or high~$\Rbal$ \mbox{(0.50--0.75)} events
versus the cross-section for events with high~$\Rcal$ ($>0.18$)
and low~$\Rbal$ ($<0.50$). 
Most events from two-photon processes fall into the former category,
whereas the latter is completely dominated by hadronic $\Zzero$~decays.
The constant cross-section from non-resonant background sources, mainly 
two-photon events, is obtained from the intercept of a straight-line fit to 
these data, which yields \mbox{$0.051\pm0.006\,{\mathrm{nb}}$.}
The error is dominated by the data statistics available.

This estimate is corrected for the small fraction,~\mbox{$(7\pm5)\%$,}
of accepted two-photon events which fall outside the defined
region of low~$\Rcal$ or high~$\Rbal$ using the
Monte Carlo generators  \mbox{PHOJET~\cite{bib-mc-PHOJET}} and
HERWIG~\cite{bib-mc-HERWIG}, which simulate  two-photon interactions.
Furthermore, the estimate based on the data contains a small bias due 
to the larger fraction of events with high-energy initial-state radiation 
photons at the off-peak energy points. These events tend to have a high value 
of~$\Rbal$ and therefore give a small contribution from signal events
to the fitted non-resonant background.
This bias is estimated using Monte Carlo simulation to be 
\mbox{$0.004\pm0.002\,{\mathrm{nb}}$.}
Overall, the non-resonant background is  
\mbox{$0.051\pm0.007\,{\mathrm{nb}}$}, which amounts to
\mbox{$(16.7\pm2.3)\times10^{-4}$} at the {\pk} energy point.

The background from four-fermion processes other than two-photon interactions
is evaluated using four-fermion event generators,
FERMISV~\cite{bib-mc-FERMISV} for $s$-channel \mbox{$\ell^+ \ell^- \, \ff$} and
PYTHIA~\cite{bib-mc-JETSET74} for $t$-channel \mbox{$\ee \, \ff$}. 
Part of the $t$-channel four-fermion events are
implicitly included in the two-photon subtraction. 
Adding the residual fraction and the $s$-channel four-fermion events
results in a small contamination at the $0.4 \times 10^{-4}$
level which is subtracted.

Other background sources are even lower;
contributions from \mbox{$\eeee$} and cosmic-ray induced events are 
estimated to be \mbox{$0.2\times10^{-4}$} each. 
There is no indication of any backgrounds induced by beam interactions
with the residual gas in the LEP vacuum or the wall of the beam pipe.

\section{\boldmath Measurement of leptonic events}
\label{sec-lept}
The branching fraction of the $\Zzero$ boson to charged leptons is 
approximately 10\%.
Consequently the
leptonic decays provide less information on $\MZ$ and $\GZ$ than the
hadronic decays of the $\Zzero$.
Since the $\Zzero$ bosons we observe at LEP are all produced through 
their coupling to electrons in the initial state, however, the measurement of
the leptonic decay channels is of particular interest.
By measuring the cross-sections for all visible $\Zzero$ decays, including
electrons, the
absolute branching fraction to invisible final states can be determined.
Assuming these states are exclusively neutrinos coupling to the $\Zzero$
according to the {\SM} then allows the 
effective number of such light neutrino species to be determined.

In contrast to the quarks in hadronic events, 
the charge of the leptons can be determined almost unambiguously, 
thereby allowing the forward-backward charge asymmetry, 
$\Afb$, to be measured. Measurements of $\Afb$ determine 
the relative strengths of 
the vector and axial-vector couplings of the $\Zzero$ to each of the 
three charged lepton species, $\gvl/\gal$. In the {\SM},
this ratio is related to the weak mixing angle, $\swsq$.
When combined with the $\elel$ cross-sections, which are 
proportional to $\gvl^2+\gal^2$, the couplings $\gal$ and $\gvl$ are
determined. 
In addition, comparisons among the partial $\Zzero$ decay widths to the three 
charged lepton species and the respective forward-backward asymmetries
provide a precise test of the lepton universality of the neutral current.

\subsection{Introduction}
  
Leptonic decays of the $\Zzero$ boson, \mbox{$\eell$}, result
in low multiplicity events. 
The different leptonic species are distinguished from each other 
mainly using the sum of the track momenta, $\ptotal$,  
and the energy deposited in the electromagnetic calorimeter,
$\Etotal$. The total momentum is calculated as a scalar sum over the
individual track momenta, $\ptotal=\Etrk$. Similarly the total energy
is the sum of the energies of the individual clusters
in the electromagnetic calorimeter, $\Etotal=\Eshw$.
{\FIG}~\ref{fig-ll} shows the distribution of ($\ptotal,\Etotal$) for a
small sample of Monte Carlo $\eell$ events.
The \mbox{$\eeee$} events have both $\Etotal$ and
$\ptotal$ concentrated around the centre-of-mass energy, $\roots$.
However, due to final-state and Bremsstrahlung photons 
the distribution of  $\ptotal$ extends to lower values.
The $\eemumu$ events are concentrated around $\ptotal=\roots$ and have
low values of $\Etotal$. Final-state radiation has the effect of producing
a small tail of events at higher values of $\Etotal$ and correspondingly 
lower values of $\ptotal$. Due to the undetected neutrinos from 
$\tau$ decays in \mbox{$\eetautau$} events the distributions of
both $\Etotal$ and $\ptotal$ are broad but well separated from the 
other leptonic decays of the $\Zzero$, with only a few events
with $\Etotal$ or $\ptotal$ close to $\roots$. 
Cuts on the global quantities $\Etotal$ and $\ptotal$ provide 
the basis for separating the different leptonic decay modes.

Detailed descriptions of the selection criteria for the 
three different \mbox{$\eell$} categories are
described in \mbox{{\SECTs}~\ref{sec-elec}-\ref{sec-tau}}.
The event selections use a series of cuts to reject background. 
The selections are exclusive, with no event allowed to be 
classified in more than one category.
The largest backgrounds in the selected \mbox{$\eeee$}, \mbox{$\mumu$}
and \mbox{$\tautau$} event samples arise from cross-contamination
between the three lepton species. 
The treatment of four-fermion final states is discussed in 
Appendix~\ref{sec-4f}. 
Backgrounds from the processes,
$\eehad$, $\eeeell$, $\eegg$ and cosmic ray events, are all relatively small. 
The background from $\eehad$ is
rejected by cuts on the number of observed tracks and the
number of clusters in the electromagnetic calorimeter. 
In two-photon processes, $\eeeell$, the scattering angles of the 
incident electron and positron tend to be close to the beam direction, 
beyond the experimental acceptance. This source of background is
rejected by lower bounds on variables such as $\Etotal$ or $\ptotal$.  
Background from cosmic rays is removed by requiring
the event to originate from the $\epem$ interaction region and to be
in time with the $\epem$ beam crossing.

\subsubsection{Corrections and systematic uncertainties}
\label{sec:lep-corrs}
Estimates of selection efficiencies and accepted background cross-sections 
are obtained from Monte Carlo samples generated at peak and
off-peak centre-of-mass energies. These efficiencies are corrected to
the $\sprime$ acceptance specified in Section~\ref{sec-sel}
and are also corrected to
account for interference between initial- and final-state radiation
diagrams (see Appendix~\ref{sec-ifi}).
Data and Monte Carlo distributions are compared to 
assess the quality of the simulation for each cut used in the 
event selection and where necessary, corrections are derived.
Depending on the exact 
nature of the systematic check, a single correction factor is often
adequate to describe all \mbox{1993--1995} data.
Otherwise, where the effect under consideration is related to the
performance of the detector, independent corrections are 
determined for each year.  
Similarly, when the effect depends on the centre-of-mass energy,
different corrections  
are derived for each of the different centre-of-mass energy points. 
The derived  corrections greatly reduce the dependence 
on the accuracy of the Monte Carlo simulation.

There are two important sources of detector-related systematic
uncertainty which affect each of the three lepton channels. 
The largest potential source of systematic bias arises 
from imperfect simulation of the detector response to charged particles 
whose paths lie near one of the wire planes of the jet chamber. 
The wire planes, which create the drift field, are situated
at $7.5^{\circ}$~intervals in azimuth around the chamber, alternating between 
the anode and cathode planes which form the 24~sectors of the jet chamber. 
The reconstruction of tracks within~$\pm0.5^{\circ}$ of an anode wire plane
can be problematic due to field distortions.
This does not cause problems for low momentum tracks, 
which bend significantly in the axial magnetic field, since only a 
small fraction of the trajectory will 
be close to a wire plane. For high momentum particles with relatively 
straight trajectories, such as muons and electrons from $\eell$, a 
serious degradation of track quality can occur in a small fraction of the
events where most of the track lies near an anode wire plane.
This can result in a degraded momentum 
measurement, single tracks being reconstructed as two tracks 
(split across a wire plane) or, in the worst instance, failure to
reconstruct the track at all, resulting in `lost' tracks.
Field distortions close to the wire planes are simulated 
in the Monte Carlo but the numbers of tracks affected is significantly
underestimated. These effects are particularly important in the selection 
of $\eemumu$. Events where the 
measured momentum of one of the muons is anomalously low, due to wire
plane effects, result in
the Monte Carlo prediction of $0.3\%$ for the selection inefficiency being 
under-estimated by approximately $0.4\%$ compared to the true value. 
Significant corrections to the Monte Carlo selection efficiencies, 
relating to these tracking problems, are therefore obtained from the data.
Since $\eemumu$ events measured to have anomalously low total momenta
are classified as $\eetautau$, the uncertainties in these corrections
lead to anti-correlated errors in the $\mumu$ and $\tautau$ selections.
Details are given in Sections~\ref{sec:muon-acc} and~\ref{sec-leptcor}. 
 
The measurement of energy in the electromagnetic calorimeter plays the
primary role in the selection of $\eeee$ events and in discriminating them
from $\eetautau$. An uncertainty in the energy scale or in the 
response of
the ECAL can therefore introduce correlated uncertainties in the
$\ee$ and $\tautau$ cross-section measurements. The Monte Carlo
simulation underestimates the fraction of $\ee$ events which 
fail the selection due to 
imperfect Monte Carlo modelling of the response of the 
electromagnetic calorimeter near the mechanical boundaries between 
calorimeter modules, \mbox{at~$\phi=\pm\,90^{\circ}$} and
\mbox{$\abscosthe=0.22$} and~0.60. In addition, there is an unsimulated
problem in the electronic gain calibration for two out of the 9440 barrel
lead glass blocks for the 1994 and 1995 data samples. As a result the 
Monte Carlo estimates of the $\eeee$ selection efficiency
need to be corrected by $0.1-0.2\%$. Details are given in 
Sections~\ref{elec-emcut} and~\ref{sec-leptcor}. 

\subsubsection[Trigger efficiency for \mbox{$\eell$} events]
{\boldmath Trigger efficiency for \mbox{$\eell$} events}

Low multiplicity events from \mbox{$\eell$} produce relatively
few energy deposits in the detector. As a result, there is the potential
that events are lost due to trigger inefficiency.
The trigger efficiency for \mbox{$\eell$} events is determined using 
independent sets of trigger signals~\cite{bib-det-trig} 
from the electromagnetic calorimeters,
from the tracking chambers, from the muon chambers and from the 
time-of-flight counters. The efficiencies are determined in bins of
polar and azimuthal angle and then combined into an overall efficiency. 
When averaged over the three years of data analysed here, the 
trigger efficiencies for the \mbox{$\eeee$}, \mbox{$\mumu$} and 
\mbox{$\tautau$} selections are 
\mbox{$(>99.99)\%$}, \mbox{$(99.96\pm0.01)\%$} and \mbox{$(99.98\pm0.01)\%$}
respectively. The online event filter~\cite{bib-det-filter} is found to 
be 100\% efficient for \mbox{$\eell$} events and no systematic error 
is assigned.

\subsection[Selection of $\eeee$ events]{\boldmath Selection of $\eeee$ events}
\label{sec-elec}
The selection of \mbox{$\eeee$} events 
is accomplished with high efficiency and purity 
by requiring low-multiplicity events with large total electromagnetic
energy and by requiring at least two 
electron\footnote{Throughout this section the word ``electron'' should be 
understood to imply ``electron or positron''. Where a distinction is required, 
the symbols~$\en$ and~$\ep$ are used.} candidates. 
In order to maintain high efficiency, an electron is identified simply
as a high energy deposit in the electromagnetic calorimeter which 
is associated with a track in the central detector. 
The main experimental systematic uncertainties are from
the energy response of the calorimeter, the edge of the polar angle
acceptance, the track reconstruction quality and the 
background from \mbox{$\eetautau$} events.

\subsubsection[$t$-channel contribution to \mbox{$\eeee$}]
{\boldmath $t$-channel contribution to \mbox{$\eeee$}}

The analysis of $s$-channel $\Zzero$~decays into $\ee$~pairs is 
complicated by indistinguishable contributions
from \mbox{$t$-channel} scattering processes, 
and interference between $s$- and \mbox{$t$-channels.}
The \mbox{$t$-channel} amplitude
is non-resonant and is dominated by photon-exchange. These contributions 
are accounted for when fitting the \mbox{$\eeee$} data 
(see Appendix~\ref{sec-tchan}), but statistical and systematic 
uncertainties on the magnitude of the \mbox{$t$-channel} contribution 
result in an overall reduction of the sensitivity of the \mbox{$\eeee$}
analysis to the~$\Zzero$ properties.
The relative size of the \mbox{$t$-channel} amplitude depends on the 
scattering angle (polar angle) of the electron, $\thelec$,
and becomes the major component at high values 
of~$\cos\thelec$. In order to enhance the \mbox{$s$-channel} component of 
the selected data sample, a cut is made, for the \mbox{$\eeee$} analysis only,
constraining the polar angle to lie well within the barrel region of the 
detector ($|\cos\thelec|<0.70$, see {\SECT}~\ref{elec-sel}). 
The relative size of the \mbox{$t$-channel} component is 
larger for the off-peak data samples than the $\sim15\%$ it represents
on-peak, due to the reduction of the \mbox{$s$-channel} contribution 
from the $\Zzero$~resonance. 
This  results in small differences in efficiencies, backgrounds and systematic 
uncertainties amongst the \mbox{$\eeee$} data samples at different energy 
points. Because of the \mbox{$t$-channel} diagram 
the differential cross-section rises steeply in the 
forward direction, resulting in an increased sensitivity
to the precise definition of the edge of the acceptance in~$\cos\thelec$. 

\subsubsection[Selection criteria for \mbox{$\eeee$}]
{\boldmath Selection criteria for \mbox{$\eeee$} }
{\label{elec-sel}}
The selection criteria for \mbox{$\eeee$} events make use of information 
from the electromagnetic calorimeter and the central tracking detectors by
requiring:
\begin{itemize}
\item Low multiplicity: 
   \subitem $2\leq\Ntra\leq 8$\  and\  $2\leq\Nclu\leq 8$,
   
   where $\Ntra$ is the number of tracks and $\Nclu$ is the
   number of clusters.
\item High energy clusters: \\
   The energies of the highest energy, $E_1$, 
   and second highest energy, $E_2$, clusters must satisfy 
   \subitem
   $E_1 > 0.2\,\roots$\  and\  $E_2>0.1\,\roots$.
\item Total electromagnetic energy: 
   \subitem $\Etotal \equiv \Eshw  >0.80 \roots$.
\item Two electrons: \\
   At least two of the three highest energy clusters of energy above
   2~GeV must be associated with a track, which is required to
   point to the cluster position to within 
   $\Delta\phi<5^{\circ}$ in azimuth and to 
   within $\Delta\theta<10^{\circ}$ in 
   polar angle.
   These clusters are identified as electron candidates.
\item Geometrical and kinematic acceptance: 
   \subitem $\abscosthelec<0.70$\  and\  $\thacol<10^{\circ}$, 

   where~$\thacol$ is the acollinearity angle of the $\ee$ pair, defined as 
   \mbox{$180^{\circ}-\alpha$}, where $\alpha$ is the opening angle 
   between the directions of the two tracks.
\end{itemize}
These criteria select \mbox{96$\,$669} events 
from the \mbox{1993--1995} data sample.
The efficiency and background are first estimated
using a sample of Monte Carlo events. Corrections to these estimates are
obtained by studying the data. 
The correction factors and their associated 
systematic errors are summarised in {\TAB}~\ref{tab-eecor} 
for the seven data samples recorded during
\mbox{1993--1995} at the three energy points.
The systematic errors for the seven data samples are strongly correlated,
as shown in {\TAB}~\ref{tab:ee_matrix}.
The main corrections and systematic uncertainties
are described below.

\subsubsection[Selection efficiency for \mbox{$\eeee$}]
{\label{elec-emcut}\boldmath Selection efficiency for \mbox{$\eeee$} }

The Monte Carlo prediction for the $\eeee$
selection efficiency within the geometrical acceptance of
$\abscosthelec<0.70$ is $(99.44\pm0.02)\%$ for peak data.
The main corrections to this efficiency  
and related systematic errors are described below.

\bigskip
\noindent {\boldmath\bf Electromagnetic energy cuts:}
The main effect of the cut on $\Etotal>0.8\roots$, 
shown in \mbox{{\FIG}~\ref{f-eeecut1}(a)}, is to 
remove background from \mbox{$\eetautau$}. From the Monte Carlo simulation
only $11\times10^{-4}$ of \mbox{$\eeee$} events are rejected by this cut. 
There is a discrepancy between data and Monte Carlo in the vicinity of the 
cut, due to imperfect Monte Carlo modelling of the electromagnetic calorimeter
response near the mechanical boundaries between calorimeter modules and
a problem with the electronic gain calibration for two lead glass blocks
in 1994 and 1995.
These two effects introduce  additional inefficiencies for the data,
estimated to be \mbox{$(9\pm 10)\times10^{-4}$,} 
\mbox{$(22\pm 7)\times10^{-4}$} and \mbox{$(17\pm 8)\times10^{-4}$} for 
1993, 1994 and 1995, respectively.
These efficiency corrections are obtained from a detailed study of 
events which fail only the cut on electromagnetic energy.
For these events, the acoplanarity,
\mbox{$\thacop\equiv\left|\left|\phi_{\en}-
\phi_{\ep}\right|-180^{\circ}\right|$,}
and the sum of the track momenta, $\ptotal$,
are used to discriminate \mbox{$\eeee$} events from the 
dominant \mbox{$\eetautau$} background. 
The distribution of~$\thacop$ for events which fail only the cut on 
electromagnetic energy is shown in \mbox{{\FIG}~\ref{f-eeecut1}(b).}
The~$\thacop$ distribution for \mbox{$\eeee$} 
events is strongly peaked at~$0^{\circ}$. The
distribution for \mbox{$\eetautau$} events 
is broader due to the momentum transverse to the tracks 
which is carried away by the unobserved neutrinos from the $\tau$~decays.
The excess of data events compared to Monte Carlo near 
\mbox{$\thacop=0^{\circ}$} indicates 
that the excess of events in the region 
of the cut
arises from \mbox{$\eeee$}. This interpretation is corroborated by 
the distribution of \mbox{$\ptotal/\roots$} for events in the region of 
small~$\thacop$, shown in \mbox{{\FIG}~\ref{f-eeecut1}(c),}
where the excess of data over Monte Carlo events is clustered  
\mbox{near~$\ptotal/\roots=1.0$}. 
The efficiency corrections are derived from
the distributions of~$\thacop$ and~$\ptotal$ in the region
\mbox{$0.6\roots<\Etotal<0.8\roots$.} 

\bigskip
\noindent {\boldmath\bf
Electron identification:}
Electron candidates are defined as electromagnetic clusters of energy  
greater than 2~GeV which are associated with a track. An  
electron can fail these requirements if the track associated with 
the cluster is of very low momentum, due to hard bremsstrahlung in 
the material in front of the tracking detectors. 
It can also fail due to the emission of a hard final state photon
or if, due to poor track reconstruction, the track fails the geometrical 
matching conditions or the track quality requirements. 

The inefficiency arising from the demand for at least two 
electron candidates is estimated from Monte Carlo simulation to be
\mbox{$(39\pm2)\times10^{-4}$} for \mbox{$\eeee$}.
The inefficiency in the data is assessed from a sample of events where 
only one electron candidate is found. This control sample is
obtained by assuming the 
electrons in the event correspond to the two highest energy clusters,
only one of which is identified as an 
electron candidate on the basis of the tracking requirements. 
The background from the non-resonant QED~process 
\mbox{$\eegg$} where one of the photons has converted to an $\ee$ pair
is reduced by removing events where two tracks are associated to the same
cluster. Within this sample 
there are more events in the data than the Monte Carlo predicts. 
The excess is
concentrated in the regions of~$\phi$ near the wire planes of the jet 
chamber. In this region the polar angle resolution for
reconstructed tracks can be degraded sufficiently such that
the track no longer points to the cluster. 
The excess of data over the Monte Carlo prediction is used to evaluate 
a correction to the Monte Carlo estimate for these inefficiencies.
The size of the correction differs for each year of data-taking.
For example in the 1994 data an efficiency correction 
of \mbox{$(26\pm5)\times10^{-4}$} is derived.

\bigskip
\noindent {\boldmath\bf Acceptance definition:}
The geometrical and kinematic acceptance for \mbox{$\eeee$} events is 
defined by cuts on $\abscosthelec$ and on~$\thacol$,
shown in 
{\FIG}~\ref{f-eeacc}. The polar angle cut is made with respect 
to~$\thelec$, the direction of the negatively charged lepton as
measured in the electromagnetic calorimeter.
This is determined from an energy-weighted 
average of the positions of the lead glass blocks which form the cluster,
which is then 
corrected for biases caused by showering in the material in front of 
the calorimeter.
The existence of any systematic offset between the reconstructed cluster
position and the actual trajectory of the electron was studied by 
measuring the displacement between well measured electron tracks 
and their associated clusters near the
critical $\costhe$ boundaries.
These studies indicated that the effective edge of the
acceptance is offset symmetrically from the nominal cut value 
towards~$\cos\thelec=0$ by \mbox{$0.0004\pm0.0006$} at 
\mbox{$\cos\thelec=\pm 0.70$.}
Consistent displacements were obtained in both
the data and Monte Carlo simulation. 
The central value of the offset is 
obtained from the high-statistics Monte Carlo samples and the uncertainty
is taken as the statistical precision of the study based on data.
For the peak data the uncertainty in the location of the
\mbox{$\cos\thelec=\pm 0.70$} boundary results in an uncertainty of
$\pm9\times10^{-4}$ in the measured cross-section.

\bigskip
\noindent {\boldmath\bf Further corrections and uncertainties:}
The inefficiency arising from the multiplicity cuts
is estimated to be~$(1\pm1)\times10^{-4}$. This estimate is obtained
from the simulation and checked by examining 
the electromagnetic calorimeter energy distribution 
for the events which fail only the multiplicity cuts. 
The trigger inefficiency is determined to be less than $5\times10^{-5}$
and no correction is applied. 
Corrections related to  
four-fermion events, determined as described in Appendix~\ref{sec-4f},
are negligible and a $2\times10^{-4}$ uncertainty is assigned.

\subsubsection[Background in the \mbox{$\eeee$} channel]
{\boldmath Background in the \mbox{$\eeee$} channel}

The expected background in the {\pk} data samples is 
\mbox{$(34\pm 6)\times10^{-4}$}, dominated by $\eetautau$. 
The energy-dependence of the \mbox{$\eeee$} cross-section is different 
from the other \mbox{$s$-channel} fermion-pair production processes
due to the \mbox{$t$-channel} photon exchange contributions.
Consequently the background fraction is dependent on centre-of-mass energy.

\bigskip
\noindent {\boldmath\bf Background from \mbox{$\eetautau$}:}
The Monte Carlo estimate of this background 
for the {\pk} energy point is \mbox{$(32\pm 1)\times10^{-4}$}. 
The majority of the 
$\eetautau$ background is rejected by the cut $\Etotal>0.8\roots$, 
shown in {\FIG}~\ref{f-eeecut1}(a).  The distribution of $\Etotal/\roots$
is well reproduced  by the Monte Carlo simulation 
in the region dominated by the \mbox{$\eetautau$} background, 
\mbox{$\Etotal<0.70\roots$}. 
To investigate the level of the background within the
$\eeee$ sample, events just above the energy cut are selected,
\mbox{$0.8<\Etotal/\roots<0.9$}. For these events the data 
and Monte Carlo distributions of~$\ptotal$ and $\thacop$ are 
again used to distinguish
$\eeee$ events from $\eetautau$. 
The Monte Carlo estimate of the \mbox{$\eetautau$} background 
has been found to be consistent with the observations from the data, 
to within the statistical errors of the comparisons made. The 
systematic uncertainty on the \mbox{$\eetautau$} background for the
{\pk} energy point is estimated to be $6\times10^{-4}$.

\bigskip
\noindent{\boldmath\bf Other backgrounds:}
The energy deposits in the electromagnetic calorimeter 
from \mbox{$\eegg$} events 
are similar to those from \mbox{$\eeee$} events. However, for
\mbox{$\eegg$} events to pass the electron identification requirements
the two electromagnetic calorimeter clusters both must have associated tracks.
The probability that both the photons convert is about~1\% and the
$\eegg$ cross-section is relatively small. Consequently, 
the background fraction in the $\eeee$ sample is small (~$\sim1\times10^{-4}$).
The background from hadronic events is 
estimated to be about the same size.
A~100\% relative uncertainty is assigned to the 
background from hadronic events and to that from \mbox{$\eegg$} events.
Backgrounds from two-photon interaction processes,
\mbox{$\eeeeff$,} and from cosmic ray events are less
than~$1\times10^{-4}$.
No correction is applied, and this estimate is taken as a systematic error.

\subsection[Selection of $\eemumu$ events]{\boldmath Selection of $\eemumu$ events}
\label{sec-muon}
Of the three $\eell$ channels, the $\mumu$ final state provides the
cleanest environment for precise measurements of the $\elel$ 
cross-sections and asymmetries. The $\mumu$ channel does not suffer from the
theoretical uncertainties associated with $t$-channel corrections in the
$\ee$ final state nor the systematic uncertainties 
arising from the less well defined experimental signature of the 
$\tautau$ final state. However, of the three lepton channels, $\mumu$
events are the most sensitive to systematic uncertainties arising
from track reconstruction.
The \mbox{$\eemumu$} events are separated from other $\Zzero$~decays and 
background processes by requiring exactly two tracks to be reconstructed 
in the central detector both of which are identified as muons.
The cross-sections are measured within the phase space 
region defined by \mbox{$m_{\mu\mu}^2/s>0.01$.}  The selection criteria
are summarised below.

\subsubsection[Selection criteria for \mbox{$\eemumu$}]
{\boldmath Selection criteria for \mbox{$\eemumu$} }
\label{muon-sel} 

The selection criteria for \mbox{$\eemumu$} events are:
\begin{itemize}
\item Two tracks: \\
Exactly two tracks are required each of which satisfies
\subitem
\ \ \ \mbox{$p_{\mathrm{track}}>6~{\mathrm{GeV}}$,}
\ \ \ \mbox{$\abscosthe<0.95$,}

where $p_{\mathrm{track}}$~is the track 
momentum and $\theta$~is the reconstructed
polar angle. Tracks identified as coming from photon conversions
are not counted.  
\item Azimuthal separation: \\
The azimuthal angular separation between the two tracks must satisfy
\mbox{$\cosdelphi<0.95$}, {\em i.e.} $\delphi>18^\circ$, 
to avoid difficulties in the muon identification of two closely separated 
tracks.  
\item Muon identification: \\
Both tracks must satisfy at least one of the 
following three muon identification criteria:
\subitem
(a) At least two muon chamber hits are associated with the track.
\subitem
(b) At least four hadronic calorimeter strips are associated with the track. 
The average number of strips in layers containing hits has to be less than 
two, in order to reject hadronic showers. For \mbox{$\abscosthe<0.65$,} 
where tracks traverse all nine layers of strips in the barrel calorimeter,
at least one hit in the last three layers of strips is required.
\subitem
(c) The track has \mbox{$p_{\mathrm{track}}>15~{\mathrm{GeV}}$} 
and the sum of the energy 
deposited in the electromagnetic calorimeter within a cone of half-angle 
63~mrad about the track is less than 3~GeV.
\item Visible energy: \\
Backgrounds from \mbox{$\eetautau$} and two-photon interaction events are 
reduced by requiring
\subitem \mbox{$\Evis>0.6\,\roots$},

where the visible energy, $\Evis$, is the scalar sum of the two track 
momenta and the energy of the highest energy cluster found in the 
electromagnetic calorimeter.
\item Despite the fact that the OPAL detector is situated in a cavern 100$\,$m 
underground there is still a large flux of cosmic ray particles.
The majority of these are muons which traverse the detector volume. 
Those which happen to pass close to the beam interaction point and 
which are synchronous with a bunch crossing can resemble \mbox{$\eemumu$}
events. Cosmic ray background is rejected by requiring that
the selected events originate from the average~$\epem$ 
interaction point and are coincident in time with the beam crossing.
\end{itemize}
These criteria select 128$\,$682~events which enter the cross-section 
analysis from the 1993--1995 data sample.
Control samples from the data are used to check and, where necessary, 
correct the Monte Carlo estimates. The dominant corrections are due to 
tracking losses and the residual background from \mbox{$\eetautau$} events.
The resulting correction factors and their systematic errors are summarised in
{\TAB}~\ref{tab-mmcor} for the seven data samples recorded during 
\mbox{1993--1995} at the three energy points.
The systematic errors of the seven data samples are strongly correlated
as shown in {\TAB}~\ref{tab:mm_matrix}. 
Unless otherwise specified, the illustrative errors quoted in the
following text refer to the 1994 sample.

\subsubsection[Selection efficiency for $\eemumu$]
{\boldmath\label{sec:muon-acc}Selection efficiency for $\eemumu$}

The Monte Carlo prediction for the $\eemumu$
selection efficiency is $(91.34\pm0.05)\%$.
This corresponds to a selection efficiency of $(98.40\pm0.03)\%$
within the geometric acceptance of $|\cos\theta_{\mu^-}|<0.95$.
Small corrections are then applied to account for the Monte Carlo events
generated below the ideal kinematic acceptance limit of \mbox{$m_\ff^2/s>0.01$},
and for interference between photons radiated in the initial and final states,
as described in Appendix~\ref{sec-ifi}.
Using the 1994 data as an example, the selection
efficiency from the Monte Carlo simulation is then corrected by 
$(-76\pm8)\times10^{-4}$ through comparisons with the data.
The main efficiency corrections 
and systematic errors are described below.

\bigskip
\noindent {\boldmath\bf Tracking losses:}
The selection of \mbox{$\eemumu$} events relies heavily on
track reconstruction in the central detector, which is required 
to measure~$\Evis$ and to associate 
tracks with activity in the outer detectors used for muon identification. 
As discussed in {\SECT}~\ref{sec:lep-corrs}, of particular concern are tracks 
whose paths lie within~$\pm0.5^{\circ}$ of an anode wire plane 
of the jet chamber. 
Figure~\ref{fig:mu_fvis}(a) shows the distribution of $\Evis$ for events 
passing all other selection cuts. There is a clear discrepancy
between data and Monte Carlo. The origin of this
discrepancy is the imperfect Monte Carlo simulation of tracking in the 
region close to the jet chamber wire planes. The discrepancy becomes
more apparent in Figure~\ref{fig:mu_fvis}(b) which
shows the corresponding $\Evis$ distribution for tracks 
within $0.5^\circ$ of the anode wire planes. 

The wire plane effect was investigated using an
alternative \mbox{$\eemumu$} selection procedure. This selection 
is independent of the central detector, relying 
instead on back-to-back signals in the muon chambers (within \mbox{30~mrad}) 
and electromagnetic calorimeters (within \mbox{50~mrad}).
By using relatively tight cuts on the acollinearity measured in the
outer detectors the background from \mbox{$\eetautau$} events is strongly 
suppressed. This is mainly due to the greater 
deflection in the magnetic field experienced by the lower momentum 
charged particles from $\tau$~decays.
Background is suppressed further by requiring at least one 
identified muon in the event. The efficiency of the tracking-independent 
selection is approximately~79\%.  

The events which are selected by the tracking-independent selection
but fail the default $\mumu$ selection are concentrated in regions where
one of the muons passes near a jet-chamber wire plane.
The numbers of ``lost'' \mbox{$\eemumu$} events, selected by these cuts, 
are corrected for the inefficiency of the tracking-independent selection. 
The corrected numbers are compared between data and Monte Carlo 
simulation in bins of $\cos\theta$ for each year of data taking. There 
is an excess of lost
events in the data, indicating that the 
Monte Carlo estimate of $30\times10^{-4}$ for the inefficiency due to tracking 
problems is too low. The difference is \mbox{$(42\pm4)\times10^{-4}$} for the
1994 data.
The uncertainties are the combined statistical errors of the data and 
Monte Carlo control samples. The correction factors obtained in this manner
are applied to the Monte Carlo inefficiency estimates.

\bigskip
\noindent {\boldmath\bf Track multiplicity cuts:}
In addition to the tracking losses discussed above,
$0.5\%$ of \mbox{$\eemumu$} events are rejected by the requirement 
of exactly two tracks due to additional tracks from converted 
final-state photons which have failed to be classified as such, 
or if a track is poorly reconstructed and split into two tracks.
A correction to the Monte Carlo prediction for the inefficiency due to 
this cut is made on the basis of a visual scan of both data and Monte 
Carlo events containing three, four or five  tracks, but which 
otherwise pass the \mbox{$\eemumu$} selection. The resulting 
corrections, typically less than $10^{-3}$, 
are listed in {\TAB}~\ref{tab-mmcor}, with the errors 
reflecting the data and Monte Carlo statistics and the uncertainty of 
the scanning procedure.

\bigskip
\noindent {\boldmath\bf Muon identification:}
Both tracks in a selected \mbox{$\eemumu$} event are required to 
be identified as  muons using data from at least one of three 
independent detector subsystems; 
the muon chambers, the hadronic calorimeter and the 
electromagnetic calorimeter. The Monte Carlo estimate of the 
inefficiency introduced by the muon identification requirement is 
$(79\pm5)\times10^{-4}$. Most of this inefficiency occurs in 
geometrical regions where either the muon chamber coverage 
or the hadronic calorimeter coverage are incomplete. 
In particular, for one 
sixth of the total azimuth in the polar angle range 
\mbox{$0.65<\abscosthe<0.85$} these gaps overlap due to 
support structures, leaving coverage only from the electromagnetic 
calorimeter. For collinear \mbox{$\eemumu$} events there is a high 
degree of correlation between the muon identification inefficiencies of 
the two tracks due to the symmetry of the detectors.

The redundancy of the three muon identification requirements 
is used to determine the single track muon identification efficiency 
in bins of azimuthal and polar angle. For each muon identification criterion,
the muon identification criteria from the other two 
outer detectors are used to define a control sample of tagged muons. 
The single track efficiencies, determined in
bins of $(\costhe,\phi)$, are used to calculate overall muon 
identification inefficiencies for \mbox{$\eemumu$} events accounting
for expected angular distributions. The total inefficiency determined 
in this way is compared between data and Monte Carlo prediction
separately for each year of data.
There is good agreement for the 1993 and 
1995 data, and no correction is applied.
For the 1994 data, it is found that the
inefficiency from the corresponding Monte Carlo sample needs to be
corrected by \mbox{($15\pm4)\times10^{-4}$}
where the errors reflect the statistical power of the checks.
The correction arises due to inadequate simulation of the response of the
hadron calorimeter in the Monte Carlo sample used to simulate data 
from the 1994 operation of the OPAL detector.

\bigskip
\noindent {\boldmath\bf Acceptance definition:}
Both muon tracks are required to lie within
the geometrical acceptance \mbox{$\abscosthe<0.95$}.
The measurement 
of~$\costhe$ in this region therefore affects the overall event 
selection inefficiency and any discrepancy between data and Monte Carlo 
simulation must be corrected. A 1~mrad bias in angle at
\mbox{$\abscosthe=0.95$} corresponds to a bias in the acceptance 
of~$5\times10^{-4}$.

For muon tracks there are generally three separate detector components 
which can be used to measure~$\costhe$; the track reconstructed in the 
central detector, the energy cluster in the electromagnetic calorimeter 
and the track found in the muon chambers. For muons close to 
\mbox{$\abscosthe=0.95$} the end-cap muon chambers have the best polar 
angle resolution, approximately 1~mrad, and this measurement
is used when available, otherwise the angle from the reconstructed
central detector track is used (about~10\% of cases). 
The changes to the numbers of \mbox{$\eemumu$} events selected when 
using alternative measurements of the polar angle are investigated 
and compared between data and Monte Carlo simulated events, 
separately for each 
year of data. Based on the scatter of these comparisons a 
systematic error of~$10\times10^{-4}$ is assigned for the 1993 data sample, 
and~$5\times10^{-4}$ for the 1994 and 1995 data samples. 

\bigskip
\noindent {\boldmath\bf Further efficiency corrections:}
The trigger inefficiency for $\eemumu$ events is estimated to be
\mbox{$(6\pm2)\times10^{-4}$,} \mbox{$(5\pm2)\times10^{-4}$}
and \mbox{$(2\pm2)\times10^{-4}$} for the 1993, 1994 and 1995 data, 
respectively.
Corrections to the selection efficiency related to 
four-fermion events are determined as 
described in Appendix \ref{sec-4f}.
Finally, the inefficiency associated with the cosmic ray veto 
is found to be less than $1\times10^{-4}$.  

\subsubsection[Background in the \mbox{$\eemumu$} channel]
{\boldmath Background in the \mbox{$\eemumu$} channel }
The background in the \mbox{$\eemumu$} event selection is
approximately~1\%, dominated by misclassified \mbox{$\eetautau$} events.
The backgrounds from two-photon interaction processes, 
such as 
$\eeeemumu$,
and from events induced by the passage 
of cosmic ray muons through 
the detector are small.
The Monte Carlo background estimates are corrected for discrepancies
between simulated and real data. These corrections are described below and
the resulting background estimates are summarised in 
{\TAB}~\ref{tab-mmcor}.

\bigskip
\noindent {\boldmath\bf $\eetautau$ background:}
The lifetime of the $\tau$~lepton is sufficiently short that only the 
decay products of the~$\tau$ are registered in the detector. For 
$\eetautau$ events to be selected as $\eemumu$ both tau decays
must result in an identified muon and the visible energy requirement
must be satisfied, \mbox{$\Evis>0.6\,\roots$}.
Approximately~17\% of $\tau$~leptons decay into a muon and neutrinos, 
consequently about~3\% of \mbox{$\eetautau$} events 
result in a visible~$\mumu$ final state. In addition, other decays of 
the~$\tau$ to single charged 
particles, in particular \mbox{$\tau\rightarrow\pi\nu$,} can be 
misidentified as muons. 

Two control samples are selected from the data to check the simulation
of the $\tautau$ background. The first is sensitive to problems with the
visible energy distribution and the second is sensitive to possible
problems with muon identification.
For the first control sample the $\Evis$ cut in the $\mumu$ selection
is relaxed to $\Evis>0.5\roots$ and the low visible energy
region considered, $\Evis<0.8\roots$. 
Events with tracks within~$0.5^{\circ}$ in azimuth of a wire plane are
rejected to remove poorly measured \mbox{$\eemumu$} events.
The $\tautau$ background is further enhanced by rejecting events, 
predominantly $\mumu$, which have low acoplanarity $<20~$mrad.
A loose cut is also made to eliminate events with high energy radiated 
photons by requiring the acollinearity to be less than \mbox{150~mrad.}
This selected control sample is predicted by Monte Carlo simulation to 
be made up of approximately~97\% \mbox{$\eetautau$} events with the remainder 
being \mbox{$\eemumu$} events. The data are in good agreement with the Monte
Carlo prediction.

A second, higher statistics control sample is selected by relaxing
the $\Evis$ requirement of the \mbox{$\eemumu$} selection
to $0.5<\Evis<0.8$, and
requiring exactly one of the tracks to be identified as a muon.
Approximately~99.2\% of Monte Carlo generated events selected in this 
way are \mbox{$\eetautau$} events, 
in the majority of which one $\tau$~decays to a 
muon and the other $\tau$~decays to a single charged particle which is 
not a muon. 
From the 1993 and 1995 samples there is good agreement 
between the numbers of events selected in the data and the
Monte Carlo expectation.  In the 1994 samples an excess of about~4\% 
of Monte Carlo over data events is observed. 
On the basis of these two studies the 
Monte Carlo prediction for the \mbox{$\eetautau$} background of
$1.00\%$ for the 
1994 \mbox{$\eemumu$} data sample is corrected to $(0.97\pm0.04)\%$.
No corrections are made for the 1993 or 1995 
samples, and a systematic uncertainty of $2\times10^{-4}$ is assigned.

\bigskip
\noindent {\boldmath\bf Other background:}
Most events from the two-photon process
\mbox{$\eeeemumu$} are rejected by the cut on visible 
energy, \mbox{$\Evis>0.6\,\roots$.} The Monte Carlo predicts 
the remaining background to be approximately~0.05\% on-peak 
and 0.1\% off-peak, at which energies the resonant $\Zzero$~production 
is less dominant.
The two-photon background is evaluated separately for each year and
systematic errors of~$1\times10^{-4}$ are assigned to these calculations.
The backgrounds from cosmic ray events, estimated from time-of-flight
and vertex information, were \mbox{$(2\pm2)\times10^{-4}$} 
in 1993 and 1994 and 
\mbox{$(3\pm2)\times10^{-4}$} in~1995.

\subsection[Selection of $\eetautau$ events]{\boldmath Selection of $\eetautau$ events}
\label{sec-tau}
Tau~leptons produced in the process \mbox{$\eetautau$} decay before
entering the sensitive volume of the detector. The branching ratio 
for the decay $\tau^-\rightarrow\ell^-\nu_\tau\overline{\nu}_\ell$ 
is about 35\%, and the resulting electron or muon has, in general,
less momentum than a directly produced fermion from 
$\Zzero\rightarrow\ell^+\ell^-$ due to the two associated neutrinos. 
In $\tau$ decays to hadrons the associated single neutrino also reduces
the total visible energy of the final state.
The $\tau$-decay branching ratio to
three or more charged hadrons is about 15\%, and electrons from 
the conversion of photons from $\pi^0$~decays further
increase the average charged multiplicity of the final state. 
The experimental signature 
for \mbox{$\eetautau$} events is therefore less well defined than that for 
either \mbox{$\eeee$} or \mbox{$\eemumu$} events. 
Consequently, the cuts used to select \mbox{$\eetautau$} events are 
relatively involved. 

\subsubsection[Selection criteria for \mbox{$\eetautau$}]
{\label{sec-tau_selection}
\boldmath Selection criteria for \mbox{$\eetautau$}  }
The selection criteria used to identify \mbox{$\eetautau$} events remain 
unchanged with respect to our previous 
publications where a more
detailed description is provided~\cite{bib-opal-ls90,bib-opal-ls91,bib-opal-ls92}. The selection is summarised below: 
\begin{itemize}
 \item Multiplicity cuts, shown in {\FIG}s~\ref{fig:tau_fig1}(a) and
       ~\ref{fig:tau_fig1}(b), 
       to reject hadronic $\Zzero$ decays:
           \subitem $2 \leq \Ntra \leq 6 $ \ \ and $\Ntra + \Nclu \leq 15$.
 \item \mbox{$\eetautau$} event topology:
           \subitem  Events are reconstructed using a cone 
                     jet-finding algorithm\cite{bib-opal-ls90} with a cone
                     half-angle of $35^\circ$. The sum of the 
                     electromagnetic calorimeter energy and  
                     scalar sum of track momenta in each cone has to be more 
                     than 1\% of the beam energy. Exactly two cones 
                     containing tracks are required, not counting cones 
                     which contain only tracks associated with 
                     photon conversions. 
      The direction of each $\tau$~candidate is taken to be the  
      total momentum vector reconstructed from the tracks 
      and electromagnetic clusters in its cone. 
 \item \mbox{$\eetautau$} event acollinearity:
           \subitem \mbox{$\thacol< 15^\circ$}

      where the acollinearity angle,~$\thacol$, is $180^{\circ}$ minus 
      the angle between the directions of the two $\tau$~candidates.
 \item Geometrical acceptance:
           \subitem $\costau<0.9$.

      where $\theta_\tau$ is the polar angle of the event axis, defined
      using the vectorial difference between the momenta of the two
      $\tau$~candidates.
 \item Rejection of \mbox{$\eemumu$} events:
       \subitem Events selected as \mbox{$\eemumu$} by the 
       criteria described in {\SECT}~\ref{muon-sel} are rejected.
 \item Rejection of \mbox{$\eeee$} events:
           \subitem $\Etotal \equiv \Eshw < 0.8 \roots$, shown in 
       {\FIG}~\ref{fig:tau_fig1}(c).
       For the region $\costau>0.7$, where there is additional
       material in front of the electromagnetic calorimeter, it is also
       required that: 
           \subitem $\Evistau < 1.05 \roots$ \ \ 
                     or \ \ $\Etotal < 0.25 \roots$,\ \ \  
                    (for $\costau>0.7$),

       where~$\Evistau$ is the total visible energy, 
       $\Evistau = \Etotal + \ptotal$.
 \item Rejection of non-resonant \mbox{$\eeeell$} events, see  
       {\FIG}~\ref{fig:tau_fig1}(d): 
           \subitem $\Evistau > 0.18 \roots$. 
 
\item Cosmic ray background is rejected by requiring that
       the selected events originate from the average~$\epem$ 
       interaction point and are coincident in time with the beam crossing.

\end{itemize}
These criteria select {107\,340} events from
the \mbox{1993--1995} data sample.
The selection efficiency and background contributions from processes other than
\mbox{$\eetautau$} are estimated using  Monte Carlo events. 
These efficiencies and backgrounds are corrected 
for the observed differences between data and Monte Carlo. 
The corrections and systematic errors are 
summarised in {\TAB} \ref{tab:ttfactor} and are described below.
The systematic errors of the seven data samples are strongly correlated,
as shown in {\TAB} \ref{tab:tt_matrix}.
Unless otherwise specified, the numbers quoted in the text 
correspond to the 1994 data. 

\subsubsection[Selection efficiency for \mbox{$\eetautau$}]
{\label{sec:taueff}
\boldmath Selection efficiency for \mbox{$\eetautau$} }

The Monte Carlo prediction for the $\eetautau$
selection efficiency is $(75.18\pm0.07)\%$.
This corresponds to a selection efficiency of $(87.70\pm0.05)\%$
within the geometric acceptance of $\costau<0.90$.
Small corrections are then applied to account for the Monte Carlo events
generated below the ideal kinematic acceptance limit of \mbox{$m_\ff^2/s>0.01$},
and for interference between photons radiated in the initial and final states,
as described in Appendix~\ref{sec-ifi} and listed in {\TAB} \ref{tab:ttfactor}.
Within the angular 
acceptance, the largest source of inefficiency arises from 
the cuts used to reject background from $\eeee$.
To estimate the efficiency for data, the inefficiency 
introduced by each cut exclusively 
is first estimated from the Monte Carlo simulation. 
These estimates are corrected by comparing the data and Monte Carlo.
The decomposition into exclusive inefficiencies
is appropriate since only 0.6\% of events fail more than one of 
the classes of selection cuts. Using the 1994 data as an example, the 
selection efficiency from the Monte Carlo simulation is corrected by 
$(-1.30\pm0.28)\%$ to give an estimated selection efficiency 
of $(73.88\pm0.29)\%$. Details of the main 
corrections to the efficiency and the main sources of systematic 
uncertainty are given below. 

\bigskip
\noindent {\boldmath\bf
Multiplicity cuts:} 
The multiplicity requirements
exclusively reject \mbox{$(1.32\pm0.02)\%$} of Monte Carlo 
\mbox{$\eetautau$} events. However, a number of effects which 
can influence multiplicity are not perfectly modelled,  
such as the simulation of the material of the detector which affects
the rate of photon conversions and the two-track resolution.
To assess the effect using the data, the multiplicity cuts are 
removed from the \mbox{$\eetautau$} selection yielding a sample 
dominated by \mbox{$\eetautau$} and \mbox{$\eehad$}. In this 
sample, \mbox{$\eetautau$} events are identified by requiring 
that one of the two $\tau$~cones is consistent
with being a \mbox{$\tau \rightarrow \mu \nu_\mu \nu_\tau$} decay.
This requirement removes essentially all \mbox{$\eehad$} events. 
Backgrounds from \mbox{$\eemumu$} and \mbox{$\eeeemumu$} 
are rejected using cuts on acoplanarity and momentum.
In this way a sample of \mbox{$\eetautau$} events is isolated
using only the multiplicity information from a single $\tau$~cone.
The other $\tau$ cone is used to provide an unbiased estimator for
the multiplicity distribution for a single $\tau$~cone.
By convolving this measured single cone multiplicity distribution
with itself, the $\eetautau$ multiplicity distribution is estimated 
using data alone. The convolution is performed 
in two dimensions, track and total multiplicity.
The multiplicity cuts are applied to the convolved distribution to determine
the exclusive inefficiency of \mbox{$(1.69\pm0.11)\%$} for the  
1994 data sample\footnote{We give here, by way of example for all these
corrections, the correspondence between this
correction to the exclusive
inefficiency and the multiplicity correction factor, $f$, 
in Table~\ref{tab:ttfactor}.
$f = 1.0 + {{(0.0169-0.0132)}/{(0.7388+0.0169})}$, where 0.7388 
is the overall corrected efficiency, {\em i.e.} $1.3536^{-1}$.}. 
The validity of the 
procedure is verified using different Monte Carlo samples. 

\bigskip
\noindent{\bf\boldmath Acollinearity and cone cuts:} 
The acollinearity cut and the requirement that there be exactly two
charged $\tau$ cones in the event reject 
$(3.29\pm0.03)\%$ of Monte Carlo \mbox{$\eetautau$} events. These cuts reject
background from $\eeee$ and $\eeqq$.
To study the effect of these cuts, the
acollinearity and cone requirements are
replaced by cuts using particle identification information such as
${\mathrm{d}}E/\mathrm{d}x$. Relatively hard cuts are 
necessary to be able to study
the acollinearity distribution for $\eetautau$ events since, in the region of
the cut, the background from $\eeee$ dominates.  
\mbox{Figure~\ref{fig:tau_fig2}(a)} shows the acollinearity distribution
for this alternative selection which has an efficiency 
for \mbox{$\eetautau$} of approximately 33\% but has little
background (less than~0.5\%). 
For Monte Carlo $\eetautau$ events it is verified that the alternative
selection does not significantly bias the acollinearity 
distribution.  
In the region \mbox{$\thacol<15^\circ$}
the good agreement between data and Monte Carlo indicates that
the modelling of the efficiency of the particle identification
cuts is reasonable. The relative normalisations of the 
data and Monte Carlo in the region \mbox{$15^\circ<\thacol<45^\circ$} 
are used to determine corrections to the Monte Carlo efficiencies for the
different centre-of-mass energies. 
A similar check of the inefficiency related to the cone requirements is
made. The corrected inefficiency due to the acollinearity and cone cuts
is $(3.55\pm0.14)\%$.

\bigskip
\noindent{\boldmath\bf
Definition of $\costau$:}
A systematic uncertainty on the selection efficiency of~$0.1\%$ is assigned
due to the uncertainty in the definition of the geometrical acceptance,
\mbox{$\costau<0.9$.} This estimate is obtained by comparing 
relative numbers of selected events in data and Monte Carlo 
using various definitions of~$\costau$, 
{\em e.g.} calculated using tracks, using clusters in the electromagnetic 
calorimeter or using both tracks and clusters. 

\bigskip
\noindent{\bf\boldmath\mbox{$\eeee$} rejection cuts:} 
\noindent
The Monte Carlo simulation predicts an exclusive inefficiency
due to the \mbox{$\eeee$} rejection cuts of \mbox{$(3.40\pm0.03)\%$}. 
To investigate potential biases, the energy-based $\eeee$ rejection
is replaced by cuts using electron identification information. 
Imperfect modelling of the detector response in the region
\mbox{$\costau>0.7$}, shown in \mbox{{\FIG}~\ref{fig:tau_fig2}(b)},
results in the Monte Carlo underestimating the true inefficiency.
The inefficiency for the data is estimated to be \mbox{$(3.92\pm0.17)\%$}. 

\bigskip
\noindent{\boldmath\bf
Further corrections and uncertainties :}
By using lepton identification information the effect of the
cuts used to reject \mbox{$\eemumu$} and \mbox{$\eeeell$} events are
found to be adequately modelled by the Monte Carlo, however corrections,
consistent with unity, and corresponding uncertainties  are obtained from
the data.
The trigger efficiency for \mbox{$\eetautau$} events is 
estimated to be $(99.98\pm0.02)\%$. During 1995, when LEP operated in 
bunch-train mode, no discernible effect on the \mbox{$\eetautau$} 
selection is observed and no systematic error is assigned. Uncertainties 
in the branching ratios of the $\tau$~lepton result in a~$0.05\%$ uncertainty 
in the \mbox{$\eetautau$} selection efficiency. 
The uncertainty on the mean $\tau$~polarisation\cite{bib-opal-taupol} 
has a negligible effect on the \mbox{$\eetautau$} selection 
efficiency ($<0.01\%$). Corrections to the selection efficiency related to 
four-fermion events are determined as 
described in Appendix \ref{sec-4f}. A 10\% uncertainty is assigned 
to the Monte Carlo expectation that 0.6\% of events fail more than one of 
the classes of selection cuts. 
  
\subsubsection[Background in the \mbox{$\eetautau$} channel]
{\label{sec:tauback}\boldmath 
Background in the \mbox{$\eetautau$} channel}

For peak data approximately 2.6\% 
of the events passing the \mbox{$\eetautau$} 
selection come from background processes. This is significantly
larger than the corresponding backgrounds in the 
\mbox{$\eeee$} and  \mbox{$\eemumu$} selections. The main backgrounds
are from \mbox{$\eeee$} (0.4\%), \mbox{$\eemumu$}  (1.1\%), 
\mbox{$\eehad$}  (0.4\%) and
$\mbox{$\eeeell$}$  (0.6\%). The background fractions from
$\eeee$
and $\mbox{$\eeeell$}$ are higher for the off-peak 
samples. To estimate the size of the background contributions 
using the data, cuts are applied to the selected 
\mbox{$\eetautau$} sample to enhance the various background sources. 
The resulting background estimations and systematic uncertainties are
given in {\TAB}~\ref{tab:ttfactor} and are summarised below (for peak data).

\bigskip
\noindent{\boldmath\bf 
Background from \mbox{$\eeee$}:}
The Monte Carlo expectation for the \mbox{$\eeee$} background fraction is 
\mbox{$(0.26\pm0.02)\%$}.  
For the reasons discussed in {\SECT}~\ref{elec-emcut}, the Monte Carlo
underestimates the \mbox{$\eeee$} background in the barrel region of 
the detector. The techniques
described in {\SECT}~\ref{elec-emcut} are used to estimate the effect
on the background level in the $\mbox{$\eetautau$}$ selection.
Similar studies indicate that the \mbox{$\eeee$} background in the
end-cap region, \mbox{($\costau>0.7$)}, is well modelled.
The corrected total background from \mbox{$\eeee$} is then
\mbox{$(0.47\pm0.07)\%$}

\bigskip
\noindent{\boldmath\bf Background from \mbox{$\eemumu$}:}
The \mbox{$\eemumu$} background is estimated 
from Monte Carlo simulation to be \mbox{$(0.98\pm0.02)\%$}. For 
\mbox{$\eemumu$} events to enter the \mbox{$\eetautau$} sample either 
one of the tracks must fail the muon identification criteria of
{\SECT}~\ref{sec-muon} or $\Evis$ must be less 
than~$0.6\roots$. The \mbox{$\eemumu$} background in the
\mbox{$\eetautau$} sample is enhanced by requiring at least one 
identified muon and by imposing loose momentum and electromagnetic energy 
cuts. The acoplanarity distribution of the surviving events is used to 
estimate corrections to the Monte Carlo expectation for the 
\mbox{$\eemumu$} background. Discrepancies between data and Monte Carlo 
are observed for tracks near the anode planes of the central jet chamber
as indicated in {\FIG} \ref{fig:tau_fig2}(c).   
As a result of these comparisons, the Monte Carlo background estimate
for the 1994 data is corrected to $(1.17\pm0.09)\%$.

\bigskip
\noindent{\boldmath\bf
Background from \mbox{$\eehad$}:}
The Lund string model as implemented in the JETSET program is used to 
describe the hadronisation process for 
the generated Monte Carlo samples. The parameters within JETSET are tuned to 
describe the properties of OPAL \mbox{$\eehad$} 
events\cite{bib-mc-OPALtune-j73,bib-mc-OPALtune-j74}.
Although these Monte Carlo samples provide a good description of the
global properties of \mbox{$\eehad$}, {\em e.g.} event shape variables,
there is no guarantee that they adequately describe the
low multiplicity region. This is illustrated by the comparison of the 
\mbox{$\eehad$} background expectations obtained from Monte Carlo samples
using two different JETSET tunes, \cite{bib-mc-OPALtune-j73} and 
\cite {bib-mc-OPALtune-j74}, which give respective background estimates of 
\mbox{$(0.40\pm0.02)\%$} and \mbox{$(0.88\pm0.02)\%$}.

To assess the \mbox{$\eehad$}  background  fraction from the data,
the multiplicity cuts of the \mbox{$\eetautau$} selection
are relaxed and events where one of the $\tau$ cones is consistent 
with a leptonic $\tau$ decay are removed from the selected 
\mbox{$\eetautau$} sample. Events in the predominant $\tau^+\tau^-$ 
track multiplicity topologies, $1-1$, $1-3$ and $3-3$, are 
also rejected. 
In this way \mbox{$\eetautau$} events are removed  
whilst retaining a large fraction of the \mbox{$\eehad$} background.
For example, \mbox{{\FIG}~\ref{fig:tau_fig2}(d)} shows the distribution of
track multiplicity after the $\eehad$ enhancement.
The resulting background enriched sample allows the \mbox{$\eehad$} 
background
fraction to be estimated from the data by  fitting the track and total
multiplicity distributions with contributions from 
\mbox{$\eetautau$} and \mbox{$\eehad$} where
the shapes are taken from Monte Carlo and the normalisations left free. 
In this manner the \mbox{$\eehad$} background is estimated to be
\mbox{$(0.41\pm0.11)\%$}. The systematic error reflects the
sensitivity to the Monte Carlo model used to describe the shape
of the multiplicity distributions.

\bigskip
\noindent{\boldmath\bf
Background from \mbox{$\eeeell$}:}
At the $\Zzero$ peak, the Monte Carlo prediction of the background from 
\mbox{$\eeeell$} two-photon processes is \mbox{$(0.53\pm0.02)\%$}.
The Monte Carlo predictions for the background from 
\mbox{$\eeeeee$} and \mbox{$\eeeemumu$} backgrounds are checked
by requiring both cones in selected \mbox{$\eetautau$} events
to be consistent with being electrons and muons respectively. 
Distributions of~$\Rvistau$ and missing transverse momentum 
are used to estimate the 
background directly from data. No deviations from the Monte Carlo
expectation are found. The statistical precision of the check
is taken as the systematic error. 
A 100\% systematic uncertainty is assigned to the background
from \mbox{$\eeeetautau$}. The resulting estimate of the background from 
two-photon processes for the 1994 data is \mbox{$(0.58\pm0.07)\%$}.

\bigskip
\noindent{\boldmath\bf
Other backgrounds:}
The background from four-fermion processes (see  
Appendix~\ref{sec-4f}) is estimated to be $(4\pm1)\times10^{-4}$ 
for the peak data. The cosmic ray background is estimated
from the data and found
to be small, $(2\pm2)\times10^{-4}$ for the 1994 data.

%
%
\subsection{Correlations among lepton species}
\label{sec-leptcor}
%
Although the cross-sections for each lepton species are measured 
independently, the requirement that there is no overlap in the selections
leads to anti-correlations between the systematic errors for 
each species. These anti-correlations reduce the precision with which  
the inter-species ratios can be measured. However, components of
the total systematic uncertainties which are anti-correlated
between lepton species do not contribute significantly to the uncertainty 
on the leptonic cross-section from all species taken together.

The separation of $\eeee$ and $\eetautau$ events 
is made primarily on the basis of the cut on the total deposited 
electromagnetic energy, $\Etotal$.
Almost all $\ee$ events failing this cut are accepted by the
$\tautau$ selection, while $\tautau$ events which exceed this cut
appear as background in the $\ee$ sample. Consequently,
the covariance between $\ee$ and $\tautau$ cross-section measurements 
is taken as the product of the systematic uncertainty in $\ee$ efficiency
due to the electromagnetic energy cut 
and the uncertainty of $\tautau$ background in the $\ee$ sample.
A second source of anti-correlation arises from $\eemumu$ events 
being explicitly excluded from the 
$\eetautau$ sample. The covariance is taken as the product of the 
uncertainty in the estimated $\mumu$ background in the $\tautau$ sample 
and the uncertainty in the loss of $\tautau$ events improperly 
identified as $\mumu$. 
The covariance between the $\mumu$ and $\tautau$ selections
and between the $\tautau$ and $\ee$ selections are
of similar size, with the exact values listed in {\TAB}~\ref{tab:ll_matrix}.
Due to their very different experimental signatures,
there is no cross-talk between the $\eeee$ and $\eemumu$ selections.

%
%
\section{Cross-section measurements}
\label{sec-meas}
Table~\ref{tab-meas-data} gives the number of selected events in each channel
used in the cross-section analysis of the entire OPAL {\LEPI} data sample.
For measuring the cross-section of a particular final state both
the detector sub-systems relevant for the selection and the 
luminosity detector are required to be fully operational.
Since the detector sub-systems used in the different event selections
are not identical, the luminosity and mean beam energy differ 
slightly for each final state. Within these `detector status' requirements, 
the number of selected final state events,
the number of luminosity events, the luminosity weighted beam energy
and the beam-energy spread are calculated for each data sample.
The cross-sections in the idealised phase space described in
{\SECT}~\ref{sec-sel} are then calculated by multiplying the number of
selected events by the correction factors given in
{\TABs}~\ref{tab-mhcor}, \ref{tab-eecor}, \ref{tab-mmcor} 
and \ref{tab:ttfactor}, and dividing by the integrated luminosity.
Details of the luminosity calculation entering these tables can be found
in~\cite{bib-lumi-siw}.

Tables~\ref{tab:hadron_xsec},~\ref{tab:ee_xsec},~\ref{tab:muon_xsec},
and~\ref{tab:tau_xsec} give the measured fermion-pair cross-sections
within the kinematic cuts described in {\SECT}~\ref{sec-sel},
for hadrons, electrons, muons, and taus respectively.
The  ``measured'' cross-sections have been corrected for all effects except
the spread of collision energies.
The ``corrected'' cross-sections have been corrected for this spread
according to the expected curvature of the cross-section as a function
of energy, and correspond to the true cross-section at the central
beam energy, as described in Appendix~\ref{sec-covmat}.
The cross-section measurements from 1990--1992 correspond to
the results published in~\cite{bib-opal-ls90,bib-opal-ls91,bib-opal-ls92},
except for two corrections.
The small effect of initial-final state
interference (see Appendix~\ref{sec-4f-ifi}) has been applied
retrospectively to the lepton cross-sections. In addition, 
the improved theoretical calculations (see {\SECT}~\ref{sec-lum})
of the small-angle Bhabha scattering 
cross-section causes a slight change of the FD luminosity 
used for the 1990--1992 cross-sections.

The SiW luminosity detector was fully operational for $94\%$ of the 1994
peak running and the energy scans in 1993 and 1995 
(but not during the 1993 and 1995 prescan periods).
During the periods when the SiW luminosity detector was not fully 
functional, the data are used to measure ratios of cross-sections.
For these measurements the integrated luminosity is
determined from the number of selected $\eeqq$ events using the
expected cross-section.
A common 10\,\% normalisation uncertainty  
is assigned to these ``pseudo-cross-sections''. This arbitrary
large uncertainty ensures that these pseudo-cross-section
measurements only contribute to the measurement of $\Rl$,
but do not affect the determination of $\MZ$, $\GZ$ or $\shadpol$ 
(see {\SECT}~\ref{sec-ewp}).
Table~\ref{tab:pseudo_xsec} gives the pseudo-cross-sections, where
the errors given are statistical only.
The systematic errors, and their correlations, are taken to be the same
as those for the peak cross-sections measured in the same year,
except for the correlated normalisation scale error.

For the 1990 data the systematic uncertainties on the absolute luminosity 
and centre-of-mass energy scales are significantly larger 
than in subsequent years
(see {\SECT}~\ref{sec-ene}). 
For this reason the 1990 cross-section measurements are treated as 
pseudo-cross-sections with a common 
luminosity as well as LEP energy scale error imposed 
across all decay channels and energy points.
Since the 1990 data includes off-peak as well as on-peak points, it
contributes to the measurement of $\Rl$ and $\GZ$,
but not to $\MZ$ or $\shadpol$.

The quoted errors on the measured cross-sections listed in 
{\TABs}~\ref{tab:hadron_xsec}--\ref{tab:tau_xsec} are the statistical
errors from the counting of signal and luminosity events.
Systematic errors, which arise from uncertainties
of the LEP {\cms} energy, the luminosity measurement and the 
event selections, have varying degrees of correlation
between the different final-state species and the various data-taking periods.
Tables~\ref{tab:ecm9092_matrix}--\ref{tab:esigrms_matrix} give the
covariance matrices for the LEP energy and its spread. 
The covariance matrix for the luminosity is specified in
{\TAB}~\ref{tab:lumi_matrix} and 
{\TABs}~\ref{tab:had_matrix}--\ref{tab:ll_matrix} show the
covariance matrices for the hadron and lepton event selections.
Appendix~\ref{sec-covmat} describes in detail how the
full covariance matrix for the cross-sections (and asymmetries)
is constructed.

Many checks were made to ensure the consistency of the event
sample over widely varying time scales. One example is shown
in {\FIG}~\ref{f-meas-tkmh}, which shows the distributions
of the number of luminosity events observed between consecutive
$\eeqq$ events and {\em vice versa}.
If OPAL had temporarily lost sensitivity to 
small angle Bhabha scattering events in the SiW detector 
or $\eeqq$ decays (but not both)
for a continuous period of a few minutes in the approximately six
months of OPAL live time spent running at the peak during
1993--1995, the failure would
be visible as a tail in one of these distributions.
Similar tests involving the lepton species also revealed no such problems.
Other checks probing the constancy of event type ratios at time scales
varying from hours to months revealed no statistically significant effect.
For the 1995 data, the cross-section for each final state was determined,
specific to each bunchlet in the bunch-train. No significant variation 
was observed.
%
%
%
\section{The asymmetry measurements}
\label{sec-afb}
The forward-backward asymmetry~$\Afbll$ is defined as 
\begin{equation}
\label{eq-afbdef}
\Afbll = \frac{\sigF-\sigB}{\sigF+\sigB},
\end{equation}
where~$\sigF$ and~$\sigB$ are the cross-sections integrated over the 
forward \mbox{($0<\costhelm<1$)} and backward \mbox{($-1<\costhelm<0$)}
hemispheres respectively, and $\thetaelm$ is the 
angle between the final state~$\ell^-$ and the~$\en$ beam direction.

In the {\SM} the $s$-channel contribution to the differential cross-section 
for the reaction \mbox{$\eell$} with unpolarised beams
is expected to have an approximately 
quadratic dependence on the cosine of 
the lepton scattering angle:
\begin{eqnarray}
  \frac{{\mathrm{d}}\sigma}{{\mathrm{d}}(\costhelm)}
  & \propto & 1 + \frac{8}{3}\Afbll\costhelm + \cos^2\!\thetaelm\ .
\end{eqnarray}
For $\eeee$ events, the angular dependence is more complex due 
to the large contributions from the $t$-channel and interference
between $s$- and $t$-channels, discussed in Appendix~\ref{sec-tchan}.
The magnitude and energy-dependence of~$\Afbll$ are sensitive to 
the vector and axial-vector couplings between leptons and the~$\Zzero$ 
(see {\SECT}~\ref{sec-ewp}). 
Within the {\SM}, 
the asymmetry is predicted to be about~0.01 at \mbox{$\roots=\MZ$} 
(after unfolding initial-state radiation)
and 
to increase with~$\roots$.

In principle the measurement of $\Afbll$ is simple and suffers
from few systematic uncertainties. The statistically most powerful method 
of extracting a measurement of the forward-backward asymmetry
is to perform an unbinned maximum likelihood fit of the differential 
cross-section 
with~$\Afbll$ as the only free parameter using 
the logarithm of the product of the event weights, $w_i$ : 
\begin{eqnarray}
     w_i & = & \frac{3}{8}(1+({\cos\theta_{\ell^-}^i})^2) 
          + \Afbll \cos\theta_{\ell^-}^i,
\end{eqnarray} 
where $\theta_{\ell^-}^i$ is the polar angle of the negatively charged
lepton in the $i$th event.  
This method, which is used for the $\eemumu$ and $\eetautau$ channels, 
has the advantage of being insensitive to any event 
selection inefficiencies, or any variation of the efficiency 
with~$\cos\theta_{\ell^-}$, provided there is no 
forward-backward charge asymmetry 
in the efficiency function itself. Technically, the inclusion of 
a selection efficiency term in the event weights results in
an additive constant when taking the logarithm of the likelihood 
function provided the efficiency can be expressed in terms of 
$|\cos\theta_{\ell^-}^i|$. This is in contrast to the 
event-counting method (see below) which must be 
corrected for selection inefficiencies. The fit yields a value 
for~$\Afbll$ over the full polar angle range. 
This can be corrected to a restricted range  
\mbox{$\abscosthe<k$} using a simple geometrical factor
\begin{eqnarray}
\label{eq:afbcos}
   \Afbll(\abscosthe<k) & = & \Afbll(\abscosthe<1)\frac{4k}{3+k^2}.
\end{eqnarray}
This region corresponds to the ideal acceptance definition 
of {\TAB} \ref{tab-sel-kine} and is chosen to  correspond closely to the
geometric acceptance of the event selection. In addition, small
corrections to the measured asymmetry are required to take account of
background and other biases.
 
The measurement of the forward-backward asymmetry in the $\eeee$
channel is more complicated. 
Due to the \mbox{$t$-channel} contributions, the angular distribution
in \mbox{$\eeee$} events can not be parametrised as a simple function 
of the forward-backward asymmetry, $\Afbee$. For this reason 
$\Afbee$ is measured from the observed numbers of events in the forward and 
backward hemispheres, $\NF$ and $\NB$:
\begin{eqnarray}
   \Afbee = \frac{\NF-\NB}{\NF+\NB}. 
\end{eqnarray}
This counting method is also used as a cross check of the unbinned maximum
likelihood method in the $\eemumu$ and $\eetautau$
channels and yields consistent results.

In each of the three leptonic channels the choice of which particle 
($\ell^+$~or~$\ell^-$) is used to define the polar angle of the event is 
varied at random from event to event. This reduces many
possible geometrical biases to the 
asymmetry measurement arising from an asymmetry in the detector. 
 In the cases where the~$\ell^+$ 
track is used, the angle~$\theta$ is measured relative to the $\ep$~beam 
direction. 
The measured asymmetries are corrected for the energy spread
of the colliding beams as described in Appendix~\ref{sec-covmat}. 
Corrections are required to account for
charge misassignment and for differences between the experimental
acceptance and the ideal acceptance of {\TAB} \ref{tab-sel-kine}.
Details of these corrections and the main systematic uncertainties are
given below.
Note that our measured asymmetries implicitly include the effects of
initial-final-state interference.
We make no correction to remove these effects from our measurements,
but calculate the predicted asymmetries accounting for this interference
using ZFITTER.
The asymmetry measurements are dominated 
by statistical rather than systematic uncertainties even when all data sets 
are combined.

\subsection[\mbox{$\eemumu$}forward-backward asymmetry]
{\boldmath\label{muon-afb}\mbox{$\eemumu$} forward-backward asymmetry}
\label{sec-mumuasy}
The forward-backward asymmetry of muon pairs, $\Afbmumu$, is a 
conceptually simple measurement. Muon pair events are selected with
high efficiency and the asymmetry measurement itself is robust
against many systematic effects. Compared to our previous
publication\cite{bib-opal-ls92} the analysis presented here represents a 
significant improvement in the understanding of the 
experimental biases.  To benefit from the resulting reduction in systematic
uncertainties, the asymmetry in the 1992 data sample has been re-evaluated.

In addition to the \mbox{$\eemumu$} selection cuts of {\SECT}~\ref{muon-sel},
the acollinearity angle between the directions of the two muon tracks 
is required to be less than~$15^{\circ}$. This removes 
about~1.3\% of the total sample by rejecting events with high-energy 
radiated photons. Such events distort the angular distribution 
and bias the asymmetry measurement due to the lower value of the effective 
centre-of-mass energy,~$\rootsprime$. Equivalent acceptance cuts are 
made inside the ZFITTER program~\cite{bib-fit-ZFITTER}
when evaluating the theoretical predictions used to fit the data, 
as discussed in {\SECT}~\ref{sec-ewp}.

About 1.2\% of the selected \mbox{$\eemumu$} events have the same sign
assigned to both candidate muon tracks. 
These events are concentrated in the region where the tracks are close to 
the jet chamber wire planes or are in the forward region, 
$|\cos\theta|>0.90$.  The probability that a track is assigned the incorrect 
charge is found to be dependent both on the polar angle of the track and on 
its direction of curvature. Both positive tracks with $\cos\theta<0$ and
negative tracks with $\cos\theta>0$ are more likely to be assigned
the wrong charge.  
Rejection of same-sign events would therefore lead 
to a selection inefficiency which is both asymmetric and charge-dependent, 
removing more events in the backward hemisphere than in the forward 
hemisphere and producing a positive bias in~$\Afbmumu$. 
The origin of this effect is not understood.
The size of this bias is 
found to be~0.0010. These ``like-charge'' events
are recovered by determining which track was the most likely to have been 
assigned the wrong charge, based on numbers of assigned track hits,
measured momentum and its uncertainty and the acoplanarity angle for
tracks close to the anode wire planes of the central detector. 
The efficiency of this procedure is determined using the
sample of events where the charge can be almost unambiguously determined
from the acoplanarity as measured in the muon chambers.   
The resulting charge measurement is estimated to be
correct for 93\% of like-charge events.
A small correction is made for the 7\% of like-charge events where
the assignment is incorrect. 
The fraction of events where both muons are assigned the wrong charge
is small, $2\times10^{-4}$, and has negligible impact on the measurement.

For Monte Carlo events the cut on $\Evis$ introduces a 
small additional bias, approximately $1\times10^{-4}$, 
in the measured asymmetry with respect to the
asymmetry within the ideal acceptance of {\TAB} \ref{tab-sel-kine}.  
However there is an additional bias for the data, related to the
detector response, from events
which fail the $\Evis$ cut due to being mismeasured. These events 
are concentrated in two regions either $|\cos\theta|>0.90$ or
within $0.5^\circ$ of an anode wire plane. Additional selection 
criteria were applied to recover these mismeasured events. The 
main background in the region $\Evis<0.6$ arises from $\eetautau$.
Cuts based on acoplanarity and the reconstructed momenta of the two muon
tracks reduced this background to a manageable level.
The inclusion of the recovered events in the event sample
changes the measured value of $\Afbmumu$ by $(6\pm2)\times10^{-4}$.

Figure~\ref{fig:mmdsdc} shows the observed angular 
distribution of \mbox{$\eemumu$} events from the \mbox{1993--1995} data 
samples, after applying small corrections for inefficiency and background.
The asymmetries are obtained from unbinned maximum 
likelihood fits to the observed distributions of
$\costhmuon$ and corrected to the restricted acceptance of $|\cos\theta|<0.95$
using Equation~\ref{eq:afbcos}.
In the event selection both reconstructed muons are required
to be within $|\cos\theta|<0.95$. In contrast, the acceptance for the
muon pair asymmetry is defined constraining only 
$\abscosthmum<0.95$, thus including a larger fraction of events with
significant initial state radiation.
The measured asymmetries are therefore corrected for the resulting bias
of about 0.0003, which was estimated using
KORALZ Monte Carlo events.
The forward-backward asymmetry measurements for the \mbox{$\eemumu$} 
channel, fully corrected to the acceptance definition of
{\TAB} \ref{tab-sel-kine}, are summarised in {\TAB}~\ref{tab:muon_afb}. 

For each of the data sets the systematic
errors are less than 10\% of the statistical uncertainties.
The total systematic uncertainty on the measured asymmetry for the highest 
statistics data sample is estimated to be 0.0004.
The largest uncertainty is related to charge
misassigned events, described above, varying between 
$\pm0.0002$ and $\pm0.0010$ depending on the data sample. 
The second largest systematic error,
$\pm0.0002$, comes from limits on the possible inadequacies of the
fitting function in the maximum likelihood fit. In addition, 
small systematic uncertainties,
each less than 0.0002, are assigned due to the definition of the angular 
acceptance, the residual tau pair background and biases from the event 
selection. The systematic errors for the different data samples and their
correlations are given in {\TAB} \ref{tab:asymm_matrix}.

\subsection[\mbox{$\eetautau$} forward-backward asymmetry]
{\boldmath \mbox{$\eetautau$} forward-backward asymmetry}

Measurement of {$\Afbtautau$} begins with the definition of a suitable
event sample.
In addition to the \mbox{$\eetautau$} selection cuts of {\SECT} 
\ref{sec-tau_selection}, events in which the sums of the charges in 
both $\tau$~cones have the same sign are rejected and 
the sum of the charges of the tracks in at least one of the
tau cones must be either~$\pm1$.  These additional requirements reject 
approximately~2\% of the selected \mbox{$\eetautau$} events. 
Unlike in the muon-pair asymmetry measurement, the softer 
momentum spectrum of charged particles from tau decays allows
like-charged events to be rejected without introducing
a significant bias.
The charge 
misassignment in $\eetautau$ events mainly arises from track reconstruction
ambiguities for tau decays producing three or more closely separated tracks.   
Another difference with respect to the muon-pair asymmetry measurement
is that the tau-pair asymmetry is subject to a potentially important
background from electron-pair events.
Due to the \mbox{$t$-channel} contribution, 
\mbox{$\eeee$} background events
have a large forward-backward asymmetry. To reduce the 
sensitivity of the $\Afbtautau$ measurement to uncertainties in the 
\mbox{$\eeee$} background, additional cuts are applied. 
These cuts, based on ${\mathrm d}E/{\mathrm d}x$, acoplanarity and 
$\Etotal$, 
reduce the \mbox{$\eeee$} background by over 90\% and remove 
only~1.5\% of \mbox{$\eetautau$} events. 
Figure~\ref{fig:tau_afb} shows the angular distributions of the 
events selected for the \mbox{$\eetautau$} asymmetry measurement.
Figure~\ref{fig:ttdsdc} shows the corresponding distributions
after corrections for background and selection efficiency are applied. 
Unbinned maximum likelihood fits to the uncorrected 
observed data distributions 
are used to measure $\Afbtautau$.  The measured 
asymmetries are then corrected for the presence of
background 
$$ \Delta\Afb = (\Afbtautau - \Afbbkg)\fbkg,$$
where~$\fbkg$ is the background fraction and $\Afbbkg$~is the asymmetry 
of the background. The resulting corrections are small, corresponding to 
less than~10\% of the statistical error. 

The measured values of $\Afbtautau$ correspond to the
asymmetry in the selected sample extrapolated to the full $\costhe$ 
acceptance. These are corrected to the acceptance 
definition {\mbox{$\costaum<0.9$}} and {\mbox{$\thacol<15^\circ$}
given in {\TAB} \ref{tab-sel-kine}. 
The corrections are obtained using
Monte Carlo $\eetautau$ events which
are treated in the same manner as data. The corrections also 
take into account other biases such as the effect
of the non-zero average polarisation of the $\tau$~lepton. 
The visible energy spectra from decays of positive and negative 
helicity $\tau$~leptons are different. The \mbox{$\eetautau$} event 
selection is approximately 1.5\% more efficient 
for the negative helicity final states than for positive helicity final 
states and a bias arises since the forward-backward asymmetries
for the two helicity states are different.
For each energy point, the measured asymmetry from
the Monte Carlo sample is compared with the true asymmetry 
within the acceptance \mbox{$\costaum<0.9$}} and {\mbox{$\thacol<15^\circ$} 
obtained from the Monte Carlo generator information.
The differences are used to obtain corrections to the 
measured asymmetries of \mbox{$-0.0027\pm0.0017$,} 
\mbox{$-0.0014\pm0.0011$} and 
\mbox{$-0.0013\pm0.0015$} for the {\pkm}, {\pk} and {\pkp} data samples, 
respectively, where the errors include statistical and systematic
components.  These corrections and uncertainties implicitly include 
effects due to possible inadequacies of the function used 
in the maximum likelihood fit.

The corrected forward-backward asymmetry measurements 
for the \mbox{$\eetautau$} 
channel are summarised in {\TAB}~\ref{tab:tau_afb}. 
The overall systematic uncertainties on~$\Afbtautau$ are~0.0018, 0.0012 
and~0.0016 for the {\pkm}, {\pk} and {\pkp} data points, respectively. 
The errors for the different data sets and their correlations are listed
in {\TAB} \ref{tab:asymm_matrix}. 
The systematic errors are dominated by the 
uncertainties on the Monte Carlo correction from the measured 
asymmetries to the asymmetry definition of {\TAB} \ref{tab-sel-kine}.
The systematic error associated with the measurement of
$\cos\theta$ is studied by using 12 different definitions of this angle,
{\em e.g.} using tracks and clusters, using tracks alone, taking the average
value from the two tau cones. 
The shifts in the measured asymmetries compared to the default result are 
reasonably well reproduced by the Monte Carlo and a systematic error
of $0.0005$ is assigned. The knowledge of the absolute angular
scale, or length-width ratio of the detector, has negligible impact on
these measurements. As a consequence of the additional 
\mbox{$\eeee$} rejection cuts, the systematic errors arising from the
uncertainties in the background fractions are small 
\mbox{(typically~0.0002).} 

\subsection[\mbox{$\eeee$} forward-backward asymmetry]
{\boldmath\label{elec-afb}\mbox{$\eeee$} forward-backward asymmetry}

$\Afbee$ is measured 
from the observed numbers of events in the forward and backward hemispheres:
\begin{eqnarray}
  \Afbee = \frac{\NF-\NB}{\NF+\NB}. 
\end{eqnarray}
Here $\NF$ and $\NB$ are the numbers of events within 
\mbox{$0.00<\cos\thelec<0.70$} 
and \mbox{$-0.70<\cos\thelec<0.00$,} respectively, and~$\thelec$ is the
$\en$ scattering angle.  In
1.5\% of the selected \mbox{$\eeee$} events the signs of reconstructed 
charges of the tracks are the same. These events are retained for the asymmetry
measurement by using the acoplanarity angle between the
electromagnetic calorimeter clusters associated with the two
electron candidates  to distinguish their charges.
This method is limited by the experimental resolution of the
acoplanarity measurement giving the correct assignment in 
about 92\% of events.
This performance is studied in both data and Monte Carlo by 
applying the cluster-based acoplanarity method to 
selected events in which the two tracks have been assigned opposite charges. 
We therefore correct the measured asymmetry for residual bias
due to the fact that
we make the wrong charge assignment for an estimated $1.2\times10^{-3}$
of all events when we recover the like-sign events using the acoplanarity
method.
The resulting corrections to $\Afbee$ are
\mbox{$0.0010\pm 0.0005$,} \mbox{$0.0004\pm 0.0003$} 
and \mbox{$0.0004\pm 0.0006$} for the
{\pkm}, {\pk} and {\pkp} energy points, respectively. 

In the process \mbox{$\eeee$}, there is a small difference in the
event selection efficiency for forward and backward events due mainly to
the cut on the electromagnetic energy. This is caused by the 
softer energy spectra of ISR photons from $s$-channel $\Zzero$ 
than from $t$-channel photon exchange processes. ISR photons tend 
to be produced along the beam direction and are therefore likely to remain
undetected. Consequently, events with harder ISR are more likely to be
rejected by the cut on the total visible electromagnetic energy 
in the event, $\Etotal>0.80\roots$.
The $t$-channel contribution is relatively more important in the forward 
direction while the contribution from the $\Zzero$ dominates in the backward
direction.
As a result a small forward-backward asymmetry in the 
\mbox{$\eeee$} event selection efficiency arises.
The resulting $\roots$-dependent effect on $\Afbee$ is evaluated using the
Monte Carlo simulation 
to be \mbox{$-0.0002\pm 0.0004$,} \mbox{$-0.0003\pm 0.0003$} 
and \mbox{$-0.0007\pm 0.0008$} for
the {\pkm}, {\pk} and {\pkp} energy points, respectively.

Unlike the unbinned maximum likelihood fit, the counting 
method used to obtain $\Afbee$ is sensitive to
variations in the selection efficiency as a function
of $|\cos\theta|$.
The efficiency variations for $\eeee$, however, are small enough to have
negligible impact on the asymmetry measurement.
The background in the \mbox{$\eeee$} sample is dominated 
by \mbox{$\eetautau$} events for which the 
expected asymmetry is different from that of \mbox{$\eeee$} events. 
Small corrections for background are obtained in the same manner 
as used for $\Afbtautau$. The corresponding systematic uncertainties
are negligible.

The offset of \mbox{$0.0004\pm 0.0005$} in the effective edge of the
acceptance in polar angle relative to
the nominal value (see {\SECT}~\ref{elec-emcut}) causes a small bias in
the measured asymmetry.
The size of this effect has been evaluated numerically, using the 
{\SM} 
prediction for the differential cross-section at the acceptance edges.
Corrections of \mbox{$0.0003\pm0.0009$,} \mbox{$0.0002\pm0.0004$} 
and \mbox{$0.0002\pm0.0005$} are obtained for the {\pkm}, {\pk} and {\pkp}
energy points, respectively. The systematic errors correspond to
the uncertainty on the edge of the acceptance. 
The accuracy of the definition of the boundary at
\mbox{$\cos\theta=0.0$} is also 
important for the asymmetry measurement, 
since this is used to separate forward 
events from backward events.
However, by randomly choosing either the~$\en$ or the~$\ep$ to classify 
each event, the effect of any possible small bias is largely cancelled 
and is reduced to a negligible level.
As a check of the asymmetry measurement, $\Afbee$ has also been 
evaluated using 
several different methods to classify events as forward or backward:
using only~$\en$ clusters and tracks;
using only~$\ep$ clusters and tracks;
using only $\theta$ measurements from tracks rather than from clusters
when a high quality track is found; and
using only events in which the two tracks have been assigned opposite charges.
There are no significant differences between these results, beyond what 
is expected from statistical fluctuations.

The forward-backward asymmetry measurements for the \mbox{$\eeee$} channel 
are summarised in {\TAB}~\ref{tab:ee_afb}.
The systematic errors for the different data samples and their
correlations are given in {\TAB} \ref{tab:asymm_matrix}.
Figure~\ref{fig:eedsdc} shows the electron differential cross-sections
at the three main energy points, fully corrected for the effects
discussed above.

%
\section{\boldmath Parametrisation of the $\Zzero$ resonance}
\label{sec-ewp-intro}
Before proceeding to interpret the measurements
we first discuss the basic formalism used to describe
cross-sections and asymmetries at the $\Zzero$ resonance.
Then we give a brief overview of radiative corrections and
the programs which we use for their calculation.
Finally, we describe the $t$-channel corrections which need to be accounted
for in $\ee$ final states.
\subsection{Lowest order formulae}
\label{sec-ewp-tree}
The process $\eetoff$ can be  mediated in the $s$-channel by two spin-1 bosons,
a massless photon and a massive $\Zzero$ boson.
In lowest order and neglecting fermion masses  the differential 
cross-section for this reaction can be written as: 
\begin{eqnarray}
\label{eq-diffxs}
\frac{2s}{\pi N_{\rm c}} \frac{{\rm d}\sff}{{\rm d}\cos\theta}
& = &  
\alfem^2 \, Q_{\rm f}^2 \,  (1+\cos^2\theta)  \\
& + &
8\,{\rm Re}\left\{ \alfem\, Q_{\rm f} \, \chi^{*}(s)
\ [ \ \Csgz(1+\cos^2\theta) + 2\Cagz\cos\theta \ ] \right\} \nonumber \\
& + & 
16\, |\chi(s)|^2
\ [ \ \Cszz(1+\cos^2\theta) + 8\Cazz\cos\theta \ ]  \nonumber
\end{eqnarray}
with
\begin{equation} 
\label{eq-z-propag}
     \chi(s) = \frac{\GF\MZbar^{2}}{8\pi\sqrt{2}}
     \frac{s}{s-\MZbar^{2} + i \, \MZbar \GZbar } 
\end{equation}
and
\begin{eqnarray}
\label{eq-cpar1}
\Csgz = & \gvhate\gvhatf ~~& ,~~  \Cagz  = \gahate\gahatf ~~,\\
\label{eq-cpar2}
\Cszz = & (\gvhate^2 + \gahate^2)(\gvhatf^2 + \gahatf^2) ~~& ,~~ 
 \Cazz =  \gvhate\gahate\gvhatf\gahatf ~~.
\end{eqnarray}
Here $\alfem$ is the electromagnetic coupling constant, $\GF$ is 
the Fermi constant
and $Q_{\mathrm{f}}$ is the electric charge of 
the final state fermion f.
The colour factor $N_{\rm c}$ is 1 for leptons and 3 for quarks,
$\MZbar$ is the mass of the $\Zzero$ boson and $\GZbar$ its total decay 
width.\footnote{The bars on $\MZbar$ and $\GZbar$ distinguish quantities
defined in terms of the 
Breit-Wigner parametrisation with an $s$--independent total width from
$\MZ$ and $\GZ$ which are defined in terms of an $s$--dependent total width,
as discussed in connection with {\EQ}~\ref{eq-z-propag2}.}
$\gahatf$ and $\gvhatf$ are the axial-vector and vector couplings 
between the participating fermions and the $\Zzero$ boson.

The first term in {\EQ}~\ref{eq-diffxs} accounts for pure photon exchange, 
the second term for $\gzif$ interference 
and the third for pure $\Zzero$ exchange.
The four coefficients $\Csgz$, $\Cagz$, $\Cszz$ and $\Cazz$ parametrise
the terms symmetric (`s') and antisymmetric (`a') in $\cos\theta$.
%
The relative size and the energy dependence of the five components of the
differential cross-section (three symmetric terms for $\gamma$, $\gzif$ 
and $Z$, and two anti-symmetric terms for $\gzif$ and $Z$)
as expected in the {\SM} are shown
in {\FIG}~\ref{fig:imb}. The symmetric terms are clearly 
dominated by pure $\Zzero$ exchange. In contrast, the antisymmetric
pure $\Zzero$ term is much smaller in magnitude 
than the corresponding contribution from 
$\gzif$ interference, except very close to the pole of the $\Zzero$ resonance, 
where the latter crosses zero.

Integrated over the full angular phase space,
the $\Zzero$ exchange term can be further expressed
as a Breit-Wigner resonance:
\begin{equation}
\sff^{\mathrm{Z}} (s) = \sfpol
            \frac{s\GZbar^2}{(s-\MZbar^2)^2 + \MZbar^2 \GZbar^2} \;\; ,
\end{equation}
where $\sigma^0_{\mathrm{f}}$ is the ``pole cross-section'', {\it i.e.}\ the total cross-section at 
$s=\MZbar^2$. It is given in 
terms of the partial widths or the $\Cszz$ parameter:
\begin{equation}
\label{eq-speak}
\sfpol =
 \frac{12\pi}{\MZbar^2}\frac{\Gee\Gff}{\GZbar^2} =
\frac{ \Cszz \,N_{\rm c} }{6 \pi} 
\left( \frac{\MZbar^2 \GF}{ \GZbar} \right)^2 \; \;  ,
\end{equation}
where $\Gee$ and $\Gff$ are the partial decay widths of the $\Zzero$ to
electron pairs and the fermion pair $\ff$, respectively. 
In terms of couplings the partial width is given by:
\begin{equation}
\label{eq-pwid1}
\Gff =  \frac{\GF N_{\rm c} \MZbar^3}{6 \pi \sqrt{2} }
        \left( {\gvhatf}^2 + {\gahatf}^2 \right ) \; \; .
\end{equation}
The anti-symmetric terms in {\EQ}~\ref{eq-diffxs} give rise to the
forward-backward asymmetry
$\Afb$ defined in {\EQ}~\ref{eq-afbdef}.
At centre-of-mass energies close to $\MZ$ it can be approximated by:
\begin{equation}
\label{eq-afb-roots}
\Afb(s) = - \frac{3 \pi \sqrt{2} \alfem  Q_{\rm f}}{\GF \MZbar^2}
       \frac{\Cagz}{\Cszz} \frac{s - \MZbar^2}{s}
       \, + \,        \frac{3 \Cazz}{\Cszz} \; .
\end{equation}
The first term arises from the $\gzif$ interference and the second
from pure $\Zzero$ exchange. The former  gives the dominant 
contribution to $\Afb$ except very close to the pole and causes a strong
$\roots$ dependence, as illustrated in {\FIG}~\ref{fig:imb} (b).
The latter is termed the
``pole forward-backward asymmetry'', $\Afbfpol$,  and
can be conveniently expressed in terms of 
the coupling parameters $\AAf$
\begin{equation}
\label{eq-afbpol}
\label{eq-cplpars}
\Afbfpol  = 
\frac{3}{4} \AAe \AAf ~~~ \mbox{with}~~~ 
\AAf = 2\,\frac{\gvhatf\gahatf}{\gvhatf^2+\gahatf^2} \; .
\end{equation}

One should note that the relations presented so far are generally
valid for the exchange of a massive spin-1 boson,
independent of the specific form or size of the couplings between the 
heavy boson and the fermions. Only the well-tested prediction of QED
for the strength of the vector-type coupling between photon and fermions
is assumed.
Within the {\SM}, however, $\gahatf$ and $\gvhatf$ are given
at tree level by the third component of the weak isospin $\Itf$, 
the electric charge
$\Qf$ and the universal electroweak mixing angle, $\swsq$:
\begin{equation}
\label{eq-couplings}
\gahatf  = \Itf ~~ , ~~ \gvhatf = \Itf - 2 \Qf \swsq \; .
\end{equation}

\subsection{Radiative corrections}
\label{sec-ewp-radcor}
Radiative corrections significantly modify the $\eetoff$ cross-sections and
forward-backward asymmetries with respect to the tree level
calculation.
One can distinguish four main categories:
\begin{itemize}
\item {\bf Photon vacuum polarisation:} Vacuum polarisation leads to
a scale dependence of the electromagnetic coupling constant $\alfem$:
\begin{equation}
\label{eq-alem-run}
 \alfem(0)  \longrightarrow
   \alfem(s) = \frac{\alfem(0)}{1 - \Delta \alfem(s)} \; .
\end{equation}
Due to uncertainties in the hadronic contribution to $\Delta \alfem(s)$
the value of $\alfem$ at $s=\MZ^2$ has a considerable uncertainty
($\alfem(\MZ^2)^{-1} = 128.886 \pm 0.090$~\cite{bib-alem-jeg}),
despite the high precision of $\alfem$ at $s = 0$ (0.04 ppm).
There is also a small imaginary component of $\Delta \alfem(s)$
which leads to an additional contribution to the $\gzif$ interference 
in conjunction with the imaginary part of the 
$\Zzero$ propagator in {\EQ}~\ref{eq-z-propag}.

\item {\bf Initial-state radiation:} Photonic radiation in the 
initial state has a profound effect on the
$\Zzero$ lineshape. It reduces the peak height by about $25\%$, 
shifts the peak upward by about 100~MeV and increases the apparent 
(FWHM) width by about 500~MeV, 
as illustrated in {\FIG}~\ref{fig:imb}~(e).
Initial-state radiative corrections can be implemented in terms of a
 radiator function, $H(s,\sprime)$, which relates the 
electroweak cross-section, $\sff$, to the observable cross-section,
$\sff^{\mathrm{obs}}$, in terms of an integral over $\sprime$, the squared
invariant mass available to the hard electroweak interaction
\begin{equation}
\sff^{\mathrm{obs}}(s) = \int^{s}_{s_{\mathrm{min}}} \sff(\sprime) H(s,\sprime) 
\,{\mathrm{d}}\sprime \; \; ,
\label{eq:radiator}
\end{equation}
where $s_{\rm min}$ is the minimum invariant mass squared of the system after
initial state radiation.

\item {\bf Final-state radiation:} The radiation of photons 
or gluons (only for $\qq$) in the final state increases
the partial widths in first order by factors
\begin{equation}
\label{eq-delqedqcd}
\delta_{\rm QED} = 1 + \frac{3}{4} \frac{Q_f^2\, \alfemmz}{\pi} \; \; , \; \;
\delta_{\rm QCD} \simeq 1 + \frac{\alfasmz}{\pi} \; ,
\end{equation}
where $\als$ is the strong coupling constant.
For hadronic final states $\delta_{\rm QCD}$ causes a sizeable correction which
allows a precise determination of $\alfasmz$ from the inclusive hadronic
width, $\Ghad$.
\item {\bf Electroweak corrections:} 
Quantum-loop effects in the $\Zzero$ 
propagator
and vertex corrections involving the Higgs and the top quark
give rise to 
radiative corrections,
which  in leading order depend quadratically on the mass of the top quark,
$\Mt$, and logarithmically on the Higgs boson mass, $\MH$.
These effects give sensitivity to physics at much larger scales than
$\MZ^2$.

\end{itemize}
When higher order corrections are considered mixed corrections arise; in particular
QED/QCD and QCD/electroweak corrections are significant.
In addition to $\gamma$ and
$\Zzero$ exchange, box diagrams also make small contributions
to the process $\eetoff$,
with a relative size $\le 10^{-4}$ close to  the $\Zzero$ resonance.

After unfolding the effects of initial and final-state radiation one can still
retain the form of the differential cross-section expressed in {\EQ}~\ref{eq-diffxs}
with a few modifications:
\begin{itemize}
\item $\alfem(0)$ must be replaced by the (complex) $\alfem(\MZ^2)$.
\item Electroweak corrections can be absorbed by replacing the
tree-level axial-vector and vector couplings,  $\gahatf$ and 
$\gvhatf$,  by the corresponding
effective couplings ${\cal G}_{A{\mathrm{f}}}$ and ${\cal G}_{V{\mathrm{f}}}$, 
which are complex numbers with small imaginary components.
In general, the effect of these imaginary components, termed ``remnants'',
on the different terms
of the differential cross-section is small. 
Most notable is their contribution to the symmetric part of the $\gzif$ 
interference, as shown in {\FIG}~\ref{fig:imb}~(c). 
\item Loop corrections to the $\Zzero$ propagator lead to an $s$--dependent
decay width~\cite{bib-s-dep-width}.
\label{pag-sdep-bw} 
This can be accounted for
by replacing
($\,\GZbar \rightarrow \GZ(s) \equiv s \GZ / \MZ^2 \,$)
in {\EQ}~\ref{eq-z-propag}, resulting in:
\begin{equation}
\label{eq-z-propag2}
     \chi(s) = \frac{\GF\MZ^{2}}{8\pi\sqrt{2}}
     \frac{s}{s-\MZ^{2} + i \, \frac{s}{\MZ} \GZ } ~~.
\end{equation}
One should note that this does not alter the
form of the resonance curve but corresponds to
a transformation which redefines both the $\Zzero$ mass,
\mbox{$ \MZ = \MZbar \, \sqrt{1 + \GZbar/\MZbar}\,$},
and width,\\ \mbox{$\GZ = \GZbar \, \sqrt{1 + \GZbar/\MZbar}$\,}.
Numerically, $\MZ$ is shifted by about $+34$ MeV and $\GZ$ by \mbox{$+1$ MeV.}
\end{itemize}
The $\Zzero$ resonance measurements are conveniently interpreted
in terms of ``model-independent'' parameters, such as mass, total width,
partial widths or pole cross-section, and pole asymmetry. 
In the presence of radiative corrections these parameters are 
no longer direct observables but depend on the specific choice of which 
radiative corrections to unfold or include and are therefore termed 
``pseudo-observables''.
In this paper we define the $\Zzero$ mass and width, $\MZ$ and $\GZ$,
in terms of the $s$--dependent Breit-Wigner of {\EQ}~\ref{eq-z-propag2}.
Following the standard convention adopted by the LEP experiments,
the partial decay 
widths absorb all final-state and electroweak corrections, such that their
sum equals the total decay width:
\begin{equation}
\label{eq-pwid-rc}
\Gff =  \frac{\GF N_{\rm c} \MZ^3}{6 \pi \sqrt{2}}
        \left( |{\cal G}_{\mathrm{V}}^{\mathrm{f}}|^2 
      R_{\mathrm{V}}^{\mathrm{f}} + |{\cal G}_{\mathrm{A}}^{\mathrm{f}}|^2 
      R_{\mathrm{A}}^{\mathrm{f}} \right )
        + \Delta_{\mathrm{ew/QCD}} \; .
\end{equation}
Here, $R_{\mathrm{V}}^{\mathrm{f}}$ and $R_{\mathrm{A}}^{\mathrm{f}}$ account for final-state corrections and fermion
masses and $\Delta_{\mathrm{ew/QCD}}$ accounts 
for non-factorisable mixed electroweak/QCD
contributions. 
After unfolding initial-state radiation the total cross-section
from $\Zzero$ exchange can then be written as:
\begin{equation}
\label{eq-breit-wig}
\sff^{\mathrm{Z}} (s) = \frac{\sfpol}{\delta_{\rm QED}}
            \frac{s\GZ^2}{(s-\MZ^2)^2 + \frac{s^2}{\MZ^2}\GZ^2} 
\mbox{~~~with~~} \sfpol = 
 \frac{12\pi}{\MZ^2}\frac{\Gee\Gff}{\GZ^2} \;\; .
\end{equation}
The factor $1/\delta_{\rm QED}$ ({\EQ}~\ref{eq-delqedqcd}) is needed to 
cancel the final-state
radiative corrections in $\Gee$ when used to describe
the couplings to the initial state.

In contrast, for the pole asymmetries  ({\EQ~\ref{eq-afbpol}),
the effects of final-state radiation and imaginary remnants
are excluded; they are interpreted in terms of the 
real parts of the effective couplings:
\begin{equation}
\label{eq-afbpol-rc}
\Afbfpol  = \frac{3}{4} \AAe \AAf ~~,~~
 \AAf = 2\,\frac{\gvf\gaf}{\gvf^2+\gaf^2}
\;\; \mbox{with} \;\; 
\gvf \equiv {\rm Re}({\cal G}_{\mathrm{V}}^{\mathrm{f}}) \, , \;
\gaf \equiv {\rm Re}({\cal G}_{\mathrm{A}}^{\mathrm{f}}) \; .
\end{equation}
The imaginary remnants 
(${\rm Im}({\cal G}_{\mathrm{A}}^{\mathrm{f}}),\,
{\rm Im}({\cal G}_{\mathrm{V}}^{\mathrm{f}})\,$)
 are evaluated in the {\SM} and the 
corresponding corrections are applied.

For the  calculation of radiative corrections we used 
the programs ZFITTER~\cite{bib-fit-ZFITTER} and TOPAZ0~\cite{bib-fit-TOPAZ0}.
Initial-state radiation is implemented  completely to  
${\cal{O}}(\alfem^2)$~\cite{bib-fit-alpha2}
with leading ${\cal{O}}(\alfem^3)$ corrections~\cite{bib-fit-alpha3}.
The radiation of fermion pairs in the initial state is also included 
to ${\cal{O}}(\alfem^3)$~\cite{bib-arbuzov}.
Photon radiation in the final state as well as
initial and final state interference is treated to ${\cal{O}}(\alfem)$.
The calculation of final-state radiation deserves a further comment.
For the asymmetries and $\eeee$ cross-section, which are measured within
tight kinematic cuts, we use $\alpha(0)$ as recommended in~\cite{bib-pcp}.
In contrast, for the other cross-sections, which are measured within
inclusive cuts, we use $\alpha(\MZ^2)$, which implicitly covers the
dominant part of small effects from final-state pair production.

In addition to the one-loop level electroweak corrections, TOPAZ0 and ZFITTER
include
the leading (${\cal{O}}(\Mt^4)$) and sub-leading (${\cal{O}}(\Mt^2 \MZ^2)$)
two-loop corrections~\cite{bib-degrassi}.
QCD corrections are calculated to ${\cal{O}}(\als^3)$~\cite{bib-alphas-2},
with mixed terms  included to ${\cal{O}}(\alfem \als)$ and the leading
${\cal{O}}(\Mt^2 \als)$ terms~\cite{bib-kniehl,bib-czak}.

\subsection[$t$-channel contributions to $\eeee$]
{\boldmath $t$-channel contributions to $\eeee$}
For $\eeee$ events not only $s$-channel annihilation contributes but also
the exchange of $\gamma$ and $\Zzero$ in the $t$-channel. Since $t$-channel
exchange is not  included in our main fitting program, ZFITTER, 
an external treatment is needed to account for it.
We use the program ALIBABA~\cite{bib-fit-ALIBABA}
to calculate the expected contribution from $t$-channel and $s$-$t$ 
interference and add them to the pure $s$ channel contributions
calculated with ZFITTER.
The details of how we treat these contributions in the
fit and the associated uncertainties are described in 
Appendix~\ref{sec-tchan}. 
The uncertainties on the fitted $\ee$ partial 
width, $\Gee$, and pole asymmetry, $\Afbpolee$, due to the 
$t$-channel are 0.11\,MeV and
0.0015, respectively, with a correlation coefficient of $0.85$.

\section{Determination of electroweak and Standard Model parameters}
\label{sec-ewp}
The 211 measurements of
cross-sections and asymmetries listed in
{\TABs}~\ref{tab:hadron_xsec} --~\ref{tab:pseudo_xsec} and
{\TABs}~\ref{tab:muon_afb} --~\ref{tab:ee_afb}
form the basis of our tests of {\SM} 
expectations, and our determination of 
electroweak parameters.
The cross-section measurements at the three energy
points of the precision scan 
determine the basic parameters of the $\Zzero$ resonance, its mass,
$\MZ$, width, $\GZ$, and its pole production cross-sections, 
$\sigma^0_{\rm f}$, for each of the four final states.
By combining the measured pole cross-sections one can 
determine directly the absolute
branching ratios ($\Gff/\GZ$) to hadrons and leptons, as can be seen
from {\EQ}~\ref{eq-breit-wig}.
The branching ratio to invisible states,
such as neutrinos, can be inferred from the difference between unity
and the sum of the visible branching ratios.
Combining the branching ratios with the total width yields the decay widths
to each species.
For leptons these widths provide a precise check of {\SM} 
predictions for the leptonic couplings.
The hadronic width, through QCD effects, provides a measurement of
$\alfas$. In contrast to almost all other measurements of
this quantity, no hadronisation corrections are needed to compare with QCD calculations,
which are available in ${\cal{O}}(\alfas^3)$. Therefore QCD uncertainties are small.
The charge asymmetries of the leptons produced at the peak allow the ratios 
of the leptonic vector and axial-vector couplings to be determined.
Finally, through radiative effects on the effective couplings, limits
can be placed on the mass of the Higgs boson.

The interpretation of our measured cross-sections and asymmetries
in terms of electroweak parameters proceeds in three stages.
First, we use a parametrisation  based on {\EQ}~\ref{eq-diffxs}, where 
the $C$ coefficients are treated as independent parameters, not imposing the 
constraints of {\EQs}~\ref{eq-cpar1} and \ref{eq-cpar2}.
Secondly, we perform the analysis in terms of 
the ``{\SLP}'', which are based on mass, 
total width, pole cross-sections and pole asymmetries of the $\Zzero$. 
The main difference with respect to the first approach is that
in deriving the  {\SLP} the 
$\gzif$ interference is constrained to the theoretical prediction.
Finally, we compare our measured 
cross-sections and asymmetries directly with calculations made in the
context of the {\SM}. 
This allows us to test the consistency of our  measurements with 
the theoretical prediction and also to determine the {\SM} parameters, 
$\MZ$, $\Mt$, $\MH$ and $\als$.

For all our fits we use ZFITTER, version 6.21, with the 
default settings\footnote{Except for $\dalh$, which we
specify directly (flag {\sf ALEM = 2}), and the correction of
\cite{bib-czak}, which is used only for the {\SM} fits in 
{\SECT}~\ref{sec-fit-sm} (flag  {\sf CZAK}) as recommended by the
authors of ZFITTER.}
to calculate the
cross-sections and asymmetries within our ideal kinematic acceptance
as functions of the fit parameters.
The fit parameters are obtained from a  $\chisq$ minimisation
based on MINUIT~\cite{bib-MINUIT}, which takes into account the full covariance
matrix of our data. In Appendix~\ref{sec-covmat} we describe
how the various sources of uncertainty are combined to construct the 
covariance matrix.

The {\SM} calculations require the full specification of the
fundamental {\SM} parameters. The main parameters are
the masses of the $\Zzero$ boson ($\MZ$), the top quark ($\Mt$) and the
Higgs boson ($\MH$), and the strong and electromagnetic coupling constants,
$\als$ and $\alpha$.
Unless explicitly specified otherwise, we use
for the calculation of imaginary remnants 
(${\rm Im}({\cal G}_{\mathrm{A}}^{\mathrm{f}}),\,
{\rm Im}({\cal G}_{\mathrm{V}}^{\mathrm{f}})\,$),
non-factorisable corrections and for comparison with the 
{\SM} predictions the following values and ranges:
\begin{eqnarray}
\label{eq-sm-pars}
\MZ \, = & 91.1856 \pm 0.0030 \; \mbox{GeV} \; & ,  \; \; 
\Mt = 175 \pm 5 \; \mbox{GeV}\; , \\ \nonumber
\MH \, = & 150^{+850}_{-60} \; \mbox{GeV} \; & ,  \; \; 
\als = 0.119 \pm 0.002 \mbox{\cite{bib-pdg98}} \; , \\ \nonumber
\dalh \, = & 0.02804 \pm 0.00065 & \; .
\end{eqnarray}
The values and ranges of $\Mt$ and $\als$ differ slightly
from the most recent evaluations~\cite{bib-top-mass,bib-alphas-siggi}.
They were chosen for consistency with the corresponding publications
of the other LEP experiments; the small differences are completely
negligible for the results presented here.
The range of $\MH$ corresponds approximately to the lower limit
from  direct searches~\cite{bib-MH-lower} and the upper limit from
theoretical considerations~\cite{bib-MH-upper}. 
The electromagnetic coupling $\alfem$ is expressed here in terms of
$\dalh$, the contribution of the five light quark flavours
to the running of $\alfem$ at $s = \MZ^2$ as described in
{\EQ}~\ref{eq-alem-run}. 
The quoted value for $\dalh$ corresponds to 
$\alfem(\MZ^2)^{-1} = 128.886 \pm 0.090$~\cite{bib-alem-jeg} 
when the contributions from the top quark and the leptons are also
included.
A further crucial parameter is the Fermi constant $\GF$, which can be 
determined precisely from the muon lifetime. 
We use the most recent value,
\mbox{$ \GF = (1.16637\pm 0.00001) \times 10^{-5} \, \mbox{GeV}^{-2}$}
\cite{bib-stuartvanritbergen}, which includes two-loop QED corrections.
Other input parameters of the {\SM}, such as 
the masses of the 
five light quarks\footnote{See~\cite{bib-fit-ZFITTER} for details on the
treatment of quark masses.}
 and the leptons, are fixed to their present values~\cite{bib-pdg98}.

\subsection[$C$-Parameter fits]{\boldmath $C$-Parameter fits}
\label{sec-ewp-cpar}
Our first analysis is based 
on a parametrisation  according to {\EQ}~\ref{eq-diffxs}, treating the
$C$-coefficients as independent fit parameters. In this ansatz, which we first introduced
in the analyses of our 1990/1991 data samples~\cite{bib-opal-ls90,bib-opal-ls91}, the
differential cross-section is described in terms of  four independent 
parameters, $\Csgz$, $\Cagz$, $\Cszz$ and $\Cazz$, which express the symmetric and 
anti-symmetric contributions from $\gzif$ interference and pure $\Zzero$
exchange, respectively.
Specifically we do not impose the constraints expressed in {\EQs}~\ref{eq-cpar1} and
{\EQs}~\ref{eq-cpar2}, which are in principle generally valid
for the exchange of a massive vector boson with arbitrary axial-vector 
and vector couplings interfering with the photon.
Contributions from new physics, however, such as an additional heavy
vector boson, would lead to additional terms in {\EQ}~\ref{eq-diffxs}.
In particular one would expect contributions from the interference between the
$\Zzero$ and the extra boson, which would have the same form
as the $\gzif$ interference. 
In the case of leptonic final states, the measurements of cross-sections
and asymmetries are sensitive to all four parameters independently.
Fitting the data with 
independent interference terms retains the sensitivity to such new physics effects.
In addition, from the independent determination of $\Csgz$ and $\Cagz$
one can distinguish $\gvf$ and $\gaf$ and determine the relative signs
of the couplings, as discussed below in {\SECT}~\ref{sec-ewp-couplings}.
This is not possible from the pure $\Zzero$ terms $\Cszz$ and $\Cazz$ alone,
which are symmetric in $\gvf$ and $\gaf$.

In this analysis we do not include any hadronic forward-backward 
asymmetry data and hence only 
the terms symmetric in $\cos \theta$, $\Csgz$ and $\Cszz$, are
accessible.
However, the sensitivity to the $\gzif$ interference ($\Csgz$) is small
at centre-of-mass energies close to $\MZ$, since a shift of the
interference term is essentially equivalent to a shift in $\MZ$. The fact that
high statistics measurements of the cross-section are available only for 
three different centre-of-mass energies further aggravates the lack of
discrimination between shifts in
$\MZ$ and the interference term. In the following analysis we therefore
use the $C$-parametrisation only for leptons.
The hadrons are parametrised in terms of an $s$--dependent Breit-Wigner 
resonance ({\EQ}~\ref{eq-breit-wig}) with $\MZ$, $\GZ$ and $\shadpol$ 
as fit parameters. 
The hadronic $\gzif$ interference term is fixed 
to the {\SM} prediction. This constraint
allows a precise determination of $\MZ$.
With the leptonic $\gzif$ interference terms left free in the fit,
$\MZ$ is determined from the hadronic data and the $\Csgz$
parameters for leptons therefore depend on the assumed interference term
for hadrons.

For the precise definition of the $C$-parameters several choices could
be made. We employ a definition such that the parameters can be interpreted
directly in terms of the real parts of the effective 
couplings, {\it i.e.}~the meaning of the 
$C$-parameters in the {\SM} corresponds to 
{\EQs}~\ref{eq-cpar1}~and~\ref{eq-cpar2}
with $\gahatf$ and $\gvhatf$ replaced by $\gaf$ and 
$\gvf$ as in {\EQ}~\ref{eq-afbpol-rc}.
We use ZFITTER to correct for initial and final-state radiation
and the running of $\alfem$ and to calculate the imaginary remnants
of the couplings. Since these remnants
are small and essentially independent of the
centre-of-mass energy in the {\LEPI}
region, this {\SM} correction does not compromise the independence
of the parametrisation from theoretical assumptions, but it preserves
a transparent interpretation of the $C$-parameters 
in terms of {\SM} couplings.

If lepton universality is not assumed, there are a total of 15 parameters:
$\MZ, \GZ, \shadpol$
and four $C$-parameters -- 
$\Csgz,\, \Cagz, \,\Cszz,\, \Cazz$ -- for each lepton species.
These fitted parameters
are shown in column two of {\TAB}~\ref{tab-ewp-cpar}.
The values obtained from the different lepton species for the $C$-parameters
are consistent with one another, compatible with lepton universality.
Column three of {\TAB}~\ref{tab-ewp-cpar} gives the results for a 
7 parameter fit when lepton universality is imposed by requiring
each of the four $C$-parameters to be equal across the three lepton species.
The  {\SM} predictions are shown in the last column.
The error correlation matrices are given in
{\TAB}s~\ref{tab-ewp-cparcor}~and~\ref{tab-ewp-cparcorlu}.

Figure~\ref{f-cparvssm} compares the parameters measured
assuming lepton universality with the
{\SM} predictions as a function of $\MH$.
We see agreement for all parameters.
The largest discrepancy, of about two standard
deviations, occurs in the parameter $\Cagz$.
The same tendency was observed in our previous analysis using only the data
collected from 1990 to 1992~\cite{bib-opal-ls92}.
This parameter corresponds to the energy dependence of the leptonic
forward-backward asymmetry.

As a check we also performed a fit releasing the {\SM} constraint
for the hadronic $\gzif$ interference by introducing a scale factor
for the $\gzif$ interference contribution ($\fgz = 1$ in the {\SM})
which was treated as an
additional free parameter. The fit resulted in
$\MZ = 91.190 \pm 0.011$ GeV and $\fgz = 0.0 \pm 3.0$,
with a correlation coefficient of $-0.96$.
This is in good agreement with the $\MZ$ result in {\TAB}~\ref{tab-ewp-cpar}
and with $\fgz = 1$.
The large increase of the $\MZ$ 
uncertainty and the high anti-correlation with $\fgz$ illustrates the fact
that {\LEPI} measurements alone give marginal
discrimination between hadronic $\gzif$ interference effects and $\MZ$ shifts.
In combination with measurements at energies further away from
the $\Zzero$ resonance, {\it e.g.},~the fermion-pair cross-sections at {\LEPII},
this problem can be resolved~\cite{bib-LEPEWWG}. 

An alternative model-independent parametrisation
of the $\Zzero$ resonance is the S-Matrix formalism~\cite{bib-smatrix}.
In practice this approach is equivalent to the $C$-parameters.
Since the S-Matrix parametrisation is the standard
framework for combined {\LEPI} and {\LEPII} cross-section and asymmetry
measurements we also give the
S-Matrix results in Appendix~\ref{sec-smat}.

\subsection{Results of the model-independent Z parameter fits}
\label{sec-ewp-leppars}
Our second  fit is based on
the {\SLP},
which consist of
 $\MZ$, $\GZ$, $\shadpol$, and $\Afblpol$ as given in 
{\EQs}~\ref{eq-breit-wig}~and~\ref{eq-afbpol-rc}, 
as well as the ratios of hadronic to leptonic widths:
\begin{equation}
\Rl  \equiv \frac{\Ghad}{\Gll} ~~~~(\ell = \mathrm{e}, \mu, \tau)~~.
\end{equation}
The partial widths include final-state and electroweak 
corrections as defined in {\EQ}~\ref{eq-pwid-rc}.
This set of pseudo-observables is closely related to the experimental 
measurements and sufficient to parametrise the $\Zzero$ properties;
correlations between the parameters are small.
For this reason it has been adopted by the four LEP collaborations
to facilitate the comparisons and averaging of the $\Zzero$ resonance
measurements.

The essential difference between this fit and the $C$-parameter fit
above is that now, in addition to the hadronic interference term,
the leptonic $\gzif$
interference terms ($\Csgz$ and $\Cagz$) are also constrained to the {\SM} 
prediction.
The expression of resonant lepton production in terms of $\Rl$, $\shadpol$
and $\Afblpol$ is  equivalent to the $\Cszz$ and $\Cazz$ formulation.

The results of the 9 parameter fit (without lepton universality) 
and the 5 parameter fit (assuming lepton universality) 
are shown in {\TAB}~\ref{tab-ewp-leppar}, together with the {\SM}
predictions in the last column.
The error correlations are given in {\TABs}~\ref{tab-ewp-lepparc5}
and~\ref{tab-ewp-lepparc9}.
For the fits where lepton universality is not imposed
the large mass of the $\tau$ is expected to reduce $\Gtautau$ by about
$0.2\%$.
Lepton universality would therefore
be reflected in $\Rtau$ being 0.047 larger than $\Ree$ and $\Rmu$.
In the 5 parameter fit, where lepton universality {\em is} imposed,
the $\tau$ mass effect is corrected;  $\Gll$ refers to the
partial decay width of the $\Zzero$ into massless charged leptons.
Figure~\ref{f-zparvssm} compares the parameters fit assuming lepton
universality with the {\SM} prediction as a function of the Higgs mass.
The agreement for all parameters is good.

In the context of this parameter set, we also make a test of the
consistency of the LEP energy calibration.
To make this test, we replace the single parameter $\MZ$
with three independent parameters for the three
periods which distinguish the stages in the
evolution of the LEP calibration; these are 1990--92, 1993--94 and 1995.
The results are given in {\TAB}~\ref{tab-ewp-mztest}.
The errors of the three $\MZ$ values are largely uncorrelated,
and the
stability of the LEP energy calibration is 
verified within the precision of our measurements.

The value of $\MZ$ we obtain in the fit to the 
{\SLP} is about 1~MeV smaller than the value we obtain in
the fit to the $C$-parameters.
This difference is due to the fact that in the 
$C$-parameter fit the leptons contribute much less to
the $\MZ$ measurement, since the leptonic $\gzif$ interference terms are 
allowed to vary freely.
One should also note that $\MZ$ shifts by $-0.5$~MeV when lepton universality
is imposed, in both the fit with $C$-parameters and the fit with
{\SLP}.
This is due to a subtle effect in the $\eeee$ $t$-channel correction, which 
gives an additional weak constraint on $\MZ$ (see Appendix~\ref{sec-tchan}).

Figures~\ref{f-fit-xsec} and~\ref{f-fit-afb} compare the cross-section and asymmetry
measurements with the results of the {\SLPnine} fit.
Figure~\ref{f-afb0vsrl} shows the contours in the $\Afblpol$\,--\, 
$\Rl$ plane, separately for the three lepton species as well as for
a universal charged lepton.
There is a large anti-correlation between $\Afbpolee$ and $\Ree$,
due to the appreciable non-$s$-channel contributions,
as discussed in Appendix~\ref{sec-tchan}.

\subsubsection[Error composition and $\chisq$]
{\boldmath Error composition and $\chisq$}
\label{sec-parerr}
Table~\ref{tab-ewp-err} shows the approximate error composition
for the {\SLP} and 
for the derived parameters which are discussed below.
Only in the case of $\shadpol$ and $\Rtau$ do systematic uncertainties exceed
the statistical errors. For $\GZ$ and the asymmetries the statistical
errors are still much larger than the systematics.
This parameter set also benefits from a division
of the major systematic uncertainties.
The LEP energy mainly affects $\MZ$ and $\GZ$,
the luminosity uncertainty enters nearly exclusively in $\shadpol$, and 
$t$-channel uncertainties are isolated in $\Ree$ and $\Afbpolee$.
Such a separation of effects greatly facilitates the combination of
the results with the other LEP experiments, since these are the
uncertainties which are common to all experiments.
For other possible parametrisations, {\it e.g.}, in terms of partial widths, 
$C$-parameters or the S-matrix, the systematic effects on the parameters are 
much more poorly separated.

The $\chisq$ values of our fits are somewhat lower than expected, 
for example in the {\SLPnine} fit 
$\chisq = 155.6$ with 194 degrees of freedom.\footnote{For
determining the number of degrees of freedom one needs to account
for the six sets of ``pseudo-cross-sections'' ({\SECT}~\ref{sec-meas}) 
and for the fact that we allow
the absolute normalisation of the cross-sections and the
energy scale to float in the 1990 data. Therefore the number of effective
measurements is reduced from 211 to 203.}
The probability to obtain   $\chisq \le 155.6$ is 2\%.
The low value of $\chisq$ can be attributed predominantly to 
anomalously low statistical fluctuations in
the data with lower
 precision taken in the early phase of LEP running (1990--1992).
It was already present in our earlier 
publications~\cite{bib-opal-ls90,bib-opal-ls91,bib-opal-ls92}. 
A fit to the 1993--1995 data alone results in $\chidof = 34.2/40$,
where the probability to obtain a lower $\chisq$ is 27\%.

\subsubsection{Theoretical uncertainties}
\label{sec-thyerr}
The relatively large effects from theoretical  uncertainties in the
luminosity determination and the $\eeee$ $t$-channel correction are discussed
elsewhere in this paper; these are already treated in the fit procedure and 
included in the quoted parameter errors.
Additional uncertainties
arise from the deconvolution
of initial-state radiation (ISR), 
possible ambiguities in the precise definition of the pseudo-observables
and residual dependence on the values of the {\SM} parameters which we
have assumed.

ISR corrections substantially affect the $\Zzero$ lineshape. 
In the fitting programs ZFITTER and TOPAZ0,
their implementation is  complete to ${\cal{O}}(\alfem^2)$ 
and includes the leading ${\cal{O}}(\alfem^3)$ terms.
In a recent 
evaluation~\cite{bib-isr-study} different schemes have been compared and
the effect of missing higher order terms estimated;
the residual uncertainties are limited to 0.004 nb for $\shadpol$,
0.1 MeV for $\MZ$ and $\GZ$, and are negligible for other parameters.
A related effect is the correction for fermion-pair radiation in the
initial state. Although this correction is about two orders
of magnitude smaller than the ISR correction it gives rise to somewhat
larger uncertainties. Comparing different schemes and 
implementations\footnote{We repeated the
fits varying the ZFITTER flags ISPP=2,3,4 which select the
parametrisations of~\cite{bib-arbuzov} and two variants 
of~\cite{bib-fit-yfspairs}.
The maximal difference is taken as the error.}
leads
to uncertainties of 0.3 MeV for $\MZ$, 0.2 MeV for $\GZ$ and
0.006 nb for $\shadpol$.

We further investigated possible differences in the definition
of pseudo-observables or their implementation by comparing the
programs ZFITTER and TOPAZ0. In a very detailed comparison~\cite{bib-pcp} 
the authors of the two packages demonstrated the good overall
consistency in their implementations of radiative corrections and
{\SM} calculations. 
In addition, we evaluated 
differences between the two programs by using one to calculate
a set of cross-sections and asymmetries and using the other to re-fit
this set.
The effective differences in terms of pseudo-observables are
0.2 MeV for $\MZ$, 0.1 MeV for $\GZ$, 0.003 nb for $\shadpol$,
0.004 for $\Rl$, and 0.0001 for $\Afbpolll$. 
The difference in $\Rl$ is the only notable effect; it corresponds
to 10\% of the experimental error.
Further effects from missing higher order electroweak corrections 
have also been studied. Such corrections are
small  and affect the pseudo-observables only through the tiny
{\SM} remnants. Numerically, the changes are found to be negligible. 

Finally, we evaluated the dependence of the fitted pseudo-observables
on the {\SM} parameters $\alfemmz$, $\als(\MZ^2)$, $\Mt$ and $\MH$.
The values assumed for these parameters affect the fitted 
pseudo-observables mainly through the $\gzif$ interference which is 
taken from the {\SM}.
The effects, however, are small; the only notable
impact is on $\MZ$, which changes by 0.2 MeV when varying $\MH$ from
90 to 1000 GeV.

The overall theoretical uncertainties for the $\Zzero$ resonance parameters 
from all these sources are given in the last column of {\TAB}~\ref{tab-ewp-err}.
These have a negligible effect on our results, since
in all cases the theoretical error is much less than the total uncertainty;
the largest
effects are for $\shadpol$ with 15\% and $\MZ$ and $\Rl$ with 10\% of the
total error.

\subsection{Interpretation}
The {\SLP} were
chosen to be closely related to the measured quantities
and have minimal correlations with each other.
These parameters can easily be transformed to other
equivalent sets of physical quantities for the interpretation
of our results.

\subsubsection[$\Zzero$ decay widths]
{\boldmath $\Zzero$ decay widths}
From the {\SLP} $\GZ$, $\shadpol$, $\Ree$, $\Rmu$,
and $\Rtau$ the 
partial $\Zzero$ decay widths 
$\Ginv$, $\Gee$, $\Gmumu$, $\Gtautau$, and $\Ghad$
can be derived, and are shown in {\TAB}~\ref{tab-ewp-pwidth}.
Here $\Ginv$ is the width of the $\Zzero$ to final states not
accounted for in the analysis of
$\eeee$, $\mumu$, $\tautau$ and $\qq$ processes discussed
in the previous sections
\begin{equation}
\Ginv \equiv \GZ - \Gee - \Gmumu - \Gtautau - \Ghad \; .
\end{equation}
The correlations between the partial
widths are large and
are given in {\TABs}~\ref{tab-ewp-pwidth5} and \ref{tab-ewp-pwidth3}.
In {\TAB}~\ref{tab-ewp-pwidth} good agreement is seen among the measured
lepton partial widths and with the {\SM} expectations.  
By assuming lepton universality, 
our measurement of $\Ginv$ becomes more precise.

A quantity  sensitive to a possible deviation of the
data from the {\SM} prediction for the invisible width
is the ratio
\begin{equation}
\Rinv \equiv \frac{\Ginv}{\Gll} \; .
\end{equation}
In this ratio experimental errors partially cancel and 
theoretical uncertainties
due to the assumed values of $\Mt$ and $\MH$ are strongly reduced.
In the  {\SM} only neutrinos contribute to
$\Ginv$. 
The first $\Zzero$ resonance measurements at SLC and LEP in
1989~\cite{bib-all-ls89}
demonstrated the existence of three generations of light neutrinos.
For three generations one expects
\[
\Rinvsm = 3 \frac{ \Gnu}{\Gll} =  5.974 \pm 0.004 \; ,
\]
where the uncertainty corresponds to variations of $\Mt = 175 \pm 5$ GeV and
$90 < \MH < 1000$ GeV.
Taking into account the correlation between $\Ginv$ and 
$\Gll$ our measured value  is 
\[
\Rinv = 5.942 \pm   0.027 \; .
\]
Dividing $\Rinv$ by the {\SM} expectation for a single generation,
${\Gnu}/{\Gll}$, gives 
\[
N_{\nu} =  2.984 \pm  0.013  \; .
\]

The measurements of the total and partial $\Zzero$ decay widths give 
important constraints for extensions of the {\SM} which predict
additional decay modes of the $\Zzero$ into new particles.
From the difference between measured widths and
the {\SM} predictions (assuming three neutrino species) 
one can derive upper limits for such contributions
from new physics.
New particles
could contribute either to one of 
the visible decay channels in the partial widths 
or to the invisible width, 
depending on their specific properties.
In order to calculate the upper limits for such extra contributions
to the widths  we added the experimental and 
theoretical errors in quadrature. For the latter we evaluated the change 
of the predicted widths when varying the {\SM} input parameters, 
$\MZ$, $\Mt$, $\als(\MZ^2)$ and $\alfem(\MZ^2)$,
within their experimental precision ({\EQ}~\ref{eq-sm-pars}).
For the mass of the Higgs boson we used 1000~GeV. This results in 
the smallest theoretical predictions for the
widths and therefore in the most conservative limits.
We obtain 
\begin{equation}
\GZ^{\rm new}  < 14.8 \, {\rm MeV} \mbox{~~and~~}
\Ginv^{\rm new}  <  3.7 \, {\rm MeV} 
\end{equation}
as one-sided upper limits at 95\% confidence level
for additional contributions to the total ($\GZ^{\rm new}$) 
and the invisible width ($\Ginv^{\rm new}$).
The limits for $\MH = 150$ GeV and for the 
visible partial widths are given in {\TAB}~\ref{tab-ewp-pwlim}. 
All limits are Bayesian with a prior probability which is uniform for
positive  $\Gamma^{\rm new}_{\rm x}$.

\subsubsection{Coupling parameters and the effective mixing angle}
From the measured pole asymmetries $\Afbpolll$ it is straightforward
({\EQ~\ref{eq-afbpol-rc}) to
determine the coupling parameters $\AAl$, which quantify 
the asymmetry of the $\Zzero$--lepton coupling for each lepton species.
The results for the three lepton species are consistent with each other and agree well
with the {\SM} predictions, as shown in {\TAB}~\ref{tab-ewp-al}.
However, since the measured $\Afbpolmumu$ and $\Afbpoltautau$
are products of  $\AAe$ with either $\AAm$ or $\AAtau$,
large anti-correlations ({\TAB}~\ref{tab-ewp-alcorr}) 
arise between $\AAe$ and $\AAm ,\; \AAtau$.

An equivalent formulation of the coupling asymmetries can be made in terms of
the effective weak mixing angle $\swsqeffl$.
Assuming lepton universality we obtain
\begin{equation}
\swsqeffl \equiv \frac{1}{4}\,\left( 1-\frac{\gvl}{\gal} \right)
 =  0.2325  \pm     0.0010  \;,
\end{equation} 
which is in good agreement with the world average
of $0.23151 \pm 0.00017$~\cite{bib-LEPEWWG}.

\subsubsection{Vector and axial-vector couplings}
\label{sec-ewp-couplings}
Using the set of the {\SLP} one can determine
the leptonic  vector and axial-vector couplings $\gvl$ and $\gal$ for
the neutral currents.
The results are given in {\TAB}~\ref{tab-ewp-gvga}.
There are strong anti-correlations between the $\ee$ couplings on one side
and the $\mumu$ and $\tautau$ couplings on the other, and also
between vector and axial-vector couplings for each lepton species.
This is shown
in {\TAB}~\ref{tab-ewp-gvga6} and illustrated in {\FIG}~\ref{f-gvvsga}.
When lepton universality is imposed the anti-correlation between
$\gvl$ and $\gal$  is reduced to $-29\%$.
Good agreement is found with the prediction of the {\SM}.
Being determined primarily from the lepton cross-sections, $\gal$
has a small relative error. 
Ratios of couplings provide a powerful test of lepton universality
for the axial-vector couplings at the 0.5\% level:
\begin{equation}
\frac{\gam}{\gae}  =      0.9989_{ -0.0058}^{+0.0033} ~~,~~~~
\frac{\gatau}{\gae}  =      1.0003_{ -0.0055}^{+0.0037}~~,~~~~
\frac{\gatau}{\gam}  =      1.0014_{ -0.0034}^{+0.0036} \; .
\end{equation}
For the vector couplings the relative uncertainties are much
larger. They are also very asymmetric and strongly correlated between lepton species.
Because the observables $\Afbpolmumu$ or $\Afbpoltautau$ are each products
of $\gvm$ or $\gvtau$ and $\gve$,
the derived values of $\gvm$ and $\gvtau$ diverge if $\gve$ 
approaches zero.
This pathology is visible in the 
strongly asymmetric errors in $\gvm$ and $\gvtau$ and the corresponding 
tails in {\FIG}~\ref{f-gvvsga}. 
When we form the vector coupling ratios the non-Gaussian behaviour of
the uncertainties is further enhanced for $\gvm/\gve$ and $\gvtau/\gve$,
while $\gvtau/\gvm$ has much smaller and rather symmetric errors:
\begin{equation}
\frac{\gvm}{\gve}  =      1.79_{ -0.64}^{+1.84} ~~,~~~~
\frac{\gvtau}{\gve}  =    1.63_{ -0.61}^{+1.72} ~~,~~~~
\frac{\gvtau}{\gvm}  =    0.91_{ -0.21}^{+0.25} \; .
\end{equation}
One should emphasise  that the non-linearities in the 
determination of $\gvl$ are driven by the
observed, statistically limited, measurement of $\Afbpolee$.
Our value of $\Afbpolee$ is small and only two standard deviations above zero, 
although compatible with the {\SM} prediction.
The problem is less apparent and the propagated errors
in the vector couplings are
significantly reduced when $\Afbpolee$ happens to be large, as is the 
case for example
in~\cite{bib-ALEPH-final}.

The  pure $\Zzero$ pseudo-observables $\Rl$ and 
$\Afbpolll$ are symmetric in $\gvl$ and $\gal$, as can be seen from
{\EQs}~\ref{eq-pwid1}~and~\ref{eq-afbpol}.
The results would not change if
the values of $\gvl$ and $\gal$ were interchanged.
Our measurements of the $C$-parameters describing the $\gzif$ 
interference ({\EQ}~\ref{eq-cpar1}) can be used to resolve
this ambiguity. $\Csgz$ parametrises the interference contribution
symmetric in $\costhe$  and $\Cagz$ the strong $\roots$ dependence
of the asymmetries ({\EQ}~\ref{eq-afb-roots}). 
The results of the fit, as given in
{\TAB}~\ref{tab-ewp-cpar}, distinguish $\gvl$ and $\gal$ unambiguously.
$\Cagz$  and $\Cazz$ also 
determine the signs of $\gam$ and $\gatau$ 
relative to $\gae$ (and the signs of $\gvm$ and $\gvtau$
relative to $\gve$).
The last remaining ambiguity, which the lineshape and asymmetry data alone cannot resolve,
is the relative sign\footnote{The absolute
sign of one of the axial-vector or vector couplings needs to be defined. 
By convention one takes $\gae$ as negative.}
 of $\gvl$ and $\gal$. This can be determined by measurements of 
the $\tau$ {\em polarisation} asymmetries~\cite{bib-tau-pol} 
or the {\em left-right} asymmetry~\cite{bib-sld-alr}, 
which determine directly the ratio $\gvl / \gal$ for a 
specific flavour.

\subsubsection[$ \als$ from the $\Zzero$ resonance parameters]
{\boldmath $ \als$ from the $\Zzero$ resonance parameters}
As discussed in {\SECT}~\ref{sec-ewp-radcor}
the hadronic partial decay  width, $\Ghad$, is increased by 
final state QCD corrections, in first order by a factor 
$1 + \frac{\als}{\pi}$. In ZFITTER corrections up to 
${\cal{O}}(\als^3)$ are included. 

The largest contribution to $\GZ$ is from $\Ghad$. As a result 
several $\Zzero$ resonance
parameters are sensitive to $\als$, namely the total width $\GZ$ directly,
the ratio of hadronic to leptonic partial width $\Rl$,
the hadronic pole cross-section $\shadpol$
and finally the leptonic pole cross-section $\sleppol$
(obtained by transforming the results in {\TAB}~\ref{tab-ewp-leppar},
$\sleppol = 1.9932 \pm 0.0043$\,nb).
From the three observables
$\Rl$, $\shadpol$, $\sleppol$ only two are independent and
the best constraint on $\als$ is 
obtained in a simultaneous fit to all parameters. This is discussed in
{\SECT}~\ref{sec-fit-sm}.
However, from the experimental point of view 
rather different effects dominate the uncertainty of each of 
these observables, and it is therefore instructive to examine the
value of $\als$ derived from each individual observable.
The measurement via $\GZ$ is statistics limited and 
free of normalisation errors, that via $\Rl$ is independent
of the luminosity and limited by lepton statistics and systematics,
while for the measurement via $\shadpol$ statistics,
selection systematics and luminosity uncertainties contribute about equally,
and finally, the measurement via $\sleppol$ is free of any hadronisation
uncertainties.

Figure~\ref{f-lsalfas} 
shows these observables and the {\SM} prediction 
as implemented in ZFITTER
as a function of $\als$. Also shown are the effects of varying the 
parameters $\Mt$, $\MH$ and $\alfemmz$. 
The resulting values for $\als$ are listed in {\TAB}~\ref{tab-ewp-alfas}.
They are in good agreement within the experimental errors. 

Since $\sleppol$, $\shadpol$ and $\Rl$ are ratios of partial or total widths,
the values of $\als$ derived through these parameters are rather
insensitive to variations of the {\SM} parameters, which affect
$\Gll$, $\Ghad$ and $\GZ$ similarly, 
 while the value
derived through $\GZ$ depends more strongly on $\Mt$ and $\MH$.
It is interesting to note that $\als$ from $\sleppol$, a measurement
relying entirely on leptonic final states,
has by far the smallest uncertainty. 
This somewhat counter-intuitive result is due to the fact that 
$\sleppol$ depends quadratically on $\GZ$ and is therefore
most sensitive to the effect of $\alfas$ on $\Ghad$, 
through the dominant contribution of the latter to $\GZ$.

QCD uncertainties on the determination of $\als(\MZ^2)$ from the
Z resonance parameters are small. The dominant effect comes from the
unknown terms of ${\cal{O}}(\als^4)$ and higher.
However, several evaluations exist in the literature, which result in rather
different error estimates.
In Reference~\cite{bib-alphas-2}, which is the basis of the ZFITTER
QCD corrections, the renormalisation scale
dependence ($1/4 < \mu^2/\MZ^2 < 4$) of the massless non-singlet terms
entering $\Ghad$ translates into \mbox{$-0.0002 < \delta \als < +0.0013$} for
$\als = 0.125$. Other contributions, such as the renormalisation scheme or
the uncertainty of the $b$ quark mass, are below $\pm 0.0005$.
In Reference~\cite{bib-alphas-siggi} the renormalisation
scale dependence of $\Rl$ is evaluated based on an effective parametrisation
of $\Rl$ as function of $\als$. Using a wider range
($1/16 < \mu^2/\MZ^2 < 16$) results in
\mbox{$-0.0004 < \delta \als < +0.0028$}.
In Reference~\cite{bib-alphas-soper} a similar scale dependence
of  $\als(\MZ^2)$ from $\Rl$ is found. However, the authors then
resum
additional terms (``$\pi^2$ terms'')
in the perturbative expansion of $\Rl$, which leads to a
slight correction of $+0.6\%$ for the resulting $\als(\MZ^2)$
and reduces the renormalisation scale uncertainty to $\pm 0.0005$.

Given the small size of this correction,
and to remain consistent with the other LEP collaborations,
we do not correct our result obtained from ZFITTER
and assign a conservative systematic error of
$\pm 0.002$ as the QCD uncertainty for $\als$ derived from the
Z resonance parameters, which is still significantly smaller than the
experimental precision.
In principle, $\als$ dependent propagator
corrections enter differently 
in the various Z resonance observables.\footnote{
The residual differences are caused by
QCD effects in terms involving the top-quark mass which
almost completely cancel in quantities that depend on
the ratio of widths.  At present, uncertainties in
these effects~\cite{bib-alphas-kuehn2} are equivalent to an error $\sim$ 1 GeV in
$\Mt$,  and are much smaller than the uncertainty on the measured top
mass.}
However, since these corrections are much smaller than the dominating
$\als$ correction to $\Ghad$, these differences can be safely
neglected for the overall QCD uncertainty.

\subsection{Standard Model fits}
\label{sec-fit-sm}
As a last step we compare our measured cross-sections and asymmetries
directly with the full {\SM} calculations implemented in
ZFITTER, using as fit parameters  $\MZ$,  $\als$, $\Mt$ and $\MH$. 
In addition, the electromagnetic coupling constant
is constrained by $\dalh = 0.02804 \pm 0.00065$~\cite{bib-alem-jeg}. 
The lineshape and asymmetry measurements alone are not sufficient to determine
simultaneously $\Mt$, $\MH$ and $\als$. The leading electroweak radiative corrections
in terms of $\Mt^2$ and $-\log(\MH)$ have the same form for all fermion 
partial widths and lepton asymmetries. 
The only exception is the $b$-quark partial width, which has
unique $\Mt$ dependent vertex corrections. This leads
to a somewhat different $\Mt$ dependence for the inclusive
 hadronic width, $\Ghad$.
However, this potential discrimination power
is absorbed by $\als$ when it can vary freely in the fit.
Therefore additional constraints on these parameters or supplementary
electroweak precision measurements are needed for a simultaneous fit
to $\Mt$, $\MH$ and $\als$.

We choose to restrict ourselves to three scenarios: 
{\it (i)} 
determination of $\als$ and $\Mt$ from the lineshape and asymmetry measurements 
alone,
restricting
$\MH$ to the range 90--1000 GeV, with a central value of 150 GeV;
{\it (ii)}
determination of $\MH$ and $\als$, using the direct
Tevatron measurement
of the top quark mass, $\Mt = 174.3 \pm 5.1$~GeV~\cite{bib-top-mass},
as an external constraint; and
{\it (iii)} determination of $\MH$ alone, using in addition to the
$\Mt$ constraint also the recent
world average of $\als = 0.1184 \pm 0.0031$~\cite{bib-alphas-siggi}.

The results of these fits are shown in {\TAB}~\ref{tab-ewp-smres}.
Our measured cross-sections and asymmetries are 
consistent with the {\SM} predictions as indicated both
by the absolute $\chidof$ and by its small change  with respect to 
the model-independent fits. In fit {\it (i)} we determine a value of
\mbox{$\Mt = 162 \pm 15 ^{+25}_{-5}\,(\MH)$ GeV}
through the indirect effect of radiative corrections, which agrees
well with the direct measurement.
This agreement is a sensitive validation of the electroweak loop corrections.
%
In fit {\it (ii)} we obtain
\begin{equation}
\als(\MZ^2) = 0.127 \pm 0.005 
\pm 0.002\,(\,{\rm QCD})
\end{equation}
for the strong coupling constant,
which is consistent within about 1.5 standard deviations with the world average.
Also in fit {\it (ii)} we find
$\MH = 390 ^{+750}_{-280}$~GeV as the mass of the Higgs boson.
In fit {\it (iii)}, when we further use the external value of $\als$,
the fitted result for the Higgs mass moves lower, to
 $\MH = 190^{+335}_{-165}$ GeV, due to the correlation between
$\als$ and $\MH$.
The correlation is illustrated in {\FIG}~\ref{f-mhas} 
which shows the 68\% C.L. contour of $\als$ and $\MH$ for fit {\it (ii)}.
Since the leading radiative corrections depend logarithmically
on $\MH$ the uncertainty of $\MH$ is very asymmetric.
But even in terms of $\log \MH$ the error is asymmetric.
This is due to the fact that the fit allows a wide
range for $\MH$ but only for $\MH \gg \MW$ is the
logarithmic dependence a good approximation.

It is important to
verify that performing a {\SM} fit on the level of  
pseudo-observables is equivalent to fitting the
cross-sections and asymmetries directly.
Using the results of the {\SLPnine} fit 
({\TABs}~\ref{tab-ewp-leppar} and ~\ref{tab-ewp-lepparc9})
as input to the {\SM} fit yields consistent results:
the central values for the {\SM}
parameters agree within 10\% of the error; uncertainties
and correlations are indistinguishable within the quoted precision.
This is illustrated in {\FIG}~\ref{f-mhas} which also shows the 
68\% contour from the pseudo-observable fit.
When these results are combined
with other electroweak measurements~\cite{bib-LEPEWWG} to determine 
{\SM} parameters, the fits are performed at the level of
pseudo-observables.

%
%
\section{Summary and conclusions}
\label{sec-sum}
We present here the OPAL analysis of the $\Zzero$ 
cross-section  and lepton forward-backward asymmetry measurements.
This analysis includes the entirety of the OPAL data sample taken near the
$\Zzero$ resonance from 1990 to 1995.
In addition to the nearly four-fold increase in statistics since our last 
publication~\cite{bib-opal-ls92} many improvements have been made in 
the experimental and theoretical systematic uncertainties of the
measurements. 
One important contribution is the reduction of the 
experimental error of the luminosity determination~\cite{bib-lumi-siw} 
by more than a factor of ten. New techniques to evaluate systematic 
effects of the event selections more than halved this
source of uncertainty for the hadronic 
cross-section.
Similar progress has been made in the precision of the external inputs
to our measurements, namely the precise determination of the LEP centre-of-mass 
energy~\cite{bib-LEPecal-98} and the theoretical calculation of the 
luminosity cross-section~\cite{bib-bward,bib-pavia}, as well as in
the theoretical tools and programs used to evaluate and interpret
our results~\cite{bib-fit-ZFITTER,bib-fit-TOPAZ0,bib-pcp}.
Overall, these coherent efforts allowed us to exploit most of the 
inherent statistical precision of
our $4.5\times10^6$ measurable $\Zzero$ decays.

In a model-independent ansatz we parametrise independently
the contributions  from pure $\Zzero$ exchange and from 
$\gzif$ interference. Our results are in good agreement  with lepton 
universality and consistent with the vector and
axial-vector couplings between the $\Zzero$ and fermions as predicted in the
{\SM}. 
Our main results in terms of $\Zzero$ resonance parameters,
assuming lepton universality, can be summarised as:
\begin{eqnarray*}
\MZ & = &       91.1852 \pm 0.0030\,\mbox{GeV} \; ,  \\
\GZ & = &        2.4948 \pm 0.0041\,\mbox{GeV} \; ,  \\
\shadpol & = &   41.501 \pm 0.055 \,\mbox{nb} \; ,  \\ 
\Rl & = &        20.823 \pm 0.044 \; ,  \\ 
\Afbpolll & = &  0.0145 \pm 0.0017\; .  
\end{eqnarray*}
Transforming these parameters yields the ratio of invisible to leptonic 
decay width,
$$
\Ginv/\Gll = 5.942 \pm 0.027 \; .
$$
Assuming  the
{\SM} couplings for leptons this can be converted into a 
measurement of the effective number of 
light neutrino species
$$
N_{\nu} =  2.984 \pm  0.013 \; .
$$
Alternatively, one can use the {\SM} prediction of $\Ginv$
for three neutrino generations and derive an upper limit for additional 
contributions from new physics to the invisible or the total Z width:
$$
\Ginv^{\rm new}  <  3.7 \, {\rm MeV} \mbox{~~~~or~~~~} \GZ^{\rm new}  < 14.8 \, {\rm MeV} \mbox{~~~at 95\,\% C.L.}
$$
Finally, we compare our measured cross-sections and asymmetries
with the full {\SM} calculations. Radiative corrections are sensitive to
the parameters $\Mt$, $\MH$ and $\als(\MZ^2)$, allowing
our measurements to determine:
\begin{eqnarray*}
\Mt & = & 162 \pm 15 ^{+25}_{-5}\,(\MH)  \mbox{~GeV}\;\; , \\
\als(\MZ^2) & = & 0.125 \pm 0.005 ^{+0.004}_{-0.001}\,(\MH)\,  
\pm 0.002\,(\,{\rm QCD}) \; \;,
\end{eqnarray*}
where we fixed $\MH = 150^{+850}_{-60}$ GeV.
Including the direct measurement of the
top quark mass ($\Mt = 174.3 \pm 5.1\mbox{~GeV}$~\cite{bib-top-mass})
as an additional constraint we obtain results  for $\als(\MZ^2)$ and 
the mass of the Higgs boson:
\begin{eqnarray*}
\als(\MZ^2) & = & 0.127 \pm 0.005 \pm 0.002\,(\,{\rm QCD}) \\
\MH & = & 390^{+750}_{-280} \mbox{~GeV}\,.
\end{eqnarray*}
%
These $\Zzero$ lineshape
and asymmetry measurements test and confirm the {\SM}
at the level of quantum loop corrections
and set tight constraints on new physics.

%
\section*{Acknowledgements}

We particularly wish to thank the SL Division for the efficient operation
of the LEP accelerator, for the painstaking calibration of the beam energy
 and for their continuing close cooperation with
our experimental group.
We would also like to extend special thanks to the community of theorists whose
contributions to precise electroweak calculations we cite, and whose efforts
have made our own meaningful.
We thank our colleagues from CEA, DAPNIA/SPP,
CE-Saclay for their efforts over the years on the time-of-flight and trigger
systems which we continue to use.  In addition to the support staff at our own
institutions we are pleased to acknowledge the  \\
Department of Energy, USA, \\
National Science Foundation, USA, \\
Particle Physics and Astronomy Research Council, UK, \\
Natural Sciences and Engineering Research Council, Canada, \\
Israel Science Foundation, administered by the Israel
Academy of Science and Humanities, \\
Minerva Gesellschaft, \\
Benoziyo Center for High Energy Physics,\\
Japanese Ministry of Education, Science and Culture (the
Monbusho) and a grant under the Monbusho International
Science Research Program,\\
Japanese Society for the Promotion of Science (JSPS),\\
German Israeli Bi-national Science Foundation (GIF), \\
Bundesministerium f\"ur Bildung und Forschung, Germany, \\
National Research Council of Canada, \\
Research Corporation, USA,\\
Hungarian Foundation for Scientific Research, OTKA T-029328, 
T023793 and OTKA F-023259.\\

%
%
\clearpage
\newpage
\begin{appendix}
%
%
\section{Four-fermion processes and radiative photon interference}
\label{sec-4f-ifi}
Small radiative corrections to fermion-pair production due to four-fermion
final states and the interference between initial- and final-state 
photon radiation are not treated by the fermion-pair Monte Carlo
generators, JETSET and KORALZ, which we used in determining our event
selection efficiencies.
We describe here, in some technical detail, how we derived the small acceptance
corrections arising from these effects.

\subsection{Treatment of four-fermion final states}
\label{sec-4f}

The selection criteria for hadronic and lepton-pair events define samples
which are primarily the result of fermion-pair production processes
$\ee\rightarrow\ff$.
Studies have been made of the much smaller contributions arising from 
four-fermion final states $\ee\rightarrow\ff\FF$, as shown in
{\FIG}~\ref{f-four-fermion}.
Some classes of four-fermion events have little connection to $\Zzero$
production, and can properly be considered as background,
for example multi-peripheral (d) or pair corrections to $t$-channel
scattering (e,f).
Other four-fermion events (b,c)
however, must be considered as radiative corrections to the fermion-pair
production (a) which interests us.
In the following we use the designation $\ff$ to refer to the primary
fermion pair of interest and $\FF$ to refer to the radiated pair,
although a rigid distinction between the pairs cannot always be made.

From the theoretical point-of-view it is desirable to design 
the selection of fermion pairs in such a way that
no stringent cuts are imposed on events with photon radiation and  
pair production.  One then benefits
from cancellations between real and virtual corrections leading to 
smaller uncertainties in the theoretical calculations.
In practice, however, the event selection efficiency for events with 
radiated pairs can be expected to be 
different from the efficiency for events without such radiation,
and this efficiency cannot be 
evaluated using the standard Monte Carlo event generators for fermion-pair 
production since they do not in general include such 
four-fermion final states.
Studies have been made following the approach described in~\cite{bib-4fcorr}
using a specially generated sample of Monte Carlo four-fermion events.

An unambiguous separation of signal and background four-fermion events
is not possible due to the interference of amplitudes which lead to the same 
final states.
We therefore adopt a practical signal definition based on kinematics.
Typically the fermion pair ($\ff$) from $\Zzero$ decay
is of high invariant mass and the radiated pair ($\FF$) is of low mass.
Four-fermion final states from $t$-channel and multi-peripheral diagrams
tend to occupy a region of phase space well separated from the signal events
of interest.
We ignore the interference between $s$-channel and $t$-channel
diagrams and generate separate four-fermion Monte Carlo samples:
four-fermion events from $s$-channel diagrams are generated using the FERMISV
program, while those from the $t$-channel diagrams are generated using 
the grc4f (version 1.11) and Pythia programs.

For channels other than $\eeee$,
($\ee\rightarrow\ff$, with $\mathrm{f}\neq\mathrm{e}$),
we identify four-fermion events
from $s$-channel diagrams as part of the fermion-pair signal
if they satisfy
$m_{\ff} > m_{\FF}$ and $m^2_{\ff}/s > 0.01$.
The second requirement is common to our definition of the fermion-pair 
signal phase space.
All other four-fermion events arising from $s$-channel diagrams 
failing these requirements, as well as those from $t$-channel diagrams
and the two photon process, are regarded as background.

In case of $\eeee$, four-fermion final states are
considered as signal if the final-state electron-pair
meets the normal requirements for the $\eeee$ signal in terms of 
electron energy, acollinearity and electron polar angle.
Here, four-fermion events from $t$-channel
diagrams are treated as signal.
In principle even contributions from {\FIG}~\ref{f-four-fermion}(e)
where the lower boson is a virtual $\Zzero$ are considered as signal.
The number of such events falling within our ideal acceptance, however,
is negligible.

Including four-fermion events leads to a correction of the effective
signal efficiency of 
\begin{equation}
\Delta_\varepsilon = - \frac{\sigma_{\ff\FF}}{\sigma}
              (\varepsilon_{\ff}-\varepsilon_{\ff\FF}) \; ,
\end{equation}
where $\sigma$ is the total 
cross-section for $\ff$ production including the effects of
pair-production, and 
$\sigma_{\ff\FF}$ is the cross-section for four-fermion
events $\ff\FF$ defined as signal.
The selection efficiency calculated using the fermion-pair Monte Carlo is
denoted by $\varepsilon_{\ff}$, and the selection efficiency calculated for
the four-fermion signal contributions is $\varepsilon_{\ff\FF}$.

We expect the largest effect in the case of the $\ee\rightarrow\mumu$ selection,
where the requirement that the number of tracks be exactly two
explicitly excludes all the visible four-fermion final states, even those which 
need to be counted as part of the signal.
Here we find that the efficiency for signal four-fermion final states is
typically 20--30\%
for $\eemumu\mumu$, $\mumu\qq$, $\mumu\tautau$, and about 80\%
for $\eemumu\ee$,
but the resulting correction to the $\eemumu$ efficiency is only about 0.1\%.
Background from the $t$-channel diagrams gives a very small contribution of
2$\times 10^{-5}$, and background four-fermion events from $s$-channel
diagrams give an even smaller contribution.
We find a correction of similar size for the $\eetautau$ selection.

In the $\eeee$ selection no tight multiplicity requirement is made.
The efficiency for signal four-fermion events is found to be high (over 90\%).
As a result, the efficiency correction is nearly cancelled by the small
background contributions, 
and no overall correction due to
four-fermion final states is necessary.
In the case of $\eehad$ the event selection is 
very inclusive so that there is no notable effect on the efficiency.
We subtract a very small four-fermion background contribution of
$0.4\times10^{-4}$.

The theoretical programs ZFITTER and TOPAZ0 which we use to compute expected
cross-sections and asymmetries treat pair radiation inclusively in the pair
mass and do not include contributions from virtual $\Zzero$ bosons.
Therefore the four-fermion signal definition as described here cannot be
exactly mapped to these calculations, e.g. the separation between signal
and background based on the mass of the pairs is not made in these programs.
However the differences are quantitatively negligible~\cite{bib-mc-LEP2yr-mcws}.

\subsection{Initial-final state interference}
\label{sec-ifi}

The effect of interference between initial and final state photon radiation 
is in general strongly suppressed at the $\Zzero$ resonance.
In view of the high precision measurements presented here, however, we studied
several possible effects.
The main effect of initial-final state interference is a change in the angular
distribution, proportional to the value of the differential cross-section,
but of opposite sign in the forward and backward hemispheres, giving an 
increasing effect towards $\cos\theta=\pm 1$.
The size of the effect depends on the cuts, mainly on $s^{\prime}$ or 
$m_{\ell\ell}$: in general the tighter the cut on the energy of radiated 
photon, the larger the effect.

Initial-final state interference is not included in the 
Monte Carlo event samples used to calculate the event selection efficiency 
for the $\eehad$, $\mumu$, or $\tautau$ processes, while it {\em is} included
in our use of ZFITTER.
We therefore do not remove its effects in our measured cross-sections and
asymmetries but need to make small corrections to our calculated selection
efficiencies to account for its absence in the Monte Carlo samples.
Since the major source of the event selection inefficiency is due to limited 
angular acceptance near $|\cos\theta|=1$ and reduced acceptance at 
small $s^{\prime}(m_{\ell\ell}$), the missing initial-final state interference
in these Monte Carlo samples can cause some bias in the acceptance
extrapolation.

The effect of this missing initial-final state interference was evaluated
using the program ZFITTER, for which this effect can be switched on and off.
To incorporate the event selection efficiency in the calculation, 
a matrix of efficiency resolved in bins of $\cos\theta$ and the invariant 
mass of the final state lepton pair, $m^2_{\ell\ell}$, was calculated using
the KORALZ Monte Carlo sample.
The expected observed cross-section was obtained by multiplying the 
efficiency in each bin by the corresponding differential cross-section 
calculated using ZFITTER, and then summing over the full phase space.
The overall efficiency was calculated as the ratio of the observed to the 
total cross-section.
Two calculations were performed, one with and one without initial-final
state interference, and the two results compared.

For both $\eemumu$ and $\eetautau$ the effect of missing
initial-final state interference is found to be small (at $10^{-4}$ level)
as a result of the large acceptance
(in both $\cos\theta$ and $m^2_{\ell\ell}$) of both these selections.
For $\eehad$, the effect of initial-final state interference 
is more complicated to calculate due to the presence of a mixed QED and
QCD parton shower in the final state.
The size of the effect is strongly suppressed, however, by the almost
complete 
acceptance of the $\eehad$ event selection in both 
$\cos\theta$ and $s^{\prime}$ (the selection inefficiency is only 
0.5\%), as well as the smaller size of the quark charges. 
Without considering the effect of the parton shower, we evaluated the
effect to be less than $10^{-5}$, which we neglect. 

In the $\eeee$ selection, the experimental acceptance and ideal 
kinematical acceptance are very close.
No appreciable phase-space
extrapolation which could introduce a significant correction
for the effect of initial-final state interference is needed.
Therefore no correction is applied to the $\eeee$ selection efficiency.

%
%
\section{\boldmath $t$-channel contributions to $\eeee$}
\label{sec-tchan}
For the reaction $\eeee$ not only $s$-channel annihilation but also the 
$t$-channel exchange of $\gamma$ and $\Zzero$ contribute.
Since our primary fitting program (ZFITTER) 
includes only $s$-channel processes
an external correction is needed to account for the contributions 
from $t$-channel exchange diagrams and their 
interference with the $s$-channel.
To ensure a consistent treatment of the $s$-channel in
all final states we continue to calculate
the $\eeee$ $s$-channel terms using ZFITTER,
but proceed as follows.
We use the program ALIBABA~\cite{bib-fit-ALIBABA}
to calculate the {\SM} prediction for both 
the pure $s$-channel and the full $s$+$t$ channel, separately
for forward and backward cross-sections.
The contributions from $t$-channel and 
$s$-$t$ interference are obtained
 by subtracting the ALIBABA $s$-channel
cross-section from the full ALIBABA $s$+$t$ cross-section,
$\sAFBt = \sAFBst - \sAFBs$.
To these $t$ and $s$-$t$ contributions we then add the 
$s$-channel cross-sections calculated with ZFITTER 
($\sZFs,\;\sZBs$) to obtain the
predictions for the total cross-section and asymmetry:
\begin{eqnarray}
\see & = & \sZFs + \sAFt + \sZBs + \sABt \nonumber \\
\Afbee & = & \frac{(\sZFs + \sAFt) - (\sZBs + \sABt)}
                  {(\sZFs + \sAFt) + (\sZBs + \sABt)} \; \; .
\end{eqnarray}
The separation into forward and backward cross-sections 
ensures a correct propagation of errors to the fitted
$s$-channel observables $\Ree$ and $\Afbpolee$. The non $s$-channel
contributions lead to a statistical correlation of $-11\,\%$ between
$\Ree$ and $\Afbpolee$.

The size of the $t$-channel corrections changes rapidly
as a function of centre-of-mass energy, since
the $s$-$t$ interference is proportional to $\sqrt{s} - \MZ$ 
in the vicinity of the $\Zzero$ pole.
We therefore parametrise the corrections $\sAFt$ and $\sABt$
as a function of $(\sqrt{s} - \MZ)$ in order to account
for their variation as  $\MZ$ converges in the fit. In this way our results 
for $\Ree$ and $\Afbpolee$ properly respect the uncertainty
of our fitted $\MZ$ and the corresponding correlations.
However, one should note a subtle side effect of this treatment.
When lepton universality is assumed in the fit the measurement of $\see$ 
at the peak contributes to the determination of $\MZ$.
The imposition of lepton universality introduces a tension in the fit
since fluctuations induce discrepancies between the measured and
predicted cross-sections for each lepton species at the peak point.
This tension can be relaxed in the case of the electron cross-section
by shifting $\MZ$ since the non-$s$ contributions in the electron
channel have a non-zero slope with energy at the pole.
This effect is responsible for the 
shift of $ 0.5$ MeV of $\MZ$ between the fits with and 
without lepton universality ({\SECT}~\ref{sec-ewp-leppars}).

Theoretical uncertainties of the $t$-channel correction have been estimated
in~\cite{bib-fit-ALIERROR} and the uncertainties in the  
forward and backward cross-section have also been determined 
separately~\cite{bib-fit-ALIERROR-fb}.
Adjusting these estimates to the smaller angular acceptance used in our
$\ee$ selection results in the uncertainties listed in 
{\TAB}~\ref{tab-ewp-eet+ti} which are separated 
into three classes according to the centre-of-mass energy
(below, at and above the peak).
For the fit these uncertainties are translated into the
corresponding uncertainties of the measured $\see$ and $\Afbee$ and included
in the covariance matrix.
It is not known to what extent the errors between the energy points and the
forward-backward regions are correlated. We tested several possibilities
-- uncorrelated, fully correlated and fully anti-correlated -- and found that
our fitted parameters are insensitive within the effective $t$-channel 
errors to the scenario chosen. For the results presented here we 
 assumed no correlation between forward and backward regions,
full correlation between data points within 
each centre-of-mass energy class and 
no correlation between data points of different classes.
The effective $t$-channel theory  uncertainties on 
$\Ree$ and $\Afbpolee$ amount to 0.027 and
0.0015, respectively, with a correlation of $-0.85$. Overall, 
statistical errors and the theoretical $t$-channel uncertainties lead to 
a correlation of $-0.20$ between $\Ree$ and $\Afbpolee$.

\section{Fit covariance matrix and energy spread corrections}
\label{sec-covmat}
For all our fits we perform a $\chisq$ minimisation using the 
MINUIT~\cite{bib-MINUIT} package. The $\chisq$ is defined as 
\begin{equation}
\chisq = \Delta^{\mathrm{T}} \, C^{-1} \, \Delta \; ,
\end{equation}
where $\Delta$ is the vector of residuals
between the measured and predicted cross-sections and asymmetries
and $C$ is the covariance matrix describing the statistical and systematic
uncertainties of the measurements and their correlations.
In total we have 211 measured points, 124 cross-sections and 87 asymmetries.
Therefore 22366 components $C_{ij}$ need to be determined
to specify the full matrix $C$.
Due to the large variety of error sources entering the measurements and 
non-trivial correlations, the construction of the covariance matrix is a rather
complex task. In general, each $C_{ij}$  is composed of
\begin{equation}
C_{ij} = 
C_{ij}^{\mathrm{\,sel,stat}} + 
C_{ij}^{\mathrm{\,lumi,stat}} + 
C_{ij}^{\mathrm{\,sel,syst}} + 
C_{ij}^{\mathrm{\,lumi,syst}} + 
C_{ij}^{\mathrm{\,t-chan}} + 
C_{ij}^{\ecm} + 
C_{ij}^{\,\delta_E}   \; .
\end{equation}
The elements $C_{ij}^{\mathrm{\,XX}}$ are in general 
constructed from `small'
covariance matrices $V^{\mathrm{\,XX}}$, which describe the year, 
energy point and
final state dependencies.
In the following we give a brief overview of how each uncertainty 
is treated. The indices $i$ and $j$ refer to the 211 measured points;
$k$ and $l$ to the specific element in the $V$ matrices. So the notation
$\sigma_i^k$ refers to the $i$-th cross-section which is related
to the row or column $k$ in the matrix $V$.

\noindent
{\large \bf Statistical errors} of the cross-sections ($\sigma_i$) 
are determined by the 
number of selected events, $N_{\rm sel}$, and corrected for estimated 
background, $N_{\rm bg}$. 
They enter only the diagonal elements
\begin{equation}
C_{ii}^{\mathrm{\,sel,stat}} = \left( 
\frac{\sqrt{N_{\rm sel}}}{N_{\rm sel} -  N_{\rm bg}} \, \sigma_i \right)^2 
\;\; . 
\end{equation}
Luminosity statistical errors 
are common to all cross-sections
for a specific running period. They are calculated as
\begin{equation}
C_{ij}^{\mathrm{\,lumi,stat}} = (\delta^{\mathrm{\,lumi,stat}}_k)^2 \,
\sigma_i^k \, \sigma_j^k \; ,
\end{equation}
where $\delta^{\mathrm{lumi,stat}}_k$ is the relative statistical luminosity
error  and $\sigma_{i,j}^k$ refers to the four cross-sections
in each running period $k$.

\noindent
{\large \bf Systematic errors} from the event selection affect mostly a specific
final state. A large fraction of this error is fully correlated but there
are also components which are independent or only partially correlated
 for different running periods or energy points. Therefore these
errors are themselves specified in a covariance matrix of relative errors,
$V^{\mathrm{\,sel}}$. We use three such matrices, one for the hadron 
cross-sections ({\TAB}~\ref{tab:had_matrix}), 
one for the three leptonic cross-sections 
({\TABs}~\ref{tab:ee_matrix}--\ref{tab:ll_matrix}), and one
for the three lepton asymmetries
({\TAB}~\ref{tab:asymm_matrix}). 
In this way  correlations among the three
lepton species are accounted for. $C_{ij}^{\mathrm{\,sel,syst}}$ 
is given by
\begin{equation}
C_{ij}^{\mathrm{\,sel,syst}} = 
V^{\mathrm{\,sel,}\sigma}_{k,l} \, \sigma_i^k \, \sigma_j^l ~~~~ \mbox{or}
~~~~
 C_{ij}^{\mathrm{\,sel,syst}} = 
V^{\mathrm{\,sel,}\Afb}_{k,l} \; .
\end{equation}
There is no correlation between the experimental errors for the cross-sections
and those for the asymmetries.

\noindent
{\large \bf Luminosity errors} include the experimental systematics 
and the theoretical uncertainties of the luminosity measurement.
They affect all cross-section measurements, but
similar to the selection errors there are components which are 
only partially correlated between running periods or energy points.
Therefore a covariance matrix of relative luminosity errors,
$V^{\mathrm{\,lumi}}$,  is also used ({\TAB}~\ref{tab:lumi_matrix}):
$$
C_{ij}^{\mathrm{\,lumi,syst}} =
V^{\mathrm{\,lumi}}_{k,l} \, \sigma_i^k \, \sigma_j^l \; .
$$

\noindent
{\large \boldmath \bf  $t$-channel  errors} refer to the theoretical
uncertainty in the $t$-channel correction for $\ee$ cross-sections
and asymmetries. These are specified in terms of forward and backward
cross-sections, $\Delta \sigF$ and $\Delta \sigB$ 
({\TAB}~\ref{tab-ewp-eet+ti}). 
Different data points are only correlated if both
energies are either below, at, or above the peak (Appendix \ref{sec-tchan}).
Then $C_{ij}^{\mathrm{\,t-chan}}$ is given by:
\begin{eqnarray}
\see^i \leftrightarrow \see^j ~: ~~~~ C_{ij}^{\mathrm{\,t-chan}} 
& = &(\Delta \sigF^k)^2 + (\Delta \sigB^k)^2 \\ \nonumber
\Afb^{{\mathrm{ee}},i} \leftrightarrow \Afb^{{\mathrm{ee}},j} ~: ~~~~ 
C_{ij}^{\mathrm{\,t-chan}} 
& = & \frac{1-\Afb^{{\mathrm{ee}},i}}{\see^i}\, \frac{1-\Afb^{{\mathrm{ee}},j}}{\see^j} \,
   (\Delta \sigF^k )^2 +
    \frac{1+\Afb^{{\mathrm{ee}},i}}{\see^i} \, \frac{1+\Afb^{{\mathrm{ee}},j}}{\see^j} \,
   (\Delta \sigB^k )^2  \\ \nonumber
\see^i \leftrightarrow \Afb^{{\mathrm{ee}},j} ~: ~~~~ C_{ij}^{\mathrm{\,t-chan}} 
& = & \frac{1-\Afb^{{\mathrm{ee}},j}}{\see^j} \,(\Delta \sigF^k) ^2 -
    \frac{1+\Afb^{{\mathrm{ee}},j}}{\see^j} \,(\Delta \sigB^k)^2 
\end{eqnarray}

\noindent
{\large \bf Centre-of-mass energy  errors} have been determined in 
\cite{bib-LEPecal-98} and are specified as a covariance matrix
$V^{\mathrm{\,\ecm}}$
(see {\TABs}~\ref{tab:ecm9092_matrix} and \ref{tab:ecm9395_matrix}). 
They are transformed into the corresponding
cross-section or asymmetry errors using the derivatives
\begin{equation}
C_{ij}^{\ecm} = V^{\mathrm{\,\ecm}}_{kl} \, \frac{{\rm d}\,O_i^k}{{\rm d}\,E} \,
\frac{{\rm d}\,O_j^l}{{\rm d}\,E} \;\;\;\;  (O = \sigma \; \mbox{or} \; \Afb) .
\end{equation}
The slopes ${\rm d}\sigma/{\rm d}E$ and ${\rm d}\Afb/{\rm d}E$ are determined numerically using 
ZFITTER ($\sigma^{\rm ZF}$) in an iteration during the fit.
For cross-sections the dependence of the luminosity on the centre-of-mass
energy must also be taken into account. Since this is a tiny effect 
it is sufficient to consider only the dominant
$1/E^2$ dependence of the low-angle Bhabha cross-section,
neglecting small distortions due to the $\gzif$ interference. 
Combining the predicted slope and the luminosity dependence  yields
\begin{equation}
\frac{{\rm d}\,\sigma_i}{{\rm d}\,E} = \frac{{\rm d}\,\sigma_i^{\rm ZF}}{{\rm d}\,E} +
2 \, \frac{\sigma_i}{E_i} \; .
\end{equation}

\noindent
{\large \bf Beam-energy spread} of the electrons and positrons in LEP 
leads to a dispersion of the centre-of-mass energy with a width 
$\delta_{\ecm} \approx 50$ MeV. Therefore the measured 
cross-sections and asymmetries
do not correspond to a sharp energy, $E_i$,  
but form a weighted average around $E_i \pm \delta_{\ecm}$ which can be shifted 
from the value exactly at $E_i$.
A similar effect is caused by the fact that many LEP fills are combined
for each data point. These fills are not at precisely the same energy but
scatter by typically 10 MeV around the average. Moreover, within a fill
the energy also varies by several MeV. These two effects need to be added
in quadrature with the intrinsic LEP $\delta_{\ecm}$. Correction terms
are determined according to
\begin{eqnarray}
\label{eq-ecm-spread}
\Delta^{\mathrm{spr}}_{ \sigma_i} & = & 
- \frac{1}{2} \left[\frac{{\rm d}^2\,\sigma}{{\rm d}\,E^2}\right]_i
 \delta_{\ecm}^2 \\ \nonumber
\Delta^{\mathrm{spr}}_{\Afb^{i}} & = &
- \left[ \frac{1}{2} \frac{{\rm d}^2\,\Afb}{{\rm d}\,E^2} + 
\frac{1}{\sigma}\, \frac{{\rm d}\,\sigma}{{\rm d}\,E}\, 
\frac{{\rm d}\,\Afb}{{\rm d}\,E}
\right]_i  \delta_{\ecm}^2  \; \,
\end{eqnarray}
and added to the measured $\sigma_i$ and $\Afb^i$.
The first and second derivatives of $\sigma$ and $\Afb$ are again determined numerically with 
ZFITTER during the fit.

The spread $\delta_{\ecm}$ has  typically an uncertainty 
of about  1.2 MeV, which is largely correlated between years and 
energy points and specified in detail in the energy spread
covariance matrix $V^{\,\delta_E}$ ({\TAB}~\ref{tab:esigrms_matrix}).
It enters  the fit covariance matrix as
\begin{equation}
C_{ij}^{\,\delta_E} = 4 \, 
\Delta^{\mathrm{spr}}_i \,
\Delta^{\mathrm{spr}}_j \, 
\frac{V^{\,\delta_E}_{kl}}{\delta_{\ecm}^k \, \delta_{\ecm}^l} \; .
\end{equation}
$\Delta^{\mathrm{spr}}_i$ refers to the corrections 
$\Delta^{\mathrm{spr}}_{\sigma_i}$ and $\Delta^{\mathrm{spr}}_{\Afb^{i}}$
in {\EQ}~\ref{eq-ecm-spread}.

\section{S-Matrix results}
\label{sec-smat}
The S-Matrix formalism~\cite{bib-smatrix} is an alternative
phenomenological approach 
to describe the $s$-channel reaction $\ee\rightarrow\ff$ by the exchange 
of two spin-1 bosons, a massless photon and a massive $Z$ boson.
The lowest-order total cross-section,
$\mathrm{\sigma^0_{tot}}$, and forward-backward asymmetry, $A^0_{\mathrm{fb}}$,
 are given
as:
\begin{eqnarray}
\sigma^0_{a}(s) & = &
                    \frac{4}{3}\pi\alpha^2
              \left[
                    \frac{\gf^a}{s}
                   +\frac{\jf^a (s-\MZbar^2) + \rf^a \, s}
                         {(s-\MZbar^2)^2 + \MZbar^2 \GZbar^2}
              \right]
              \qquad\qquad\mathrm{for} \,\, \mathit{a} = \mathrm{tot,fb}\\
A^0_{\mathrm{fb}}(s) & = & \frac{3}{4}
                           \frac{\sigma^0_{\mathrm{fb}}(s)}
                                {\sigma^0_{\mathrm{tot}}(s)}\,,
\end{eqnarray}
where $\sqrt{s}$ is the centre-of-mass energy.  The S-Matrix ansatz uses
 a Breit-Wigner denominator with $s$--{\em in}dependent width for the Z
resonance, \hbox{$s-\MZbar^2+i\MZbar\GZbar$}. 
As discussed in {\SECT}~\ref{sec-ewp-radcor} the two Breit-Wigner forms with
$s$--independent width and $s$--dependent width, respectively, are equivalent.
They differ only in that the definition of the mass and width,
$\MZbar$ and $\GZbar$ are shifted with respect to 
$\MZ$ and $\GZ$.
Note that in order to avoid confusion with different definitions we 
quote values for $\MZ$ and $\GZ$, rather than $\MZbar$
and $\GZbar$, by applying the respective transformations.
 
The S-Matrix parameters $\rf$, $\jf$ and $\gf$, which are real numbers,
describe
the Z exchange, $\gamma$Z interference and photon exchange contributions,
respectively. For the latter ($\gf$) the QED prediction is in general used.
The parameters $\rf$ and $\jf$  are identical at tree-level to the
$C$-parameters
introduced in {\SECT}~\ref{sec-ewp-tree}, apart from constant factors:
\begin{equation}
\begin{array}{lllllll}
\rtotf & = & \kappa^2 \, \Cszz    ~~~~ & , & ~~~~  \rfbf & = & 4 \kappa^2 \, \Cazz   \\
\jtotf & = & 2 \kappa \, \Csgz  ~~~~ & , & ~~~~  \jfbf & = & 2 \kappa \, \Cagz \; ,
\end{array}
\end{equation}
where $\kappa = \frac{{\mathrm{G_F}}\MZ^2}{2\sqrt{2}\pi\alpha}~ \approx ~ 1.50$.
When radiative corrections are considered the relation is less direct;
the S-Matrix parameters
absorb by definition all electroweak and final state corrections.
Further differences are caused by the treatment of the imaginary
components in $\alpha(\MZ^2)$ and the couplings
${\cal G}_{A{\mathrm{f}}}$ and ${\cal G}_{V{\mathrm{f}}}$.

In the S-Matrix approach the $\qq$ final states are also parametrised
in terms of
$r_{\rm q}$ and $j_{\rm q}$. For the inclusive hadronic final state these
are summed over all colours and open quark flavours to yield
the corresponding parameters $\rhad$ and $\jhad$.
The results of the full fit with 16 parameters are given in
{{\TAB}~\ref{tab-ewp-smat}}.
The 12 parameters describing the three lepton flavours are
consistent with lepton universality
and we find overall good agreement with the {\SM} expectations. 
The interpretation suffers, however, from the large correlations between
the fitted parameters
({\TAB}~\ref{tab-ewp-smatcor16}).
As discussed in 
{\SECT}~\ref{sec-ewp-cpar} the precision of the fitted $\MZ$ is much
reduced when the
hadronic interference, $\jhad$,  is treated as a fit parameter; 
the correlation coefficient between $\MZ$ and $\jhad$ is $-0.96$.
For comparison, we performed a fit with the hadronic interference
fixed to the {\SM}
prediction. The results are shown in {{\TAB}~\ref{tab-ewp-smat}}.
This parametrisation
is in practice equivalent to the $C$-parameters ({{\TAB}~\ref{tab-ewp-cpar}});
we transformed the $C$-parameters into S-Matrix parameters, accounting
for the differences in the treatment of radiative corrections,  
and obtained consistent results at the level of 1\,--\,2\,\% of the errors.

\end{appendix}
\clearpage \newpage

\begin{table}[htb]
\begin{center}
\begin{tabular}{|cl|c|r|r|r|r|r|}
\hline
 Year & Data & $\roots$ & $\int{{\cal{L}}\,{\mathrm{d}}t}$ 
 & $N_{\mathrm{had}}$ & $N_{\mathrm{ee}}$ & $N_{\mu\mu}$ & $N_{\tau\tau}$  \\
 & sample & (GeV) & (${\mathrm{pb}}^{-1}$) &&&& \\
\hline
     &{\pkmt}  &$  88.22 $&$   0.5 $&$    2229 $&$     169 $&$     109 $&$      81 $ \\
     &{\pkmd}  &$  89.23 $&$   0.6 $&$    5322 $&$     306 $&$     231 $&$     214 $ \\
     &{\pkmu}  &$  90.23 $&$   0.4 $&$    7045 $&$     320 $&$     316 $&$     221 $ \\
1990 &{\pk}    &$  91.22 $&$   3.5 $&$  103664 $&$    3363 $&$    4834 $&$    3563 $ \\
     &{\pkpu}  &$  92.21 $&$   0.5 $&$   10412 $&$     271 $&$     527 $&$     364 $ \\
     &{\pkpd}  &$  93.22 $&$   0.6 $&$    6848 $&$     203 $&$     308 $&$     260 $ \\
     &{\pkpt}  &$  94.22 $&$   0.6 $&$    4373 $&$     128 $&$     202 $&$     161 $ \\
\hline
     & prescan &$  91.25 $&$   5.1 $&$  156592 $&$    5624 $&$    7563 $&$    6059 $ \\
     &{\pkmt}  &$  88.48 $&$   0.7 $&$    3646 $&$     297 $&$     176 $&$     166 $ \\
     &{\pkmd}  &$  89.47 $&$   0.8 $&$    7991 $&$     451 $&$     363 $&$     289 $ \\
     &{\pkmu}  &$  90.23 $&$   0.9 $&$   16011 $&$     683 $&$     744 $&$     569 $ \\
1991 &{\pk}    &$  91.22 $&$   3.0 $&$   92025 $&$    3365 $&$    4422 $&$    3603 $ \\
     &{\pkpu}  &$  91.97 $&$   0.8 $&$   20353 $&$     566 $&$     916 $&$     734 $ \\
     &{\pkpd}  &$  92.97 $&$   0.6 $&$    8356 $&$     325 $&$     478 $&$     436 $ \\
     &{\pkpt}  &$  93.72 $&$   0.9 $&$    9404 $&$     284 $&$     404 $&$     359 $ \\
\hline
1992 &{\pk}    &$  91.30 $&$  24.9 $&$  733059 $&$   23998 $&$   32492 $&$   27036 $ \\
\hline
\hline
1990--1992&Total&$        $&$  44.4 $&$ 1187330 $&$   40353 $&$   54085 $&$   44115 $ \\
\hline
\hline
     & prescan(a)
               &$  91.14 $&$   0.3 $&$    9905 $&$     345 $&$     454 $&$     370 $ \\
     & prescan(b)
               &$  91.32 $&$   5.3 $&$  162218 $&$    5256 $&$    7139 $&$    6002 $ \\
1993 &{\pkm}   &$  89.45 $&$   8.5 $&$   85727 $&$    4595 $&$    3884 $&$    3336 $ \\
     &{\pk}    &$  91.21 $&$   8.8 $&$  265494 $&$    8766 $&$   10871 $&$    9712 $ \\
     &{\pkp}   &$  93.04 $&$   9.0 $&$  125320 $&$    3549 $&$    5521 $&$    4612 $ \\
\hline
     &{\pk}(ab)&$  91.22 $&$  50.1 $&$ 1520277 $&$   49142 $&$   67791 $&$   55886 $ \\
1994 &{\pk}(c) &$  91.43 $&$   0.4 $&$   11255 $&$     345 $&$     500 $&$     381 $ \\
     &{\pk}(d) &$  91.22 $&$   2.3 $&$   69062 $&$    2170 $&$    3060 $&$    2478 $ \\
\hline
     & prescan(a)
               &$  91.80 $&$   0.2 $&$    5941 $&$     178 $&$     259 $&$     191 $ \\
     & prescan(b)
               &$  91.30 $&$   9.9 $&$  300676 $&$    9642 $&$   13401 $&$   11049 $ \\
1995 &{\pkm}   &$  89.44 $&$   8.4 $&$   84236 $&$    4407 $&$    3768 $&$    3185 $ \\
     &{\pk}    &$  91.28 $&$   4.6 $&$  140749 $&$    4623 $&$    6338 $&$    5262 $ \\
     &{\pkp}   &$  92.97 $&$   8.9 $&$  127707 $&$    3651 $&$    5696 $&$    4876 $ \\
\hline
\hline
1993--1995&Total&$        $&$ 116.7 $&$ 2908566 $&$   96669 $&$  128682 $&$  107340 $ \\
\hline
\hline
Grand &Total    &$        $&$ 161.1 $&$ 4095896 $&$  137022 $&$  182767 $&$  151455 $ \\
\hline
\end{tabular}
\caption[Summary of the data samples used for the cross-section measurements]{
Summary of the data samples used for the cross-section measurements,
showing the numbers of selected events for each final state at each energy 
point and the integrated luminosities~($\int{{\cal{L}}\,{\mathrm{d}}t}$)

for the \mbox{$\eehad$} analyses.
The integrated luminosities for the other final states vary within 
about 1\% due to different requirements on the 
status of the detector performance for the various event selections.
For the leptonic forward-backward asymmetry measurements the data samples 
available for analysis are generally larger since there is no reliance 
on the operation of the luminometers. The peak data from 1994, and 
from the 1993 and 1995 prescan data samples, have been divided into 
subsets corresponding to data-taking periods characterised by 
significantly different mean values of~$\roots$.
\label{tab-meas-data}
}
\end{center}
\end{table}

%
%
\begin{table}[htbp]  \begin{center}
 \begin{tabular} {|l|l|c|c|c|c|c|}
\hline
        Measurement & \multicolumn{1}{|c|}{process} &
       $E_{\mathrm{min}}$(GeV)  & $\thacol^{\mathrm{max}}$  &
       $\abscosthlm_{\mathrm{max}}$  & $(s'/s)_{\mathrm{min}}$  &
       $(m_\ff^2/s)_{\mathrm{min}}$  \\
\hline
               & $\eeqq$      & $-$  &    $-$     & 1.00 & 0.01 &  $-$  \\
cross-sections & $\eemumu$    & $-$  &    $-$     & 1.00 & $-$  & 0.01  \\
               & $\eetautau$  & $-$  &    $-$     & 1.00 & $-$  & 0.01  \\
               & $\eeee$      & 0.2  & $10^\circ$ & 0.70 & $-$  &  $-$  \\
\hline
               & $\eeee$      & 0.2  & $10^\circ$ & 0.70 & $-$  &  $-$  \\
asymmetries    & $\eemumu$    & 6.0  & $15^\circ$ & 0.95 & $-$  &  $-$  \\
               & $\eetautau$  & 6.0  & $15^\circ$ & 0.90 & $-$  &  $-$  \\
\hline
\end{tabular}
\caption[Ideal kinematic cuts for cross-sections and asymmetries]
{\label{tab-sel-kine}
The measured cross-sections and asymmetries are corrected to correspond
to ideal regions of phase-space adapted to theoretical calculations.
The phase-space is defined by the maximum $\abscosthlm$
in which the fermion must fall, and, either the minimum final-state
energy fraction, or a combined requirement on the minimum fermion energy,
$E_{\mathrm{min}}$(GeV), and $\thacol^{\mathrm{max}}$
where~$\thacol$ is the acollinearity angle of the fermion pair, defined as
\mbox{$180^{\circ}-\alpha$}, where $\alpha$ is the opening angle
between the directions of the two fermions.
The final-state energy fraction is defined in terms of either the
squared centre-of-mass energy available after initial-state radiation,
$(\sprime/s)$, or the final-state fermion pair mass squared, $(m_\ff^2/s)$.
The symbol~($-$) indicates that no requirements are made on the indicated
quantity.
}
\end{center}  \end{table}

%



\begin{table}[htb] 
\begin{sideways}
\begin{minipage}[b]{\textheight}
\begin{center}

{\footnotesize

\begin{tabular}{|l|c|c|c|c|c|c|c|c|c|c|c|c|c|c|}
\hline
   & \multicolumn{6}{|c|}{1993} & \multicolumn{2}{|c|}{1994} &
     \multicolumn{6}{|c|}{1995} \\
\cline{2-15}
                    & \multicolumn{2}{|c|}{\pkm}
                    & \multicolumn{2}{|c|}{\pk}  
                    & \multicolumn{2}{|c|}{\pkp}
                    & \multicolumn{2}{|c|}{\pk}  
                    & \multicolumn{2}{|c|}{\pkm}
                    & \multicolumn{2}{|c|}{\pk}  
                    & \multicolumn{2}{|c|}{\pkp}  \\
\cline{2-15}
 & $f$ & $\df$ & $f$ & $\df$ & $f$ & $\df$
 & $f$ & $\df$
 & $f$ & $\df$ & $f$ & $\df$ & $f$ & $\df$ \\
                    & & (\%) 
                    & & (\%) 
                    & & (\%) 
                    & & (\%) 
                    & & (\%) 
                    & & (\%) 
                    & & (\%) \\ 
\hline
$\eehad$ Monte Carlo
              &  1.00481  &  0.011  &  1.00481  &  0.011  &  1.00481  &  0.011  
                                   &  1.00481  &  0.011  
              &  1.00481  &  0.011  &  1.00481  &  0.011  &  1.00481  &  0.011 \\
ISR effects   &  1.00037 &  0.030  &  0.99997 &  0.003  &  1.00007 &  0.010
                                   &  0.99997 &  0.003 
              &  1.00037 &  0.030  &  0.99997 &  0.003  &  1.00007 &  0.010 \\
\hline
{\bf Acceptance}   & & & & & & & & & & & & & & \\
Hadronisation &  1.00055 &  0.048  &  1.00055 &  0.048  &  1.00055 &  0.048
                                   &  1.00055 &  0.048
              &  1.00055 &  0.048  &  1.00055 &  0.048  &  1.00055 &  0.048 \\
Detector simulation 
              &  0.99960 &  0.040  &  0.99960 &  0.040  &  0.99960 &  0.040
                                   &  0.99960 &  0.040
              &  0.99960 &  0.040  &  0.99960 &  0.040  &  0.99960 &  0.040  \\
Detector performance 
              & 1.00008  &  0.020  & 1.00008  &  0.020  & 1.00008  &  0.020  
                                   & 1.00008  &  0.020  
              & 1.00058  &  0.047  & 1.00058  &  0.047  & 1.00058  &  0.047 \\
\hline\hline
{\bf Corrected Acceptance}
              & 1.00542  &  0.071  & 1.00501  &  0.066  & 1.00511  &  0.067
                                   &  1.00501 &  0.066  
              & 1.00593  &  0.083  &  1.00551 &  0.079  & 1.00561  &  0.079 \\
\hline
\hline
{\bf Backgrounds}   & & & & & & & & & & & & & & \\
$\eetautau$   & 0.99841  &  0.020  & 0.99841  &  0.020  & 0.99841  &  0.020  
                                   & 0.99841  &  0.020  
              & 0.99841  &  0.020  & 0.99841  &  0.020  & 0.99841  &  0.020  \\
Non-resonant   & & & & & & & & & & & & & & \\
($0.051 \pm 0.007$ nb)
              & 0.99490   &  0.070  & 0.99833   &  0.023  & 0.99637   &  0.050 
                                    & 0.99833   &  0.023 
              & 0.99490   &  0.070  & 0.99833   &  0.023  & 0.99637   &  0.050 \\
Four fermion 
              & 0.99994  &  0.004  & 0.99996  &  0.003  & 0.99992  &  0.006 
                                   & 0.99996  &  0.003 
              & 0.99994  &  0.004  & 0.99996  &  0.003  & 0.99992  &  0.006 \\
$\eeee$ plus   & & & & & & & & & & & & & & \\
Cosmics 
              & 0.99988  &  0.009  & 0.99996  &  0.003  & 0.99992  &  0.006 
                                   & 0.99996  &  0.003 
              & 0.99988  &  0.009  & 0.99996  &  0.003  & 0.99992  &  0.006 \\
\hline\hline
{\bf Background Sum}
              & 0.99314  &  0.073  & 0.99666  &  0.031  & 0.99463  &  0.055 
                                   & 0.99666  &  0.031  
              & 0.99314  &  0.073  & 0.99666  &  0.031  & 0.99463  &  0.055 \\
\hline
\hline
{\bf Total Correction}
              & 0.99852  &  0.102  &1.00166   &  0.073  & 0.99971  &  0.086
                                   &1.00166   &  0.073
              & 0.99902  &  0.111  &1.00216   &  0.085  & 1.00021  &  0.096 \\
\hline

\end{tabular}
}

\caption[Correction factors for \mbox{$\eehad$} cross-section]
{\label{tab-mhcor}
Summary of the correction factors,~$f$, and their relative systematic 
errors,~$\df$, for the \mbox{$\eehad$} cross-section measurements.
These numbers, when multiplied by the number of events actually selected,
give the number of signal events which would have been observed in the ideal
acceptance described in {\TAB}~\protect\ref{tab-sel-kine}.
ISR effects encompass the off-peak acceptance change due to initial-state radiation 
and the contamination from events with $\sprime / s < 0.01$.
Hadronisation refers to the full correction and uncertainty resulting
from the acceptance hole emulation ({\TAB}~\ref{tab-had-hole}).
The error correlation between the energy points and data-taking years 
is specified in {\TAB}~\ref{tab:had_matrix}.
}
\end{center}
\end{minipage}
\end{sideways}
\end{table}

\begin{table}[htbp]  \begin{center}
 \begin{tabular} {|l|r|}
\hline
                            & Uncertainty ($\times 10^{-4}$) \\
\hline
Residual hadronisation model dependence:  & \\
~~~Inefficiency correction                 & $ 2.3 $  \\
~~~Inefficiency from barrel region         & $ 2.2 $  \\
Adjustment of cut variables                & $ 1.5 $  \\
Change of ECAL acceptance radius           & $ 0.4 $  \\
Barrel detector simulation                 & $ 3.3 $  \\
\hline
\bf{Total}                                 & $\bf{4.8} $  \\
\hline
\end{tabular}
\caption[Systematic errors of the acceptance hole emulation]
{\label{tab-had-hole}
Systematic errors on the selection inefficiency for \mbox{$\eehad$} 
events arising from uncertainties of the acceptance hole emulation.
}
\end{center}  \end{table}

%
\begin{table}[htb] 
\begin{sideways}
\begin{minipage}[b]{\textheight}
\begin{center}
\setlength{\tabcolsep}{1.6mm}

\begin{tabular}{|l|c|c|c|c|c|c|c|c|c|c|c|c|c|c|}
\hline
   & \multicolumn{6}{|c|}{1993} & \multicolumn{2}{|c|}{1994} &
     \multicolumn{6}{|c|}{1995} \\
\cline{2-15}
                    & \multicolumn{2}{|c|}{\pkm}
                    & \multicolumn{2}{|c|}{\pk}  
                    & \multicolumn{2}{|c|}{\pkp}
                    & \multicolumn{2}{|c|}{\pk}  
                    & \multicolumn{2}{|c|}{\pkm}
                    & \multicolumn{2}{|c|}{\pk}  
                    & \multicolumn{2}{|c|}{\pkp}  \\
\cline{2-15}
 & $f$ & $\df$ & $f$ & $\df$ & $f$ & $\df$
 & $f$ & $\df$
 & $f$ & $\df$ & $f$ & $\df$ & $f$ & $\df$ \\
                    & & (\%) 
                    & & (\%) 
                    & & (\%) 
                    & & (\%) 
                    & & (\%) 
                    & & (\%) 
                    & & (\%) \\ 
\hline

{\bf Monte Carlo} & & & & & & & & & & & & & & \\
$\eeee$ Monte Carlo     & 1.0063 & 0.06 & 1.0056 & 0.02 & 1.0061 & 0.04
                        & 1.0056 & 0.02 
                        & 1.0063 & 0.06 & 1.0056 & 0.02 & 1.0061 & 0.04 \\ 
\hline
{\bf Acceptance Correction}  & & & & & & & & & & & & & & \\
Electromagnetic energy  & 1.0009 & 0.10 & 1.0009 & 0.10 & 1.0009 & 0.10
                        & 1.0022 & 0.07
                        & 1.0017 & 0.09 & 1.0017 & 0.08 & 1.0017 & 0.09 \\
Electron identification & 1.0025 & 0.08 & 1.0025 & 0.08 & 1.0025 & 0.08
                        & 1.0026 & 0.05
                        & 1.0031 & 0.08 & 1.0031 & 0.08 & 1.0031 & 0.08 \\
Acceptance definition   & 1.0000 & 0.14 & 1.0000 & 0.09 & 1.0000 & 0.10
                        & 1.0000 & 0.09
                        & 1.0000 & 0.14 & 1.0000 & 0.09 & 1.0000 & 0.10 \\
Low multiplicity        & 1.0001 & 0.01 & 1.0001 & 0.01 & 1.0001 & 0.01
                        & 1.0001 & 0.01
                        & 1.0001 & 0.01 & 1.0001 & 0.01 & 1.0001 & 0.01 \\
\hline
{\bf Other Corrections} & & & & & & & & & & & & & & \\
Four-fermion events     & 1.0000 & 0.03 & 1.0000 & 0.02 & 1.0000 & 0.03
                        & 1.0000 & 0.02
                        & 1.0000 & 0.03 & 1.0000 & 0.02 & 1.0000 & 0.03 \\

\hline\hline
{\bf Signal Correction} & 1.0098 & 0.20 & 1.0091 & 0.16 & 1.0096 & 0.17
                        & 1.0105 & 0.13
                        & 1.0112 & 0.20 & 1.0105 & 0.15 & 1.0110 & 0.17 \\

\hline\hline
{\bf Backgrounds}  & & & & & & & & & & & & & & \\  
$\eetautau$             & 0.9982 & 0.04 & 0.9968 & 0.06 & 0.9965 & 0.07
                        & 0.9968 & 0.06
                        & 0.9982 & 0.04 & 0.9968 & 0.06 & 0.9965 & 0.07 \\
$\eegg$                 & 0.9999 & 0.01 & 0.9999 & 0.01 & 0.9999 & 0.02
                        & 0.9999 & 0.01
                        & 0.9999 & 0.01 & 0.9999 & 0.01 & 0.9999 & 0.02 \\
$\eehad$                & 0.9999 & 0.01 & 0.9999 & 0.02 & 0.9998 & 0.02
                        & 0.9999 & 0.02
                        & 0.9999 & 0.01 & 0.9999 & 0.02 & 0.9998 & 0.02 \\
$\eeeell$               & 1.0000 & 0.01 & 1.0000 & 0.01 & 1.0000 & 0.01
                        & 1.0000 & 0.01
                        & 1.0000 & 0.01 & 1.0000 & 0.01 & 1.0000 & 0.01 \\

\hline\hline
{\bf Background Correction} & 0.9979 & 0.04 & 0.9966 & 0.06 & 0.9961 & 0.08
                        & 0.9966 & 0.06
                        & 0.9979 & 0.04 & 0.9966 & 0.06 & 0.9961 & 0.08 \\

\hline
\hline
{\bf Total Correction Factor}  & 1.0078 & 0.21 & 1.0057 & 0.17 & 1.0057 & 0.19
                        & 1.0071 & 0.14  
                        & 1.0091&  0.20&  1.0070 & 0.16 & 1.0070 & 0.18 \\
\hline
\end{tabular}

\caption[Correction factors for \mbox{$\eeee$} cross-section]
{\label{tab-eecor}
Summary of the correction factors,~$f$, and their relative systematic 
errors,~$\df$, for the \mbox{$\eeee$} cross-section measurements.
The Monte Carlo correction factor corresponds to the efficiency for
events within the ideal phase space definition.
The factors listed under acceptance corrections take into account the
observed discrepancies between the data and Monte Carlo.
The total correction factor, when multiplied by the number
of events actually selected,
gives the number of signal events which would have been observed in the ideal
acceptance described in {\TAB}~\protect\ref{tab-sel-kine}.
The error correlation matrix is given in {\TAB}~\protect\ref{tab:ee_matrix}.
}
\end{center}

\end{minipage}
\end{sideways}
\end{table}

%
\begin{table}[p] 
\begin{sideways}
\begin{minipage}[b]{\textheight}
\begin{center}
\setlength{\tabcolsep}{1.6mm}
\begin{tabular}{|l|c|c|c|c|c|c|c|c|c|c|c|c|c|c|}
\hline
   & \multicolumn{6}{|c|}{1993} & \multicolumn{2}{|c|}{1994} &
     \multicolumn{6}{|c|}{1995} \\
\cline{2-15}
                    & \multicolumn{2}{|c|}{\pkm}
                    & \multicolumn{2}{|c|}{\pk}  
                    & \multicolumn{2}{|c|}{\pkp}
                    & \multicolumn{2}{|c|}{\pk}  
                    & \multicolumn{2}{|c|}{\pkm}
                    & \multicolumn{2}{|c|}{\pk}  
                    & \multicolumn{2}{|c|}{\pkp}  \\
\cline{2-15}
 & $f$ & \small$\df$ & $f$ & \small$\df$ & $f$ & \small$\df$
 & $f$ & \small$\df$
 & $f$ & \small$\df$ & $f$ & \small$\df$ & $f$ & \small$\df$ \\
                    & & (\%) 
                    & & (\%) 
                    & & (\%) 
                    & & (\%) 
                    & & (\%) 
                    & & (\%) 
                    & & (\%) \\ 
\hline
{\bf Monte Carlo} & & & & & & & & & & & & & & \\
 $\epem \, \rightarrow \mpmm$ Monte Carlo
                & 1.0995 & 0.10 & 1.0955 & 0.07 & 1.0986 & 0.10
                & 1.0948 & 0.04
                & 1.1032 & 0.12 & 1.0970 & 0.05 & 1.1001 & 0.10 \\
  $s^{\prime}$ cut correction
                & 0.9971 & --  & 0.9990 & --  & 0.9980 & --
                & 0.9990 & -- 
                & 0.9971 & --  & 0.9990 & --  & 0.9980 & --  \\
  Initial/final state interference
                & 1.0003 & --  & 1.0002 & --  & 1.0001 & --
                & 1.0002 & -- 
                & 1.0003 & --  & 1.0002 & --  & 1.0001 & --  \\ \hline

{\bf Acceptance Correction}  & & & & & & & & & & & & & & \\
 Tracking losses
                & 1.0046 & 0.06 & 1.0046 & 0.06 & 1.0046 & 0.06
                & 1.0042 & 0.04
                & 1.0043 & 0.06 & 1.0043 & 0.06 & 1.0043 & 0.06 \\
 Track multiplicity cuts
                & 0.9999 & 0.05 & 1.0007 & 0.04 & 1.0000 & 0.04
                & 1.0004 & 0.02
                & 1.0007 & 0.09 & 1.0010 & 0.04 & 1.0013 & 0.08 \\
 Muon identification
                & 1.0000 & 0.05 & 1.0000 & 0.05 & 1.0000 & 0.05
                & 1.0015 & 0.04
                & 1.0000 & 0.06 & 1.0000 & 0.06 & 1.0000 & 0.06 \\
 Acceptance definition
                & 1.0000 & 0.10 & 1.0000 & 0.10 & 1.0000 & 0.10
                & 1.0000 & 0.05
                & 1.0000 & 0.05 & 1.0000 & 0.05 & 1.0000 & 0.05 \\
\hline
{\bf Other Corrections} & & & & & & & & & & & & & & \\
 Trigger efficiency
                & 1.0006 & 0.02 & 1.0006 & 0.02 & 1.0006 & 0.02
                & 1.0005 & 0.02
                & 1.0002 & 0.02 & 1.0002 & 0.02 & 1.0002 & 0.02 \\
 Four-fermion events
                & 1.0009 & 0.01 & 1.0011 & 0.01 & 1.0011 & 0.01
                & 1.0011 & 0.01
                & 1.0009 & 0.01 & 1.0011 & 0.01 & 1.0011 & 0.01 \\
\hline
\hline
{\bf Signal Correction}  
                & 1.1032 & 0.17 & 1.1022 & 0.15 & 1.1034 & 0.17
                                & 1.1024 & 0.09
                & 1.1071 & 0.18 & 1.1034 & 0.12 & 1.1056 & 0.16 \\
\hline
\hline
{\bf Backgrounds}  & & & & & & & & & & & & & & \\
 $\epem \, \rightarrow \tptm$
                & 0.9914 & 0.02 & 0.9914 & 0.02 & 0.9914 & 0.02
                & 0.9903 & 0.04
                & 0.9905 & 0.02 & 0.9905 & 0.02 & 0.9905 & 0.02 \\
 $\epem \, \rightarrow \epem \mpmm$
                & 0.9988 & 0.01 & 0.9995 & 0.01 & 0.9991 & 0.01
                & 0.9996 & 0.01
                & 0.9987 & 0.01 & 0.9995 & 0.01 & 0.9990 & 0.01 \\
 Cosmic rays 
                & 0.9998 & 0.02 & 0.9998 & 0.02 & 0.9998 & 0.02
                & 0.9998 & 0.02
                & 0.9997 & 0.02 & 0.9997 & 0.02 & 0.9997 & 0.02 \\
\hline\hline
{\bf Background Correction} 
                & 0.9900 & 0.03 & 0.9907 & 0.03 & 0.9903 & 0.03
                                & 0.9897 & 0.05
                & 0.9889 & 0.03 & 0.9897 & 0.03 & 0.9892 & 0.03 \\
\hline
\hline
{\bf Total Correction Factor     }
                & 1.0922 & 0.17 & 1.0920 & 0.16 & 1.0927 & 0.17
                & 1.0910 & 0.10
                & 1.0948 & 0.18 & 1.0920 & 0.12 & 1.0937 & 0.17 \\
\hline
\end{tabular}

\caption[Correction factors for \mbox{$\eemumu$} cross-section]
{\label{tab-mmcor}
Summary of the correction factors,~$f$, and their relative systematic 
errors,~$\df$, for the \mbox{$\eemumu$} cross-section measurements.
These numbers, when multiplied by the number of events actually selected,
give the number of signal events which would have been observed in the ideal
acceptance described in {\TAB}~\protect\ref{tab-sel-kine}.
The effects tracking losses, track multiplicity cuts and muon identification
were, in principle, simulated by the Monte Carlo. The quoted
corrections were introduced to take into account the observed discrepancies
between the data and Monte Carlo for these effects.
The error correlation matrix is given in {\TAB}~\protect\ref{tab:mm_matrix}.
}
\end{center}

\end{minipage}
\end{sideways}
\end{table}

%

\begin{table}[htb] 
\begin{sideways}
\begin{minipage}[b]{\textheight}
\begin{center}
\setlength{\tabcolsep}{1.6mm}

\begin{tabular}{|l|c|c|c|c|c|c|c|c|c|c|c|c|c|c|}

\hline
   & \multicolumn{6}{|c|}{1993} & \multicolumn{2}{|c|}{1994} &
     \multicolumn{6}{|c|}{1995} \\
\cline{2-15}
                    & \multicolumn{2}{|c|}{\pkm}
                    & \multicolumn{2}{|c|}{\pk}  
                    & \multicolumn{2}{|c|}{\pkp}
                    & \multicolumn{2}{|c|}{\pk}  
                    & \multicolumn{2}{|c|}{\pkm}
                    & \multicolumn{2}{|c|}{\pk}  
                    & \multicolumn{2}{|c|}{\pkp}  \\
\cline{2-15}
 & $f$ & \small$\df$ & $f$ & \small$\df$ & $f$ & \small$\df$
 & $f$ & \small$\df$
 & $f$ & \small$\df$ & $f$ & \small$\df$ & $f$ & \small$\df$ \\
                    & & (\%) 
                    & & (\%) 
                    & & (\%) 
                    & & (\%) 
                    & & (\%) 
                    & & (\%) 
                    & & (\%) \\ 
\hline
{\bf Monte Carlo} & & & & & & & & & & & & & & \\
 $\eetautau$ Monte Carlo 
                    & 1.3384 & 0.22 & 1.3302 & 0.09 & 1.3388 & 0.19 
                                    & 1.3302 & 0.09 
                    & 1.3384 & 0.22 & 1.3302 & 0.09 & 1.3388 & 0.19 \\ 
$\sprime$ cut correction
                    & 0.9976 & --   & 0.9992 &  --  & 0.9984 &  -- 
                                    & 0.9992 &  --
                    & 0.9976 & --   & 0.9992 &  --  & 0.9984 &  --  \\
Initial/final state interference
                    & 1.0005 &  --  & 1.0004 &  --  & 1.0001 &  --
                                    & 1.0004 &  --
                    & 1.0005 &  --  & 1.0004 &  --  & 1.0001 &  --  \\
\hline
{\bf Acceptance Correction}    & & & & & & & & & & & & & & \\
Multiplicity cuts  
                    & 1.0018 & 0.16 & 1.0017 & 0.16 & 1.0019 & 0.17
                                    & 1.0049 & 0.14
                    & 1.0021 & 0.16 & 1.0021 & 0.16 & 1.0022 & 0.17 \\
Acollinearity and cone cuts
                    & 0.9999 & 0.25 & 1.0002 & 0.23 & 1.0007 & 0.28
                                    & 1.0034 & 0.19
                    & 1.0008 & 0.25 & 1.0008 & 0.23 & 1.0012 & 0.28 \\
Definition of $\costau$    
                    & 1.0000 & 0.10 & 1.0000 & 0.10 & 1.0000 & 0.10
                                    & 1.0000 & 0.10
                    & 1.0000 & 0.10 & 1.0000 & 0.10 & 1.0000 & 0.10 \\
$\eeee$ rejection 
                    & 1.0037 & 0.25 & 1.0038 & 0.26 & 1.0037 & 0.25
                                    & 1.0068 & 0.23
                    & 1.0044 & 0.25 & 1.0045 & 0.26 & 1.0043 & 0.25 \\
$\eemumu$ rejection 
                    & 0.9997 & 0.08 & 0.9997 & 0.08 & 0.9997 & 0.08
                                    & 0.9998 & 0.05
                    & 0.9993 & 0.08 & 0.9994 & 0.08 & 0.9993 & 0.08 \\
$\eeeell$ rejection 
                    & 1.0019 & 0.12 & 1.0018 & 0.11 & 1.0019 & 0.12
                                    & 1.0014 & 0.07
                    & 1.0000 & 0.10 & 1.0000 & 0.10 & 1.0000 & 0.10 \\
Cosmic ray cuts  
                    & 1.0001 & 0.01 & 1.0001 & 0.01 & 1.0001 & 0.01
                                    & 1.0001 & 0.01
                    & 1.0001 & 0.01 & 1.0001 & 0.01 & 1.0001 & 0.01 \\
Combinations of cuts
                    & 1.0000 & 0.09 & 1.0000 & 0.08 & 1.0000 & 0.10
                                    & 1.0000 & 0.08
                    & 1.0000 & 0.09 & 1.0000 & 0.08 & 1.0000 & 0.10 \\
\hline
{\bf Other Corrections} & & & & & & & & & & & & & & \\
Trigger efficiency  
                    & 1.0002 & 0.01 & 1.0002 & 0.01 & 1.0002 & 0.01
                                    & 1.0002 & 0.01
                    & 1.0002 & 0.01 & 1.0002 & 0.01 & 1.0002 & 0.01 \\
Tau branching ratios    
                    & 1.0000 & 0.05 & 1.0000 & 0.05 & 1.0000 & 0.05
                                    & 1.0000 & 0.05
                    & 1.0000 & 0.05 & 1.0000 & 0.05 & 1.0000 & 0.05 \\
Four-fermion events 
                    & 1.0012 & 0.04 & 1.0013 & 0.04 & 1.0018 & 0.04
                                    & 1.0013 & 0.04                 
                    & 1.0012 & 0.04 & 1.0013 & 0.04 & 1.0018 & 0.04 \\
\hline
\hline
{\bf Signal Correction}  
                    & 1.3473 & 0.50 & 1.3414 & 0.45 & 1.3502 & 0.51
                                    & 1.3536 & 0.39
                    & 1.3467 & 0.50 & 1.3409 & 0.44 & 1.3490 & 0.50 \\
\hline
\hline
{\bf Backgrounds }   & & & & & & & & & & & & & & \\
$\eeee$             & 0.9921 & 0.16 & 0.9966 & 0.07 & 0.9963 & 0.08
                                    & 0.9953 & 0.07
                    & 0.9906 & 0.16 & 0.9959 & 0.07 & 0.9957 & 0.08 \\
$\eemumu$           & 0.9900 & 0.12 & 0.9902 & 0.11 & 0.9902 & 0.11
                                    & 0.9886 & 0.09
                    & 0.9892 & 0.13 & 0.9893 & 0.12 & 0.9894 & 0.13 \\
$\eehad$            & 0.9961 & 0.10 & 0.9960 & 0.11 & 0.9961 & 0.10
                                    & 0.9960 & 0.11
                    & 0.9961 & 0.10 & 0.9960 & 0.11 & 0.9961 & 0.10 \\
$\eeeell$           & 0.9848 & 0.22 & 0.9948 & 0.07 & 0.9893 & 0.16
                                    & 0.9943 & 0.07
                    & 0.9838 & 0.22 & 0.9947 & 0.07 & 0.9887 & 0.16 \\
Four-fermion        & 0.9991 & 0.02 & 0.9994 & 0.02 & 0.9993 & 0.02
                                    & 0.9994 & 0.02
                    & 0.9991 & 0.02 & 0.9994 & 0.02 & 0.9993 & 0.02 \\
Cosmic rays         & 0.9991 & 0.05 & 0.9997 & 0.02 & 0.9994 & 0.04
                                    & 0.9996 & 0.02
                    & 0.9993 & 0.06 & 0.9998 & 0.02 & 0.9996 & 0.04 \\
\hline
\hline
{\bf Background Correction}& 0.9618 & 0.32 & 0.9768 & 0.19 & 0.9708 & 0.24
                                    & 0.9733 & 0.18
                    & 0.9587 & 0.32 & 0.9752 & 0.19 & 0.9690 & 0.25 \\
\hline
\hline
{\bf Total Correction Factor}
                    & 1.2960 & 0.59 & 1.3105 & 0.48 & 1.3108 & 0.56
                                    & 1.3178 & 0.42
                    & 1.2913 & 0.59 & 1.3077 & 0.48 & 1.3074 & 0.56 \\
\hline
\end{tabular}

\caption[Correction factors for \mbox{$\eetautau$} cross-section]
{\label{tab:ttfactor}
Summary of the correction factors,~$f$, and their relative systematic 
errors,~$\df$, for the \mbox{$\eetautau$} cross-section 
measurements.
These numbers, when multiplied by the number of events actually selected,
give the number of signal events which would have been observed in the ideal
acceptance described in {\TAB}~\protect\ref{tab-sel-kine}.
The error correlation matrix is given in {\TAB}~\ref{tab:ll_matrix}.
}
\end{center}

\end{minipage}
\end{sideways}
\end{table}

%
\begin{table}                                                                                                                       
\begin{center}                                                                                                                      
\begin{tabular}{|c l|c|c|r|r|r|}                                                                                                    
\hline                                                                                                                              
 & & \multicolumn{2}{|c|}{$\roots$(GeV)}                                                                                            
 &\multicolumn{3}{|c|}{$\eeqq$ cross-section (nb)} \\                                                                               
\cline{3-7}                                                                                                                         
 \multicolumn{2}{|c|}{Sample}& mean & rms & \multicolumn{1}{|c|}{measured}                                                          
& \multicolumn{1}{|c|}{corrected} & \multicolumn{1}{|c|}{fit} \\                                                                    
\cline{3-7}                                                                                                                         
\hline
       & peak-3     &$  88.2510 $&$   0.0481 $&$    4.669 \pm   0.110 $&$    4.667 $&$    4.605 $ \\
       & peak-2     &$  89.2510 $&$   0.0490 $&$    8.501 \pm   0.130 $&$    8.494 $&$    8.687 $ \\
       & peak-1     &$  90.2490 $&$   0.0500 $&$   18.899 \pm   0.281 $&$   18.890 $&$   18.713 $ \\
  1990 & peak       &$  91.2440 $&$   0.0510 $&$   30.445 \pm   0.130 $&$   30.488 $&$   30.500 $ \\
       & peak+1     &$  92.2350 $&$   0.0520 $&$   21.400 \pm   0.271 $&$   21.394 $&$   21.529 $ \\
       & peak+2     &$  93.2380 $&$   0.0529 $&$   12.434 \pm   0.180 $&$   12.427 $&$   12.450 $ \\
       & peak+3     &$  94.2350 $&$   0.0539 $&$    7.947 \pm   0.130 $&$    7.944 $&$    8.045 $ \\
\hline
       & prescan    &$  91.2540 $&$   0.0471 $&$   30.355 \pm   0.099 $&$   30.391 $&$   30.510 $ \\
       & peak-3     &$  88.4810 $&$   0.0441 $&$    5.326 \pm   0.095 $&$    5.323 $&$    5.258 $ \\
       & peak-2     &$  89.4720 $&$   0.0451 $&$   10.087 \pm   0.126 $&$   10.080 $&$   10.210 $ \\
       & peak-1     &$  90.2270 $&$   0.0461 $&$   18.243 \pm   0.171 $&$   18.234 $&$   18.398 $ \\
  1991 & peak       &$  91.2230 $&$   0.0471 $&$   30.370 \pm   0.129 $&$   30.407 $&$   30.469 $ \\
       & peak+1     &$  91.9690 $&$   0.0481 $&$   24.603 \pm   0.215 $&$   24.605 $&$   24.836 $ \\
       & peak+2     &$  92.9680 $&$   0.0490 $&$   14.058 \pm   0.178 $&$   14.051 $&$   14.304 $ \\
       & peak+3     &$  93.7170 $&$   0.0500 $&$    9.916 \pm   0.115 $&$    9.912 $&$    9.949 $ \\
\hline
  1992 & peak       &$  91.2990 $&$   0.0520 $&$   30.566 \pm   0.045 $&$   30.609 $&$   30.514 $ \\
\hline
       & peak-2     &$  89.4505 $&$   0.0564 $&$   10.053 \pm   0.037 $&$   10.042 $&$   10.048 $ \\
  1993 & peak       &$  91.2063 $&$   0.0570 $&$   30.352 \pm   0.070 $&$   30.407 $&$   30.433 $ \\
       & peak+2     &$  93.0351 $&$   0.0570 $&$   13.856 \pm   0.043 $&$   13.847 $&$   13.808 $ \\
\hline
       & peak(ab)   &$  91.2199 $&$   0.0565 $&$   30.379 \pm   0.029 $&$   30.433 $&$   30.463 $ \\
  1994 & peak(c)    &$  91.4287 $&$   0.0562 $&$   30.308 \pm   0.339 $&$   30.353 $&$   30.163 $ \\
       & peak(d)    &$  91.2195 $&$   0.0557 $&$   30.598 \pm   0.138 $&$   30.650 $&$   30.462 $ \\
\hline
       & peak-2     &$  89.4415 $&$   0.0568 $&$    9.989 \pm   0.037 $&$    9.978 $&$    9.981 $ \\
  1995 & peak       &$  91.2829 $&$   0.0578 $&$   30.559 \pm   0.097 $&$   30.614 $&$   30.520 $ \\
       & peak+2     &$  92.9715 $&$   0.0581 $&$   14.282 \pm   0.044 $&$   14.272 $&$   14.278 $ \\
\hline                                                                                                                              
\end{tabular}                                                                                                                       
\caption[The $\eeqq$ cross-section]{ The $\eeqq$ production                                                                         
cross-section near the $\Zzero$ resonance.                                                                                          
The cross-section is corrected to the simple kinematic acceptance region                                                            
defined by \mbox{$\sprime/s>0.01$.}                                                                                                 
For each data sample, we list here the mean $\roots$ of the                                                                         
colliding beams, its root-mean-square (rms) spread,                                                                                 
and the observed $\eeqq$ cross-section.                                                                                             
The errors shown are statistical only.                                                                                              
The cross-section measurements are also shown after being corrected                                                                 
for the beam energy spread to correspond to the physical cross-section                                                              
at the central value of $\roots$.                                                                                                   
The fit values are the result of the \SLPnine\ fit.                                                                                 
   \label{tab:hadron_xsec}                                                                                                          
}                                                                                                                                   
\end{center}                                                                                                                        
\end{table}                                                                                                                         

\begin{table}                                                                                                                       
\begin{center}                                                                                                                      
\begin{tabular}{|c l|c|c|r|r|r|}                                                                                                    
\hline                                                                                                                              
 & & \multicolumn{2}{|c|}{$\roots$(GeV)}                                                                                            
 &\multicolumn{3}{|c|}{$\eeee$ cross-section (nb)} \\                                                                               
\cline{3-7}                                                                                                                         
 \multicolumn{2}{|c|}{Sample}& mean & rms & \multicolumn{1}{|c|}{measured}                                                          
& \multicolumn{1}{|c|}{corrected} & \multicolumn{1}{|c|}{fit} \\                                                                    
\cline{3-7}                                                                                                                         
\hline
       & peak-3     &$  88.2510 $&$   0.0481 $&$   0.3520 \pm  0.0281 $&$   0.3519 $&$   0.3403 $ \\
       & peak-2     &$  89.2510 $&$   0.0490 $&$   0.4850 \pm  0.0281 $&$   0.4848 $&$   0.4795 $ \\
       & peak-1     &$  90.2490 $&$   0.0500 $&$   0.8110 \pm  0.0451 $&$   0.8109 $&$   0.7762 $ \\
  1990 & peak       &$  91.2420 $&$   0.0510 $&$   1.0120 \pm  0.0180 $&$   1.0132 $&$   1.0054 $ \\
       & peak+1     &$  92.2350 $&$   0.0520 $&$   0.6010 \pm  0.0371 $&$   0.6007 $&$   0.6307 $ \\
       & peak+2     &$  93.2380 $&$   0.0529 $&$   0.3650 \pm  0.0261 $&$   0.3648 $&$   0.3594 $ \\
       & peak+3     &$  94.2350 $&$   0.0539 $&$   0.2310 \pm  0.0210 $&$   0.2309 $&$   0.2429 $ \\
\hline
       & prescan    &$  91.2540 $&$   0.0471 $&$   0.9875 \pm  0.0140 $&$   0.9886 $&$   1.0036 $ \\
       & peak-3     &$  88.4810 $&$   0.0441 $&$   0.3623 \pm  0.0239 $&$   0.3622 $&$   0.3636 $ \\
       & peak-2     &$  89.4720 $&$   0.0451 $&$   0.5628 \pm  0.0280 $&$   0.5626 $&$   0.5282 $ \\
       & peak-1     &$  90.2270 $&$   0.0461 $&$   0.7617 \pm  0.0299 $&$   0.7616 $&$   0.7677 $ \\
  1991 & peak       &$  91.2230 $&$   0.0471 $&$   1.0093 \pm  0.0190 $&$   1.0104 $&$   1.0079 $ \\
       & peak+1     &$  91.9690 $&$   0.0481 $&$   0.6885 \pm  0.0300 $&$   0.6884 $&$   0.7422 $ \\
       & peak+2     &$  92.9680 $&$   0.0490 $&$   0.4165 \pm  0.0270 $&$   0.4163 $&$   0.4114 $ \\
       & peak+3     &$  93.7170 $&$   0.0500 $&$   0.3020 \pm  0.0180 $&$   0.3019 $&$   0.2920 $ \\
\hline
  1992 & peak       &$  91.2990 $&$   0.0520 $&$   1.0062 \pm  0.0065 $&$   1.0074 $&$   0.9958 $ \\
\hline
       & peak-2     &$  89.4505 $&$   0.0564 $&$   0.5414 \pm  0.0080 $&$   0.5411 $&$   0.5231 $ \\
  1993 & peak       &$  91.2063 $&$   0.0570 $&$   1.0064 \pm  0.0109 $&$   1.0080 $&$   1.0098 $ \\
       & peak+2     &$  93.0351 $&$   0.0570 $&$   0.3945 \pm  0.0067 $&$   0.3942 $&$   0.3973 $ \\
\hline
       & peak(ab)   &$  91.2197 $&$   0.0565 $&$   1.0047 \pm  0.0046 $&$   1.0063 $&$   1.0083 $ \\
  1994 & peak(c)    &$  91.4286 $&$   0.0562 $&$   0.9422 \pm  0.0512 $&$   0.9434 $&$   0.9635 $ \\
       & peak(d)    &$  91.2195 $&$   0.0557 $&$   0.9676 \pm  0.0210 $&$   0.9691 $&$   1.0083 $ \\
\hline
       & peak-2     &$  89.4415 $&$   0.0568 $&$   0.5260 \pm  0.0080 $&$   0.5257 $&$   0.5209 $ \\
  1995 & peak       &$  91.2829 $&$   0.0578 $&$   1.0074 \pm  0.0150 $&$   1.0089 $&$   0.9988 $ \\
       & peak+2     &$  92.9715 $&$   0.0581 $&$   0.4088 \pm  0.0068 $&$   0.4085 $&$   0.4107 $ \\
\hline                                                                                                                              
\end{tabular}                                                                                                                       
\caption[The $\eeee$ cross-section]{ The $\eeee$ production                                                                         
cross-section near the $\Zzero$ resonance.                                                                                          
The cross-section is corrected to the simple kinematic acceptance region                                                            
defined by \mbox{$\abscosthelec<0.70$} and \mbox{$\thacol<10^{\circ}$.}                                                             
                                                                                                                                    
For each data sample, we list here the mean $\roots$ of the                                                                         
colliding beams, its root-mean-square (rms) spread,                                                                                 
and the observed $\eeee$ cross-section.                                                                                             
The errors shown are statistical only.                                                                                              
The cross-section measurements are also shown after being corrected                                                                 
for the beam energy spread to correspond to the physical cross-section                                                              
at the central value of $\roots$.                                                                                                   
The fit values are the result of the \SLPnine\ fit.                                                                                 
   \label{tab:ee_xsec}                                                                                                              
}                                                                                                                                   
\end{center}                                                                                                                        
\end{table}                                                                                                                         

\begin{table}                                                                                                                       
\begin{center}                                                                                                                      
\begin{tabular}{|c l|c|c|r|r|r|}                                                                                                    
\hline                                                                                                                              
 & & \multicolumn{2}{|c|}{$\roots$(GeV)}                                                                                            
 &\multicolumn{3}{|c|}{$\eemumu$ cross-section (nb)} \\                                                                             
\cline{3-7}                                                                                                                         
 \multicolumn{2}{|c|}{Sample}& mean & rms & \multicolumn{1}{|c|}{measured}                                                          
& \multicolumn{1}{|c|}{corrected} & \multicolumn{1}{|c|}{fit} \\                                                                    
\cline{3-7}                                                                                                                         
\hline
       & peak-3     &$  88.2510 $&$   0.0481 $&$   0.2450 \pm  0.0243 $&$   0.2449 $&$   0.2345 $ \\
       & peak-2     &$  89.2510 $&$   0.0490 $&$   0.4181 \pm  0.0284 $&$   0.4178 $&$   0.4302 $ \\
       & peak-1     &$  90.2490 $&$   0.0500 $&$   0.8432 \pm  0.0477 $&$   0.8427 $&$   0.9116 $ \\
  1990 & peak       &$  91.2440 $&$   0.0510 $&$   1.4833 \pm  0.0221 $&$   1.4854 $&$   1.4779 $ \\
       & peak+1     &$  92.2350 $&$   0.0520 $&$   1.0841 \pm  0.0477 $&$   1.0838 $&$   1.0467 $ \\
       & peak+2     &$  93.2380 $&$   0.0529 $&$   0.5961 \pm  0.0345 $&$   0.5958 $&$   0.6102 $ \\
       & peak+3     &$  94.2350 $&$   0.0539 $&$   0.3970 \pm  0.0284 $&$   0.3969 $&$   0.3983 $ \\
\hline
       & prescan    &$  91.2540 $&$   0.0471 $&$   1.4851 \pm  0.0182 $&$   1.4869 $&$   1.4784 $ \\
       & peak-3     &$  88.4810 $&$   0.0441 $&$   0.2319 \pm  0.0202 $&$   0.2318 $&$   0.2657 $ \\
       & peak-2     &$  89.4720 $&$   0.0451 $&$   0.5171 \pm  0.0273 $&$   0.5167 $&$   0.5034 $ \\
       & peak-1     &$  90.2270 $&$   0.0461 $&$   0.9082 \pm  0.0353 $&$   0.9078 $&$   0.8965 $ \\
  1991 & peak       &$  91.2230 $&$   0.0471 $&$   1.4859 \pm  0.0232 $&$   1.4877 $&$   1.4764 $ \\
       & peak+1     &$  91.9690 $&$   0.0481 $&$   1.2447 \pm  0.0424 $&$   1.2448 $&$   1.2057 $ \\
       & peak+2     &$  92.9680 $&$   0.0490 $&$   0.6836 \pm  0.0354 $&$   0.6833 $&$   0.6994 $ \\
       & peak+3     &$  93.7170 $&$   0.0500 $&$   0.4794 \pm  0.0242 $&$   0.4792 $&$   0.4899 $ \\
\hline
  1992 & peak       &$  91.2990 $&$   0.0520 $&$   1.4781 \pm  0.0083 $&$   1.4802 $&$   1.4785 $ \\
\hline
       & peak-2     &$  89.4505 $&$   0.0564 $&$   0.4964 \pm  0.0081 $&$   0.4959 $&$   0.4956 $ \\
  1993 & peak       &$  91.2072 $&$   0.0568 $&$   1.4563 \pm  0.0142 $&$   1.4589 $&$   1.4748 $ \\
       & peak+2     &$  93.0351 $&$   0.0570 $&$   0.6681 \pm  0.0091 $&$   0.6677 $&$   0.6755 $ \\
\hline
       & peak(ab)   &$  91.2200 $&$   0.0565 $&$   1.4752 \pm  0.0058 $&$   1.4778 $&$   1.4761 $ \\
  1994 & peak(c)    &$  91.4286 $&$   0.0562 $&$   1.4786 \pm  0.0674 $&$   1.4808 $&$   1.4617 $ \\
       & peak(d)    &$  91.2195 $&$   0.0557 $&$   1.4762 \pm  0.0272 $&$   1.4787 $&$   1.4761 $ \\
\hline
       & peak-2     &$  89.4415 $&$   0.0568 $&$   0.4912 \pm  0.0081 $&$   0.4907 $&$   0.4923 $ \\
  1995 & peak       &$  91.2827 $&$   0.0578 $&$   1.5085 \pm  0.0193 $&$   1.5111 $&$   1.4789 $ \\
       & peak+2     &$  92.9715 $&$   0.0581 $&$   0.6894 \pm  0.0093 $&$   0.6889 $&$   0.6981 $ \\
\hline                                                                                                                              
\end{tabular}                                                                                                                       
\caption[The $\eemumu$ cross-section]{ The $\eemumu$ production                                                                     
cross-section near the $\Zzero$ resonance.                                                                                          
The cross-section is corrected to the simple kinematic acceptance region                                                            
defined by \mbox{$m_\ff^2/s>0.01$.}                                                                                                 
For each data sample, we list here the mean $\roots$ of the                                                                         
colliding beams, its root-mean-square (rms) spread,                                                                                 
and the observed $\eemumu$ cross-section.                                                                                           
The errors shown are statistical only.                                                                                              
The cross-section measurements are also shown after being corrected                                                                 
for the beam energy spread to correspond to the physical cross-section                                                              
at the central value of $\roots$.                                                                                                   
The fit values are the result of the \SLPnine\ fit.                                                                                 
   \label{tab:muon_xsec}                                                                                                            
}                                                                                                                                   
\end{center}                                                                                                                        
\end{table}                                                                                                                         

\begin{table}                                                                                                                       
\begin{center}                                                                                                                      
\begin{tabular}{|c l|c|c|r|r|r|}                                                                                                    
\hline                                                                                                                              
 & & \multicolumn{2}{|c|}{$\roots$(GeV)}                                                                                            
 &\multicolumn{3}{|c|}{$\eett$ cross-section (nb)} \\                                                                               
\cline{3-7}                                                                                                                         
 \multicolumn{2}{|c|}{Sample}& mean & rms & \multicolumn{1}{|c|}{measured}                                                          
& \multicolumn{1}{|c|}{corrected} & \multicolumn{1}{|c|}{fit} \\                                                                    
\cline{3-7}                                                                                                                         
\hline
       & peak-3     &$  88.2510 $&$   0.0481 $&$   0.2152 \pm  0.0245 $&$   0.2151 $&$   0.2343 $ \\
       & peak-2     &$  89.2510 $&$   0.0490 $&$   0.4304 \pm  0.0296 $&$   0.4301 $&$   0.4299 $ \\
       & peak-1     &$  90.2490 $&$   0.0500 $&$   0.9328 \pm  0.0633 $&$   0.9323 $&$   0.9109 $ \\
  1990 & peak       &$  91.2440 $&$   0.0510 $&$   1.4566 \pm  0.0245 $&$   1.4587 $&$   1.4767 $ \\
       & peak+1     &$  92.2350 $&$   0.0520 $&$   1.0292 \pm  0.0552 $&$   1.0289 $&$   1.0459 $ \\
       & peak+2     &$  93.2380 $&$   0.0529 $&$   0.6518 \pm  0.0409 $&$   0.6515 $&$   0.6097 $ \\
       & peak+3     &$  94.2350 $&$   0.0539 $&$   0.4095 \pm  0.0327 $&$   0.4094 $&$   0.3980 $ \\
\hline
       & prescan    &$  91.2540 $&$   0.0471 $&$   1.4325 \pm  0.0194 $&$   1.4343 $&$   1.4771 $ \\
       & peak-3     &$  88.4810 $&$   0.0441 $&$   0.2769 \pm  0.0235 $&$   0.2768 $&$   0.2655 $ \\
       & peak-2     &$  89.4720 $&$   0.0451 $&$   0.4845 \pm  0.0297 $&$   0.4841 $&$   0.5029 $ \\
       & peak-1     &$  90.2270 $&$   0.0461 $&$   0.8331 \pm  0.0368 $&$   0.8327 $&$   0.8957 $ \\
  1991 & peak       &$  91.2230 $&$   0.0471 $&$   1.4384 \pm  0.0256 $&$   1.4402 $&$   1.4751 $ \\
       & peak+1     &$  91.9690 $&$   0.0481 $&$   1.1892 \pm  0.0450 $&$   1.1893 $&$   1.2047 $ \\
       & peak+2     &$  92.9680 $&$   0.0490 $&$   0.6953 \pm  0.0409 $&$   0.6950 $&$   0.6988 $ \\
       & peak+3     &$  93.7170 $&$   0.0500 $&$   0.4989 \pm  0.0276 $&$   0.4987 $&$   0.4895 $ \\
\hline
  1992 & peak       &$  91.2990 $&$   0.0520 $&$   1.4734 \pm  0.0092 $&$   1.4755 $&$   1.4773 $ \\
\hline
       & peak-2     &$  89.4505 $&$   0.0564 $&$   0.5074 \pm  0.0092 $&$   0.5069 $&$   0.4951 $ \\
  1993 & peak       &$  91.2060 $&$   0.0570 $&$   1.5048 \pm  0.0158 $&$   1.5074 $&$   1.4734 $ \\
       & peak+2     &$  93.0351 $&$   0.0570 $&$   0.6660 \pm  0.0102 $&$   0.6656 $&$   0.6750 $ \\
\hline
       & peak(ab)   &$  91.2198 $&$   0.0565 $&$   1.4812 \pm  0.0065 $&$   1.4838 $&$   1.4749 $ \\
  1994 & peak(c)    &$  91.4286 $&$   0.0562 $&$   1.3708 \pm  0.0724 $&$   1.3730 $&$   1.4605 $ \\
       & peak(d)    &$  91.2195 $&$   0.0557 $&$   1.4477 \pm  0.0301 $&$   1.4502 $&$   1.4748 $ \\
\hline
       & peak-2     &$  89.4415 $&$   0.0568 $&$   0.4892 \pm  0.0091 $&$   0.4887 $&$   0.4919 $ \\
  1995 & peak       &$  91.2827 $&$   0.0578 $&$   1.4988 \pm  0.0213 $&$   1.5014 $&$   1.4776 $ \\
       & peak+2     &$  92.9715 $&$   0.0581 $&$   0.7089 \pm  0.0105 $&$   0.7084 $&$   0.6976 $ \\
\hline                                                                                                                              
\end{tabular}                                                                                                                       
\caption[The $\eett$ cross-section]{ The $\eett$ production                                                                         
cross-section near the $\Zzero$ resonance.                                                                                          
The cross-section is corrected to the simple kinematic acceptance region                                                            
defined by \mbox{$m_\ff^2/s>0.01$.}                                                                                                 
For each data sample, we list here the mean $\roots$ of the                                                                         
colliding beams, its root-mean-square (rms) spread,                                                                                 
and the observed $\eett$ cross-section.                                                                                             
The errors shown are statistical only.                                                                                              
The cross-section measurements are also shown after being corrected                                                                 
for the beam energy spread to correspond to the physical cross-section                                                              
at the central value of $\roots$.                                                                                                   
The fit values are the result of the \SLPnine\ fit.                                                                                 
   \label{tab:tau_xsec}                                                                                                             
}                                                                                                                                   
\end{center}                                                                                                                        
\end{table}                                                                                                                         

\begin{table}                                                                                                                       
\setlength{\tabcolsep}{1.0mm}                                                                                                       
\begin{center}                                                                                                                      
\begin{tabular}{|l|c|c|r|r|r|r|}                                                                                                    
\hline                                                                                                                              
 & \multicolumn{2}{|c|}{$\roots$ (GeV)}                                                                                             
 & \multicolumn{1}{|c|}{input (nb)}                                                                                                 
 & \multicolumn{3}{|c|}{pseudo-cross-sections (nb)}  \\                                                                             
\cline{2-7}                                                                                                                         
 \multicolumn{1}{|c|}{Sample} & mean & rms & \multicolumn{1}{|c|}{$\eeqq$}&                                                         
 \multicolumn{1}{|c|}{$\eeee$}                                                                                                      
 & \multicolumn{1}{|c|}{$\eemumu$} & \multicolumn{1}{|c|}{$\eett$}  \\                                                              
\hline
  1993 & & & & & & \\
 \small prescan(a) &$  91.1386 $&$   0.0867 $&$   30.176 \pm   0.306 $&$    1.056 \pm   0.057 $&$ 
   1.499 \pm   0.071 $&$    1.444 \pm   0.078 $ \\
 \small prescan(b) &$  91.3211 $&$   0.0566 $&$   30.479 \pm   0.076 $&$    0.988 \pm   0.014 $&$ 
   1.461 \pm   0.018 $&$    1.471 \pm   0.020 $ \\
 \small peak       &$  91.1964 $&$   0.0576 $&$   30.383 \pm   0.198 $&$    1.024 \pm   0.036 $&$ 
   1.420 \pm   0.045 $&$    1.435 \pm   0.050 $ \\
\hline
  1994 & & & & & & \\
 \small peak(ab)   &$  91.2183 $&$   0.0568 $&$   30.437 \pm   0.122 $&$    1.051 \pm   0.023 $&$ 
   1.469 \pm   0.028 $&$    1.426 \pm   0.031 $ \\
\hline
  1995 & & & & & & \\
 \small prescan(a) &$  91.7990 $&$   0.0568 $&$   26.910 \pm   0.352 $&$    0.810 \pm   0.061 $&$ 
   1.278 \pm   0.080 $&$    1.129 \pm   0.084 $ \\
 \small prescan(b) &$  91.3030 $&$   0.0576 $&$   30.501 \pm   0.056 $&$    0.983 \pm   0.010 $&$ 
   1.481 \pm   0.013 $&$    1.463 \pm   0.014 $ \\
\hline                                                                                                                              
\end{tabular}                                                                                                                       
\caption[The pseudo cross-sections]{ For some periods of running at the                                                             
$\Zzero$ peak a precise luminosity measurement is not available.                                                                    
We nevertheless use this data to measure inter-species cross-section ratios.                                                        
To make these data compatible with other cross-section measurements,                                                                
we fix each $\eeqq$ cross-section to its                                                                                            
expected value, and normalise the cross-sections for the three lepton species                                                       
measured in the same period to this arbitrary hadron cross-section.                                                                 
We term such measurements pseudo-cross-sections, and allow the absolute                                                             
scale of each set of four to float by 10\% in the fit using appropriate                                                             
error matrices.                                                                                                                     
For each data sample we list here the mean $\roots$ of the                                                                          
colliding beams, its rms spread, and the observed pseudo-cross-sections.                                                            
The errors shown are statistical only.                                                                                              
   \label{tab:pseudo_xsec}                                                                                                          
}                                                                                                                                   
\end{center}                                                                                                                        
\end{table}                                                                                                                         

%
\clearpage
%
\begin{table}                                                                                                                       
\begin{center}                                                                                                                      
\begin{tabular}{|r|rrrrrrr|rrrrrrrr|}                                                                                                    
\hline                                                                                                                              
%
 \multicolumn{16}{|c|}{Centre-of-mass energy errors}     \\
\hline
   & \multicolumn{7}{|c|}{1990} & \multicolumn{8}{|c|}{1991} \\
   & $-3$& $-2$& $-1$& 0& $+1$& $+2$& $+3$& 0& $-3$& $-2$& $-1$& 0& $+1$& $+2$& $+3$ \\
\hline
$-3$      &  193 &194 &195 &196 &197 &198 &199 &  6.1 & 7.4 &7.0 &6.6 &6.1 &5.7 &5.1& 4.7  \\
$-2$	  &  194 &195 &196 &197 &198 &199 &200 &  5.8 & 6.7 &6.4 &6.1 &5.8 &5.5 &5.2& 4.9  \\
$-1$	  &  195 &196 &197 &198 &199 &200 &201 &  5.5 & 5.9 &5.8 &5.6 &5.5 &5.4 &5.2& 5.0  \\
1990 ~~0  &  196 &197 &198 &199 &200 &202 &203 &  5.1 & 5.0 &5.1 &5.1 &5.1 &5.2 &5.2& 5.2 \\
$+1$	  &  197 &198 &199 &200 &202 &203 &204 &  4.8 & 4.0 &4.3 &4.5 &4.8 &5.0 &5.2& 5.4  \\
$+2$	  &  198 &199 &200 &202 &203 &204 &205 &  4.4 & 2.4 &3.3 &3.8 &4.4 &4.8 &5.2& 5.6  \\
$+3$	  &  199 &200 &201 &203 &204 &205 &206 &  4.0&$-$1.9&1.8 &2.9 &3.9 &4.6 &5.3& 5.8  \\ \hline
     pre  &    6.1 &  5.8 &  5.5 &  5.1 &  4.8 &  4.4 &  4.0 & 20.3 & 6.6 &6.3 &6.1 &5.8 &5.6 &5.2& 5.0  \\
$-3$	  &    7.4 &  6.7 &  5.9 &  5.0 &  4.0 &  2.4 &$-$1.9&  6.6 & 9.7 &7.8 &7.3 &6.6 &6.0 &5.2& 4.4  \\
$-2$	  &    7.0 &  6.4 &  5.8 &  5.1 &  4.3 &  3.3 &  1.8 &  6.3 & 7.8 &8.8 &6.9 &6.3 &5.9 &5.2& 4.6  \\
$-1$	  &    6.6 &  6.1 &  5.6 &  5.1 &  4.5 &  3.8 &  2.9 &  6.1 & 7.3 &6.9 &8.2 &6.1 &5.7 &5.2& 4.8  \\
1991 ~~0  &    6.1 &  5.8 &  5.5 &  5.1 &  4.8 &  4.4 &  3.9 &  5.8 & 6.6 &6.3 &6.1 &7.6 &5.6 &5.2& 4.9 \\
$+1$	  &    5.7 &  5.5 &  5.4 &  5.2 &  5.0 &  4.8 &  4.6 &  5.6 & 6.0 &5.9 &5.7 &5.6 &7.3 &5.2& 5.1  \\
$+2$	  &    5.1 &  5.2 &  5.2 &  5.2 &  5.2 &  5.2 &  5.3 &  5.2 & 5.2 &5.2 &5.2 &5.2 &5.2 &7.3& 5.3  \\
$+3$	  &    4.7 &  4.9 &  5.0 &  5.2 &  5.4 &  5.6 &  5.8 &  5.0 & 4.4 &4.6 &4.8 &4.9 &5.1 &5.3& 7.4  \\ \hline
\end{tabular}                                                                                                                       
\caption[The LEP centre-of-mass energy covariance matrix for 1990--1992]
{
The signed square-root of the covariance matrix elements for systematic errors
in the LEP centre-of-mass energy calibration for 1990--1991.
The column and row headings label the centre-of-mass energies relative to
the peak in GeV.
The entries are the absolute errors in units of MeV.
The energy measurement in 1992 is uncorrelated with other years.  Its error is
 18.0 MeV.
   \label{tab:ecm9092_matrix}                                                                                                          
}                                                                                                                                   
\end{center}                                                                                                                        
\end{table}                                                                                                                         

\begin{table}                                                                                                                       
\begin{center}                                                                                                                      
\begin{tabular}{|l|rrrrrrr|}                                                                                                    
\hline                                                                                                                              
\multicolumn{8}{|c|}{ Centre-of-mass energy errors} \\
\hline
               &   1993  &    1993  &    1993  &    1994  &    1995  &    1995  &    1995  \\
               & pk$-2$  &  peak    &  pk$+2$  &  peak    &  pk$-2$  &  peak    &  pk$+2$  \\
\hline
     1993   pk$-2$&$    3.55 $&$    2.88 $&$    2.73 $&$    2.36 $&$    1.31 $&$    1.21 $&$    1.22 $  \\                          
     1993   peak  &$    2.88 $&$    6.76 $&$    2.77 $&$    2.49 $&$    1.16 $&$    1.22 $&$    1.17 $  \\                          
     1993   pk$+2$&$    2.73 $&$    2.77 $&$    3.09 $&$    2.28 $&$    1.25 $&$    1.27 $&$    1.35 $  \\                          
     1994   peak  &$    2.36 $&$    2.49 $&$    2.28 $&$    3.78 $&$    1.25 $&$    1.32 $&$    1.26 $  \\                          
     1995   pk$-2$&$    1.31 $&$    1.16 $&$    1.25 $&$    1.25 $&$    1.83 $&$    1.28 $&$    1.26 $  \\                          
     1995   peak  &$    1.21 $&$    1.22 $&$    1.27 $&$    1.32 $&$    1.28 $&$    5.41 $&$    1.38 $  \\                          
     1995   pk$+2$&$    1.22 $&$    1.17 $&$    1.35 $&$    1.26 $&$    1.26 $&$    1.38 $&$    1.74 $  \\                          
\hline                                                                                                                              
\end{tabular}                                                                                                                       
\caption[The LEP centre-of-mass energy covariance matrix for 1993--1995]
{
The signed square-root of the covariance matrix elements for systematic
errors in the LEP centre-of-mass energy in 1993 -- 1995.
These errors are specific to the OPAL interaction point.
The entries are the absolute errors in units of MeV.
   \label{tab:ecm9395_matrix}                                                                                                          
}                                                                                                                                   
\end{center}                                                                                                                        
\end{table}                                                                                                                         

\begin{table}                                                                                                                       
\begin{center}                                                                                                                      
\begin{tabular}{|l|rrrrrrrrrr|}                                                                                                    
\hline                                                                                                                              
\multicolumn{11}{|c|}{Centre-of-mass energy spread errors} \\
\hline
               &   1990  &    1991  &    1992  &    1993  &    1993  &    1993  &    1994  &    1995  &    1995  &    1995  \\
               &         &          &          &  pk$-2$  &  peak    &  pk$+2$  &  peak    &  pk$-2$  &  peak    &  pk$+2$  \\
\hline
     1990         &$  3.0 $&$  3.0 $&$  0.0 $&$  0.0 $&$  0.0 $&$  0.0 $&$  0.0 $&$  0.0 $&$  0.0 $&$  0.0 $  \\                    
     1991         &$  3.0 $&$  3.0 $&$  0.0 $&$  0.0 $&$  0.0 $&$  0.0 $&$  0.0 $&$  0.0 $&$  0.0 $&$  0.0 $  \\                    
     1992         &$  0.0 $&$  0.0 $&$  3.0 $&$  1.8 $&$  1.8 $&$  1.8 $&$  1.8 $&$  1.8 $&$  1.8 $&$  1.8 $  \\                    
     1993   pk$-2$&$  0.0 $&$  0.0 $&$  1.8 $&$  1.1 $&$  1.1 $&$  1.1 $&$  1.1 $&$  1.1 $&$  1.1 $&$  1.1 $  \\                    
     1993   peak  &$  0.0 $&$  0.0 $&$  1.8 $&$  1.1 $&$  1.1 $&$  1.1 $&$  1.1 $&$  1.1 $&$  1.1 $&$  1.1 $  \\                    
     1993   pk$+2$&$  0.0 $&$  0.0 $&$  1.8 $&$  1.1 $&$  1.1 $&$  1.1 $&$  1.1 $&$  1.1 $&$  1.1 $&$  1.1 $  \\                    
     1994   peak  &$  0.0 $&$  0.0 $&$  1.8 $&$  1.1 $&$  1.1 $&$  1.1 $&$  1.1 $&$  1.1 $&$  1.1 $&$  1.1 $  \\                    
     1995   pk$-2$&$  0.0 $&$  0.0 $&$  1.8 $&$  1.1 $&$  1.1 $&$  1.1 $&$  1.1 $&$  1.3 $&$  1.3 $&$  1.3 $  \\                    
     1995   peak  &$  0.0 $&$  0.0 $&$  1.8 $&$  1.1 $&$  1.1 $&$  1.1 $&$  1.1 $&$  1.3 $&$  1.3 $&$  1.3 $  \\                    
     1995   pk$+2$&$  0.0 $&$  0.0 $&$  1.8 $&$  1.1 $&$  1.1 $&$  1.1 $&$  1.1 $&$  1.3 $&$  1.3 $&$  1.3 $  \\                    
\hline                                                                                                                              
\end{tabular}                                                                                                                       
\caption[The LEP energy spread covariance matrix]
{
The signed square-root of the covariance matrix elements for systematic
errors in the LEP centre-of-mass energy spread.
The entries are the absolute errors in units of MeV.
   \label{tab:esigrms_matrix}                                                                                                          
}                                                                                                                                   
\end{center}                                                                                                                        
\end{table}                                                                                                                         

\begin{table}                                                                                                                       
\begin{center}                                                                                                                      
\begin{tabular}{|l|rrrrrrrrrr|}                                                                                                    
\hline                                                                                                                              
 \multicolumn{11}{|c|}{Luminosity errors} \\
\hline
               &   1990  &    1991  &    1992  &    1993  &    1993  &    1993  &    1994  &    1995  &    1995  &    1995  \\
               &         &          &          &  pk$-2$  &  peak    &  pk$+2$  &  peak    &  pk$-2$  &  peak    &  pk$+2$  \\
\hline
     1990         &$ 3000.00 $&$   30.91 $&$   30.91 $&$    5.40 $&$    5.40 $&
$    5.40 $&$    5.40 $&$    5.40 $&$    5.40 $&$    5.40 $  \\                                                                     
     1991         &$   30.91 $&$   60.25 $&$   30.91 $&$    5.40 $&$    5.40 $&                                                    
$    5.40 $&$    5.40 $&$    5.40 $&$    5.40 $&$    5.40 $  \\                                                                     
     1992         &$   30.91 $&$   30.91 $&$   41.82 $&$    5.40 $&$    5.40 $&                                                    
$    5.40 $&$    5.40 $&$    5.40 $&$    5.40 $&$    5.40 $  \\                                                                     
     1993   pk$-2$&$    5.40 $&$    5.40 $&$    5.40 $&$    6.52 $&$    6.27 $&                                                    
$    6.09 $&$    6.26 $&$    6.44 $&$    6.26 $&$    6.08 $  \\                                                                     
     1993   peak  &$    5.40 $&$    5.40 $&$    5.40 $&$    6.27 $&$    6.52 $&                                                    
$    6.27 $&$    6.44 $&$    6.26 $&$    6.44 $&$    6.26 $  \\                                                                     
     1993   pk$+2$&$    5.40 $&$    5.40 $&$    5.40 $&$    6.09 $&$    6.27 $&                                                    
$    6.52 $&$    6.26 $&$    6.08 $&$    6.26 $&$    6.44 $  \\                                                                     
     1994   peak  &$    5.40 $&$    5.40 $&$    5.40 $&$    6.26 $&$    6.44 $&                                                    
$    6.26 $&$    6.49 $&$    6.26 $&$    6.44 $&$    6.26 $  \\                                                                     
     1995   pk$-2$&$    5.40 $&$    5.40 $&$    5.40 $&$    6.44 $&$    6.26 $&                                                    
$    6.08 $&$    6.26 $&$    6.59 $&$    6.33 $&$    6.15 $  \\                                                                     
     1995   peak  &$    5.40 $&$    5.40 $&$    5.40 $&$    6.26 $&$    6.44 $&                                                    
$    6.26 $&$    6.44 $&$    6.33 $&$    6.58 $&$    6.33 $  \\                                                                     
     1995   pk$+2$&$    5.40 $&$    5.40 $&$    5.40 $&$    6.08 $&$    6.26 $&                                                    
$    6.44 $&$    6.26 $&$    6.15 $&$    6.33 $&$    6.58 $  \\                                                                     
\hline                                                                                                                              
\end{tabular}                                                                                                                       
\caption[The luminosity systematic covariance matrix]
{
The signed square-root of the covariance matrix elements for systematic
errors in the luminosity measurement.
The entries are the relative errors in units of $10^{-4}$.
The theoretical error in the calculated luminometer acceptance is included.
The large error for the 1990 luminosity has been artificially inflated,
as explained in the text.
   \label{tab:lumi_matrix}                                                                                                          
}                                                                                                                                   
\end{center}                                                                                                                        
\end{table}                                                                                                                         

\begin{table}                                                                                                                       
\begin{center}                                                                                                                      
\begin{tabular}{|l|rrrrrrrrrr|}                                                                                                    
\hline                                                                                                                              
\multicolumn{11}{|c|}{$\eeqq$ cross-section errors} \\
\hline
               &   1990  &    1991  &    1992  &    1993  &    1993  &    1993  &    1994  &    1995  &    1995  &    1995  \\
               &         &          &          &  pk$-2$  &  peak    &  pk$+2$  &  peak    &  pk$-2$  &  peak    &  pk$+2$  \\
\hline
     1990         &$   40.00 $&$   19.00 $&$   19.00 $&$   10.51 $&$   10.51 $&                                                     
$   10.51 $&$   10.51 $&$   10.51 $&$   10.51 $&$   10.51 $  \\                                                                     
     1991         &$   19.00 $&$   19.00 $&$   19.00 $&$    7.27 $&$    7.27 $&                                                     
$    7.27 $&$    7.27 $&$    7.27 $&$    7.27 $&$    7.27 $  \\                                                                     
     1992         &$   19.00 $&$   19.00 $&$   19.00 $&$    7.27 $&$    7.27 $&                                                     
$    7.27 $&$    7.27 $&$    7.27 $&$    7.27 $&$    7.27 $  \\                                                                     
     1993   pk$-2$&$   10.51 $&$    7.27 $&$    7.27 $&$   10.24 $&$    8.05 $&                                                     
$    9.28 $&$    7.80 $&$   10.04 $&$    7.80 $&$    9.07 $  \\                                                                     
     1993   peak  &$   10.51 $&$    7.27 $&$    7.27 $&$    8.05 $&$    7.32 $&                                                     
$    7.75 $&$    7.04 $&$    7.80 $&$    7.04 $&$    7.49 $  \\                                                                     
     1993   pk$+2$&$   10.51 $&$    7.27 $&$    7.27 $&$    9.28 $&$    7.75 $&                                                     
$    8.65 $&$    7.49 $&$    9.07 $&$    7.49 $&$    8.41 $  \\                                                                     
     1994   peak  &$   10.51 $&$    7.27 $&$    7.27 $&$    7.80 $&$    7.04 $&                                                     
$    7.49 $&$    7.32 $&$    7.80 $&$    7.04 $&$    7.49 $  \\                                                                     
     1995   pk$-2$&$   10.51 $&$    7.27 $&$    7.27 $&$   10.04 $&$    7.80 $&                                                     
$    9.07 $&$    7.80 $&$   11.08 $&$    9.10 $&$   10.21 $  \\                                                                     
     1995   peak  &$   10.51 $&$    7.27 $&$    7.27 $&$    7.80 $&$    7.04 $&                                                     
$    7.49 $&$    7.04 $&$    9.10 $&$    8.46 $&$    8.84 $  \\                                                                     
     1995   pk$+2$&$   10.51 $&$    7.27 $&$    7.27 $&$    9.07 $&$    7.49 $&                                                     
$    8.41 $&$    7.49 $&$   10.21 $&$    8.84 $&$    9.63 $  \\                                                                     
\hline                                                                                                                              
\end{tabular}                                                                                                                       
\caption[The $\eeqq$ cross-section systematic covariance matrix]
{
The signed square-root of the covariance matrix elements for systematic
errors in the measurement of the $\eeqq$ cross-section.
The hadrons are not appreciably correlated with any of the leptons.
The entries are the relative errors in units of $10^{-4}$.
They do not include errors in the luminosity measurement, which are specified
separately in {\TAB}~\ref{tab:lumi_matrix}.
   \label{tab:had_matrix}                                                                                                          
}                                                                                                                                   
\end{center}                                                                                                                        
\end{table}                                                                                                                         

\begin{table}                                                                                                                       
\begin{center}                                                                                                                      
\begin{tabular}{|l|rrrrrrrrrr|}                                                                                                    
\hline
\multicolumn{11}{|c|}{$\eeee$ cross-section errors} \\
\hline
               &   1990  &    1991  &    1992  &    1993  &    1993  &    1993  &    1994  &    1995  &    1995  &    1995  \\
               &         &          &          &  pk$-2$  &  peak    &  pk$+2$  &  peak    &  pk$-2$  &  peak    &  pk$+2$  \\
\hline
     1990         &$ 70.0 $&$ 22.0 $&$ 22.0 $&$ 14.0 $&$ 14.0 $&$ 14.0 $&$ 14.0 $&$ 14.0 $&$ 14.0 $&$ 14.0 $ \\
     1991         &$ 22.0 $&$ 22.0 $&$ 22.0 $&$ 14.0 $&$ 14.0 $&$ 14.0 $&$ 14.0 $&$ 14.0 $&$ 14.0 $&$ 14.0 $ \\
     1992         &$ 22.0 $&$ 22.0 $&$ 23.0 $&$ 14.0 $&$ 14.0 $&$ 14.0 $&$ 14.0 $&$ 14.0 $&$ 14.0 $&$ 14.0 $ \\
     1993   pk$-2$&$ 14.0 $&$ 14.0 $&$ 14.0 $&$ 20.0 $&$ 17.0 $&$ 18.0 $&$ 15.0 $&$ 18.0 $&$ 15.0 $&$ 16.0 $ \\
     1993   peak  &$ 14.0 $&$ 14.0 $&$ 14.0 $&$ 17.0 $&$ 17.0 $&$ 17.0 $&$ 14.0 $&$ 15.0 $&$ 14.0 $&$ 15.0 $ \\
     1993   pk$+2$&$ 14.0 $&$ 14.0 $&$ 14.0 $&$ 18.0 $&$ 17.0 $&$ 19.0 $&$ 15.0 $&$ 16.0 $&$ 15.0 $&$ 17.0 $ \\
     1994   peak  &$ 14.0 $&$ 14.0 $&$ 14.0 $&$ 15.0 $&$ 14.0 $&$ 15.0 $&$ 14.0 $&$ 15.0 $&$ 14.0 $&$ 15.0 $ \\
     1995   pk$-2$&$ 14.0 $&$ 14.0 $&$ 14.0 $&$ 18.0 $&$ 15.0 $&$ 16.0 $&$ 15.0 $&$ 19.0 $&$ 17.0 $&$ 17.0 $ \\
     1995   peak  &$ 14.0 $&$ 14.0 $&$ 14.0 $&$ 15.0 $&$ 14.0 $&$ 15.0 $&$ 14.0 $&$ 17.0 $&$ 16.0 $&$ 16.0 $ \\
     1995   pk$+2$&$ 14.0 $&$ 14.0 $&$ 14.0 $&$ 16.0 $&$ 15.0 $&$ 17.0 $&$ 15.0 $&$ 17.0 $&$ 16.0 $&$ 18.0 $ \\
\hline                                                                                                                              
\end{tabular}                                                                                                                       
\caption[The $\eeee$ cross-section systematic covariance matrix]
{
The signed square-root of the covariance matrix elements for systematic
errors in the measurement of the $\eeee$ cross-section.
Additional terms arise between electrons and taus, which are given
in {\TAB}~\ref{tab:ll_matrix}.
The entries are the relative errors in units of $10^{-4}$.
They do not include errors in the luminosity measurement, which are specified
separately in {\TAB}~\ref{tab:lumi_matrix}.
   \label{tab:ee_matrix}                                                                                                          
}                                                                                                                                   
\end{center}                                                                                                                        
\end{table}                                                                                                                         

\begin{table}                                                                                                                       
\begin{center}                                                                                                                      
\begin{tabular}{|l|rrrrrrrrrr|}                                                                                                    
\hline                                                                                                                              
\multicolumn{11}{|c|}{$\eemumu$ cross-section errors} \\
\hline
               &   1990  &    1991  &    1992  &    1993  &    1993  &    1993  &    1994  &    1995  &    1995  &    1995  \\
               &         &          &          &  pk$-2$  &  peak    &  pk$+2$  &  peak    &  pk$-2$  &  peak    &  pk$+2$  \\
\hline
     1990         &$ 50.0 $&$ 16.0 $&$ 16.0 $&$  9.0 $&$  9.0 $&$  9.0 $&$  9.0 $&$  9.0 $&$  9.0 $&$  9.0 $ \\
     1991         &$ 16.0 $&$ 25.0 $&$ 16.0 $&$  9.0 $&$  9.0 $&$  9.0 $&$  9.0 $&$  9.0 $&$  9.0 $&$  9.0 $ \\
     1992         &$ 16.0 $&$ 16.0 $&$ 16.0 $&$  9.0 $&$  9.0 $&$  9.0 $&$  9.0 $&$  9.0 $&$  9.0 $&$  9.0 $ \\
     1993   pk$-2$&$  9.0 $&$  9.0 $&$  9.0 $&$ 17.0 $&$ 13.3 $&$ 13.3 $&$  8.4 $&$  9.8 $&$  9.8 $&$  9.8 $ \\
     1993   peak  &$  9.0 $&$  9.0 $&$  9.0 $&$ 13.3 $&$ 16.0 $&$ 13.3 $&$  8.4 $&$  9.8 $&$  9.8 $&$  9.8 $ \\
     1993   pk$+2$&$  9.0 $&$  9.0 $&$  9.0 $&$ 13.3 $&$ 13.3 $&$ 17.0 $&$  8.4 $&$  9.8 $&$  9.8 $&$  9.8 $ \\
     1994   peak  &$  9.0 $&$  9.0 $&$  9.0 $&$  8.4 $&$  8.4 $&$  8.4 $&$ 10.0 $&$  8.5 $&$  8.5 $&$  8.5 $ \\
     1995   pk$-2$&$  9.0 $&$  9.0 $&$  9.0 $&$  9.8 $&$  9.8 $&$  9.8 $&$  8.5 $&$ 18.0 $&$ 10.6 $&$ 10.6 $ \\
     1995   peak  &$  9.0 $&$  9.0 $&$  9.0 $&$  9.8 $&$  9.8 $&$  9.8 $&$  8.5 $&$ 10.6 $&$ 12.0 $&$ 10.6 $ \\
     1995   pk$+2$&$  9.0 $&$  9.0 $&$  9.0 $&$  9.8 $&$  9.8 $&$  9.8 $&$  8.5 $&$ 10.6 $&$ 10.6 $&$ 17.0 $ \\
\hline                                                                                                                              
\end{tabular}                                                                                                                       
\caption[The $\eemumu$ cross-section systematic covariance matrix]
{
The signed square-root of the covariance matrix elements for systematic
errors in the measurement of the $\eemumu$ cross-section.
Additional terms arise between muons and taus, which are given
in {\TAB}~\ref{tab:ll_matrix}.
The entries are the relative errors in units of $10^{-4}$.
They do not include errors in the luminosity measurement, which are specified
separately in {\TAB}~\ref{tab:lumi_matrix}.
   \label{tab:mm_matrix}                                                                                                          
}                                                                                                                                   
\end{center}                                                                                                                        
\end{table}                                                                                                                         

\begin{table}                                                                                                                       
\begin{center}                                                                                                                      
\begin{tabular}{|l|rrrrrrrrrr|}                                                                                                    
\hline                                                                                                                              
\multicolumn{11}{|c|}{$\eetautau$ cross-section errors} \\
\hline
               &   1990  &    1991  &    1992  &    1993  &    1993  &    1993  &    1994  &    1995  &    1995  &    1995  \\
               &         &          &          &  pk$-2$  &  peak    &  pk$+2$  &  peak    &  pk$-2$  &  peak    &  pk$+2$  \\
\hline
     1990         &$130. $&$ 40. $&$ 40. $&$ 20. $&$ 20. $&$ 20. $&$ 20. $&$ 20. $&$ 20. $&$ 20. $ \\
     1991         &$ 40. $&$ 76. $&$ 40. $&$ 20. $&$ 20. $&$ 20. $&$ 20. $&$ 20. $&$ 20. $&$ 20. $ \\
     1992         &$ 40. $&$ 40. $&$ 43. $&$ 20. $&$ 20. $&$ 20. $&$ 20. $&$ 20. $&$ 20. $&$ 20. $ \\
     1993   pk$-2$&$ 20. $&$ 20. $&$ 20. $&$ 60. $&$ 48. $&$ 54. $&$ 33. $&$ 45. $&$ 34. $&$ 34. $ \\
     1993   peak  &$ 20. $&$ 20. $&$ 20. $&$ 48. $&$ 49. $&$ 47. $&$ 35. $&$ 34. $&$ 37. $&$ 34. $ \\
     1993   pk$+2$&$ 20. $&$ 20. $&$ 20. $&$ 54. $&$ 47. $&$ 57. $&$ 32. $&$ 34. $&$ 33. $&$ 43. $ \\
     1994   peak  &$ 20. $&$ 20. $&$ 20. $&$ 33. $&$ 35. $&$ 32. $&$ 42. $&$ 33. $&$ 35. $&$ 33. $ \\
     1995   pk$-2$&$ 20. $&$ 20. $&$ 20. $&$ 45. $&$ 34. $&$ 34. $&$ 33. $&$ 60. $&$ 47. $&$ 55. $ \\
     1995   peak  &$ 20. $&$ 20. $&$ 20. $&$ 34. $&$ 37. $&$ 33. $&$ 35. $&$ 47. $&$ 48. $&$ 47. $ \\
     1995   pk$+2$&$ 20. $&$ 20. $&$ 20. $&$ 34. $&$ 34. $&$ 43. $&$ 33. $&$ 55. $&$ 47. $&$ 58. $ \\
\hline                                                                                                                              
\end{tabular}                                                                                                                       
\caption[The $\eetautau$ cross-section systematic covariance matrix]
{
The signed square-root of the covariance matrix elements for systematic
errors in the measurement of the $\eetautau$ cross-sections.
The $\tau\tau-$ee terms and $\tau\tau-\mu\mu$ terms are given
in {\TAB}~\ref{tab:ll_matrix}.
The entries are the relative errors in units of $10^{-4}$.
They do not include errors in the luminosity measurement, which are specified
separately in {\TAB}~\ref{tab:lumi_matrix}.
   \label{tab:tt_matrix}                                                                                                          
}                                                                                                                                   
\end{center}                                                                                                                        
\end{table}                                                                                                                         

\begin{table}                                                                                                                       
\begin{center}                                                                                                                      
\begin{tabular}{|l|c|rrrrrrrrrr|}                                                                                                    
\hline
\multicolumn{12}{|c|}{inter-species cross-section errors} \\                                                                                                                             
\hline
               & &   1990  &    1991  &    1992  &    1993  &    1993  &    1993  &    1994  &    1995  &    1995  &    1995  \\
               & &         &          &          &  pk$-2$  &  peak    &  pk$+2$  &  peak    &  pk$-2$  &  peak    &  pk$+2$  \\
\hline
               & & $\tau\tau$& $\tau\tau$ & $\tau\tau$ & $\tau\tau$ & $\tau\tau$
                & $\tau\tau$ & $\tau\tau$ & $\tau\tau$ & $\tau\tau$ & $\tau\tau$  \\
\hline
     1990         &ee      &$-20.0 $&$ -4.8 $&$ -4.4 $&$  0.0 $&$  0.0 $&$  0.0 $&$  0.0 $&$  0.0 $&$  0.0 $&$  0.0 $ \\
     1991         &ee      &$ -4.8 $&$-18.0 $&$ -4.1 $&$  0.0 $&$  0.0 $&$  0.0 $&$  0.0 $&$  0.0 $&$  0.0 $&$  0.0 $ \\
     1992         &ee      &$ -4.4 $&$ -4.1 $&$-15.0 $&$  0.0 $&$  0.0 $&$  0.0 $&$  0.0 $&$  0.0 $&$  0.0 $&$  0.0 $ \\
     1993   pk$-2$&ee      &$  0.0 $&$  0.0 $&$  0.0 $&$-14.0 $&$ -9.0 $&$ -9.4 $&$ -2.9 $&$ -3.1 $&$ -2.9 $&$ -3.0 $ \\
     1993   peak  &ee      &$  0.0 $&$  0.0 $&$  0.0 $&$ -9.0 $&$-10.0 $&$ -7.9 $&$ -2.4 $&$ -2.6 $&$ -2.4 $&$ -2.5 $ \\
     1993   pk$+2$&ee      &$  0.0 $&$  0.0 $&$  0.0 $&$ -9.4 $&$ -7.9 $&$-11.0 $&$ -2.5 $&$ -2.8 $&$ -2.5 $&$ -2.6 $ \\
     1994   peak  &ee      &$  0.0 $&$  0.0 $&$  0.0 $&$ -2.9 $&$ -2.4 $&$ -2.5 $&$ -9.0 $&$ -2.5 $&$ -2.3 $&$ -2.4 $ \\
     1995   pk$-2$&ee      &$  0.0 $&$  0.0 $&$  0.0 $&$ -3.1 $&$ -2.6 $&$ -2.8 $&$ -2.5 $&$-11.0 $&$ -7.5 $&$ -7.9 $ \\
     1995   peak  &ee      &$  0.0 $&$  0.0 $&$  0.0 $&$ -2.9 $&$ -2.4 $&$ -2.5 $&$ -2.3 $&$ -7.5 $&$ -9.0 $&$ -7.1 $ \\
     1995   pk$+2$&ee      &$  0.0 $&$  0.0 $&$  0.0 $&$ -3.0 $&$ -2.5 $&$ -2.6 $&$ -2.4 $&$ -7.9 $&$ -7.1 $&$-10.0 $ \\
\hline
     1990         &$\mu\mu$&$-40.0 $&$ -8.0 $&$ -7.3 $&$  0.0 $&$  0.0 $&$  0.0 $&$  0.0 $&$  0.0 $&$  0.0 $&$  0.0 $ \\
     1991         &$\mu\mu$&$ -8.0 $&$-24.0 $&$ -5.3 $&$  0.0 $&$  0.0 $&$  0.0 $&$  0.0 $&$  0.0 $&$  0.0 $&$  0.0 $ \\
     1992         &$\mu\mu$&$ -7.3 $&$ -5.3 $&$-18.0 $&$  0.0 $&$  0.0 $&$  0.0 $&$  0.0 $&$  0.0 $&$  0.0 $&$  0.0 $ \\
     1993   pk$-2$&$\mu\mu$&$  0.0 $&$  0.0 $&$  0.0 $&$-11.0 $&$ -8.3 $&$ -8.3 $&$ -2.4 $&$ -3.1 $&$ -2.9 $&$ -3.0 $ \\
     1993   peak  &$\mu\mu$&$  0.0 $&$  0.0 $&$  0.0 $&$ -8.3 $&$-11.0 $&$ -8.3 $&$ -2.4 $&$ -3.1 $&$ -2.9 $&$ -3.0 $ \\
     1993   pk$+2$&$\mu\mu$&$  0.0 $&$  0.0 $&$  0.0 $&$ -8.3 $&$ -8.3 $&$-11.0 $&$ -2.4 $&$ -3.1 $&$ -2.9 $&$ -3.0 $ \\
     1994   peak  &$\mu\mu$&$  0.0 $&$  0.0 $&$  0.0 $&$ -2.4 $&$ -2.4 $&$ -2.4 $&$ -8.0 $&$ -2.8 $&$ -2.5 $&$ -2.6 $ \\
     1995   pk$-2$&$\mu\mu$&$  0.0 $&$  0.0 $&$  0.0 $&$ -3.1 $&$ -3.1 $&$ -3.1 $&$ -2.8 $&$-14.0 $&$ -9.8 $&$-10.1 $ \\
     1995   peak  &$\mu\mu$&$  0.0 $&$  0.0 $&$  0.0 $&$ -2.9 $&$ -2.9 $&$ -2.9 $&$ -2.5 $&$ -9.8 $&$-12.0 $&$ -9.4 $ \\
     1995   pk$+2$&$\mu\mu$&$  0.0 $&$  0.0 $&$  0.0 $&$ -3.0 $&$ -3.0 $&$ -3.0 $&$ -2.6 $&$-10.1 $&$ -9.4 $&$-13.0 $ \\
\hline                                                                                                                              
\end{tabular}                                                                                                                       
\caption[The inter-species cross-section systematic covariance matrix]
{
The signed square-root of the covariance matrix elements for
systematic errors correlated between species in the measurement of the
lepton cross-sections.
The entries are the relative errors in units of $10^{-4}$.
They do not include errors in the luminosity measurement, which are specified
separately in {\TAB}~\ref{tab:lumi_matrix}.
   \label{tab:ll_matrix}                                                                                                          
}                                                                                                                                   
\end{center}                                                                                                                        
\end{table}                                                                                                                         

\begin{table}                                                                                                                       
\begin{center}                                                                                                                      
\begin{tabular}{|c l|c|c|r|r|r|r|}                                                                                                  
\hline                                                                                                                              
 & & \multicolumn{2}{|c|}{$\roots$(GeV)}                                                                                            
 & &\multicolumn{3}{|c|}{$\eemumu$ asymmetry} \\                                                                                    
\cline{3-4} \cline{6-8}                                                                                                             
 \multicolumn{2}{|c|}{Sample}& mean & rms & $N_{\mu\mu}$                                                                            
& \multicolumn{1}{|c|}{measured}                                                                                                    
& \multicolumn{1}{|c|}{ corrected}  & \multicolumn{1}{|c|}{fit} \\                                                                  
\hline
       & peak-3     &$  88.2600 $&$   0.0481 $&$     130 $&$  -0.1590 \pm  0.0830 $&$  -0.1591 $&$  -0.2720 $ \\
       & peak-2     &$  89.2550 $&$   0.0490 $&$     258 $&$  -0.2780 \pm  0.0570 $&$  -0.2782 $&$  -0.1781 $ \\
       & peak-1     &$  90.2570 $&$   0.0500 $&$     391 $&$  -0.0770 \pm  0.0480 $&$  -0.0772 $&$  -0.0825 $ \\
  1990 & peak       &$  91.2430 $&$   0.0510 $&$    4963 $&$   0.0100 \pm  0.0130 $&$   0.0100 $&$   0.0047 $ \\
       & peak+1     &$  92.2420 $&$   0.0520 $&$     576 $&$   0.0490 \pm  0.0400 $&$   0.0491 $&$   0.0735 $ \\
       & peak+2     &$  93.2450 $&$   0.0529 $&$     354 $&$   0.0940 \pm  0.0510 $&$   0.0941 $&$   0.1218 $ \\
       & peak+3     &$  94.2420 $&$   0.0539 $&$     231 $&$   0.0830 \pm  0.0610 $&$   0.0830 $&$   0.1561 $ \\
\hline
       & prescan    &$  91.2540 $&$   0.0471 $&$    7563 $&$   0.0020 \pm  0.0110 $&$   0.0020 $&$   0.0056 $ \\
       & peak-3     &$  88.4800 $&$   0.0441 $&$     176 $&$  -0.2280 \pm  0.0700 $&$  -0.2281 $&$  -0.2515 $ \\
       & peak-2     &$  89.4720 $&$   0.0451 $&$     363 $&$  -0.1060 \pm  0.0500 $&$  -0.1061 $&$  -0.1573 $ \\
       & peak-1     &$  90.2270 $&$   0.0461 $&$     744 $&$  -0.0690 \pm  0.0340 $&$  -0.0691 $&$  -0.0853 $ \\
  1991 & peak       &$  91.2240 $&$   0.0471 $&$    4422 $&$  -0.0220 \pm  0.0140 $&$  -0.0220 $&$   0.0032 $ \\
       & peak+1     &$  91.9690 $&$   0.0481 $&$     916 $&$   0.0020 \pm  0.0310 $&$   0.0021 $&$   0.0571 $ \\
       & peak+2     &$  92.9680 $&$   0.0490 $&$     478 $&$   0.1520 \pm  0.0420 $&$   0.1521 $&$   0.1101 $ \\
       & peak+3     &$  93.7170 $&$   0.0500 $&$     404 $&$   0.0850 \pm  0.0460 $&$   0.0851 $&$   0.1393 $ \\
\hline
  1992 & peak       &$  91.2989 $&$   0.0520 $&$   33733 $&$   0.0088 \pm  0.0051 $&$   0.0088 $&$   0.0092 $ \\
\hline
       & prescan(a) &$  91.1354 $&$   0.0868 $&$     449 $&$  -0.0439 \pm  0.0442 $&$  -0.0439 $&$  -0.0041 $ \\
       & prescan(b) &$  91.3209 $&$   0.0566 $&$    7325 $&$   0.0081 \pm  0.0109 $&$   0.0081 $&$   0.0109 $ \\
  1993 & peak-2     &$  89.4502 $&$   0.0564 $&$    3937 $&$  -0.1503 \pm  0.0146 $&$  -0.1505 $&$  -0.1594 $ \\
       & peak       &$  91.2063 $&$   0.0570 $&$   12066 $&$  -0.0020 \pm  0.0085 $&$  -0.0020 $&$   0.0018 $ \\
       & peak+2     &$  93.0348 $&$   0.0570 $&$    5628 $&$   0.1001 \pm  0.0124 $&$   0.1002 $&$   0.1130 $ \\
\hline
       & peak(ab)   &$  91.2198 $&$   0.0565 $&$   70493 $&$   0.0040 \pm  0.0035 $&$   0.0040 $&$   0.0028 $ \\
  1994 & peak(c)    &$  91.4285 $&$   0.0561 $&$     637 $&$   0.0182 \pm  0.0371 $&$   0.0183 $&$   0.0193 $ \\
       & peak(d)    &$  91.2194 $&$   0.0558 $&$    3157 $&$   0.0096 \pm  0.0166 $&$   0.0096 $&$   0.0028 $ \\
\hline
       & prescan(a) &$  91.7994 $&$   0.0568 $&$     256 $&$   0.0523 \pm  0.0585 $&$   0.0524 $&$   0.0460 $ \\
       & prescan(b) &$  91.3032 $&$   0.0575 $&$   14619 $&$   0.0094 \pm  0.0077 $&$   0.0094 $&$   0.0095 $ \\
  1995 & peak-2     &$  89.4414 $&$   0.0568 $&$    3739 $&$  -0.1414 \pm  0.0151 $&$  -0.1416 $&$  -0.1603 $ \\
       & peak       &$  91.2826 $&$   0.0578 $&$    6510 $&$   0.0198 \pm  0.0115 $&$   0.0198 $&$   0.0079 $ \\
       & peak+2     &$  92.9716 $&$   0.0581 $&$    5688 $&$   0.1170 \pm  0.0123 $&$   0.1171 $&$   0.1103 $ \\
\hline                                                                                                                              
\end{tabular}                                                                                                                       
\caption[The $\eemumu$ asymmetry]{ The $\eemumu$ forward--backward                                                                  
charge asymmetry near the $\Zzero$ resonance.                                                                                       
The measured asymmetry is corrected to the simple kinematic acceptance region                                                       
defined by \mbox{$\abscosthlm<  0.95$} and \mbox{$\thacol<    15^{\circ}$,}                                                         
with the energy of each fermion required to be greater than \mbox{$   6.0$ GeV.}                                                    
For each data sample we list here the mean $\roots$ of the                                                                          
colliding beams, its root-mean-square (rms) spread,                                                                                 
and the observed $\eemumu$ asymmetry.                                                                                               
The errors shown are statistical only.                                                                                              
The asymmetries are also shown after being corrected for                                                                            
the beam energy spread to correspond to the physical asymmetry                                                                      
at the central value of $\roots$.                                                                                                   
The fit values are the result of the \SLPnine\ fit.                                                                                 
   \label{tab:muon_afb}                                                                                                             
}                                                                                                                                   
\end{center}                                                                                                                        
\end{table}                                                                                                                         

\begin{table}                                                                                                                       
\begin{center}                                                                                                                      
\begin{tabular}{|c l|c|c|r|r|r|r|}                                                                                                  
\hline                                                                                                                              
 & & \multicolumn{2}{|c|}{$\roots$(GeV)}                                                                                            
 & &\multicolumn{3}{|c|}{$\eett$ asymmetry} \\                                                                                      
\cline{3-4} \cline{6-8}                                                                                                             
 \multicolumn{2}{|c|}{Sample}& mean & rms & $N_{\tau\tau}$                                                                          
& \multicolumn{1}{|c|}{measured}                                                                                                    
& \multicolumn{1}{|c|}{ corrected}  & \multicolumn{1}{|c|}{fit} \\                                                                  
\hline
       & peak-3     &$  88.2610 $&$   0.0481 $&$     102 $&$  -0.3467 \pm  0.0800 $&$  -0.3468 $&$  -0.2651 $ \\
       & peak-2     &$  89.2540 $&$   0.0490 $&$     224 $&$  -0.0667 \pm  0.0620 $&$  -0.0669 $&$  -0.1743 $ \\
       & peak-1     &$  90.2630 $&$   0.0500 $&$     281 $&$  -0.0954 \pm  0.0550 $&$  -0.0956 $&$  -0.0810 $ \\
  1990 & peak       &$  91.2460 $&$   0.0510 $&$    3722 $&$  -0.0084 \pm  0.0150 $&$  -0.0084 $&$   0.0034 $ \\
       & peak+1     &$  92.2420 $&$   0.0520 $&$     416 $&$   0.0686 \pm  0.0460 $&$   0.0687 $&$   0.0700 $ \\
       & peak+2     &$  93.2480 $&$   0.0529 $&$     302 $&$   0.1277 \pm  0.0550 $&$   0.1278 $&$   0.1171 $ \\
       & peak+3     &$  94.2400 $&$   0.0539 $&$     183 $&$   0.0887 \pm  0.0720 $&$   0.0887 $&$   0.1503 $ \\
\hline
       & prescan    &$  91.2540 $&$   0.0471 $&$    6059 $&$   0.0146 \pm  0.0120 $&$   0.0146 $&$   0.0040 $ \\
       & peak-3     &$  88.4800 $&$   0.0441 $&$     166 $&$  -0.2547 \pm  0.0680 $&$  -0.2548 $&$  -0.2454 $ \\
       & peak-2     &$  89.4720 $&$   0.0451 $&$     289 $&$  -0.1017 \pm  0.0540 $&$  -0.1018 $&$  -0.1541 $ \\
       & peak-1     &$  90.2270 $&$   0.0461 $&$     569 $&$  -0.0714 \pm  0.0390 $&$  -0.0715 $&$  -0.0843 $ \\
  1991 & peak       &$  91.2240 $&$   0.0471 $&$    3603 $&$  -0.0004 \pm  0.0160 $&$  -0.0004 $&$   0.0017 $ \\
       & peak+1     &$  91.9690 $&$   0.0481 $&$     734 $&$   0.0386 \pm  0.0350 $&$   0.0387 $&$   0.0540 $ \\
       & peak+2     &$  92.9680 $&$   0.0490 $&$     436 $&$   0.0947 \pm  0.0440 $&$   0.0948 $&$   0.1056 $ \\
       & peak+3     &$  93.7170 $&$   0.0500 $&$     359 $&$   0.1697 \pm  0.0480 $&$   0.1697 $&$   0.1340 $ \\
\hline
  1992 & peak       &$  91.2990 $&$   0.0520 $&$   28037 $&$   0.0152 \pm  0.0056 $&$   0.0152 $&$   0.0075 $ \\
\hline
       & prescan(a) &$  91.1361 $&$   0.0867 $&$     347 $&$  -0.0191 \pm  0.0508 $&$  -0.0191 $&$  -0.0053 $ \\
       & prescan(b) &$  91.3211 $&$   0.0566 $&$    5745 $&$  -0.0084 \pm  0.0136 $&$  -0.0084 $&$   0.0092 $ \\
  1993 & peak-2     &$  89.4504 $&$   0.0564 $&$    3255 $&$  -0.1497 \pm  0.0167 $&$  -0.1499 $&$  -0.1561 $ \\
       & peak       &$  91.2052 $&$   0.0570 $&$   10374 $&$  -0.0060 \pm  0.0094 $&$  -0.0060 $&$   0.0002 $ \\
       & peak+2     &$  93.0348 $&$   0.0570 $&$    4551 $&$   0.1106 \pm  0.0142 $&$   0.1107 $&$   0.1085 $ \\
\hline
       & peak(ab)   &$  91.2196 $&$   0.0565 $&$   56230 $&$   0.0008 \pm  0.0040 $&$   0.0008 $&$   0.0013 $ \\
  1994 & peak(c)    &$  91.4277 $&$   0.0561 $&$     451 $&$  -0.0926 \pm  0.0438 $&$  -0.0925 $&$   0.0172 $ \\
       & peak(d)    &$  91.2194 $&$   0.0558 $&$    2471 $&$  -0.0193 \pm  0.0189 $&$  -0.0193 $&$   0.0013 $ \\
\hline
       & prescan(a) &$  91.7987 $&$   0.0568 $&$     195 $&$   0.1095 \pm  0.0673 $&$   0.1096 $&$   0.0432 $ \\
       & prescan(b) &$  91.3032 $&$   0.0575 $&$   11765 $&$   0.0006 \pm  0.0088 $&$   0.0006 $&$   0.0078 $ \\
  1995 & peak-2     &$  89.4413 $&$   0.0568 $&$    3060 $&$  -0.1334 \pm  0.0171 $&$  -0.1336 $&$  -0.1569 $ \\
       & peak       &$  91.2823 $&$   0.0578 $&$    5242 $&$   0.0227 \pm  0.0132 $&$   0.0227 $&$   0.0062 $ \\
       & peak+2     &$  92.9715 $&$   0.0581 $&$    4711 $&$   0.0938 \pm  0.0139 $&$   0.0939 $&$   0.1058 $ \\
\hline                                                                                                                              
\end{tabular}                                                                                                                       
\caption[The $\eett$ asymmetry]{ The $\eett$ forward--backward                                                                      
charge asymmetry near the $\Zzero$ resonance.                                                                                       
The measured asymmetry is corrected to the simple kinematic acceptance region                                                       
defined by \mbox{$\abscosthlm<  0.90$} and \mbox{$\thacol<    15^{\circ}$,}                                                         
with the energy of each fermion required to be greater than \mbox{$   6.0$ GeV.}                                                    
For each data sample we list here the mean $\roots$ of the                                                                          
colliding beams, its root-mean-square (rms) spread,                                                                                 
and the observed $\eett$ asymmetry.                                                                                                 
The errors shown are statistical only.                                                                                              
The asymmetries are also shown after being corrected for                                                                            
the beam energy spread to correspond to the physical asymmetry                                                                      
at the central value of $\roots$.                                                                                                   
The fit values are the result of the \SLPnine\ fit.                                                                                 
   \label{tab:tau_afb}                                                                                                              
}                                                                                                                                   
\end{center}                                                                                                                        
\end{table}                                                                                                                         

\begin{table}                                                                                                                       
\begin{center}                                                                                                                      
\begin{tabular}{|c l|c|c|r|r|r|r|}                                                                                                  
\hline                                                                                                                              
 & & \multicolumn{2}{|c|}{$\roots$(GeV)}                                                                                            
 & &\multicolumn{3}{|c|}{$\eeee$ asymmetry} \\                                                                                      
\cline{3-4} \cline{6-8}                                                                                                             
 \multicolumn{2}{|c|}{Sample}& mean & rms & $N_{\rm{ee}}$                                                                           
& \multicolumn{1}{|c|}{measured}                                                                                                    
& \multicolumn{1}{|c|}{ corrected}  & \multicolumn{1}{|c|}{fit} \\                                                                  
\hline
       & peak-3     &$  88.2610 $&$   0.0481 $&$     194 $&$   0.3730 \pm  0.0670 $&$   0.3731 $&$   0.3877 $ \\
       & peak-2     &$  89.2550 $&$   0.0490 $&$     321 $&$   0.3290 \pm  0.0530 $&$   0.3291 $&$   0.2977 $ \\
       & peak-1     &$  90.2610 $&$   0.0500 $&$     384 $&$   0.2260 \pm  0.0500 $&$   0.2261 $&$   0.1911 $ \\
  1990 & peak       &$  91.2460 $&$   0.0510 $&$    3719 $&$   0.0850 \pm  0.0160 $&$   0.0849 $&$   0.1004 $ \\
       & peak+1     &$  92.2450 $&$   0.0520 $&$     308 $&$   0.0790 \pm  0.0570 $&$   0.0789 $&$   0.0691 $ \\
       & peak+2     &$  93.2470 $&$   0.0529 $&$     236 $&$  -0.0070 \pm  0.0650 $&$  -0.0070 $&$   0.0994 $ \\
       & peak+3     &$  94.2420 $&$   0.0539 $&$     160 $&$   0.2180 \pm  0.0780 $&$   0.2181 $&$   0.1588 $ \\
\hline
       & prescan    &$  91.2540 $&$   0.0471 $&$    5624 $&$   0.0900 \pm  0.0130 $&$   0.0899 $&$   0.0999 $ \\
       & peak-3     &$  88.4790 $&$   0.0441 $&$     297 $&$   0.4700 \pm  0.0510 $&$   0.4701 $&$   0.3697 $ \\
       & peak-2     &$  89.4690 $&$   0.0451 $&$     451 $&$   0.2750 \pm  0.0450 $&$   0.2751 $&$   0.2759 $ \\
       & peak-1     &$  90.2270 $&$   0.0461 $&$     683 $&$   0.1820 \pm  0.0380 $&$   0.1821 $&$   0.1948 $ \\
  1991 & peak       &$  91.2200 $&$   0.0471 $&$    3365 $&$   0.1070 \pm  0.0170 $&$   0.1069 $&$   0.1022 $ \\
       & peak+1     &$  91.9690 $&$   0.0481 $&$     566 $&$   0.1180 \pm  0.0420 $&$   0.1179 $&$   0.0705 $ \\
       & peak+2     &$  92.9680 $&$   0.0490 $&$     325 $&$   0.0780 \pm  0.0550 $&$   0.0780 $&$   0.0866 $ \\
       & peak+3     &$  93.7170 $&$   0.0500 $&$     284 $&$   0.0580 \pm  0.0590 $&$   0.0580 $&$   0.1256 $ \\
\hline
  1992 & peak       &$  91.2990 $&$   0.0520 $&$   25280 $&$   0.0996 \pm  0.0063 $&$   0.0995 $&$   0.0969 $ \\
\hline
       & prescan(a) &$  91.1447 $&$   0.0857 $&$     343 $&$   0.2141 \pm  0.0528 $&$   0.2139 $&$   0.1076 $ \\
       & prescan(b) &$  91.3213 $&$   0.0566 $&$    5254 $&$   0.0930 \pm  0.0137 $&$   0.0928 $&$   0.0955 $ \\
  1993 & peak-2     &$  89.4502 $&$   0.0565 $&$    4750 $&$   0.2619 \pm  0.0140 $&$   0.2621 $&$   0.2778 $ \\
       & peak       &$  91.2055 $&$   0.0571 $&$    9628 $&$   0.1052 \pm  0.0101 $&$   0.1051 $&$   0.1032 $ \\
       & peak+2     &$  93.0346 $&$   0.0570 $&$    3664 $&$   0.0958 \pm  0.0164 $&$   0.0958 $&$   0.0894 $ \\
\hline
       & peak(ab)   &$  91.2196 $&$   0.0565 $&$   51939 $&$   0.1046 \pm  0.0044 $&$   0.1045 $&$   0.1022 $ \\
  1994 & peak(c)    &$  91.4282 $&$   0.0562 $&$     462 $&$   0.1312 \pm  0.0461 $&$   0.1310 $&$   0.0893 $ \\
       & peak(d)    &$  91.2195 $&$   0.0558 $&$    2281 $&$   0.0999 \pm  0.0208 $&$   0.0998 $&$   0.1022 $ \\
\hline
       & prescan(a) &$  91.7987 $&$   0.0567 $&$     261 $&$   0.0894 \pm  0.0617 $&$   0.0892 $&$   0.0740 $ \\
       & prescan(b) &$  91.3032 $&$   0.0575 $&$   10741 $&$   0.0943 \pm  0.0096 $&$   0.0941 $&$   0.0967 $ \\
  1995 & peak-2     &$  89.4417 $&$   0.0568 $&$    4451 $&$   0.2840 \pm  0.0144 $&$   0.2842 $&$   0.2787 $ \\
       & peak       &$  91.2823 $&$   0.0578 $&$    4812 $&$   0.0736 \pm  0.0144 $&$   0.0735 $&$   0.0980 $ \\
       & peak+2     &$  92.9715 $&$   0.0581 $&$    3706 $&$   0.0878 \pm  0.0164 $&$   0.0878 $&$   0.0867 $ \\
\hline                                                                                                                              
\end{tabular}                                                                                                                       
\caption[The $\eeee$ asymmetry]{ The $\eeee$ forward--backward                                                                      
charge asymmetry near the $\Zzero$ resonance.                                                                                       
The measured asymmetry is corrected to the simple kinematic acceptance region                                                       
defined by \mbox{$\abscosthlm<  0.70$} and \mbox{$\thacol<    10^{\circ}$,}                                                         
with the energy of each fermion required to be greater than \mbox{$   0.2$ GeV.}                                                    
For each data sample we list here the mean $\roots$ of the                                                                          
colliding beams, its root-mean-square (rms) spread,                                                                                 
and the observed $\eeee$ asymmetry.                                                                                                 
The errors shown are statistical only.                                                                                              
The asymmetries are also shown after being corrected for                                                                            
the beam energy spread to correspond to the physical asymmetry                                                                      
at the central value of $\roots$.                                                                                                   
The fit values are the result of the \SLPnine\ fit.                                                                                 
   \label{tab:ee_afb}                                                                                                               
}                                                                                                                                   
\end{center}                                                                                                                        
\end{table}                                                                                                                         

\begin{table}                                                                                                                       
\begin{center}                                                                                                                      
\begin{tabular}{|l|c|rrrrrrrrrr|}                                                                                                    
\hline                                                                                                                              
\multicolumn{12}{|c|}{lepton asymmetry errors} \\
\hline
               & &   1990  &    1991  &    1992  &    1993  &    1993  &    1993  &    1994  &    1995  &    1995  &    1995  \\
               & &         &          &          &  pk$-2$  &  peak    &  pk$+2$  &  peak    &  pk$-2$  &  peak    &  pk$+2$  \\
\hline
               & & ee      &  ee      &  ee      &  ee      &  ee      &  ee      &  ee      &  ee      &  ee      &  ee      \\
\hline
     1990         &  ee    &$     5.0 $&$     1.0 $&$     1.0 $&$     1.0 $&$     1.0 $&                                            
$     1.0 $&$     1.0 $&$     1.0 $&$     1.0 $&$     1.0 $  \\                                                                     
     1991         &  ee    &$     1.0 $&$     3.0 $&$     1.0 $&$     1.0 $&$     1.0 $&                                            
$     1.0 $&$     1.0 $&$     1.0 $&$     1.0 $&$     1.0 $  \\                                                                     
     1992         &  ee    &$     1.0 $&$     1.0 $&$     2.0 $&$     1.0 $&$     1.0 $&                                            
$     1.0 $&$     1.0 $&$     1.0 $&$     1.0 $&$     1.0 $  \\                                                                     
     1993   pk$-2$&  ee    &$     1.0 $&$     1.0 $&$     1.0 $&$     1.1 $&$     1.0 $&                                            
$     1.0 $&$     1.0 $&$     1.0 $&$     1.0 $&$     1.0 $  \\                                                                     
     1993   peak  &  ee    &$     1.0 $&$     1.0 $&$     1.0 $&$     1.0 $&$     1.0 $&                                            
$     1.0 $&$     1.0 $&$     1.0 $&$     1.0 $&$     1.0 $  \\                                                                     
     1993   pk$+2$&  ee    &$     1.0 $&$     1.0 $&$     1.0 $&$     1.0 $&$     1.0 $&                                            
$     1.2 $&$     1.0 $&$     1.0 $&$     1.0 $&$     1.0 $  \\                                                                     
     1994   peak  &  ee    &$     1.0 $&$     1.0 $&$     1.0 $&$     1.0 $&$     1.0 $&                                            
$     1.0 $&$     1.0 $&$     1.0 $&$     1.0 $&$     1.0 $  \\                                                                     
     1995   pk$-2$&  ee    &$     1.0 $&$     1.0 $&$     1.0 $&$     1.0 $&$     1.0 $&                                            
$     1.0 $&$     1.0 $&$     1.1 $&$     1.0 $&$     1.0 $  \\                                                                     
     1995   peak  &  ee    &$     1.0 $&$     1.0 $&$     1.0 $&$     1.0 $&$     1.0 $&                                            
$     1.0 $&$     1.0 $&$     1.0 $&$     1.0 $&$     1.0 $  \\                                                                     
     1995   pk$+2$&  ee    &$     1.0 $&$     1.0 $&$     1.0 $&$     1.0 $&$     1.0 $&                                            
$     1.0 $&$     1.0 $&$     1.0 $&$     1.0 $&$     1.2 $  \\                                                                     
\hline
               & &$\mu\mu$ &$\mu\mu$  &$\mu\mu$  &$\mu\mu$  &$\mu\mu$  &$\mu\mu$  &$\mu\mu$  &$\mu\mu$  &$\mu\mu$  &$\mu\mu$  \\
\hline
     1990         &$\mu\mu$&$     1.0 $&$     0.4 $&$     0.4 $&$     0.4 $&$     0.4 $&                                            
$     0.4 $&$     0.4 $&$     0.4 $&$     0.4 $&$     0.4 $  \\                                                                     
     1991         &$\mu\mu$&$     0.4 $&$     1.0 $&$     0.4 $&$     0.4 $&$     0.4 $&                                            
$     0.4 $&$     0.4 $&$     0.4 $&$     0.4 $&$     0.4 $  \\                                                                     
     1992         &$\mu\mu$&$     0.4 $&$     0.4 $&$     0.5 $&$     0.4 $&$     0.4 $&                                            
$     0.4 $&$     0.4 $&$     0.4 $&$     0.4 $&$     0.4 $  \\                                                                     
     1993   pk$-2$&$\mu\mu$&$     0.4 $&$     0.4 $&$     0.4 $&$     1.1 $&$     0.4 $&                                            
$     0.4 $&$     0.4 $&$     0.4 $&$     0.4 $&$     0.4 $  \\                                                                     
     1993   peak  &$\mu\mu$&$     0.4 $&$     0.4 $&$     0.4 $&$     0.4 $&$     0.7 $&                                            
$     0.4 $&$     0.4 $&$     0.4 $&$     0.4 $&$     0.4 $  \\                                                                     
     1993   pk$+2$&$\mu\mu$&$     0.4 $&$     0.4 $&$     0.4 $&$     0.4 $&$     0.4 $&                                            
$     1.0 $&$     0.4 $&$     0.4 $&$     0.4 $&$     0.4 $  \\                                                                     
     1994   peak  &$\mu\mu$&$     0.4 $&$     0.4 $&$     0.4 $&$     0.4 $&$     0.4 $&                                            
$     0.4 $&$     0.4 $&$     0.4 $&$     0.4 $&$     0.4 $  \\                                                                     
     1995   pk$-2$&$\mu\mu$&$     0.4 $&$     0.4 $&$     0.4 $&$     0.4 $&$     0.4 $&                                            
$     0.4 $&$     0.4 $&$     1.2 $&$     0.4 $&$     0.4 $  \\                                                                     
     1995   peak  &$\mu\mu$&$     0.4 $&$     0.4 $&$     0.4 $&$     0.4 $&$     0.4 $&                                            
$     0.4 $&$     0.4 $&$     0.4 $&$     0.9 $&$     0.4 $  \\                                                                     
     1995   pk$+2$&$\mu\mu$&$     0.4 $&$     0.4 $&$     0.4 $&$     0.4 $&$     0.4 $&                                            
$     0.4 $&$     0.4 $&$     0.4 $&$     0.4 $&$     0.9 $  \\                                                                     
\hline
               & &$\tau\tau$     &$\tau\tau$  &$\tau\tau$  &$\tau\tau$  &$\tau\tau$  &$\tau\tau$  &$\tau\tau$  &$\tau\tau$  &$\tau\tau$  &$\tau\tau$  \\
\hline
     1990         &$\tau\tau$&$     3.0 $&$     3.0 $&$     2.0 $&$     1.8 $&$     1.2 $&                                            
$     1.6 $&$     1.2 $&$     1.8 $&$     1.2 $&$     1.6 $  \\                                                                     
     1991         &$\tau\tau$&$     3.0 $&$     3.0 $&$     2.0 $&$     1.8 $&$     1.2 $&                                            
$     1.6 $&$     1.2 $&$     1.8 $&$     1.2 $&$     1.6 $  \\                                                                     
     1992         &$\tau\tau$&$     2.0 $&$     2.0 $&$     2.0 $&$     1.8 $&$     1.2 $&                                            
$     1.6 $&$     1.2 $&$     1.8 $&$     1.2 $&$     1.6 $  \\                                                                     
     1993   pk$-2$&$\tau\tau$&$     1.8 $&$     1.8 $&$     1.8 $&$     1.8 $&$     1.2 $&                                            
$     1.6 $&$     1.2 $&$     1.8 $&$     1.2 $&$     1.6 $  \\                                                                     
     1993   peak  &$\tau\tau$&$     1.2 $&$     1.2 $&$     1.2 $&$     1.2 $&$     1.2 $&                                            
$     1.2 $&$     1.2 $&$     1.2 $&$     1.2 $&$     1.2 $  \\                                                                     
     1993   pk$+2$&$\tau\tau$&$     1.6 $&$     1.6 $&$     1.6 $&$     1.6 $&$     1.2 $&                                            
$     1.6 $&$     1.2 $&$     1.6 $&$     1.2 $&$     1.6 $  \\                                                                     
     1994   peak  &$\tau\tau$&$     1.2 $&$     1.2 $&$     1.2 $&$     1.2 $&$     1.2 $&                                            
$     1.2 $&$     1.2 $&$     1.2 $&$     1.2 $&$     1.2 $  \\                                                                     
     1995   pk$-2$&$\tau\tau$&$     1.8 $&$     1.8 $&$     1.8 $&$     1.8 $&$     1.2 $&                                            
$     1.6 $&$     1.2 $&$     1.8 $&$     1.2 $&$     1.6 $  \\                                                                     
     1995   peak  &$\tau\tau$&$     1.2 $&$     1.2 $&$     1.2 $&$     1.2 $&$     1.2 $&                                            
$     1.2 $&$     1.2 $&$     1.2 $&$     1.2 $&$     1.2 $  \\                                                                     
     1995   pk$+2$&$\tau\tau$&$     1.6 $&$     1.6 $&$     1.6 $&$     1.6 $&$     1.2 $&                                            
$     1.6 $&$     1.2 $&$     1.6 $&$     1.2 $&$     1.6 $  \\                                                                     
\hline
\end{tabular}                                                                                                                       
\caption[The lepton asymmetry systematic error covariance matrix]
{
The signed square-root of the covariance matrix elements for systematic
errors in the measurement of the lepton asymmetries.
There are no appreciable inter-species correlations in the asymmetry measurements.
The entries are the absolute errors in units of $10^{-3}$.
   \label{tab:asymm_matrix}                                                                                                          
}                                                                                                                                   
\end{center}                                                                                                                        
\end{table}                                                                                                                         

%
\begin{table}[htbp]  \begin{center}                                                                                     
\renewcommand{\arraystretch}{1.2}                                                                                       
\begin{tabular}{|l|l|c|c|c|}                                                                                            
 \hline                                                                                                                 
Parameter &\multicolumn{1}{|c|}{Interpretation in} & Without Lepton & With Lepton  & Standard Model  \\                 
          &\multicolumn{1}{|c|}{terms of couplings}& Universality   & Universality & Prediction       \\                
 \hline                                                                                                                 
$\MZ$ (GeV) &               & $91.1866 \pm   0.0031$ & $91.1861 \pm   0.0030$ & input \\        
$\GZ$ (GeV) &               & $ 2.4942 \pm   0.0042$ & $ 2.4940 \pm   0.0041$ &                 
                             $ 2.4949^{+  0.0021}_{ -0.0074}$ \\                                      
$\shadpol$ (nb)&            & $ 41.505 \pm    0.055$ & $ 41.505 \pm    0.055$ &                 
                             $ 41.480^{+   0.012}_{  -0.011}$ \\                                      
\hline                                                                                                                  
$\Cszz(\ee)$    & $(\gagve)(\gagve)$        & $ 0.06330 \pm  0.00031$ & & \\                                
$\Cszz(\mumu)$  & $(\gagve)(\gagvm)$        & $ 0.06359 \pm  0.00025$ & & \\                                
$\Cszz(\tautau)$& $(\gagve)(\gagvtau)$      & $ 0.06366 \pm  0.00033$ & & \\                                
 \hline                                                                                                                 
$\Cszz(\elel)$  & $(\gagvl)^2$            & & $ 0.06353 \pm  0.00022$ &                                     
                                            $ 0.06382^{+ 0.00009}_{-0.00031}$ \\                      
 \hline                                                                                                                 
$\Cazz(\ee)$    & $\gae\gve\gae\gve$        & $ 0.000189 \pm  0.000094$ & & \\                                
$\Cazz(\mumu)$  & $\gae\gve\gam\gvm$        & $ 0.000320 \pm  0.000049$ & & \\                                
$\Cazz(\tautau)$& $\gae\gve\gatau\gvtau$    & $ 0.000293 \pm  0.000064$ & & \\                                
 \hline                                                                                                                 
$\Cazz(\elel)$  & $(\gal\gvl)^2$          & & $ 0.000294 \pm  0.000036$ &                                     
                                            $ 0.000336^{+ 0.000014}_{-0.000039}$ \\                      
 \hline                                                                                                                 
$\Cagz(\ee)$    & $\gae\gae$                & $ 0.242 \pm  0.022$ & & \\                                
$\Cagz(\mumu)$  & $\gae\gam$                & $ 0.232 \pm  0.011$ & & \\                                
$\Cagz(\tautau)$& $\gae\gatau$              & $ 0.234 \pm  0.013$ & & \\                                
 \hline                                                                                                                 
$\Cagz(\elel)$  & $\gal^2$                & & $ 0.2350 \pm  0.0080$ &                                     
                                            $  0.25130^{+  0.00013}_{ -0.00047}$ \\                      
 \hline                                                                                                                 
$\Csgz(\ee)$    & $\gve\gve$                &$ -0.0262 \pm   0.0139$ & & \\                                 
$\Csgz(\mumu)$  & $\gve\gvm$                &$ -0.0029 \pm   0.0093$ & & \\                                 
$\Csgz(\tautau)$& $\gve\gvtau$              &$ -0.0011 \pm   0.0106$ & & \\                                 
 \hline                                                                                                                 
$\Csgz(\elel)$  & $\gvl^2$                & &$  -0.0075 \pm    0.0064$ &                                      
                                           $  0.00134^{+  0.00006}_{ -0.00015}$ \\                       
  \hline                                                                                                                
$\chi^2/$d.o.f & & 146.6 /  188 &  151.8 /  196 & \\                                      
 \hline                                                                                                                 
\end{tabular}                                                                                                           
\renewcommand{\arraystretch}{1.0}                                                                                       
\end{center}                                                                                                            
\caption[Results of the C-parameter fits                                                                                
with and without assuming lepton universality]                                                                          
{\label{tab-ewp-cpar}                                                                                                   
Results of the C-parameter fits                                                                                         
to the measured cross-section and lepton asymmetry data                                                                 
with and without imposing lepton universality (7 and 15 parameters).                                                    
Theory uncertainties, other than those                                                                                  
on the $t$-channel for the electrons and the luminosity,                                                   
are not included in the errors (see section~\ref{sec-thyerr}).                                                          
In the last column we give the values predicted by the {\SM}                                                            
assuming the parameter variations given in {\EQ}~\ref{eq-sm-pars}.                                                      
}                                                                                                                       
\end{table}                                                                                                             

\clearpage                                                                                                                          
\begin{table}[htbp]                                                                                                                 
\begin{sideways}                                                                                                                    
\begin{minipage}[b]{\textheight}                                                                                                    
\begin{center}\begin{tabular}{|rl|rrrrrrrrrrrrrrr|}                                                                                 
 \hline                                                                                                                             
 \multicolumn{2}{|l|}{Parameters} & 1 & 2 & 3 & 4 & 5 & 6 & 7 & 8 & 9 & 10                                                          
 & 11 & 12 & 13 & 14 & 15 \\                                                                                                        
 \hline                                                                                                                             
  1 & $\MZ$            & $ 1.000$ & $  .043$ & $  .034$ & $ -.092$ & $ -.008$ & $ -.005$ & $ -.054$
    & $  .067$ & $  .052$ & $  .009$ & $  .009$ & $  .009$ & $ -.088$ & $ -.142$ & $ -.122$\\
  2 & $\GZ$            & $  .043$ & $ 1.000$ & $ -.354$ & $  .572$ & $  .710$ & $  .542$ & $  .003$
    & $  .024$ & $  .017$ & $  .000$ & $  .049$ & $  .043$ & $ -.024$ & $  .022$ & $  .021$\\
  3 & $\shadpol$       & $  .034$ & $ -.354$ & $ 1.000$ & $ -.098$ & $ -.120$ & $ -.090$ & $  .006$
    & $  .008$ & $  .007$ & $ -.004$ & $ -.007$ & $ -.006$ & $ -.011$ & $ -.018$ & $ -.016$\\
  4 & $\Cszz(\ee)$     & $ -.092$ & $  .572$ & $ -.098$ & $ 1.000$ & $  .457$ & $  .338$ & $  .171$
    & $ -.007$ & $ -.007$ & $ -.003$ & $  .029$ & $  .024$ & $  .055$ & $  .033$ & $  .030$\\
  5 & $\Cszz(\mumu)$   & $ -.008$ & $  .710$ & $ -.120$ & $  .457$ & $ 1.000$ & $  .413$ & $  .009$
    & $  .032$ & $  .014$ & $ -.003$ & $  .059$ & $  .032$ & $ -.018$ & $  .158$ & $  .020$\\
  6 & $\Cszz(\tautau)$ & $ -.005$ & $  .542$ & $ -.090$ & $  .338$ & $  .413$ & $ 1.000$ & $  .005$
    & $  .014$ & $  .027$ & $ -.001$ & $  .028$ & $  .075$ & $ -.012$ & $  .017$ & $  .127$\\
  7 & $\Cazz(\ee)$     & $ -.054$ & $  .003$ & $  .006$ & $  .171$ & $  .009$ & $  .005$ & $ 1.000$
    & $ -.014$ & $ -.011$ & $  .080$ & $ -.001$ & $ -.001$ & $ -.004$ & $  .010$ & $  .008$\\
  8 & $\Cazz(\mumu)$   & $  .067$ & $  .024$ & $  .008$ & $ -.007$ & $  .032$ & $  .014$ & $ -.014$
    & $ 1.000$ & $  .016$ & $  .001$ & $  .208$ & $  .003$ & $ -.008$ & $  .008$ & $ -.010$\\
  9 & $\Cazz(\tautau)$ & $  .052$ & $  .017$ & $  .007$ & $ -.007$ & $  .014$ & $  .027$ & $ -.011$
    & $  .016$ & $ 1.000$ & $  .001$ & $  .002$ & $  .184$ & $ -.006$ & $ -.009$ & $  .008$\\
 10 & $\Cagz(\ee)$     & $  .009$ & $  .000$ & $ -.004$ & $ -.003$ & $ -.003$ & $ -.001$ & $  .080$
    & $  .001$ & $  .001$ & $ 1.000$ & $  .000$ & $  .000$ & $  .217$ & $ -.001$ & $ -.001$\\
 11 & $\Cagz(\mumu)$   & $  .009$ & $  .049$ & $ -.007$ & $  .029$ & $  .059$ & $  .028$ & $ -.001$
    & $  .208$ & $  .002$ & $  .000$ & $ 1.000$ & $  .002$ & $ -.003$ & $ -.084$ & $  .000$\\
 12 & $\Cagz(\tautau)$ & $  .009$ & $  .043$ & $ -.006$ & $  .024$ & $  .032$ & $  .075$ & $ -.001$
    & $  .003$ & $  .184$ & $  .000$ & $  .002$ & $ 1.000$ & $ -.002$ & $  .000$ & $ -.089$\\
 13 & $\Csgz(\ee)$     & $ -.088$ & $ -.024$ & $ -.011$ & $  .055$ & $ -.018$ & $ -.012$ & $ -.004$
    & $ -.008$ & $ -.006$ & $  .217$ & $ -.003$ & $ -.002$ & $ 1.000$ & $  .025$ & $  .020$\\
 14 & $\Csgz(\mumu)$   & $ -.142$ & $  .022$ & $ -.018$ & $  .033$ & $  .158$ & $  .017$ & $  .010$
    & $  .008$ & $ -.009$ & $ -.001$ & $ -.084$ & $  .000$ & $  .025$ & $ 1.000$ & $  .033$\\
 15 & $\Csgz(\tautau)$ & $ -.122$ & $  .021$ & $ -.016$ & $  .030$ & $  .020$ & $  .127$ & $  .008$
    & $ -.010$ & $  .008$ & $ -.001$ & $  .000$ & $ -.089$ & $  .020$ & $  .033$ & $ 1.000$\\
 \hline                                                                                                                             
\end{tabular}\end{center}                                                                                                           
\caption[Correlation matrix for the C-parameter fit                                                                                 
which does not assume lepton universality]                                                                                          
{\label{tab-ewp-cparcor}                                                                                                            
Error correlation matrix for the C-parameters                                                                                       
in {\TAB}~\ref{tab-ewp-cpar} which do not assume lepton universality.                                                               
}                                                                                                                                   
\end{minipage}                                                                                                                      
\end{sideways}                                                                                                                      
\end{table}                                                                                                                         

\begin{table}[htbp]                                                                                                                 
\begin{center}\begin{tabular}{|rl|rrrrrrr|}                                                                                         
 \hline                                                                                                                             
 \multicolumn{2}{|l|}{Parameters} & 1 & 2 & 3 & 4 & 5 & 6 & 7 \\                                                                    
 \hline                                                                                                                             
  1 & $\MZ$            & $ 1.000$ & $  .011$ & $  .059$ & $ -.052$ & $  .064$ & $  .004$ & $ -.207$\\
  2 & $\GZ$            & $  .011$ & $ 1.000$ & $ -.353$ & $  .796$ & $  .028$ & $  .064$ & $  .025$\\
  3 & $\shadpol$       & $  .059$ & $ -.353$ & $ 1.000$ & $ -.135$ & $  .009$ & $ -.010$ & $ -.032$\\
  4 & $\Cszz(\elel)$   & $ -.052$ & $  .796$ & $ -.135$ & $ 1.000$ & $  .051$ & $  .078$ & $  .119$\\
  5 & $\Cazz(\elel)$   & $  .064$ & $  .028$ & $  .009$ & $  .051$ & $ 1.000$ & $  .185$ & $  .005$\\
  6 & $\Cagz(\elel)$   & $  .004$ & $  .064$ & $ -.010$ & $  .078$ & $  .185$ & $ 1.000$ & $ -.031$\\
  7 & $\Csgz(\elel)$   & $ -.207$ & $  .025$ & $ -.032$ & $  .119$ & $  .005$ & $ -.031$ & $ 1.000$\\
 \hline                                                                                                                             
\end{tabular}\end{center}                                                                                                           
\caption[Correlation matrix for the C-parameter fit                                                                                 
which assumes lepton universality]                                                                                                  
{\label{tab-ewp-cparcorlu}                                                                                                          
Error correlation matrix for the C-parameters                                                                                       
in {\TAB}~\ref{tab-ewp-cpar}                                                                                                        
which assume lepton universality                                                                                                    
}                                                                                                                                   
\end{table}                                                                                                                         

\begin{table}[htbp]  \begin{center}                                                                                                 
\renewcommand{\arraystretch}{1.4}                                                                                                   
\begin{tabular}{|l|r|r|r|}                                                                                                          
\hline                                                                                                                              
  & Without lepton  &  With lepton  &    {\SM}      \\                                                                              
  & universality    &  universality &    prediction \\                                                                              
\hline                                                                                                                              
 $\MZ \gev       $ & 91.1858 $\pm$ 0.0030 & 91.1852 $\pm$ 0.0030 & input \\                                                         
 $\GZ \gev       $ &  2.4948 $\pm$ 0.0041 &  2.4948 $\pm$ 0.0041 &                                                                  
 $  2.4949^{+ 0.0021}_{-0.0074}$ \\                                                                                                 
 $\shadpol  \nb  $ &  41.501 $\pm$ 0.055  &  41.501 $\pm$ 0.055  &                                                                  
 $  41.480^{+  0.012}_{ -0.011}$ \\                                                                                                 
 \hline
 $\Ree           $ &  20.902 $\pm$ 0.084 &                                                                                          
                                         &$  20.738^{+ 0.015}_{-0.023}$ \\                                                          
 $\Rmu           $ &  20.811 $\pm$ 0.058 &                                                                                          
                                         &$  20.738^{+ 0.015}_{-0.023}$ \\                                                          
 $\Rtau          $ &  20.832 $\pm$ 0.091 &                                                                                          
                                         &$  20.785^{+ 0.015}_{-0.023}$ \\                                                          
 $\Rl            $ &                      &  20.823 $\pm$ 0.044 &                                                                   
                                         $  20.738^{+ 0.015}_{-0.023}$ \\                                                           
 \hline
 $\Afbpolee      $ &   0.0089 $\pm$ 0.0044 &  &                             \\                                                      
 $\Afbpolmumu    $ &   0.0159 $\pm$ 0.0023 &  &                             \\                                                      
 $\Afbpoltautau  $ &   0.0145 $\pm$ 0.0030 &  &                             \\                                                      
 $\Afbpolll      $ &                      &   0.0145 $\pm$ 0.0017 &                                                                 
                                         $   0.0158^{+ 0.0007}_{-0.0018}$ \\                                                        
 \hline
 $\chi^2/$d.o.f.      & 155.6/ 194 &  158.3/ 198 &  \\                                                                              
\hline                                                                                                                              
\end{tabular}                                                                                                                       
\end{center}                                                                                                                        
\caption[Results for the {\SLP}]                                                                                                    
{\label{tab-ewp-leppar}                                                                                                             
Results of fitting the {\SLP} to the                                                                                                
measured cross-sections and asymmetries.                                                                                            
The theory uncertainties given in {\TAB}~\ref{tab-ewp-err}, other than those                                                              
on the $t$-channel for the electrons and the luminosity,                                                               
are not included in the errors.                                                                                                     
In the last column we give the values                                                                                               
calculated in the context of the {\SM} assuming                                                                                     
the parameter variations given in {\EQ}~\ref{eq-sm-pars}.                                                                           
}                                                                                                                                   
\end{table}                                                                                                                         

\begin{table}[htbp]  \begin{center}                                                                                                 
\begin{tabular}{|l|rrrrr|}                                                                                                          
\hline                                                                                                                              
\makebox{\rule[-1.5ex]{0cm}{4.5ex}}

  & $\MZ$             & $\GZ$             & $\shadpol$       
 & $\Rl$             & $\Afbpolll$        \\
 \hline
 $\MZ$             &$  1.00$ &$   0.05$ &$   0.03$ &$   0.04$ &$   0.07$ \\                                                         
 $\GZ$             &$   0.05$ &$  1.00$ &$  -0.35$ &$   0.02$ &$  -0.01$ \\                                                         
 $\shadpol$        &$   0.03$ &$  -0.35$ &$  1.00$ &$   0.29$ &$   0.02$ \\                                                         
 $\Rl$             &$   0.04$ &$   0.02$ &$   0.29$ &$  1.00$ &$  -0.01$ \\                                                         
 $\Afbpolll$       &$   0.07$ &$  -0.01$ &$   0.02$ &$  -0.01$ &$  1.00$ \\                                                         
\hline                                                                                                                              
\end{tabular}                                                                                                                       
\end{center}                                                                                                                        
\caption[5 parameter fit correlation matrix]                                                                                        
{\label{tab-ewp-lepparc5}                                                                                                           
Error correlation matrix for the 5 parameter fit assuming lepton                                                                    
universality given in {\TAB}~\ref{tab-ewp-leppar}.                                                                                  
}                                                                                                                                   
\end{table}                                                                                                                         

\begin{table}[htbp]  \begin{center}                                                                                                 
\begin{tabular}{|l|rrrrrrrrr|}                                                                                                      
\hline                                                                                                                              
\makebox{\rule[-1.5ex]{0cm}{4.5ex}}                                                                                                 
  & $\MZ$             & $\GZ$             & $\shadpol$       
 & $\Ree$            & $\Rmu$            & $\Rtau$          
 & $\Afbpolee$       & $\Afbpolmumu$     & $\Afbpoltautau$    \\
 \hline
 $\MZ$             &$  1.00$ &$   0.05$ &$   0.03$ &$   0.11$ &$   0.00$ &$   0.00$ &$  -0.05$ &$   0.08$ &$   0.06$ \\             
 $\GZ$             &$   0.05$ &$  1.00$ &$  -0.35$ &$   0.01$ &$   0.02$ &$   0.01$ &$   0.00$ &$   0.00$ &$   0.00$ \\             
 $\shadpol$        &$   0.03$ &$  -0.35$ &$  1.00$ &$   0.15$ &$   0.22$ &$   0.14$ &$   0.01$ &$   0.01$ &$   0.01$ \\             
 $\Ree$            &$   0.11$ &$   0.01$ &$   0.15$ &$  1.00$ &$   0.09$ &$   0.04$ &$  -0.20$ &$   0.03$ &$   0.02$ \\             
 $\Rmu$            &$   0.00$ &$   0.02$ &$   0.22$ &$   0.09$ &$  1.00$ &$   0.06$ &$   0.00$ &$   0.01$ &$   0.00$ \\             
 $\Rtau$           &$   0.00$ &$   0.01$ &$   0.14$ &$   0.04$ &$   0.06$ &$  1.00$ &$   0.00$ &$   0.00$ &$   0.01$ \\             
 $\Afbpolee$       &$  -0.05$ &$   0.00$ &$   0.01$ &$  -0.20$ &$   0.00$ &$   0.00$ &$  1.00$ &$  -0.02$ &$  -0.01$ \\             
 $\Afbpolmumu$     &$   0.08$ &$   0.00$ &$   0.01$ &$   0.03$ &$   0.01$ &$   0.00$ &$  -0.02$ &$  1.00$ &$   0.02$ \\             
 $\Afbpoltautau$   &$   0.06$ &$   0.00$ &$   0.01$ &$   0.02$ &$   0.00$ &$   0.01$ &$  -0.01$ &$   0.02$ &$  1.00$ \\             
\hline                                                                                                                              
\end{tabular}                                                                                                                       
\end{center}                                                                                                                        
\caption[9 parameter fit correlation matrix]                                                                                        
{\label{tab-ewp-lepparc9}                                                                                                           
Error correlation matrix for the 9 parameter fit                                                                                    
without assuming lepton                                                                                                             
universality given in {\TAB}~\ref{tab-ewp-leppar}.                                                                                  
}                                                                                                                                   
\end{table}                                                                                                                         

\begin{table}[htbp]  
\begin{center}
\renewcommand{\arraystretch}{1.4}
\begin{tabular}{|l|c||ccc|}
\hline
                   & $\MZ \gev  $           &  1990--2           &  1993--4           &  1995     \\
\hline
  1990--2          & 91.1851 $\pm$ 0.0091 &$ 1.00$ &$  0.02$&$  0.01$ \\
  1993--4          & 91.1870 $\pm$ 0.0046 &$  0.02$&$ 1.00$ &$  0.09$ \\
  1995             & 91.1851 $\pm$ 0.0039 &$  0.01$&$  0.09$&$  1.00$ \\
\hline
\end{tabular}
\end{center}                                                                                                                        
\caption[Results for the $\MZ$ consistency test]                                                                                
{\label{tab-ewp-mztest}
Results of the {\SLP} fit with independent $\MZ$ for the three main
data-taking phases, 1990--2, 1993--4 and 1995. 
The correlations are given in columns 3--5.
}                                                                                                                                   
\end{table}

\begin{table}[htbp]  
\renewcommand{\arraystretch}{1.2}
\begin{center} 
\begin{tabular}{|l|l|l|l|l|l||l|} \hline
%
& \multicolumn{6}{|c|}{Error sources} \\ \cline{2-7}
Parameter   & statis- & \multicolumn{4}{|c||}{included systematics} & addtl.  \\
\cline{3-6}  
            & ~tics & selection & ~lumi & $t$-chan. & $E_{\mathrm{beam}}$  & theory \\ \hline
$\MZ$ (GeV) & 0.0023 & 0.0001 & 0.0003 & 0.0001 & 0.0018 & 0.0003 \\ 
$\GZ$ (GeV) & 0.0036 & 0.0010 & 0.0007 & 0.0000 & 0.0013 & 0.0002 \\ 
$\shadpol$ (nb) & 0.031 & 0.033  & 0.029  & 0.0000 & 0.011  & 0.008  \\ 
$\Ree$ & 0.067  & 0.040  & 0.009  & 0.027 & 0.014        & 0.004  \\  
$\Rmu$ & 0.050  & 0.027  & 0.008  & 0.000 & 0.000        & 0.004  \\
$\Rtau$ & 0.055 & 0.071  & 0.010  & 0.000 & 0.000        & 0.004  \\   
$\Afbpolee$ & 0.0038 & 0.0016 & 0.0000 & 0.0018 & 0.0004    & 0.0001 \\ 
$\Afbpolmumu$ & 0.0022 & 0.0004 & 0.0000 & 0.0000 & 0.0003    & 0.0001 \\ 
$\Afbpoltautau$ & 0.0026 & 0.0014 & 0.0000 & 0.0000 & 0.0003    & 0.0001 \\ 
\hline
$\Rl$    & 0.034  & 0.027  & 0.008  & 0.005  & 0.003     & 0.004  \\ 
$\Afbpolll$ & 0.0015 & 0.0006 & 0.0000 & 0.0003 & 0.0002    & 0.0001 \\ 
\hline
$\sleppol$ (nb) & 0.0032 & 0.0021 & 0.0018 & 0.0005 & 0.0004 & 0.0005 \\ 
$\Ghad$ (MeV) & 2.8 & 1.8 & 0.7 & 0.3 & 0.8                & 0.3    \\ 
$\Gll$   (MeV) & 0.12 & 0.05 & 0.03 & 0.01 & 0.04           & 0.01   \\ 
$\Ginv$ (MeV) & 1.9 & 1.6 & 0.7 & 0.2 & 0.5                & 0.3 \\ 
$\Ginv/\Gll$ & 0.017 & 0.018 & 0.010 & 0.002 & 0.004        & 0.003\\ 
\hline
$\alpha_s$ & 0.0035 & 0.0026 & 0.0026 & 0.0008 & 0.0006    & \\ 
$\log_{10} (\MH /$GeV) & 0.41  & 0.06  & 0.13  & 0.05 & 0.07      & \\ 
\hline
\end{tabular}
\caption[Error sources for the $\Zzero$ resonance parameters]                        
{\label{tab-ewp-err}
Contribution of the various error sources to the uncertainty
in the direct and derived $\Zzero$ resonance parameters.
``Statistics'' include the event counting and luminosity point-to-point 
statistical errors; ``selection'' are the systematic errors
associated with the event selection; ``lumi'' errors contain the experimental
systematic uncertainties and the theoretical error of the luminosity
cross-section;
``$t$-channel'' are the theoretical uncertainties of the $t$-channel correction
for $\eeee$; ``$E_{\mathrm{beam}}$'' includes the uncertainties in the LEP
centre-of-mass energy and centre-of-mass energy spread;
``addtl.\ theory'' gives the uncertainties related to the determination
of the fit parameters from the measured cross-sections and asymmetries which are
{\bf not} included in the tables.
}
\end{center}
\end{table}

\begin{table}[htbp]  \begin{center}                                                                                                 
                                                                                                                                    
\renewcommand{\arraystretch}{1.4}                                                                                                   
\begin{tabular}{|l|c|c|c|}                                                                                                          
\hline                                                                                                                              
  & Without lepton  &  With lepton  & Standard Model\\                                                                              
  & universality    &  universality &    prediction \\                                                                              
\hline                                                                                                                              
 $\Ginv   $(MeV) &   494.4 $\pm$ 4.2 &   498.1 $\pm$ 2.6 &                                                                          
                               $  501.64^{+ 0.26}_{-0.93}$ \\                                                                       
 $\Gee    $(MeV) &   83.66 $\pm$  0.20 & &                                                                                          
                               $   83.977^{+ 0.058}_{-0.206}$ \\                                                                    
 $\Gmumu  $(MeV) &   84.03 $\pm$  0.30 & &                                                                                          
                               $   83.976^{+ 0.058}_{-0.206}$ \\                                                                    
 $\Gtautau$(MeV) &   83.94 $\pm$  0.41 & &                                                                                          
                               $   83.786^{+ 0.058}_{-0.206}$ \\                                                                    
 $\Gll    $(MeV)                   &&   83.82 $\pm$  0.15 &                                                                         
                               $   83.977^{+ 0.058}_{-0.206}$ \\                                                                    
 $\Ghad   $(MeV) &  1748.8 $\pm$ 4.6 &  1745.4 $\pm$ 3.5 &                                                                          
                               $  1741.5^{+ 1.8}_{-5.9}$ \\                                                                         
\hline                                                                                                                              
\end{tabular}                                                                                                                       
\renewcommand{\arraystretch}{1.0}                                                                                                   
\caption[Results for the $\Zzero$ partial decay widths]                                                                             
{\label{tab-ewp-pwidth}                                                                                                             
$\Zzero$ partial decay widths obtained from a parameter                                                                             
transformation from the fitted {\SLP}                                                                                               
given in {\TAB}~\ref{tab-ewp-leppar}.                                                                                               
In the last column we give the values of the widths                                                                                 
calculated in the context of the {\SM} assuming                                                                                     
the parameter variations given in {\EQ}{~\ref{eq-sm-pars}}.                                                                         
}                                                                                                                                   
\end{center}                                                                                                                        
\end{table}                                                                                                                         

\begin{table}[htbp]  \begin{center}                                                                                                 
\begin{tabular}{|l|rrrrr|}                                                                                                          
\hline                                                                                                                              
 & $\Ginv   $ & $\Gee    $ & $\Gmumu  $ & $\Gtautau$ & $\Ghad   $ \\                                                                
\hline                                                                                                                              
 $\Ginv   $ &$  1.00$  &$   0.81$  &$  -0.40$  &$  -0.35$  &$  -0.63$  \\                                                           
 $\Gee    $ &$   0.81$  &$  1.00$  &$  -0.19$  &$  -0.15$  &$  -0.25$  \\                                                           
 $\Gmumu  $ &$  -0.40$  &$  -0.19$  &$  1.00$  &$   0.33$  &$   0.64$  \\                                                           
 $\Gtautau$ &$  -0.35$  &$  -0.15$  &$   0.33$  &$  1.00$  &$   0.47$  \\                                                           
 $\Ghad   $ &$  -0.63$  &$  -0.25$  &$   0.64$  &$   0.47$  &$  1.00$  \\                                                           
\hline                                                                                                                              
\end{tabular}                                                                                                                       
\end{center}                                                                                                                        
\caption[5x5 partial width correlation matrix]                                                                                      
{\label{tab-ewp-pwidth5}                                                                                                            
Error correlation matrix for the measurements of                                                                                    
the partial widths,                                                                                                                 
without assuming lepton                                                                                                             
universality, presented in {\TAB}~\ref{tab-ewp-pwidth}.                                                                             
}                                                                                                                                   
\end{table}                                                                                                                         

\begin{table}[htbp]  \begin{center}                                                                                                 
\begin{tabular}{|l|rrr|}                                                                                                            
\hline                                                                                                                              
 & $\Ginv   $ & $\Gll    $ & $\Ghad   $ \\                                                                                          
\hline                                                                                                                              
 $\Ginv   $ &$  1.00 $ &$   0.60 $ &$  -0.27 $ \\                                                                                   
 $\Gll    $ &$   0.60 $ &$  1.00 $ &$   0.36 $ \\                                                                                   
 $\Ghad   $ &$  -0.27 $ &$   0.36 $ &$  1.00 $ \\                                                                                   
\hline                                                                                                                              
\end{tabular}                                                                                                                       
\end{center}                                                                                                                        
\caption[3x3 partial width correlation matrix]                                                                                      
{\label{tab-ewp-pwidth3}                                                                                                            
Error correlation matrix for the measurements of                                                                                    
the partial widths,                                                                                                                 
assuming lepton                                                                                                                     
universality, presented in {\TAB}~\ref{tab-ewp-pwidth}.                                                                             
}                                                                                                                                   
\end{table}                                                                                                                         

%
%
%
%
\begin{table}[htbp]  
\renewcommand{\arraystretch}{1.2}
\begin{center} 
%
\begin {tabular}{|l|r|r|rr||r|r|}
\hline
& \multicolumn{2}{|c|}{$\Gamma^{\rm exp} - \Gamma^{\rm SM}$ (MeV)}
&  \multicolumn{2}{|c||}{errors (MeV)}
& \multicolumn{2}{|c|}{$\Gamma^{\rm new}_{95}$ (MeV)} \\
\hline
$\MH$(GeV) &150 & 1000 & & & 150 & 1000 \\
\hline
$\GZ$     &$ -0.1 $ &$ 7.1  $ & $ \pm4.1$  & $ \pm2.1$  &$ 9.0 $ &$ 14.8$  \\
$\Ginv$   &$ -3.5 $ &$ -2.6 $ & $ \pm2.6$  & $ \pm0.2$  &$ 3.4 $ &$ 3.7 $  \\
$\Ghad$   &$  3.9 $ &$ 9.6  $ & $ \pm3.5$  & $ \pm1.9$  &$ 10.8$ &$ 16.2$  \\
$\Gll$    &$ -0.15$ &$ 0.05 $ & $ \pm0.15$ & $ \pm0.05$ &$ 0.22$ &$ 0.34$  \\ 
$\Gee$    &$ -0.31$ &$-0.11 $ & $ \pm0.20$ & $ \pm0.05$ &$ 0.26$ &$ 0.35$  \\
$\Gmumu$  &$  0.05$ &$ 0.25 $ & $ \pm0.30$ & $ \pm0.05$ &$ 0.63$ &$ 0.78$  \\
$\Gtautau$&$  0.16$ &$ 0.36 $ & $ \pm0.41$ & $ \pm0.05$ &$ 0.93$ &$ 1.08$  \\ \hline
\end{tabular}
\caption[Limits on partial widths]                        
{\label{tab-ewp-pwlim}
Upper limits  for total and partial $\Zzero$ widths.
The second column gives the difference between the measured width and 
the expected width in the {\SM} for $\MH = 150$ GeV, and the third column
for $\MH = 1000$ GeV.
These two results share the same errors, which are given in the fourth
column.
The first error is  experimental and the second error reflects parametric
uncertainties in the SM input 
parameters $\als ,\;\Mt$ and $\dalh$ as specified in {\EQ}~\ref{eq-sm-pars}. 
The last two columns show the corresponding upper limits 
(one-sided Bayesian limits at 95\,\% C.L.)
for new contributions, beyond the {\SM}.
For $\Gll$, $\Ginv$ and $\Ghad$ the experimental results with lepton universality imposed
have been used.
Both experimental and theoretical uncertainties are correlated between the different widths,
therefore the limits cannot be used simultaneously.

}
\end{center}
\end{table}

\begin{table}[htbp]  \begin{center}                                                                                                 
\renewcommand{\arraystretch}{1.4}                                                                                                   
\begin{tabular}{|l|c|c|c|}                                                                                                            
\hline                                                                                                                              
  & Without lepton  &  With lepton  & {\SM}\\                                                                              
  & universality    &  universality &    prediction \\                                                                              
\hline

 $\AAe$       & $  0.109^{+ 0.025}_{-0.032}$  &&\\                                                                              
 $\AAm$       & $  0.194^{+ 0.086}_{-0.044}$  &&\\                                                                              
 $\AAtau$     & $  0.177^{+ 0.083}_{-0.047}$  &&\\                                                                              
 $\AAl$       &
            & $  0.1392^{+ 0.0078}_{-0.0082}$& $  0.1450^{+ 0.0030}_{-0.0084}$  \\                                           
\hline                                                                                                                              
\end{tabular}                                                                                                                       
\caption[Results for the coupling parameters]                                                                                
{\label{tab-ewp-al}                                                                                                               
The leptonic coupling parameters obtained from a parameter                                                                         
transformation from the {\SLP}
given in {\TAB}~\ref{tab-ewp-leppar}.                                                                                                
In the last column we give the value of the parameter
calculated in the context of the {\SM} assuming                                                                            
the parameter variations given in {\EQ}{~\ref{eq-sm-pars}}.                                                                      
}                                                                                                                                   
\end{center}                                                                                                                        
\end{table}                                                                                                                         

\begin{table}[htbp]  \begin{center}                                                                                                 
\begin{tabular}{|l|rrr|}                                                                                                            
\hline                                                                                                                              
 & $\AAe   $ & $\AAm    $ & $\AAtau   $ \\                                                                                          
\hline                                                                                                                              
 $\AAe    $ &$  1.000  $ &$  -0.869 $ &$ -0.777 $ \\                                                                                   
 $\AAm    $ &$ -0.869  $ &$   1.000 $ &$  0.681 $ \\                                                                                   
 $\AAtau  $ &$ -0.777  $ &$   0.681 $ &$  1.000 $ \\                                                                                   
\hline                                                                                                                              
\end{tabular}                                                                                                                       
\end{center}                                                                                                                        
\caption[Coupling parameter correlation matrix]                                                                                      
{\label{tab-ewp-alcorr}                                                                                                            
Error correlation matrix for the measurements of                                                                                          
the leptonic coupling parameters,                                                                                                                 
presented in {\TAB}~\ref{tab-ewp-al}.                                                                              
}                                                                                                                                   
\end{table}                                                                                                                         

\begin{table}[htbp]  \begin{center}                                                                                                 
\renewcommand{\arraystretch}{1.4}                                                                                                   
\begin{tabular}{|l|c|c|c|}                                                                                                          
\hline                                                                                                                              
  & Without lepton  &  With lepton  & Standard Model\\                                                                              
  & universality    &  universality &    prediction \\                                                                              
\hline

 $\gae$       & $  -0.5009^{+ 0.0007}_{-0.0007}$  &&\\                                                                              
 $\gam$       & $  -0.5004^{+ 0.0026}_{-0.0013}$  &&\\                                                                              
 $\gatau$     & $  -0.5011^{+ 0.0024}_{-0.0016}$  &&\\                                                                              
 $\gal$       &
           & $  -0.50095^{+ 0.00046}_{-0.00046}$& $  -0.50130^{+ 0.00047}_{-0.00013}$  \\

 $\gve$       & $   -0.027^{+  0.008}_{ -0.006}$  &&\\                                                                              
 $\gvm$       & $   -0.049^{+  0.011}_{ -0.022}$  &&\\                                                                              
 $\gvtau$     & $   -0.045^{+  0.012}_{ -0.021}$  &&\\                                                                              
 $\gvl$       &
           & $  -0.0350^{+ 0.0021}_{-0.0020}$& $  -0.0365^{+ 0.0022}_{-0.0008}$  \\                                                 
\hline                                                                                                                              
\end{tabular}                                                                                                                       
\caption[Results for the axial and vector couplings]                                                                                
{\label{tab-ewp-gvga}                                                                                                               
Axial-vector and vector couplings obtained from a parameter                                                                         
transformation from the standard LEP parameter set                                                                                  
given in {\TAB}~\ref{tab-ewp-leppar}.                                                                                               
In the last column we give the values of the couplings                                                                              
calculated in the context of the {\SM} assuming                                                                                     
the parameter variations given in {\EQ}{~\ref{eq-sm-pars}}.                                                                         
}                                                                                                                                   
\end{center}                                                                                                                        
\end{table}                                                                                                                         

\begin{table}[htbp]  \begin{center}                                                                                                 
\begin{tabular}{|l|rrrrrr|}                                                                                                         
\hline                                                                                                                              
\makebox{\rule[-1.5ex]{0cm}{3.5ex}}

  & $\gae$       & $\gam$       & $\gatau$     & $\gve$       & $\gvm$       & $\gvtau$     \\
 \hline
 $\gae$       &$  1.00$ &$  -0.37$ &$  -0.29$ &$  -0.41$ &$   0.34$ &$   0.31$ \\                                                   
 $\gam$       &$  -0.37$ &$  1.00$ &$   0.49$ &$   0.70$ &$  -0.83$ &$  -0.54$ \\                                                   
 $\gatau$     &$  -0.29$ &$   0.49$ &$  1.00$ &$   0.52$ &$  -0.45$ &$  -0.70$ \\                                                   
 $\gve$       &$  -0.41$ &$   0.70$ &$   0.52$ &$  1.00$ &$  -0.87$ &$  -0.77$ \\                                                   
 $\gvm$       &$   0.34$ &$  -0.83$ &$  -0.45$ &$  -0.87$ &$  1.00$ &$   0.68$ \\                                                   
 $\gvtau$     &$   0.31$ &$  -0.54$ &$  -0.70$ &$  -0.77$ &$   0.68$ &$  1.00$ \\                                                   
\hline                                                                                                                              
\end{tabular}                                                                                                                       
\end{center}                                                                                                                        
\caption[6x6 leptonic coupling correlation matrix]                                                                                  
{\label{tab-ewp-gvga6}                                                                                                              
Error correlation matrix for the measurements of                                                                                    
the axial vector and vector couplings,                                                                                              
without assuming lepton                                                                                                             
universality, presented in {\TAB}~\ref{tab-ewp-gvga}.                                                                               
}                                                                                                                                   
\end{table}                                                                                                                         

%
%
%
%
\begin{table}[htbp]  
\renewcommand{\arraystretch}{1.4}
\begin{center} 
\begin{tabular}{|l|c|} \hline
Observable   & $\als$ \\ \hline
~~$\Rl$        & $0.132  \pm  0.007 _{-0.001}^{  + 0.003}$\\
~~$\GZ$        & $0.119  \pm  0.008 _{-0.004}^{  + 0.017}$\\
~~$\shadpol$   & $0.114  \pm  0.010 _{-0.001}^{  + 0.002}$\\
~~$\sleppol$   & $0.127  \pm  0.005 _{-0.001}^{  + 0.003}$\\
\hline
\end{tabular}
\caption[$\als$ from the $\Zzero$ resonance parameters]                           
{\label{tab-ewp-alfas}
Determination of $\als$ from the $\Zzero$ resonance parameters.
The central value is obtained with the {\SM} parameters
$\Mt ,\; \MH$ and $\dalh$ as specified in {\EQ}~\ref{eq-sm-pars}. 
The second error reflects
the effect on $\als$ when these are varied within the given ranges. 
In all cases an additional uncertainty of $\pm 0.002$ arises from
QCD uncertainties on $\Ghad$, and must be included.
}
\end{center}
\end{table} 

%
%
%
%
%
%
\begin{table}[htbp]  
\renewcommand{\arraystretch}{1.3}
\begin{center} 
\begin{tabular}{|l|r@{$\pm$}l|r@{$\pm$}l|r@{$\pm$}l|} \hline
& \multicolumn{2}{|c|}{OPAL $\Zzero$ resonance} 
& \multicolumn{2}{|c|}{external} 
& \multicolumn{2}{|c|}{external}  \\[-0.3ex]
& \multicolumn{2}{|c|}{measurements alone} 
& \multicolumn{2}{|c|}{$\Mt = (174.3\pm5.1)$GeV} 
& \multicolumn{2}{|c|}{$\Mt = (174.3\pm5.1)$GeV} \\
& \multicolumn{2}{|c|}{ } 
& \multicolumn{2}{|c|}{ } 
& \multicolumn{2}{|c|}{$\als = 0.1184\pm0.0031$}  \\ 
\hline
$\MZ$ (GeV)& $91.1851$ & $ 0.0030$  & $91.1851$ & $0.0030$  & $91.1852$ & $0.0030 $  \\
$\Mt$ (GeV)& $162 $ & $ 15 ^{+25}_{-5}$& $174.3 $ & $ 5.1$      & $173.4 $ & $ 5.1$ \\
$\dalh \; (\times 10^2)$      
           & $2.803 $ & $ 0.064 ^{-0.003}_{+0.001}$ 
                                     & $2.802 $ & $ 0.065$  & $2.800$ & $ 0.064$ \\
$\als(\MZ^2)$     & $0.125 $ & $ 0.005 ^{+0.004}_{-0.001}$ 
                                     & $0.127 $ & $ 0.005$  & $0.121 $ & $ 0.003$ \\
$\log_{10} (\MH/{\rm GeV})$
& \multicolumn{2}{|c|}{--}          
& \multicolumn{2}{|c|}{$2.59^{+0.46}_{-0.55} $}    
& \multicolumn{2}{|c|}{$2.28^{+0.44}_{-0.89}  $~} \\ \hline
$\MH$ (GeV)
& \multicolumn{2}{|c|}{150 (fixed) } 
& \multicolumn{2}{|c|}{$390 ^{+750}_{-280}$} 
& \multicolumn{2}{|c|}{$190^{+335}_{-165} $~}\\ \hline
$\chidof$   
& \multicolumn{2}{|c|}{159.7/200}
& \multicolumn{2}{|c|}{159.7/200}
& \multicolumn{2}{|c|}{161.7/201~} \\ \hline
\end{tabular}
\caption[Results of the {\SM} fits]                                                                                
{\label{tab-ewp-smres}                                                                                                               
Results of the full {\SM} fit to the measured cross-sections and asymmetries.
In the second column $\MH$ is fixed to 150 GeV.
The second errors show the variation for
 $\MH = 90$ GeV {\it (lower)} and $\MH = 1000$ GeV {\it (upper)}.
In the remaining columns $\MH$ is determined from the data, with
additional external constraints, as indicated.
In all cases the electromagnetic coupling constant $\dalh$
was used as additional fit parameter with the constraint
given in {\EQ}~\ref{eq-sm-pars}.
}                                                                                                                                   
\end{center}                                                                                                                        
\end{table}                                                                                                                         




%
\begin{table}[htbp]
\begin{center}\begin{tabular}{|l|c|}
 \hline
     & $\Delta \sigma (\mathrm{pb})$  \\
 \hline
$\seeF (\rm{pk}~-)$ & 1.28 \\
$\seeF (\rm{pk}~0)$ & 1.12 \\
$\seeF (\rm{pk}~+)$ & 1.20 \\
$\seeB (\rm{pk}~-)$ & 0.32 \\
$\seeB (\rm{pk}~0)$ & 0.32 \\
$\seeB (\rm{pk}~+)$ & 0.32 \\
 \hline
\end{tabular}\end{center}
\caption[Uncertainties in the electron t+ti cross-sections]
{\label{tab-ewp-eet+ti}
Uncertainties in the forward
($\seeF$) and backward
($\seeB$) electron cross-section for $t$-channel plus $s$-$t$ 
interference diagrams.
The designations
$(\rm{pk}~-)$, $(\rm{pk}~0)$,
and $(\rm{pk}~+)$ refer to the energy points respectively below, at and above
the $\Zzero$ resonance, where the peak region is taken
to lie within $\pm 0.9$ GeV of $\MZ$.
}
\end{table}

\clearpage
\begin{table}[htbp]  \begin{center}                                                                                                 
\renewcommand{\arraystretch}{1.4}                                                                                                   
\begin{tabular}{|l|r|r|r|}                                                                                                          
 \hline                                                                                                                             
Parameter  & \multicolumn{1}{|c|}{fitting $\jtoth$}&\multicolumn{1}{|c|}{fixing $\jtoth$}&                                          
        \multicolumn{1}{|c|}{{\SM}} \\                                                                                              
  & & & \multicolumn{1}{|c|}{Prediction}     \\                                                                                     
 \hline                                                                                                                             
$\MZ$ (GeV)                & $91.1901 \pm   0.0115$ & $91.1866 \pm   0.0031$ & input \\                                             
$\GZ$ (GeV)                & $ 2.4936 \pm   0.0047$ & $ 2.4943 \pm   0.0041$ &                                                      
                             $ 2.4949^{+  0.0021}_{ -0.0074}$ \\                                                                    
  \hline                                                                                                                            
$\rtoth$                   & $  2.962 \pm    0.010$ & $  2.963 \pm    0.009$ &                                                      
                             $ 2.9627^{+  0.0051}_{ -0.0173}$ \\                                                                    
$\jtoth$                   & $    0.01 \pm    0.650$ &   &                                                                          
                             $  0.2181^{+  0.0048}_{ -0.0139}$ \\                                                                   
\hline                                                                                                                              
$\rtote$                   & $  0.14122 \pm   0.00085$ & $  0.14138 \pm   0.00069$ & \\                                             
$\rtotm$                   & $  0.14205 \pm   0.00061$ & $  0.14212 \pm   0.00056$ &                                                
                             $  0.14260^{+  0.00020}_{ -0.00070}$ \\                                                                
$\rtott$                   & $  0.14221 \pm   0.00078$ & $  0.14229 \pm   0.00074$ & \\                                             
\hline                                                                                                                              
$\jtote$                   & $  -0.085 \pm    0.052$ & $  -0.076 \pm    0.044$ & \\                                                 
$\jtotm$                   & $  -0.013 \pm    0.042$ & $  -0.003 \pm    0.030$ &                                                    
                             $  0.0043^{+  0.0002}_{ -0.0005}$ \\                                                                   
$\jtott$                   & $  -0.007 \pm    0.045$ & $   0.003 \pm    0.034$ & \\                                                 
\hline                                                                                                                              
$\rfbe$                    & $  0.00134 \pm   0.00086$ & $  0.00140 \pm   0.00084$ & \\                                             
$\rfbm$                    & $  0.00265 \pm   0.00046$ & $  0.00261 \pm   0.00044$ &                                                
                             $  0.00300^{+  0.00013}_{ -0.00035}$ \\                                                                
$\rfbt$                    & $  0.00238 \pm   0.00059$ & $  0.00234 \pm   0.00057$ & \\                                             
\hline                                                                                                                              
$\jfbe$                    & $   0.763 \pm    0.070$ & $   0.763 \pm    0.070$ & \\                                                 
$\jfbm$                    & $   0.732 \pm    0.036$ & $   0.732 \pm    0.036$ &                                                    
                             $  0.7985^{+  0.0007}_{ -0.0016}$ \\                                                                   
$\jfbt$                    & $   0.740 \pm    0.042$ & $   0.740 \pm    0.042$ & \\                                                 
\hline                                                                                                                              
$\chi^2/$d.o.f.   & 146.6 /  187 &  146.7 /  188 & \\                                                                               
 \hline                                                                                                                             
\end{tabular}                                                                                                                       
\renewcommand{\arraystretch}{1.0}                                                                                                   
\end{center}                                                                                                                        
\caption[Results of the 16 and 15 parameter S-Matrix fits]                                                                          
{\label{tab-ewp-smat}                                                                                                               
Results of the 16 and 15 parameter S-Matrix fits to the measured                                                                    
cross-section and lepton asymmetry data.                                                                                            
The uncertainties on the LEP energy are included in the errors quoted.                                                              
In the last column we give the predictions of                                                                                       
the {\SM} assuming the parameters and variations given in                                                                           
{\TAB}~\ref{eq-sm-pars}.                                                                                                            
}                                                                                                                                   
\end{table}                                                                                                                         

\clearpage                                                                                                                          
\begin{table}[htbp]                                                                                                                 
\begin{sideways}                                                                                                                    
\begin{minipage}[b]{\textheight}                                                                                                    
\begin{center}\begin{tabular}{|rl|rrrrrrrrrrrrrrrr|}                                                                                
 \hline                                                                                                                             
 \multicolumn{2}{|l|}{Parameters} & 1 & 2 & 3 & 4 & 5 & 6 & 7 & 8 & 9 & 10                                                          
 & 11 & 12 & 13 & 14 & 15 & 16\\                                                                                                    
 \hline                                                                                                                             
  & & & & & & & & & & & & & & & & & \\                                                                                              
  1 &      $\MZ$       & $ 1.000$ & $ -.441$ & $ -.428$ & $ -.964$ & $ -.578$ & $ -.357$ & $ -.281$ & $ -.537$
    & $ -.709$ & $ -.658$ & $ -.252$ & $  .295$ & $  .237$ & $  .024$ & $  .002$ & $  .007$\\
  2 &      $\GZ$       & $ -.441$ & $ 1.000$ & $  .934$ & $  .466$ & $  .684$ & $  .757$ & $  .595$ & $  .232$
    & $  .344$ & $  .320$ & $  .116$ & $ -.117$ & $ -.096$ & $ -.011$ & $  .043$ & $  .035$\\
  3 &      $\rtoth$    & $ -.428$ & $  .934$ & $ 1.000$ & $  .457$ & $  .679$ & $  .755$ & $  .594$ & $  .222$
    & $  .333$ & $  .310$ & $  .117$ & $ -.110$ & $ -.090$ & $ -.013$ & $  .043$ & $  .036$\\
  4 &      $\jtoth$    & $ -.964$ & $  .466$ & $  .457$ & $ 1.000$ & $  .585$ & $  .377$ & $  .296$ & $  .536$
    & $  .708$ & $  .657$ & $  .247$ & $ -.289$ & $ -.232$ & $ -.023$ & $  .000$ & $ -.005$\\
  5 &      $\rtote$    & $ -.578$ & $  .684$ & $  .679$ & $  .585$ & $ 1.000$ & $  .561$ & $  .433$ & $  .338$
    & $  .431$ & $  .400$ & $  .277$ & $ -.175$ & $ -.141$ & $ -.027$ & $  .024$ & $  .017$\\
  6 &      $\rtotm$    & $ -.357$ & $  .757$ & $  .755$ & $  .377$ & $  .561$ & $ 1.000$ & $  .475$ & $  .186$
    & $  .359$ & $  .259$ & $  .098$ & $ -.079$ & $ -.075$ & $ -.011$ & $  .059$ & $  .028$\\
  7 &      $\rtott$    & $ -.281$ & $  .595$ & $  .594$ & $  .296$ & $  .433$ & $  .475$ & $ 1.000$ & $  .147$
    & $  .218$ & $  .277$ & $  .076$ & $ -.073$ & $ -.043$ & $ -.008$ & $  .027$ & $  .074$\\
  8 &      $\jtote$    & $ -.537$ & $  .232$ & $  .222$ & $  .536$ & $  .338$ & $  .186$ & $  .147$ & $ 1.000$
    & $  .395$ & $  .365$ & $  .130$ & $ -.161$ & $ -.129$ & $  .189$ & $ -.002$ & $ -.004$\\
  9 &      $\jtotm$    & $ -.709$ & $  .344$ & $  .333$ & $  .708$ & $  .431$ & $  .359$ & $  .218$ & $  .395$
    & $ 1.000$ & $  .482$ & $  .181$ & $ -.190$ & $ -.170$ & $ -.017$ & $ -.041$ & $ -.003$\\
 10 &      $\jtott$    & $ -.658$ & $  .320$ & $  .310$ & $  .657$ & $  .400$ & $  .259$ & $  .277$ & $  .365$
    & $  .482$ & $ 1.000$ & $  .168$ & $ -.197$ & $ -.138$ & $ -.016$ & $  .000$ & $ -.051$\\
 11 &      $\rfbe$     & $ -.252$ & $  .116$ & $  .117$ & $  .247$ & $  .277$ & $  .098$ & $  .076$ & $  .130$
    & $  .181$ & $  .168$ & $ 1.000$ & $ -.084$ & $ -.068$ & $  .054$ & $ -.001$ & $ -.002$\\
 12 &      $\rfbm$     & $  .295$ & $ -.117$ & $ -.110$ & $ -.289$ & $ -.175$ & $ -.079$ & $ -.073$ & $ -.161$
    & $ -.190$ & $ -.197$ & $ -.084$ & $ 1.000$ & $  .082$ & $  .008$ & $  .181$ & $  .004$\\
 13 &      $\rfbt$     & $  .237$ & $ -.096$ & $ -.090$ & $ -.232$ & $ -.141$ & $ -.075$ & $ -.043$ & $ -.129$
    & $ -.170$ & $ -.138$ & $ -.068$ & $  .082$ & $ 1.000$ & $  .006$ & $  .002$ & $  .162$\\
 14 &      $\jfbe$     & $  .024$ & $ -.011$ & $ -.013$ & $ -.023$ & $ -.027$ & $ -.011$ & $ -.008$ & $  .189$
    & $ -.017$ & $ -.016$ & $  .054$ & $  .008$ & $  .006$ & $ 1.000$ & $  .000$ & $  .000$\\
 15 &      $\jfbm$     & $  .002$ & $  .043$ & $  .043$ & $  .000$ & $  .024$ & $  .059$ & $  .027$ & $ -.002$
    & $ -.041$ & $  .000$ & $ -.001$ & $  .181$ & $  .002$ & $  .000$ & $ 1.000$ & $  .002$\\
 16 &      $\jfbt$     & $  .007$ & $  .035$ & $  .036$ & $ -.005$ & $  .017$ & $  .028$ & $  .074$ & $ -.004$
    & $ -.003$ & $ -.051$ & $ -.002$ & $  .004$ & $  .162$ & $  .000$ & $  .002$ & $ 1.000$\\
 &  & & & & & & & & & & & & & & & & \\                                                                                              
 \hline                                                                                                                             
\end{tabular}\end{center}                                                                                                           
\caption[Correlation matrix for the 16 parameter S-Matrix fit]                                                                      
{\label{tab-ewp-smatcor16}                                                                                                          
Error correlation matrix for the S-Matrix fit in {\TAB}~\ref{tab-ewp-smat}
without assuming                                                                      
lepton universality, and without fixing $\jtoth$, which controls                                                                    
the $\gamma Z$ interference in the hadron channel.                                                                                  
}                                                                                                                                   
\end{minipage}                                                                                                                      
\end{sideways}                                                                                                                      
\end{table}                                                                                                                         

%
%
\clearpage \newpage
\clearpage
\newpage
\begin{figure}[htb]
  \setlength{\unitlength}{1mm}
\begin{center}
\mbox{
\begin{picture}(170,180)
\put(0,    0){\epsfig{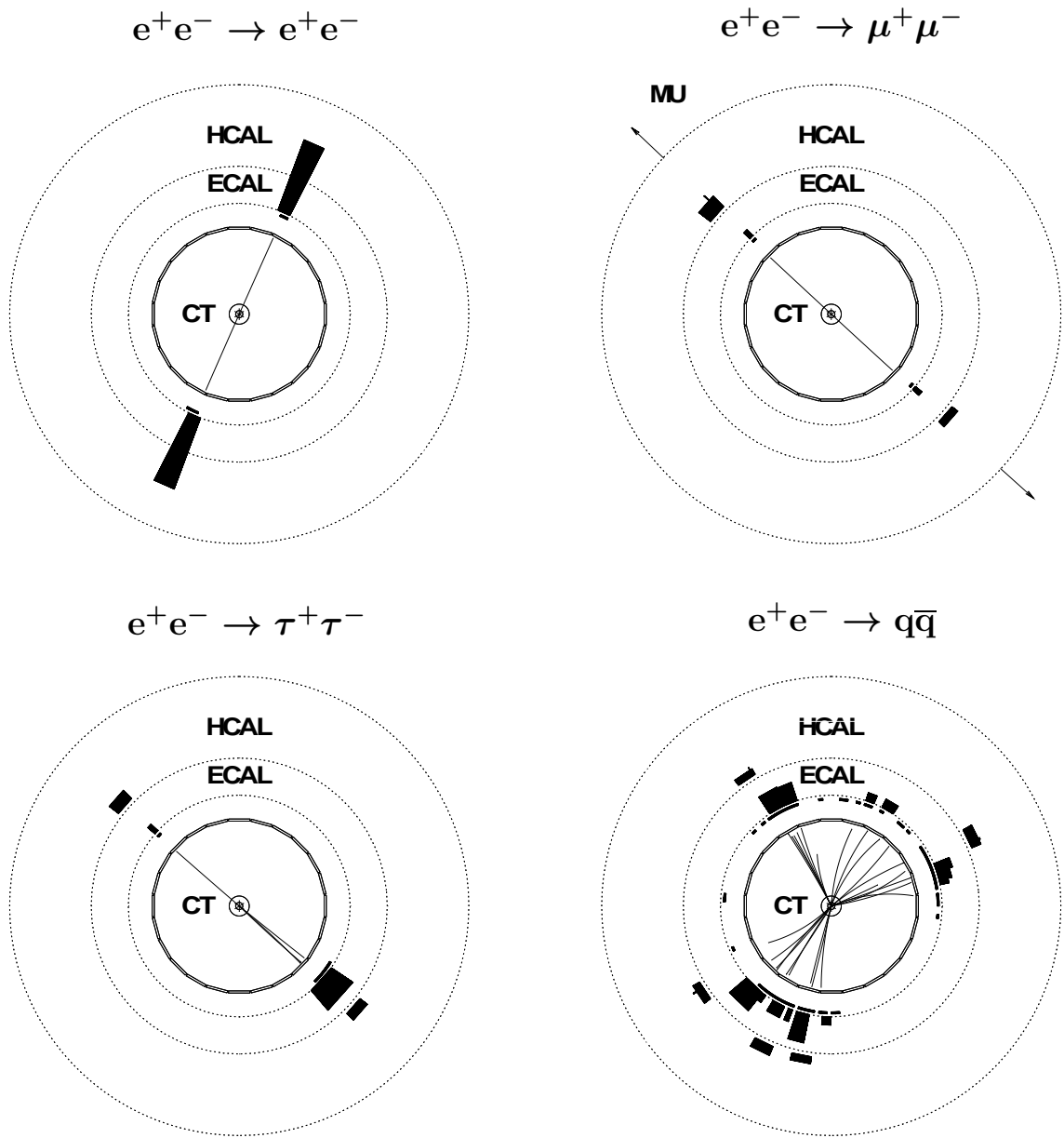}} 
\put(0,  168){\makebox(85,1){\Large \boldmath $\eeee$}}
\put(85, 168){\makebox(85,1){\Large \boldmath $\eemumu$}}
\put(0,   83){\makebox(85,1){\Large \boldmath $\eetautau$}}
\put(85,  83){\makebox(85,1){\Large \boldmath $\eehad$}}
\end{picture}
}
\caption[OPAL event examples]{Typical examples 
for the four event categories.
These views of the four final states measured in this analysis
all show the detector projected along the beam axis, parallel to
the magnetic field generated by the solenoid located between
CT and ECAL.
The approximately radial lines within the volume of the central
tracker (CT) represent the reconstructed tracks of inonising particles.
The dark trapezoids in the volumes of the electromagnetic calorimeter
(ECAL) and hadronic calorimeter (HCAL) represent corresponding
observed energy deposits.
The arrows within the volume of the multiple-layer muon chambers (MU)
represent reconstructed track segments of penetrating particles.
}
\label{f-opal-ev}
\end{center}
\end{figure}

\clearpage
\newpage
\begin{figure}[htb]
\begin{center}
\begin{minipage}{0.3\textwidth}
  \mbox{\epsfxsize1.0\textwidth\epsffile{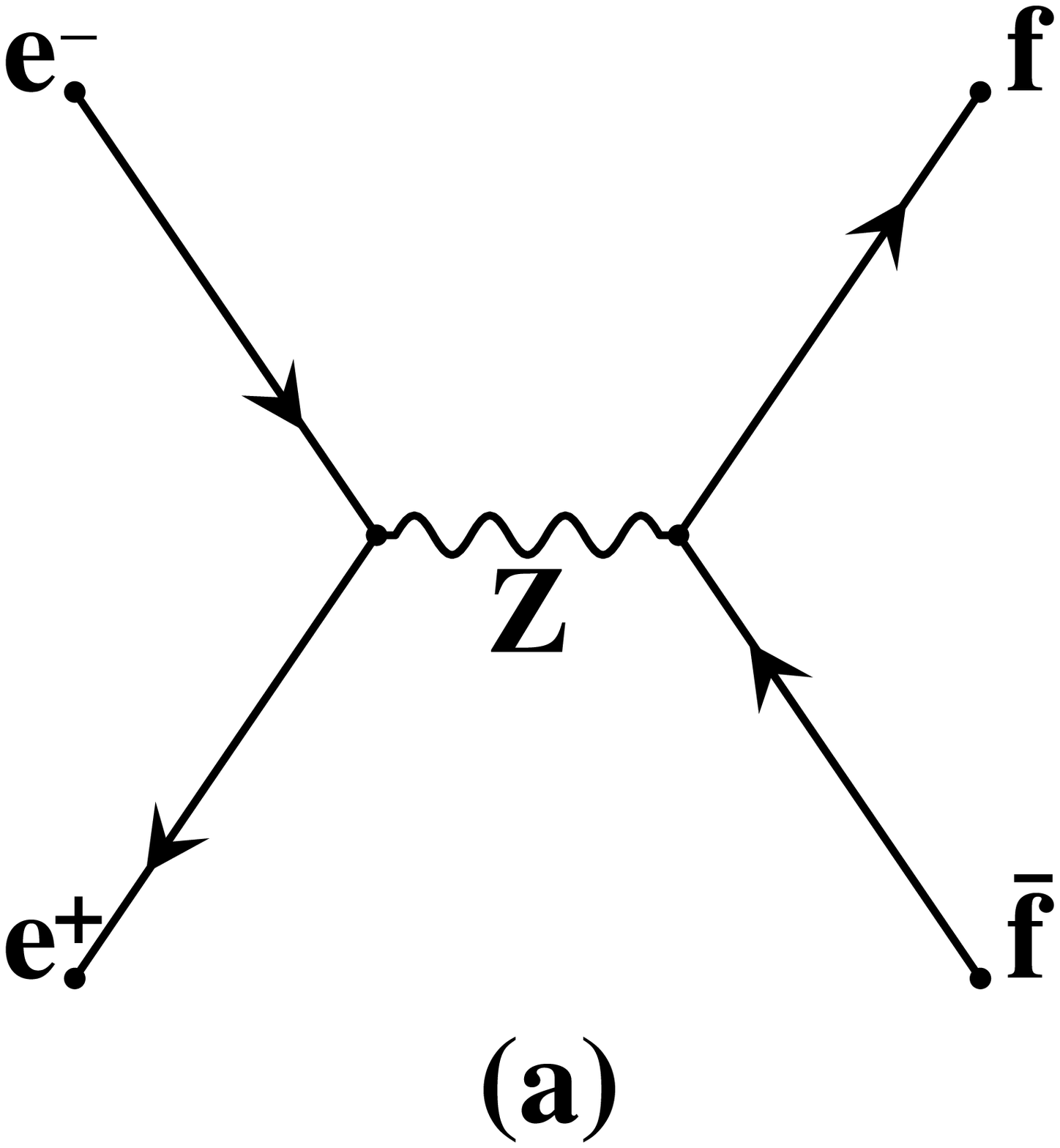}}
\end{minipage}~~
\begin{minipage}{0.3\textwidth}
  \mbox{\epsfxsize1.0\textwidth\epsffile{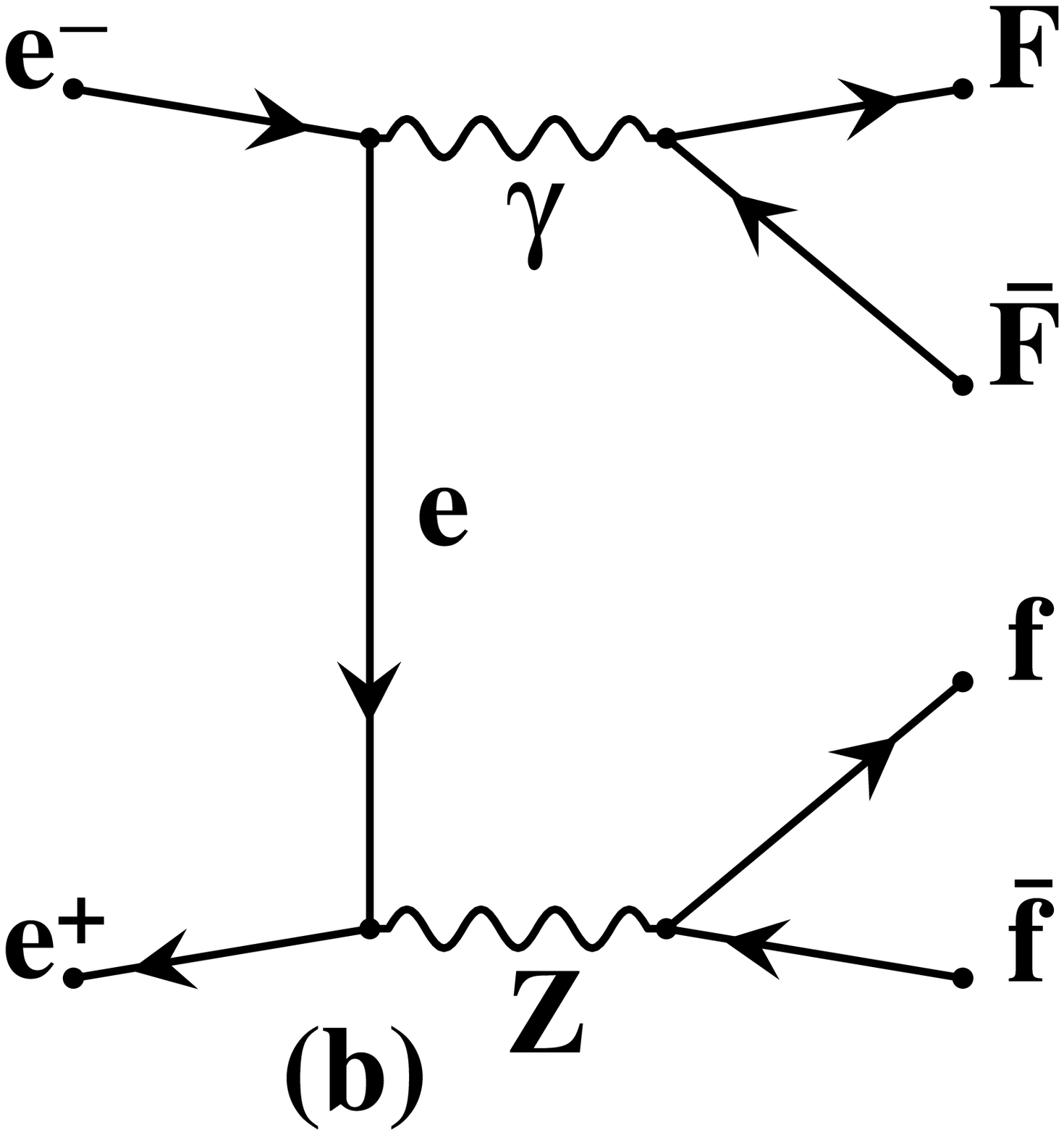}}
\end{minipage}~~
\begin{minipage}{0.3\textwidth}
  \mbox{\epsfxsize1.0\textwidth\epsffile{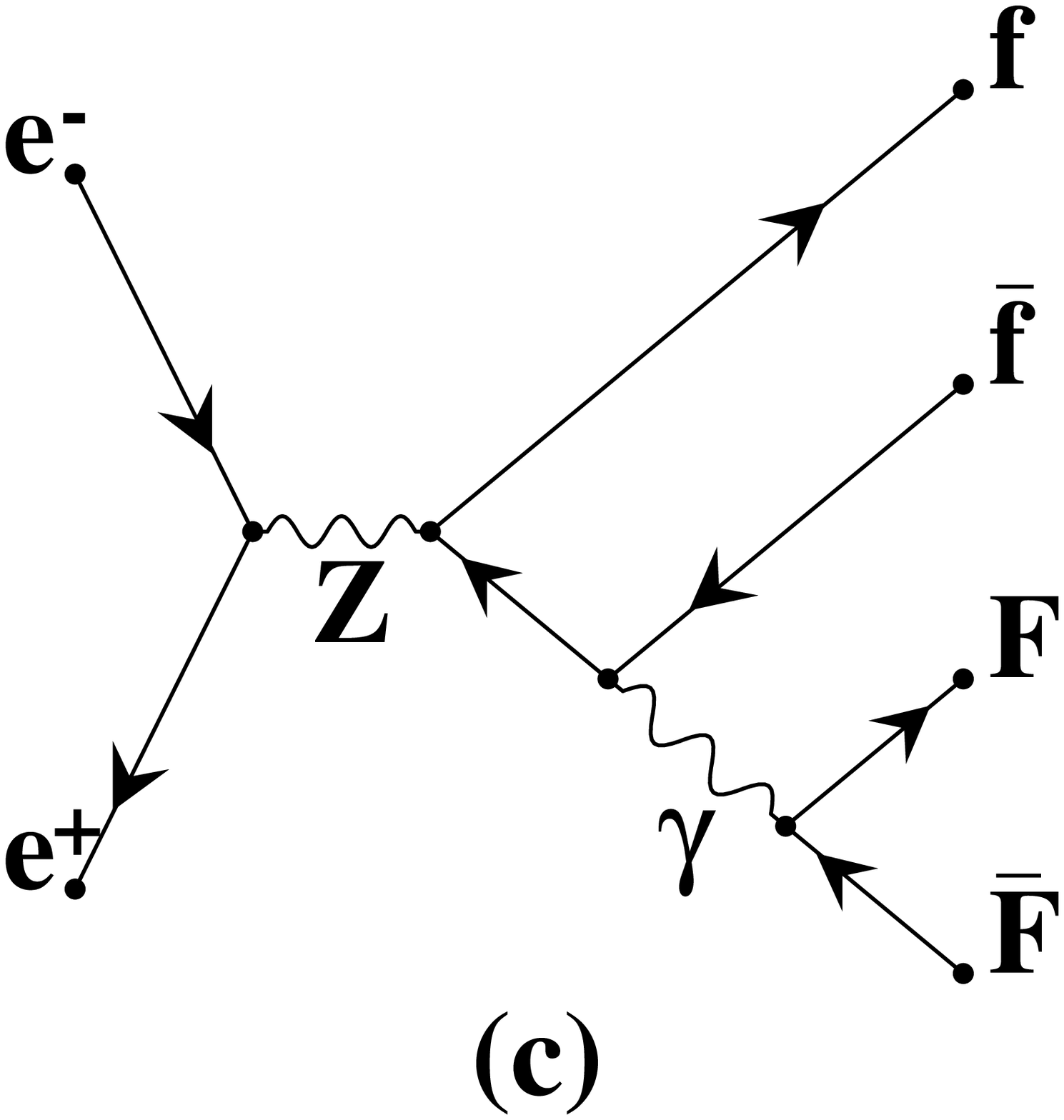}}
\end{minipage}

\begin{minipage}{0.3\textwidth}
  \mbox{\epsfxsize1.0\textwidth\epsffile{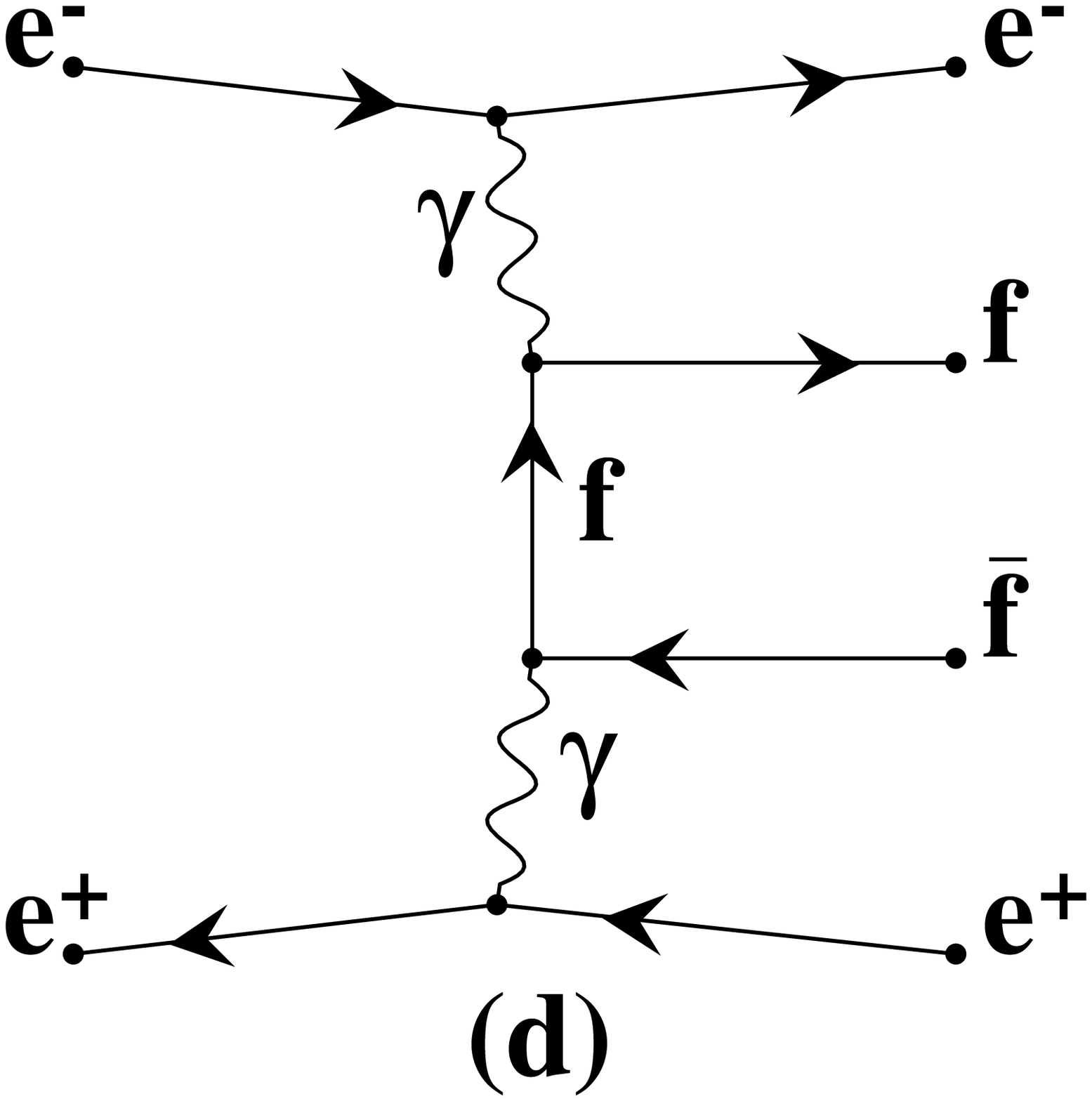}}
\end{minipage}~~
\begin{minipage}{0.3\textwidth}
  \mbox{\epsfxsize1.0\textwidth\epsffile{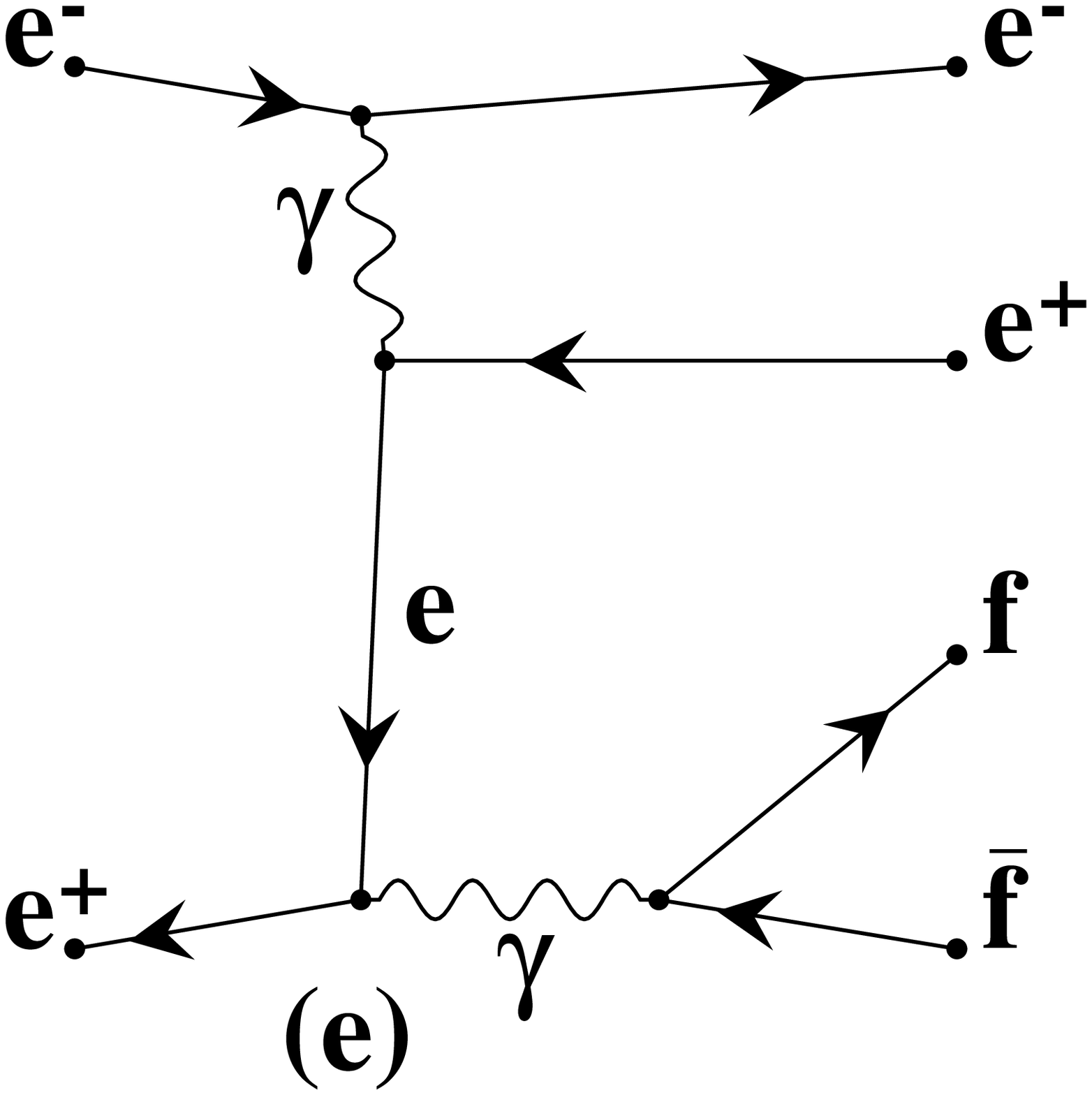}}
\end{minipage}~~
\begin{minipage}{0.3\textwidth}
  \mbox{\epsfxsize1.0\textwidth\epsffile{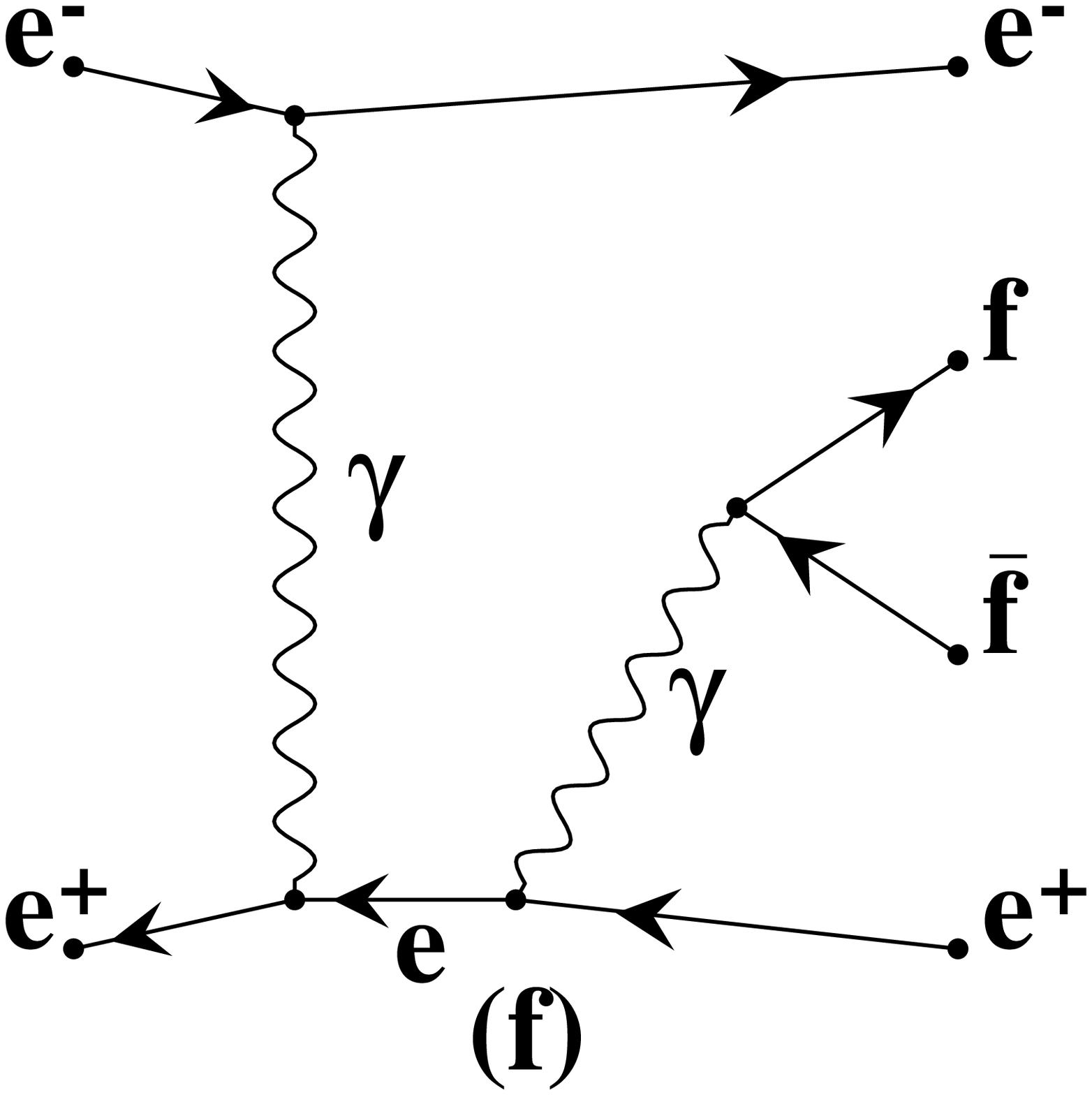}}
\end{minipage}
\caption[Two- and four-fermion diagrams]{
Two- and four-fermion diagrams. \\
(a) The basic fermion-pair $s$-channel diagram which is responsible
for almost all of the $\ff$ signal.\\
(b) Radiative correction to (a) with a fermion pair
$\FF$ in the initial state. Treated as signal.\\
(c) Radiative correction to (a) with a fermion pair
$\FF$ in the final state. Treated as signal.\\
(d) Multiperipheral (two-photon) diagram. Treated as background.\\
(e) Initial-state pair production in the $t$-channel, treated as background
except for $\eeee$.\\
(f) Final-state pair production in the $t$-channel, treated as background
except for $\eeee$.\\
Only the dominant boson is indicated. Additional contributions arise by
substituting $\Zzero$ and $\gamma$.
}
\label{f-four-fermion}
\end{center}
\end{figure}

%
%
\clearpage
\newpage
%
\begin{figure}[htb]
\begin{center}
\mbox{\epsfxsize1.0\textwidth\epsffile{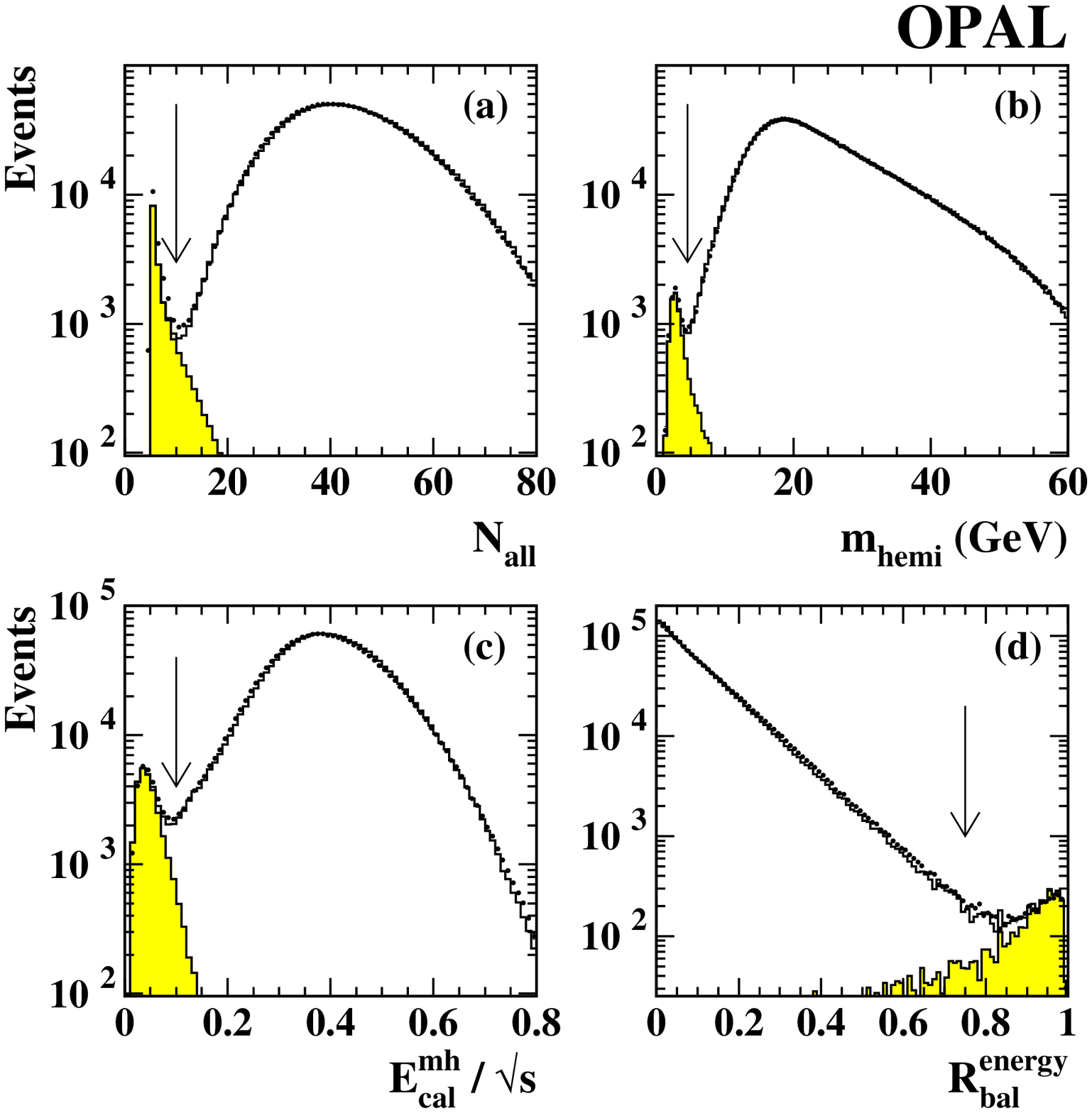}}
\end{center}
\caption[Distributions of the fundamental hadronic selection variables]{
The \mbox{$\eehad$} event selection.
A comparison of the cut variables between data and Monte Carlo 
simulation:
(a)~the total multiplicity,~$\Nall$,
(b)~the hemisphere invariant mass sum,~$\Mhemi$, 
(c)~the visible energy,~$\Rcal$,
(d)~the energy imbalance along the beam direction,~$\Rbal$. 
The points are the data and the open histograms the \mbox{$\eehad$} 
Monte Carlo simulation.
The shaded histograms show the background Monte Carlo prediction 
which is dominated in the case of ~(a) and~(b) by \mbox{$\eell$} and
in the case of ~(c) and~(d) by \mbox{$\eeeeff$}.
The cuts are indicated by the arrows. 
In each case events are plotted only if they pass all the other selection cuts.
\label{mhcuts} }
\end{figure}

\clearpage
\newpage
\begin{figure}[htb]
\begin{center}
\mbox{\epsfxsize1.00\textwidth\epsffile{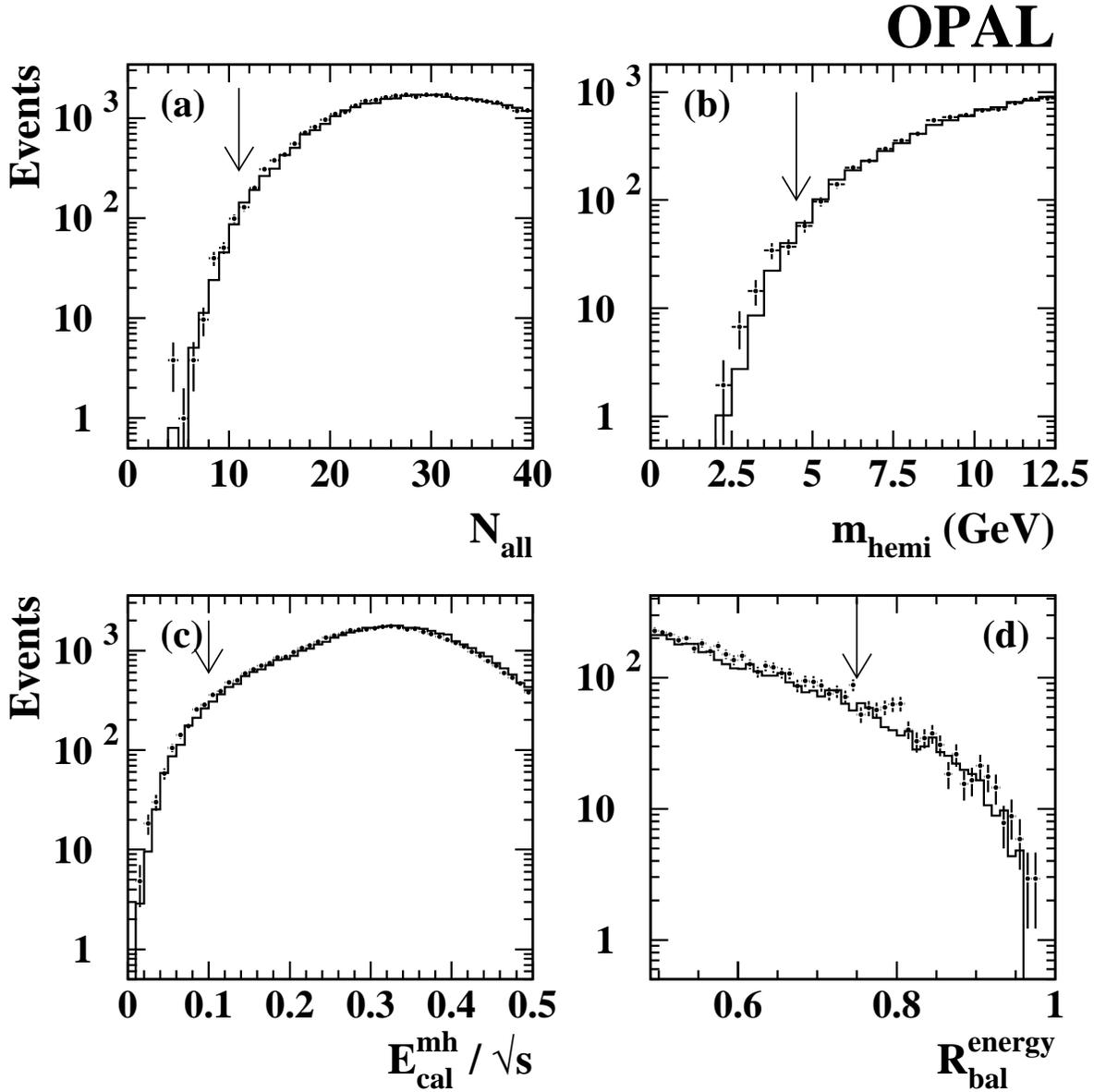}}
\end{center}
\caption[Comparison of data and MC simulation near the $x$-axis hole]{
Distributions of the emulated cut variables for \mbox{$\eehad$}
events at small angles to the 
\mbox{$x$-axis,} which have been processed by the acceptance hole 
emulation program, as described in the text.
The data~(points) and the JETSET Monte Carlo events~(histogram) are compared.
For these plots events are required to pass the 
standard \mbox{$\eehad$} selection before the hole emulation.
\label{mhhole2} }
\end{figure}

\clearpage
\newpage
\begin{figure}[htb]
\begin{center}
\mbox{\epsfxsize1.00\textwidth\epsffile{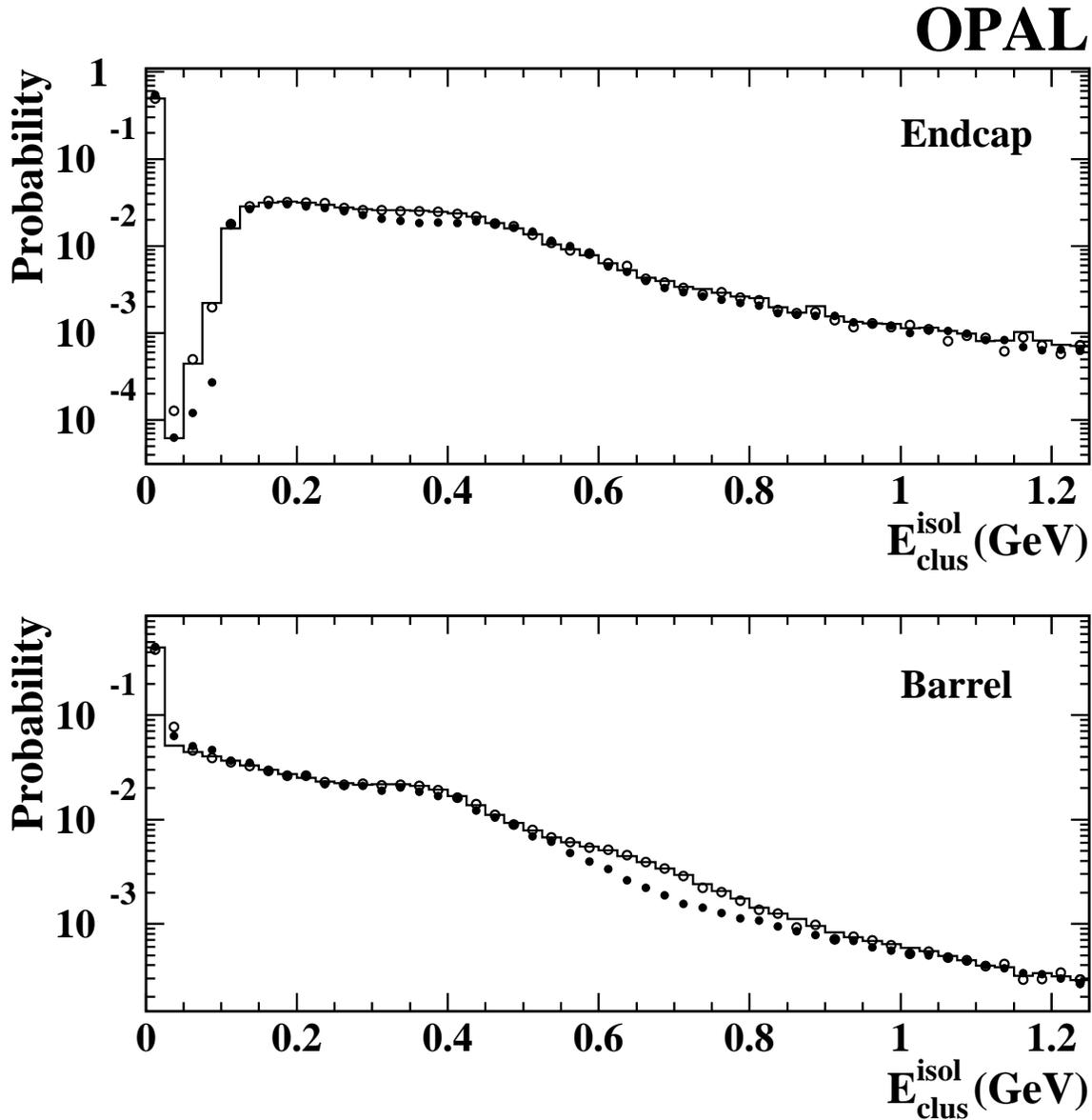}}
\end{center}
\caption[Rescaling of the MC simulated cluster energies]{
Detector simulation study for the \mbox{$\eehad$} selection.
The plots show the energy spectra (normalised to the number of 
tracks) for single ECAL~clusters near isolated tracks 
for data~(histogram),
for the standard Monte Carlo simulation~(closed circles) and
for the corrected Monte Carlo simulation~(open circles). 
The upper plot is for tracks pointing into the endcaps of the 
electromagnetic calorimeter.
The lower plot is for tracks pointing into the barrel region. 
Both plots are summed over all momenta. 
A clear improvement in the modelling of the distributions can
be seen after correction.
\label{clenergy} }
\end{figure}

\clearpage
\newpage
\begin{figure}[htb]
\begin{center}
\mbox{\epsfxsize1.00\textwidth\epsffile{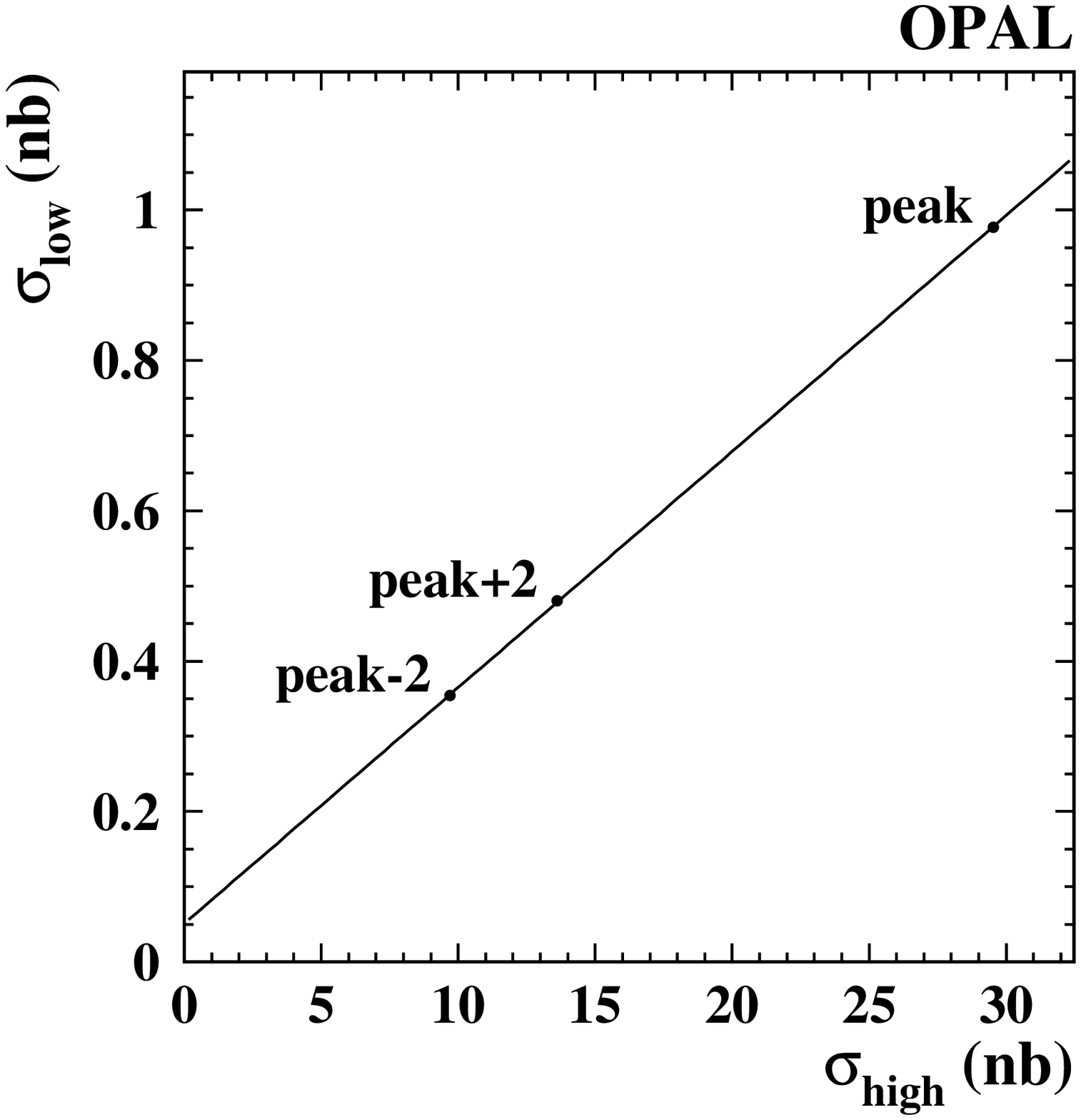}}
\end{center}
\caption[Non-resonant background in the $\eehad$ channel]{
The non-resonant background evaluation for the 
\mbox{$\eehad$} selection.
The plot shows the cross-section for events having 
a low visible energy,~\mbox{$0.10<\Rcal<0.18$}, or 
a large energy imbalance along the beam direction,~\mbox{$0.50<\Rbal<0.75$}, 
($\sigma_{\rm low}$) 
versus the cross-section for events having a high~$\Rcal$ ($>0.18$)
and a small~$\Rbal$ ($<0.50$) ($\sigma_{\rm high}$). 
The intercept of the fitted straight line yields the non-resonant background
estimate. 
The distributions for $\Rcal$ and $\Rbal$ are shown 
in {\FIG}~\ref{mhcuts} (c) and (d), respectively.
\label{nrb} }
\end{figure}

\clearpage \newpage
%

\clearpage
\newpage
\begin{figure}[htbp]
 \epsfxsize=\textwidth
 \epsffile{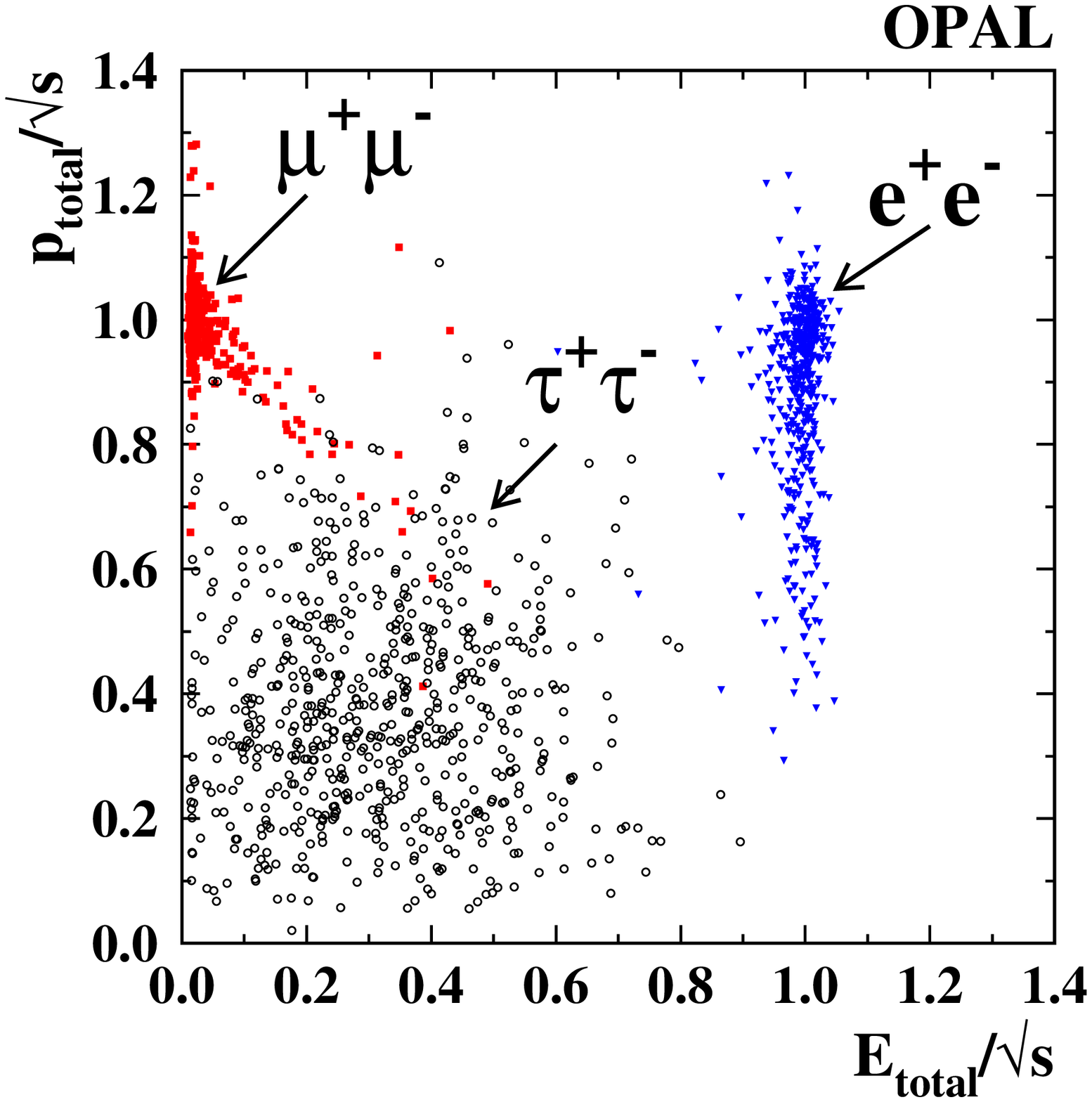}
 \caption[Separation of $\eell$ events in $\Etotal / \roots$ 
         {\em vs.} $\ptotal / \roots$.]
          {The separation of Monte Carlo $\eell$ events using 
          the energy deposited in the electromagnetic
          calorimeter summed over all clusters, $\Etotal$ 
          versus the scalar sum of the momenta of the
          reconstructed tracks in the event, $\ptotal$. 
          Both quantities are scaled to the centre-of-mass energy
          $\roots$. The triangles show
          $\eeee$ events, the solid squares show $\eemumu$ events and
          the open circles show $\eetautau$ events.}
 \label{fig-ll}
\end{figure}

\clearpage
\newpage
\begin{figure}[p]
\begin{center}\hspace*{-0.5cm}
\mbox{\epsfxsize1.0\textwidth\epsffile{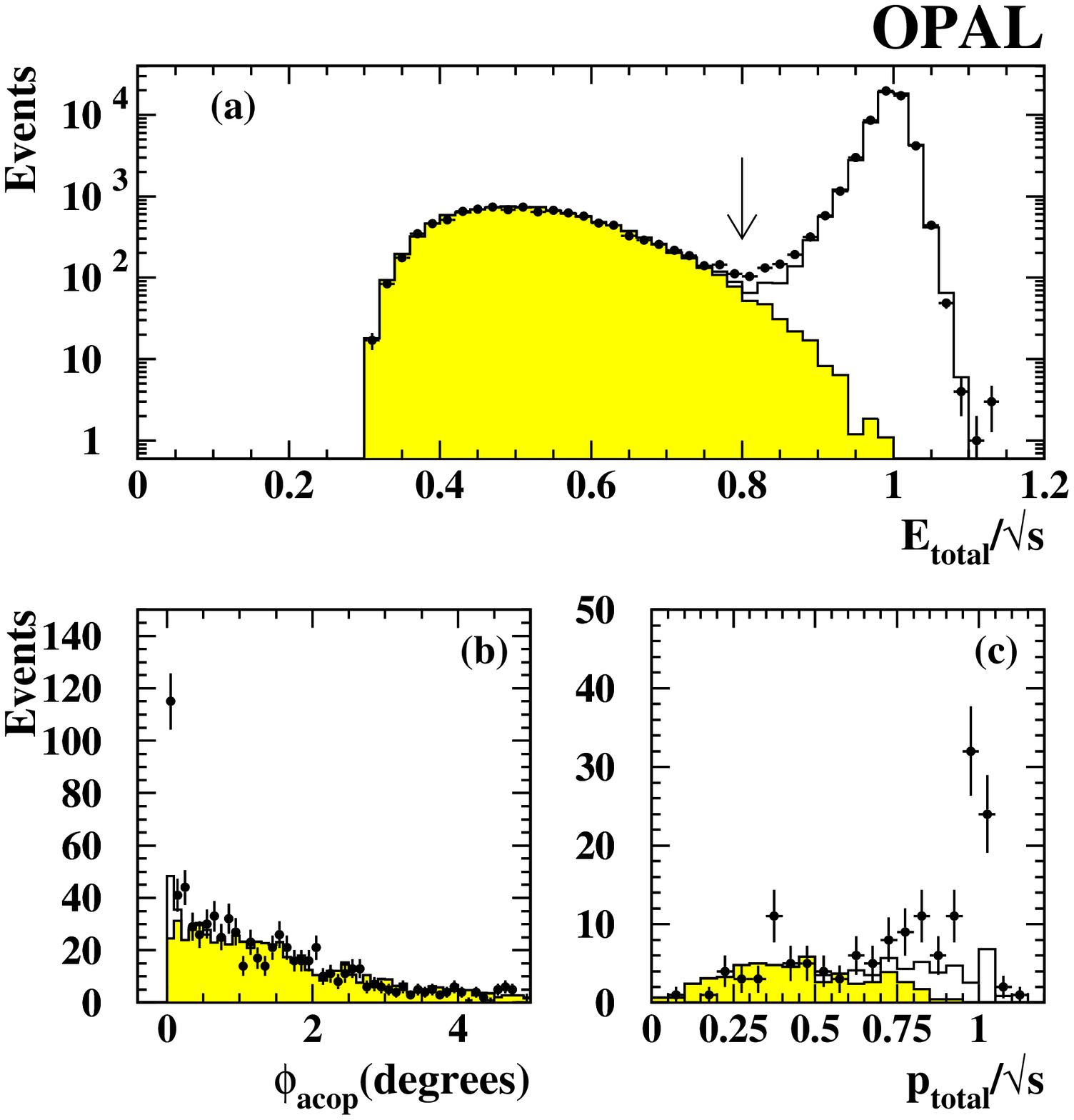}}
\caption[Fundamental distributions of the \mbox{$\eeee$} event selection
variables]{
The \mbox{$\eeee$} event selection.\newline
(a) Distribution of the sum of electromagnetic energy,~$\Etotal/\roots$,
after all the other cuts have been applied,
in the angular range \mbox{$\abscosthelec<0.70$.}
The arrow indicates the selection cut used.\newline
(b) Distribution of the acoplanarity angle of~$\en$ and~$\ep$ tracks for 
events satisfying \mbox{$0.7 < \Etotal/\roots < 0.8$.}\newline
(c) Distribution of the scalar sum of the track momenta,~$\ptotal/\roots$, 
for the events satisfying \mbox{$0.7 < \Etotal/\roots < 0.8$} 
and \mbox{$\thacop < 0.2^\circ$.}
In each case the points are the on-peak data, 
the open histogram shows the Monte Carlo expectation and 
the shaded histogram shows the contribution from background processes.
}
\label{f-eeecut1}
\end{center}
\end{figure}


\clearpage\newpage
\begin{figure}[p]
\begin{center}
\mbox{\epsfxsize1.0\textwidth\epsffile{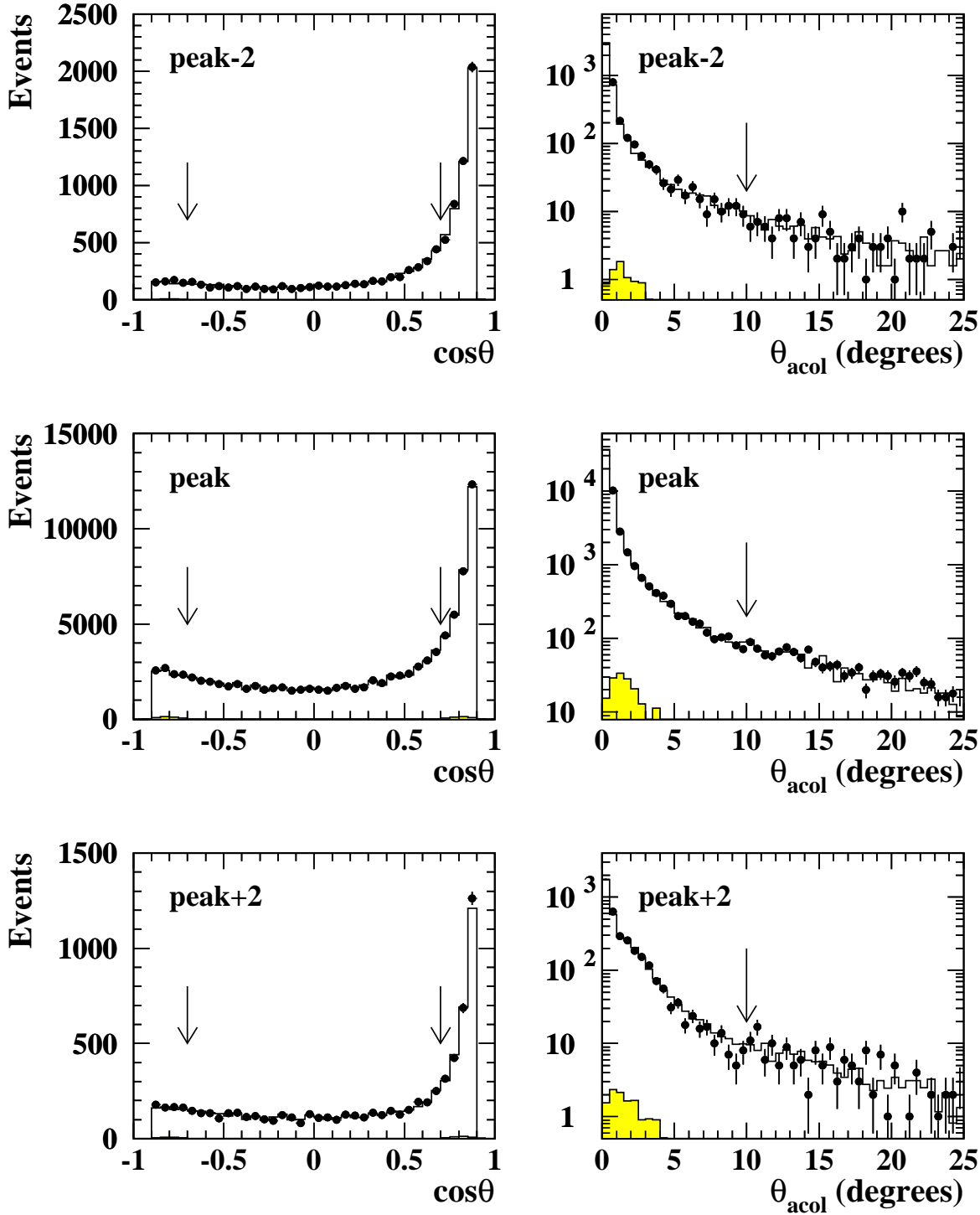}}
\caption[Angular and acollinearity distributions of the {$\eeee$} event sample]{
The \mbox{$\eeee$} event acceptance.
Angular distributions and acollinearity distributions from
data samples at three different centre-of-mass energies.
In each case the points are the data, 
the open histogram shows the Monte Carlo expectation and 
the shaded histogram shows the contribution from background processes.
The arrows indicate the acceptance cuts used.
}
\label{f-eeacc}
\end{center}
\end{figure}

%

\clearpage
\newpage
\begin{figure}[htbp]
 \epsfxsize=\textwidth
 \epsffile{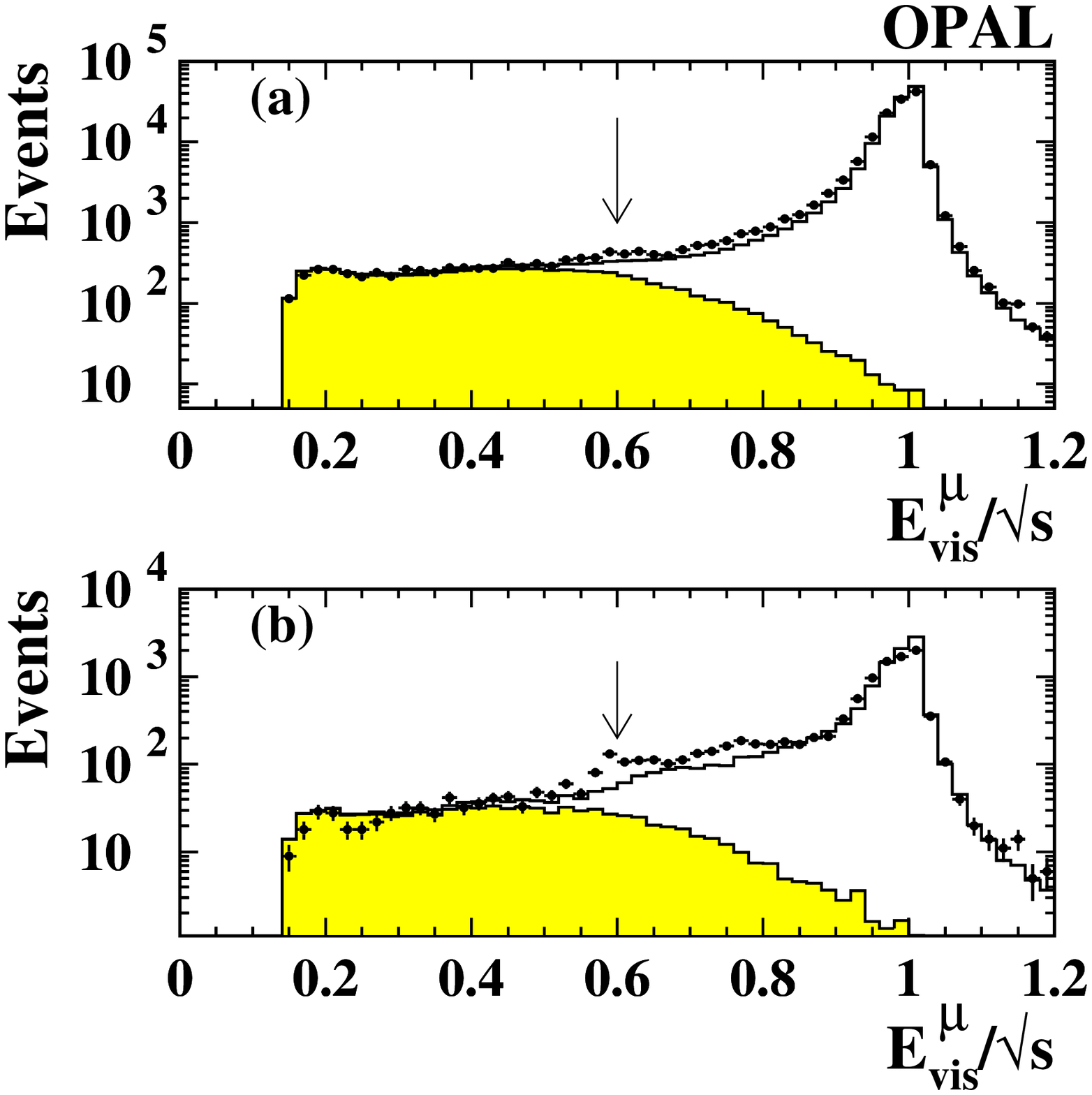}
 \caption[$\Evis$ distributions for $\eemumu$]{    
    Distribution of $\Evis$ for events passing all of the $\eemumu$
    selection cuts with the exception of $\Evis>0.6\roots$.  
    The 1993--1995 data are shown by the points and
    the histograms indicate the Monte Carlo expectation.
    The shaded histograms represent the background (mainly  $\eetautau$).
    Figure (a) gives the distribution for all events, while (b) 
    shows those events with tracks within $0.5^\circ$ of the
    anode wire planes of the central tracking chamber.
 \label{fig:mu_fvis}
}
\end{figure}

%

\clearpage
\newpage
\begin{figure}[htbp]
 \epsfxsize=\textwidth
 \epsffile{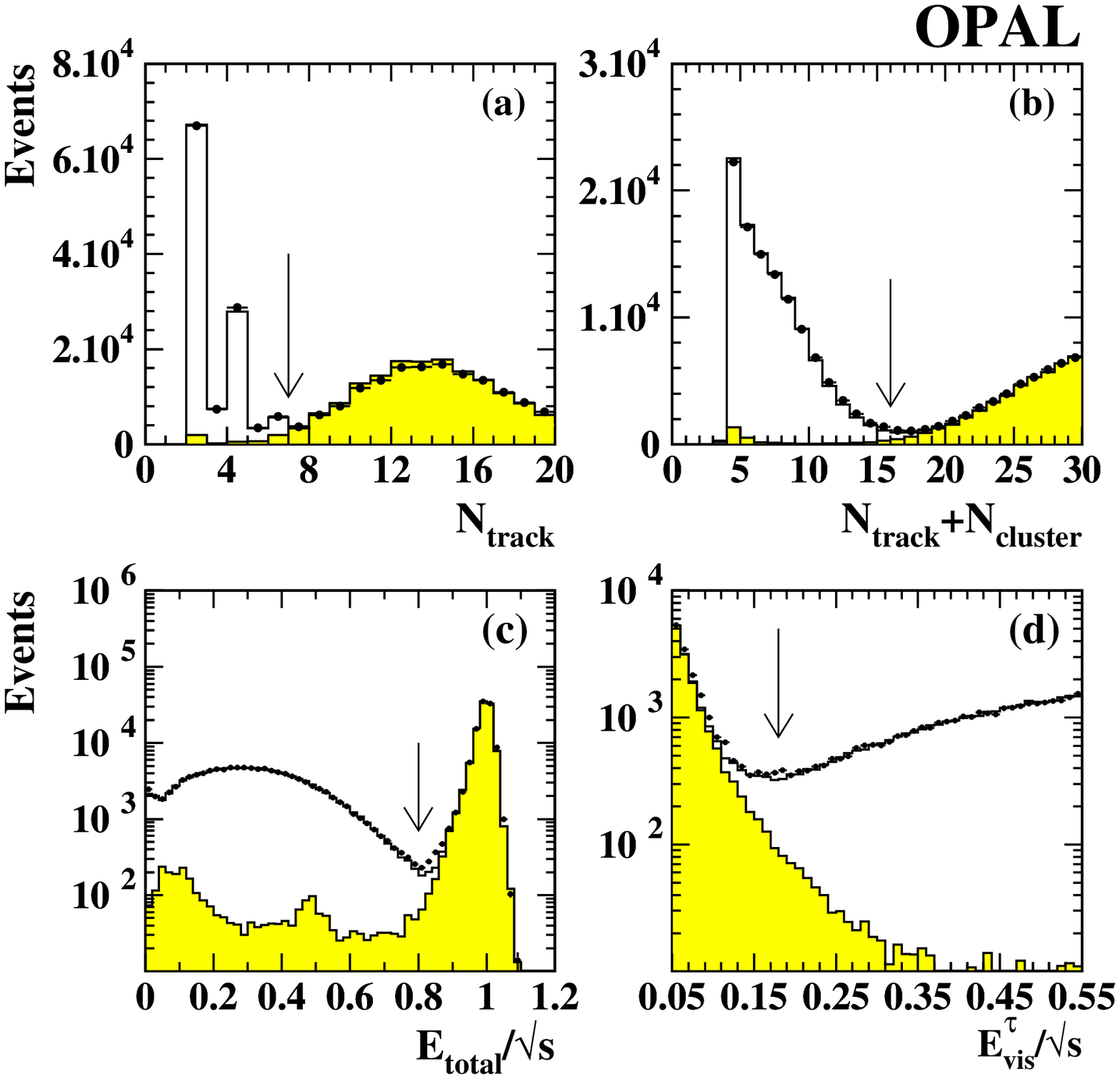}
 \caption[Distributions of the principal selection cut variables
 for $\eetautau$]
 {    
 Cuts used in the \mbox{$\eetautau$} event selection.
 Distributions of
 (a) and (b) the charged and total multiplicity, $\Ntra$ and
  $\Ntra+\Nclu$ used to reject background from $\eehad$,
 (c)~$\Etotal$ for events with \mbox{$\costau<0.7$} showing the cut used to 
 reject the background from \mbox{$\eeee$} in this barrel region of the 
 detector and
 (d)~$\Rvistau$ showing the cut used to reject two-photon interaction events.
 In all cases all other selection cuts have been applied.
 The points represent the \mbox{1993--1995} data (on-peak and off-peak), 
 the histograms show the Monte 
 Carlo expectation and the shaded histograms indicate the background 
 component, predominantly \mbox{$\eehad$} in~(a) and (b), $\eeee$ in (c)
 and  \mbox{$\eeeell$} in~(d). 
 The cuts are indicated by the arrows.}
 \label{fig:tau_fig1}
\end{figure}


\clearpage
\newpage
\begin{figure}[htbp]
 \epsfxsize=\textwidth
 \epsffile{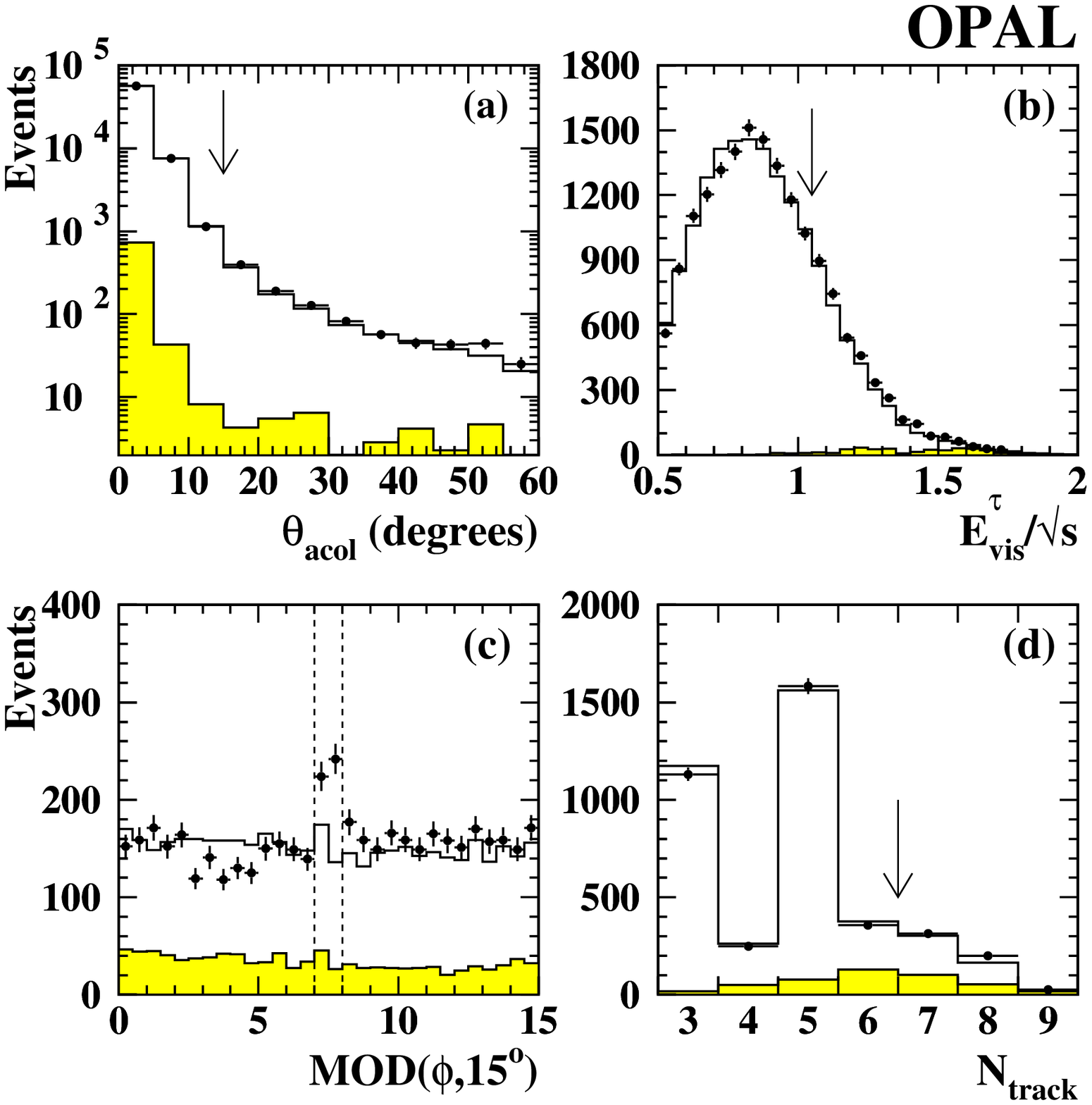}
 \caption[Systematic checks of the $\eetautau$ selection]{    
 Systematic checks of the \mbox{$\eetautau$} event selection.\newline
 Each plot corresponds to one of the
 control samples used to check the efficiency and background 
 of the $\eetautau$ event selection. Plot a) corresponds to the sample used
 to assess the effect of the acollinearity cut, b) corresponds to the
 check of the inefficiency due to the cut on $\Evistau$ in the region 
 \mbox{$\costau>0.7$}, c) shows the excess of $\eemumu$ background events in 
 the region of the jet chamber anode planes, and d) shows the distribution
 of $\Ntra$ for the control sample used to assess the $\eeqq$ background. 
 In all plots the data are shown by the points, the total 
 Monte Carlo expectations
 are shown by the histograms and the contributions from events other than
 $\eetautau$ are shown by the shaded histograms.
  Details are given in the text.}
 \label{fig:tau_fig2}
\end{figure}

%

\clearpage
\newpage
\begin{figure}[p]
\begin{center}\hspace*{-0.5cm}
\mbox{\epsfxsize1.0\textwidth\epsffile{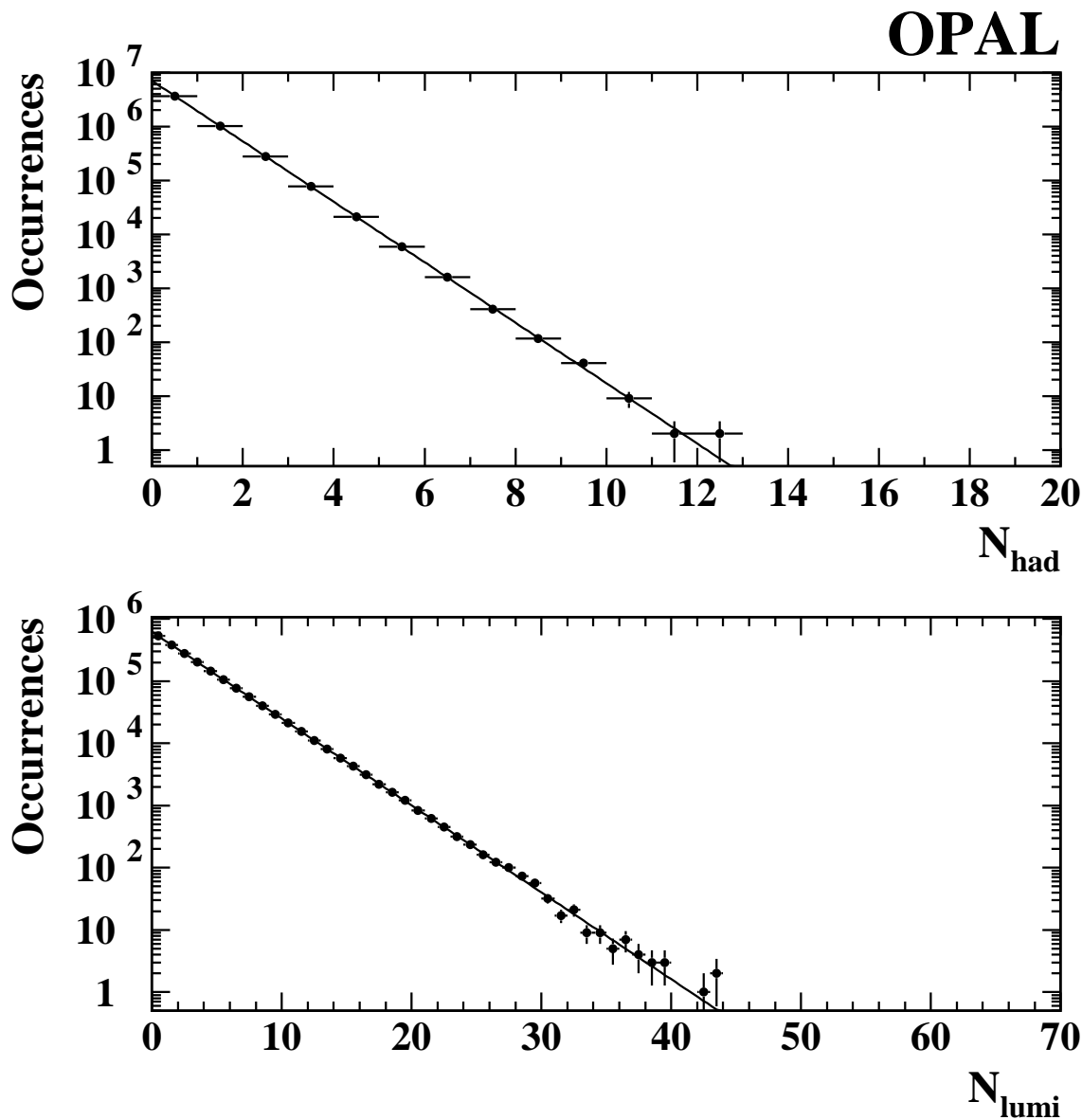}}
\caption[$eeqq$ and luminosity event sensitivity checks]{
The distribution of the number of events of type (a) observed
between adjacent events of type (b) forms a sensitive test for the
constancy of the a/b rate ratio.
When the ratio truly remains constant, the resulting
distribution is a pure exponential, indicated by the line,
whose logarithmic slope depends on the event type ratio.
The upper plot shows the number of $\eeqq$ events observed between
adjacent luminosity events in the 1993--1995 peak data.
The lower plot shows the reverse: the number of luminosity events observed
between adjacent $\eeqq$ events.
The tails of these distributions are particularly sensitive to any interruption
in the experimental sensitivity to events of type (b), at the level of a
few minutes in the sample of about six months of livetime shown.
There were no overflows in either of these distributions.
}
\label{f-meas-tkmh}
\end{center}
\end{figure}

%

\clearpage
\newpage
\begin{figure}[htbp]
 \epsfxsize=\textwidth
 \epsffile{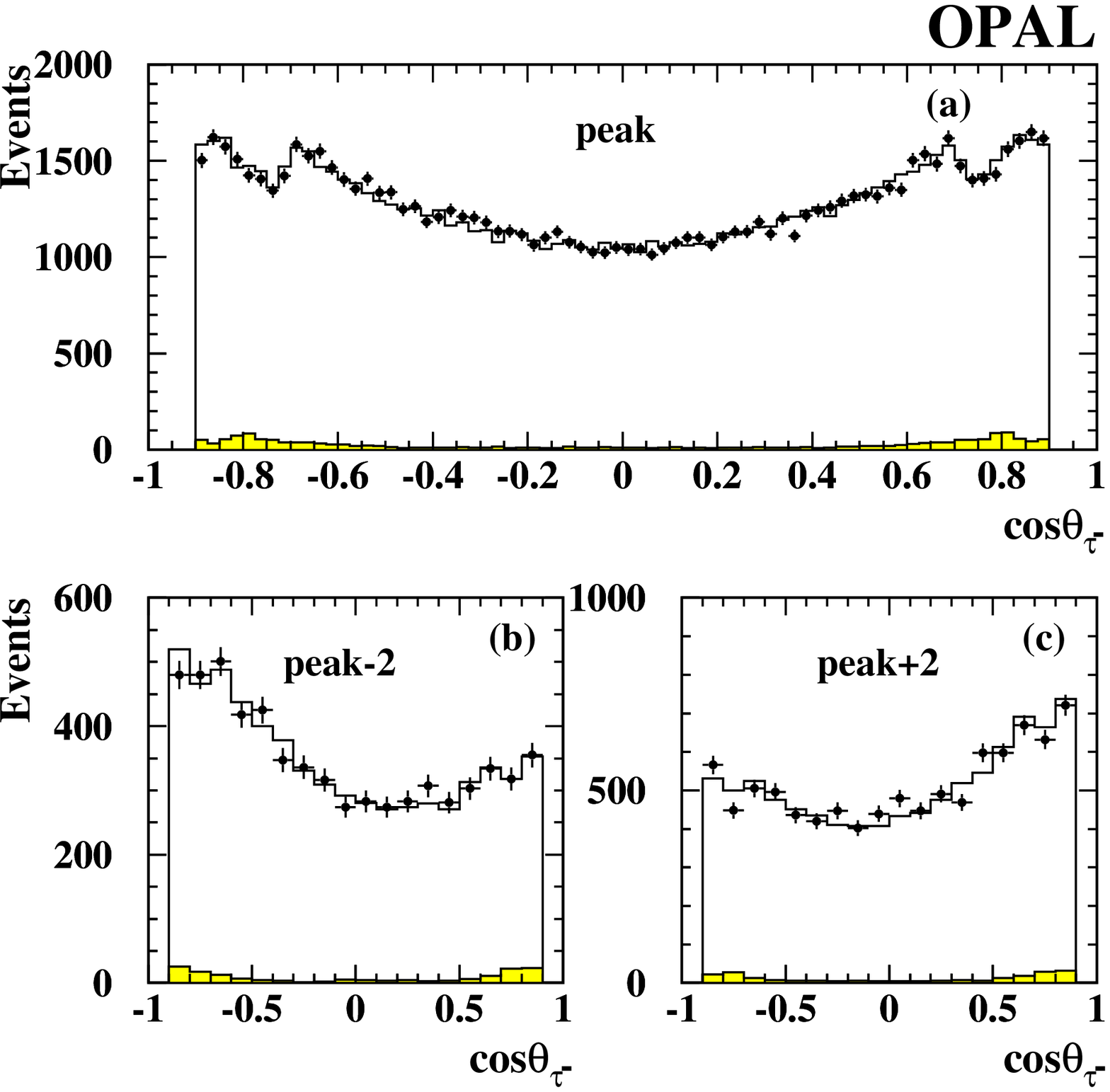}
\caption[Tau angular distributions]{
         The \mbox{$\eetautau$} distribution of polar angle.
         Distributions of~$\costaums$ for the combined \mbox{1993--1995} data 
         (points) separated into {\pkm}, {\pk} and {\pkp} energy points.
         For these plots~$\costaum$ is determined from the average
         of the polar angles of the positive and negative $\tau$~cones
         determined using tracks and electromagnetic clusters.  
         The additional cuts used for the asymmetry analysis have been 
         applied. The Monte Carlo expectation is shown by the histogram 
         and the background contribution is shown as the shaded component.}
 \label{fig:tau_afb}
\end{figure}


\clearpage
\newpage
\begin{figure}[p]
\begin{center}\hspace*{-0.5cm}
\mbox{\epsfxsize1.0\textwidth\epsffile{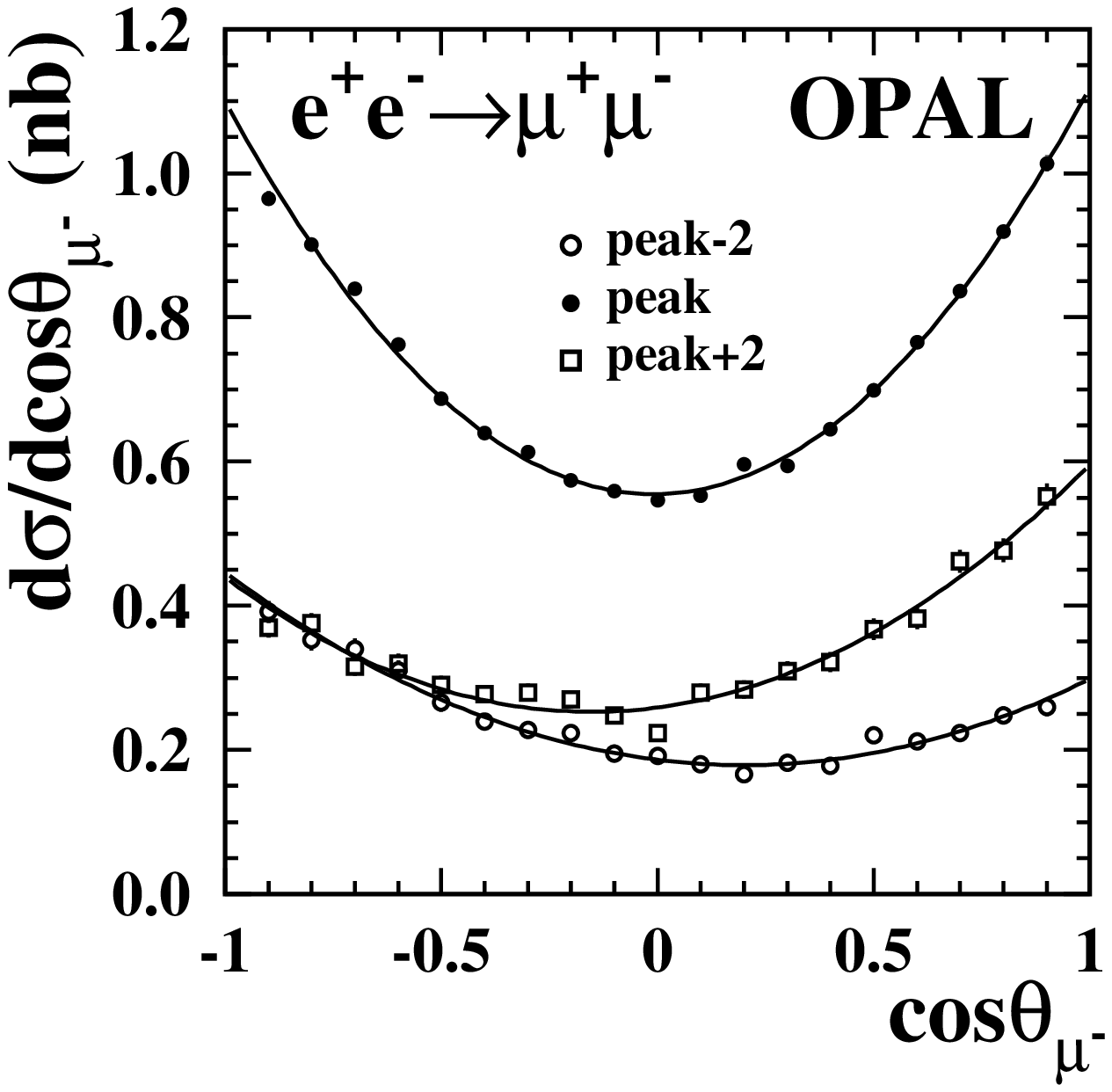}}
\caption[The differential cross-section in $\cos\theta$ for $\eemumu$]{
Observed differential cross-sections as a function of $\cos\theta$ 
for the process $\eemumu$ at the three centre-of-mass energies
in the 1993--1995 data. 
Corrections have been applied for inefficiency and background.
Only statistical errors are shown, 
bin-by-bin systematic uncertainties are not included.
The curves correspond to fits to a simple parametrisation of the form 
$a(1 + \cos^2\theta) + b\cos\theta$.
}
\label{fig:mmdsdc}
\end{center}
\end{figure}


\clearpage
\newpage
\begin{figure}[p]
\begin{center}\hspace*{-0.5cm}
\mbox{\epsfxsize1.0\textwidth\epsffile{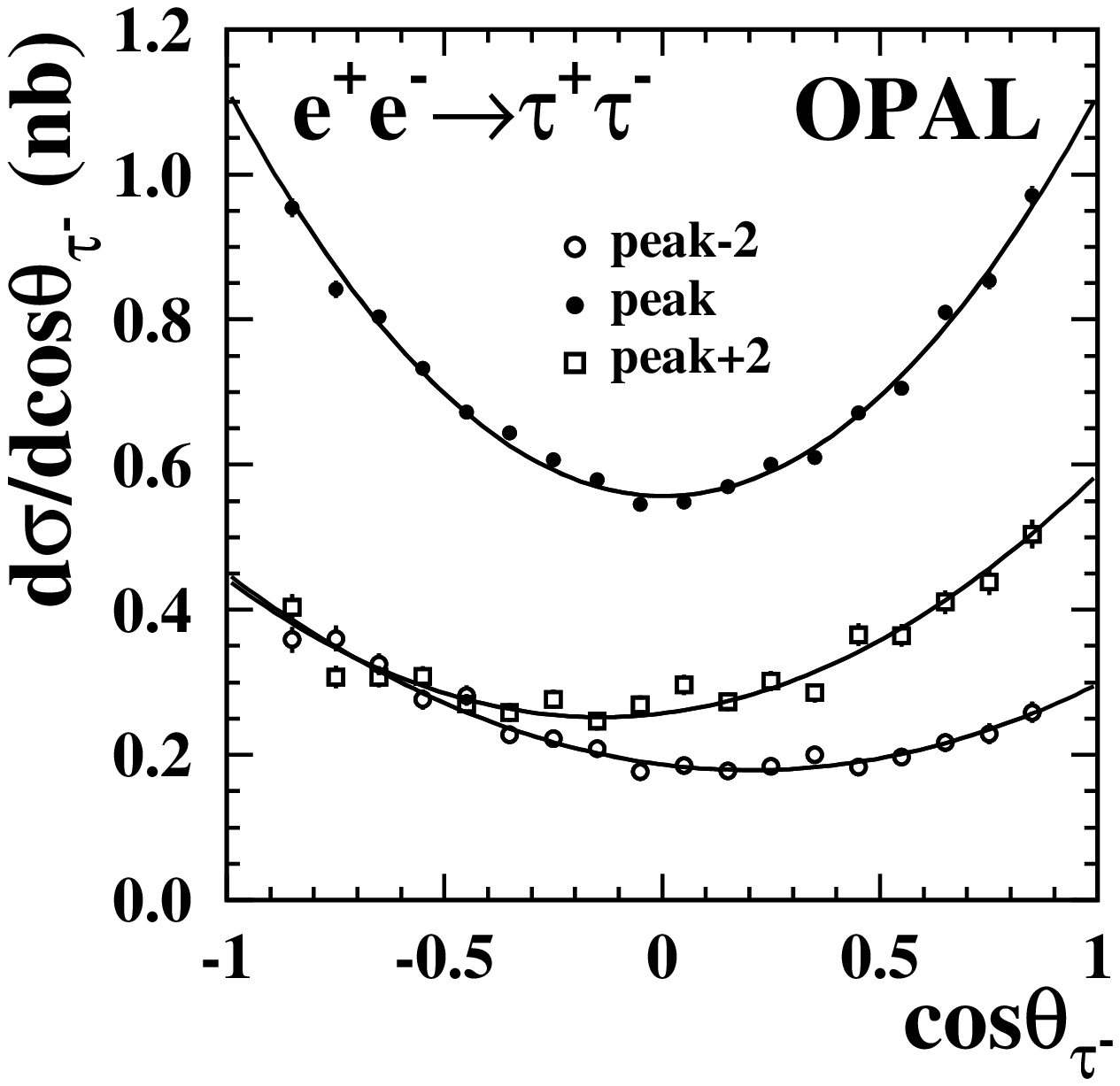}}
\caption[The differential cross-section in $\cos\theta$ for $\eetautau$]{
Observed differential cross-sections as a function of $\cos\theta$ 
for the process $\eetautau$ at the three centre-of-mass energies
in the 1993--1995 data. 
Corrections have been applied for inefficiency and background.
Only statistical errors are shown, 
bin-by-bin systematic uncertainties are not included.
The curves correspond to fits to a simple parametrisation of the form 
$a(1 + \cos^2\theta)  + b\cos\theta$.
}
\label{fig:ttdsdc}
\end{center}
\end{figure}


\clearpage
\newpage
\begin{figure}[p]
\begin{center}\hspace*{-0.5cm}
\mbox{\epsfxsize1.0\textwidth\epsffile{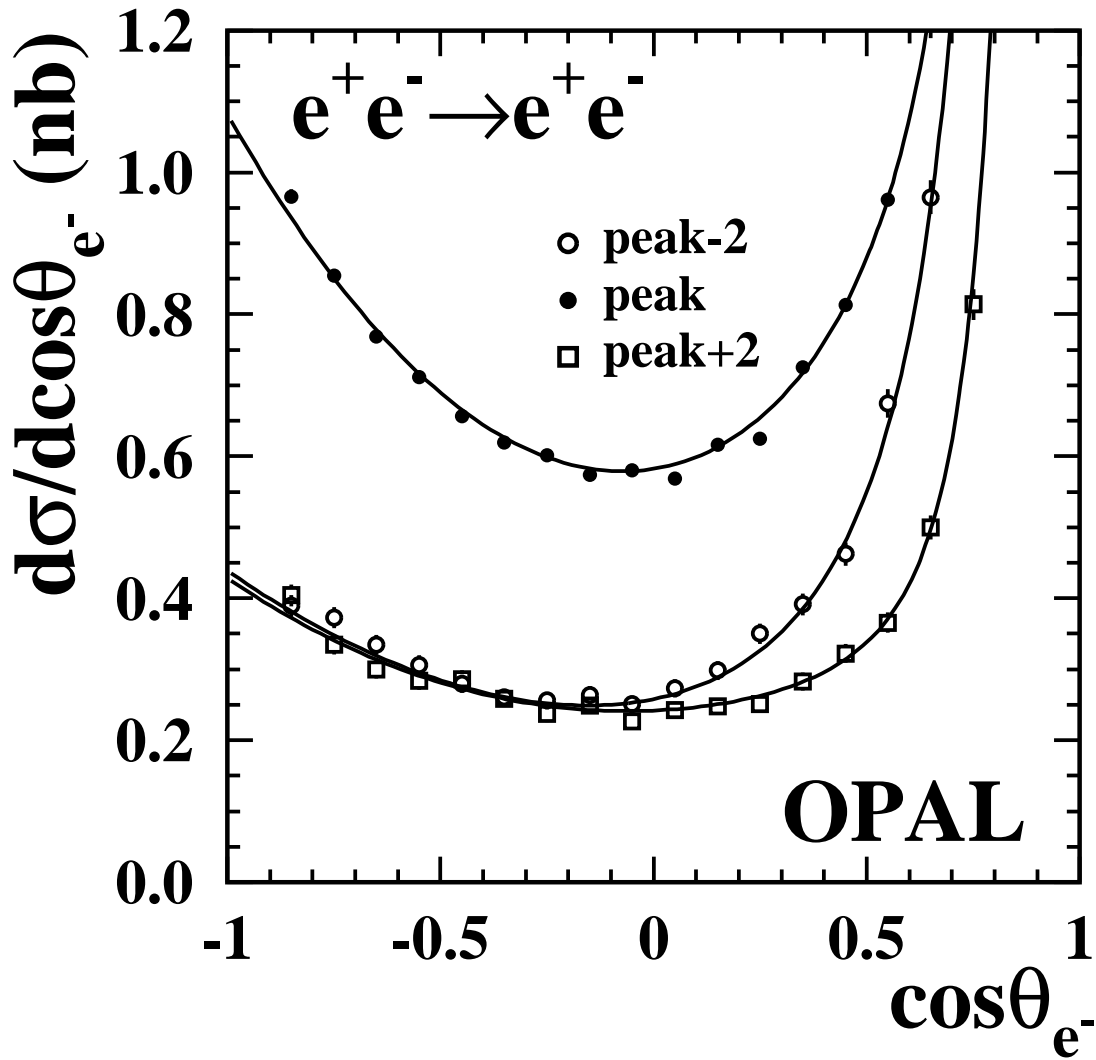}}
\caption[The differential cross-section in $\cos\theta$ for $\eeee$]{
Observed differential cross-sections as a function of $\cos\theta$ 
for the process $\eeee$ at the three centre-of-mass energies
in the 1993--1995 data.
Corrections have been applied for inefficiency and background.
Only statistical errors are shown, 
bin-by-bin systematic uncertainties are not included. The curves show
the predictions of ALIBABA.}
\label{fig:eedsdc}
\end{center}
\end{figure}

%
%
\clearpage
\newpage
\begin{figure}[bht]
\begin{center}
\mbox{\epsfxsize0.95\textwidth\epsffile{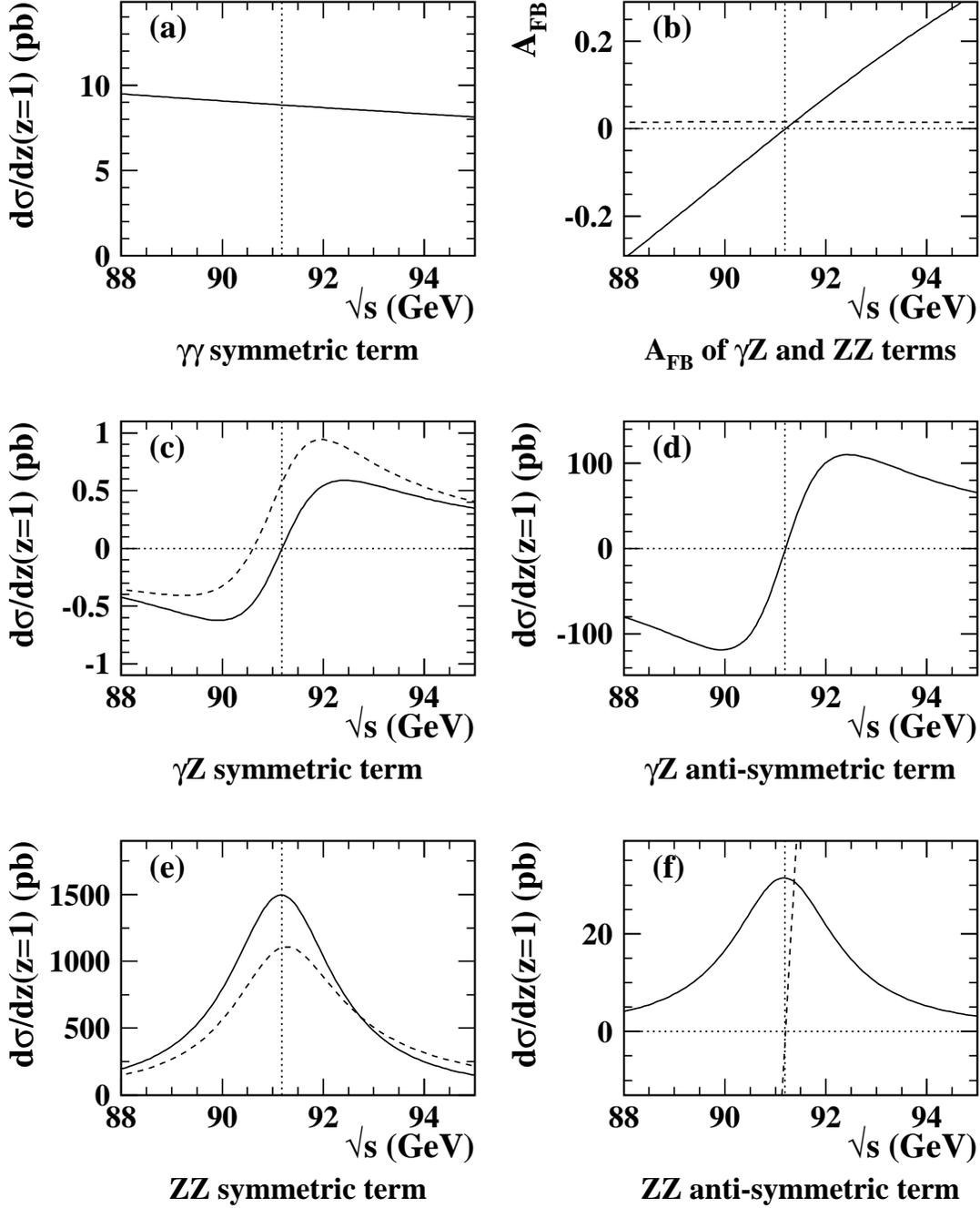}}
\caption[Contribution of individual terms to the differential cross-section]{
The contribution at tree level of the five terms of
equation~\ref{eq-diffxs} {(solid lines)} to the
differential cross-section of $\eemumu$ 
~~(${\mathrm{d}}\sigma\,/\, {\mathrm{d}}z\,(z=1),\,z\equiv \costhe$, in (pb))~~
as a function of $\roots$.
Plots {(a), (c)}  and {(e)} give the symmetric terms from 
 $\gamma$, $\gzif$ and $Z$ exchange respectively;
{(d)} and {(f)} the antisymmetric $\gzif$ and $Z$ exchange terms.
%
The dashed line in {(c)} demonstrates the 
effect when the imaginary parts of the couplings are taken into account.
 The dashed line in {(e)} illustrates the profound change of the lineshape
due to initial-state
radiation.
The dashed curve in {(f)} shows the $\gzif$ term of  {(d)}
superposed to illustrate how rapidly it dominates the more interesting $ZZ$
term as the energy moves away from the peak.
{(b)} shows the forward-backward asymmetry which results from 
the $\gzif$ {(solid)} and $ZZ$ {(dashed)} terms 
when the cross-sections are integrated over $-1 < \costhe < +1$.
The dotted vertical line in each plot indicates $\roots = \MZ$.
}
\label{fig:imb}
\end{center}
\end{figure}


\clearpage
\newpage
\begin{figure}[bht]
\begin{center}
\mbox{\epsfxsize1.0\textwidth\epsffile{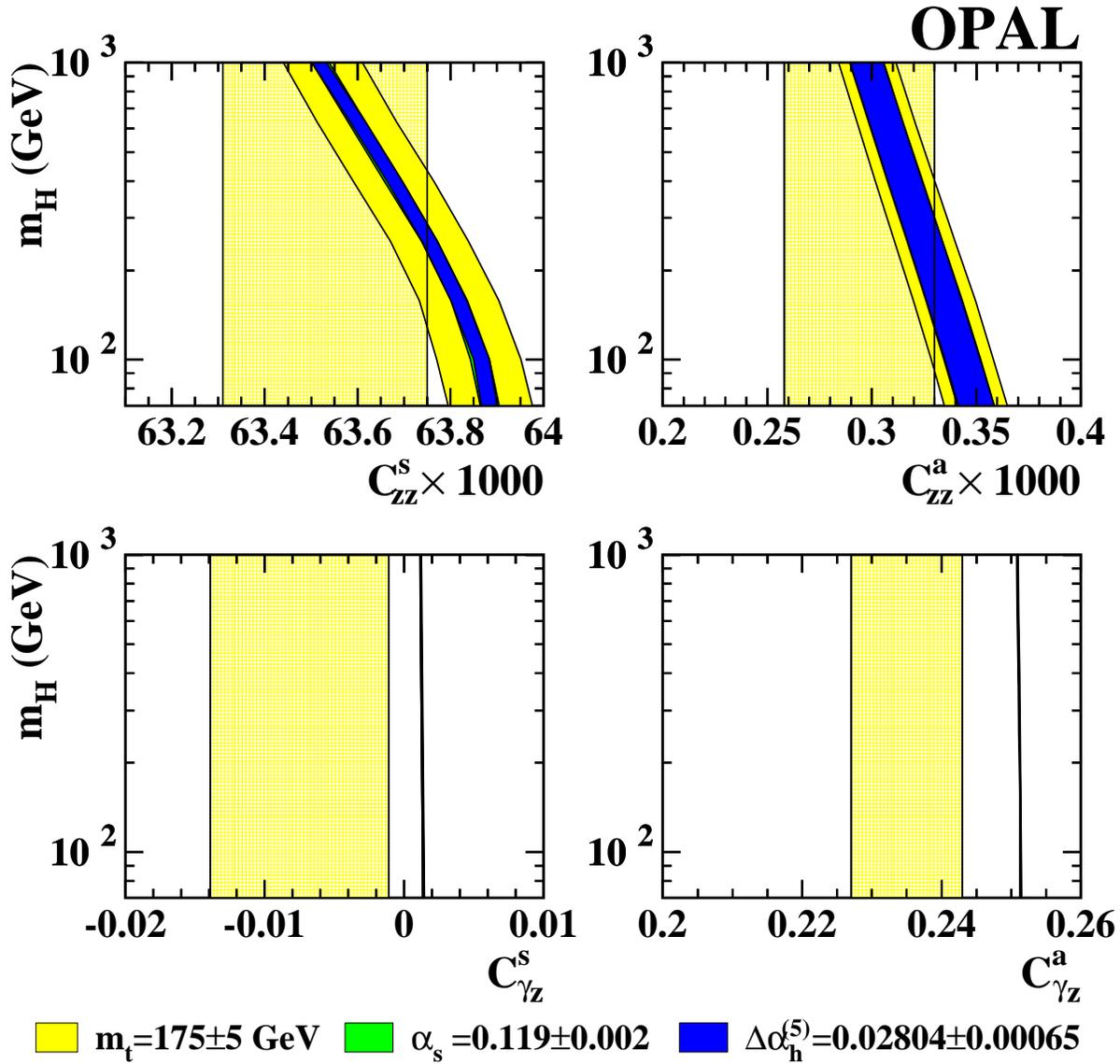}}
\caption[$C$-parameters as a function of the Higgs mass]{
Comparison of the results from the C-parameter fit with
the {\SM} prediction as a function of Higgs mass $\MH$.
The vertical bands indicate the fit results and their horizontal widths
correspond to one standard deviation error intervals.
The shaded area shows (linearly) the variation of the {\SM}
prediction for $\Mt$, $\alfas$, and $\dalh$ in the indicated ranges.
These parameters are insensitive to $\als$, and its variation band is
therefore invisible.
}
\label{f-cparvssm}
\end{center}
\end{figure}


\clearpage
\newpage
\begin{figure}[bht]
\begin{center}
\mbox{\epsfxsize1.0\textwidth\epsffile{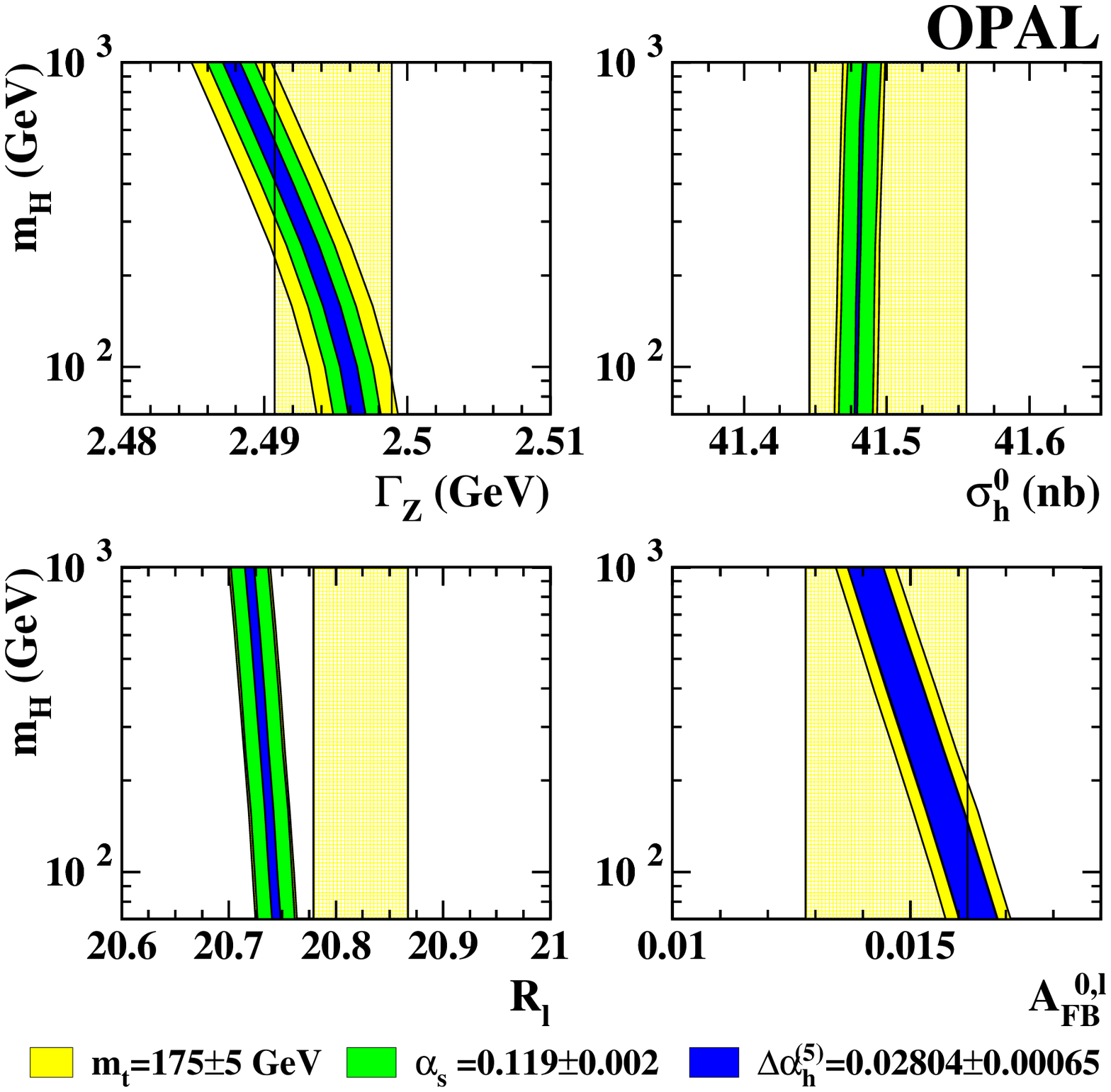}}
\caption[Model-independent Z parameters as a function of the Higgs mass]{
Comparison of the results of the {\SLPfive} fit with
the {\SM} prediction as a function of Higgs mass $\MH$.
The vertical bands indicate the fit results and their horizontal widths
correspond to one standard deviation error intervals.
The shaded area shows (linearly) the variation of the {\SM}
prediction for $\Mt$, $\alfas$, and $\dalh$ in the indicated ranges.
}
\label{f-zparvssm}
\end{center}
\end{figure}

\clearpage
\newpage
  \setlength{\unitlength}{1mm}
\begin{figure}[bht]
\begin{center}
\mbox{
\begin{picture}(170,220)
\put(0,  110){\epsfig{file=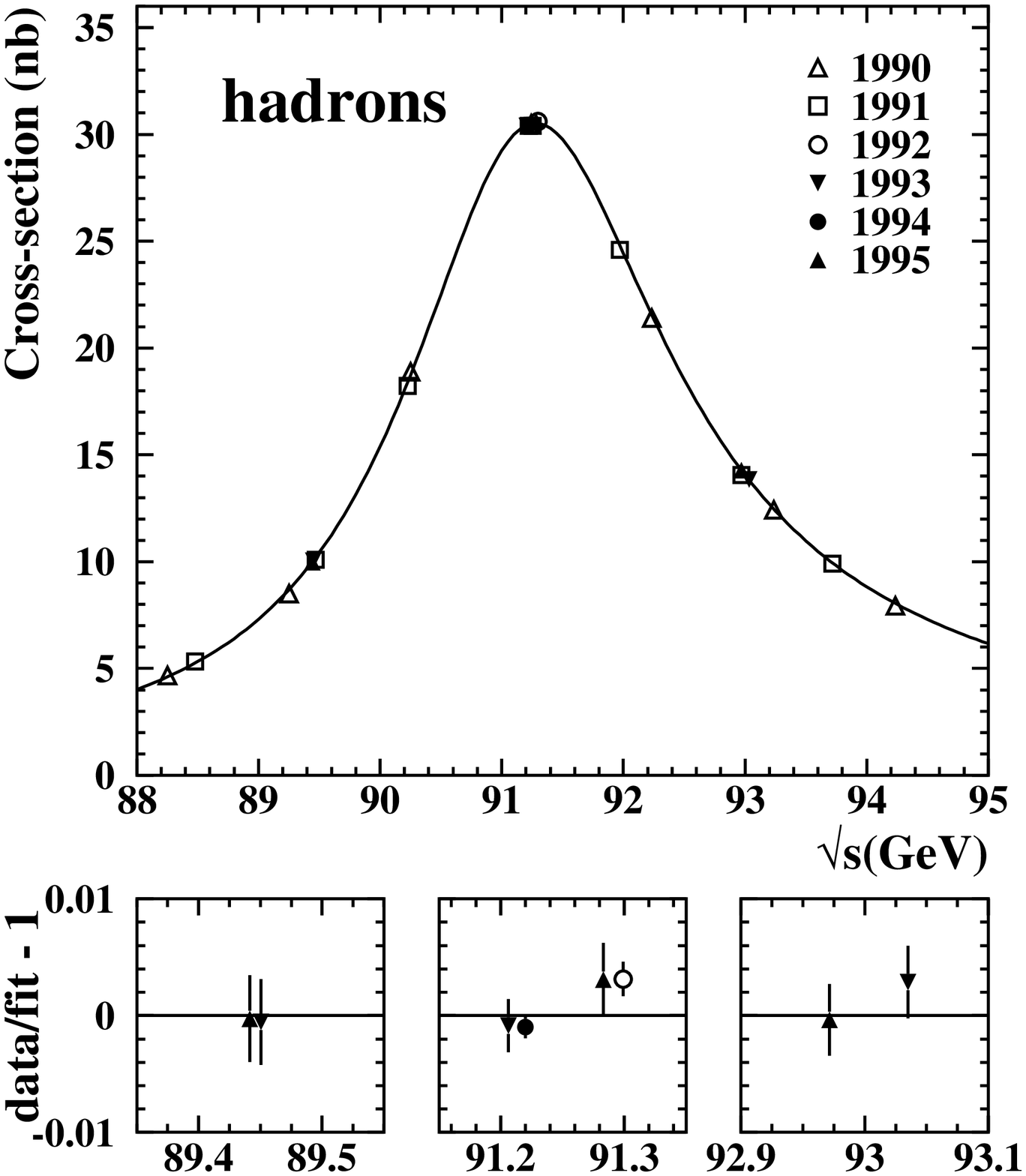,width=85mm}} 
\put(85, 110){\epsfig{file=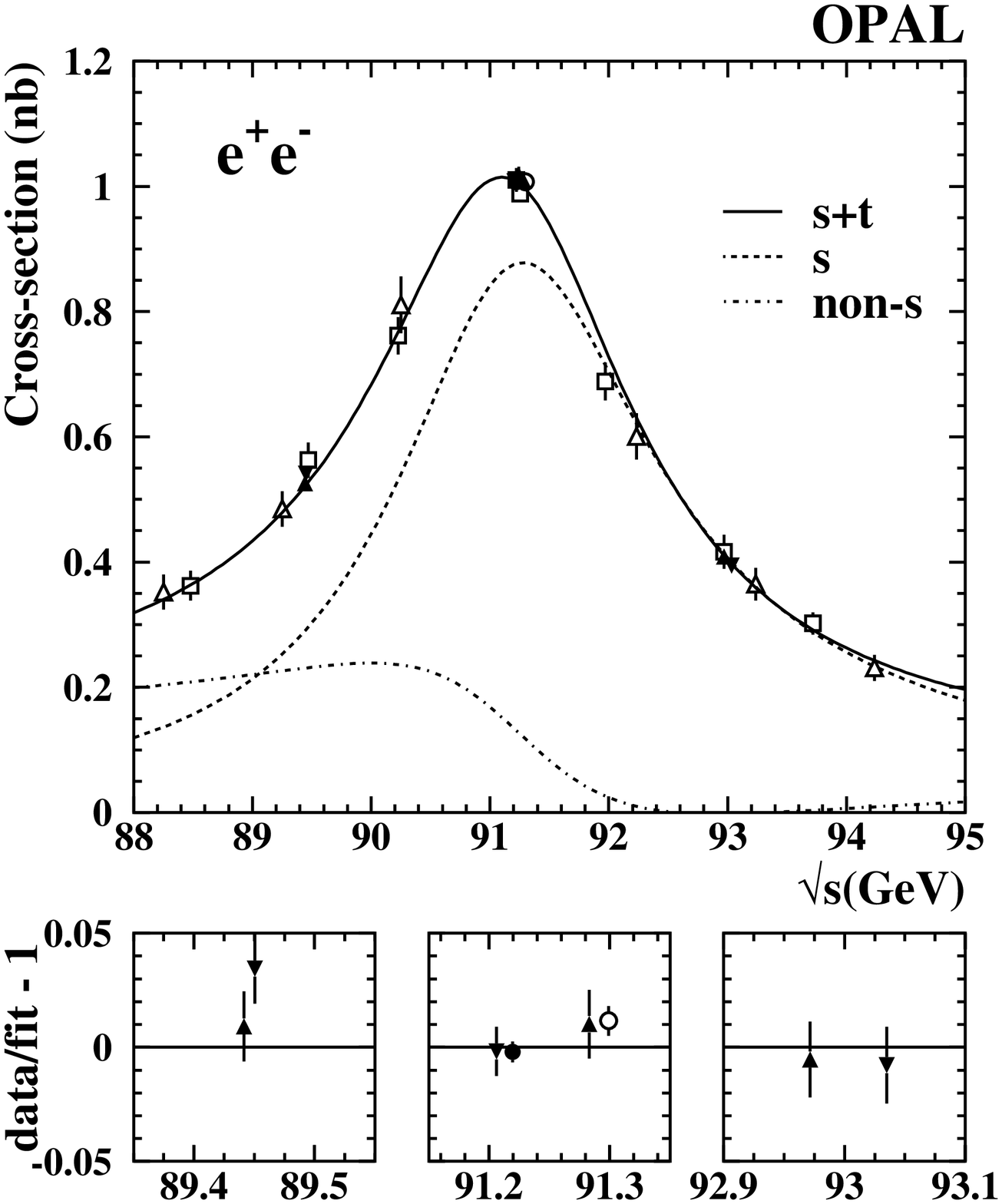,width=85mm}} 
\put(0,    0){\epsfig{file=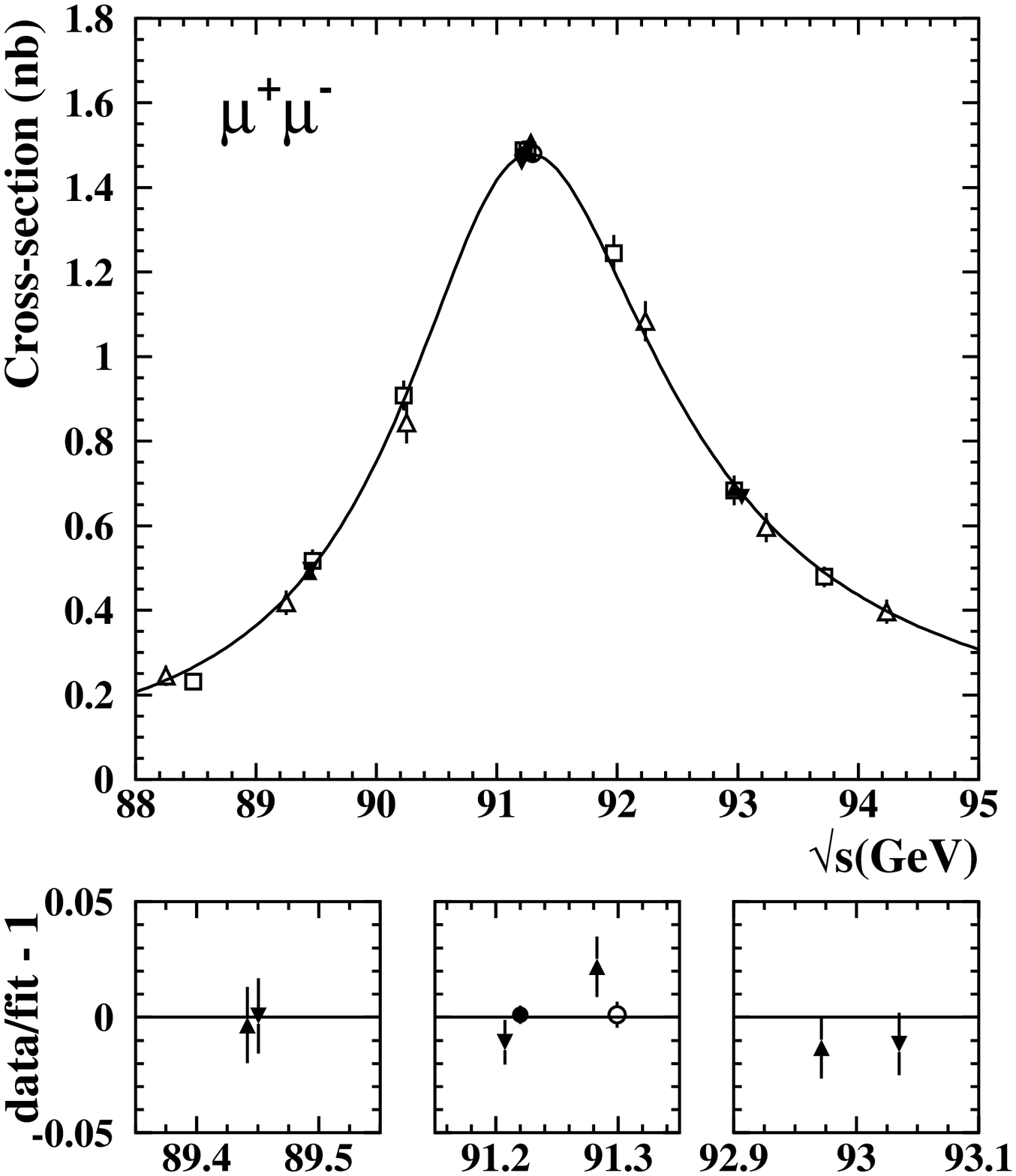,width=85mm}} 
\put(85,   0){\epsfig{file=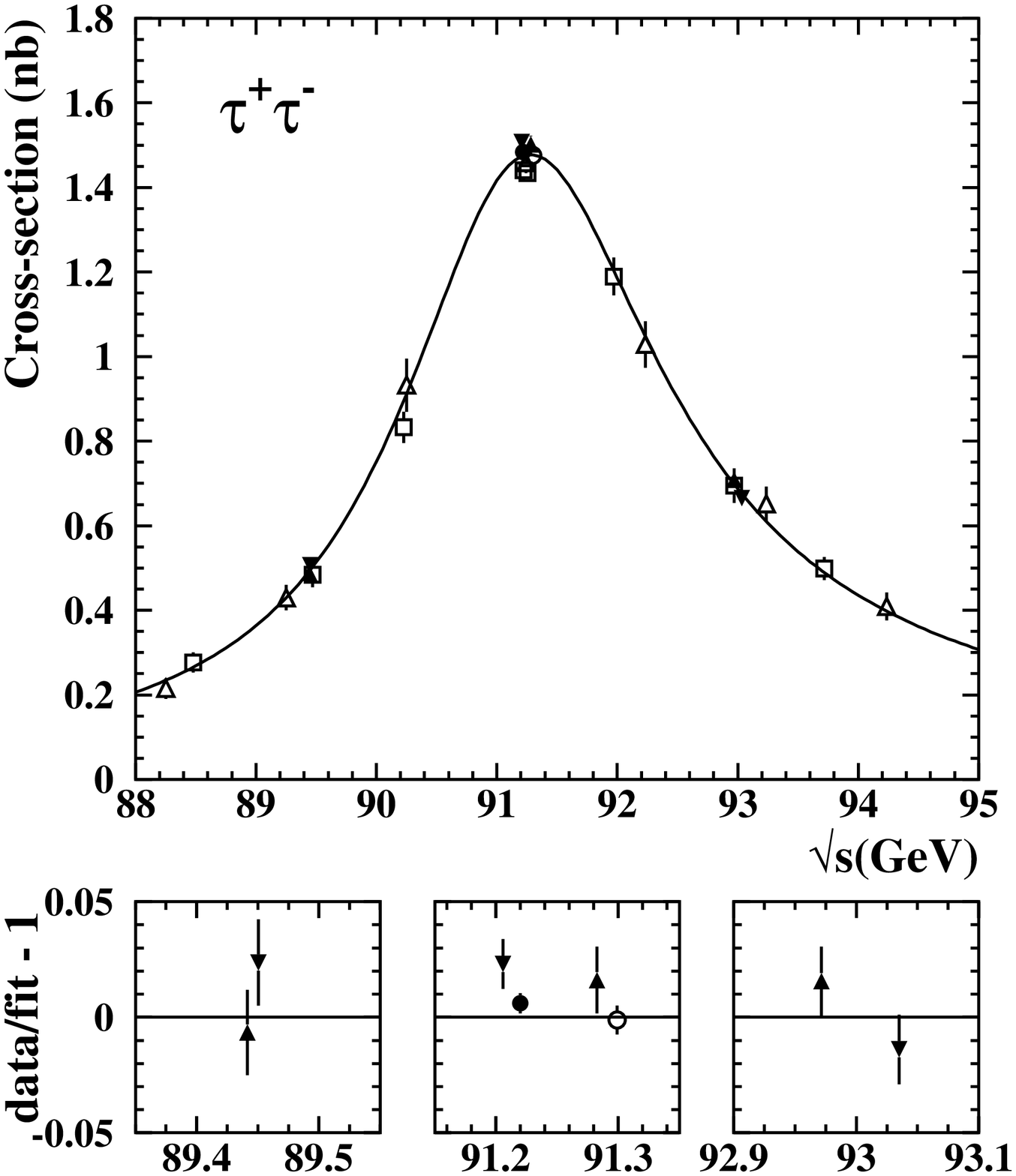,width=85mm}} 
\end{picture}
}
\caption[$\Zzero$ cross-sections]{
The measured cross-sections for hadronic and leptonic final states
as a function of centre-of-mass energy.
The errors shown are statistical only.
The solid line is the result of the {\SLPnine} fit
to the combined leptonic and 
hadronic data (without assuming lepton universality) described
in {\SECT}~\ref{sec-ewp-leppars}.
The lower plots show the residuals to the fit.
For the electrons, the dashed curves show the contributions of
the pure $s$-channel, and $t$-channel plus $s-t$ interference, respectively.
}
\label{f-fit-xsec}
\end{center}
\end{figure}

\clearpage
\newpage
  \setlength{\unitlength}{1mm}
\begin{figure}[bht]
\begin{center}
\mbox{
\begin{picture}(170,220)
\put(0,  110){\epsfig{file=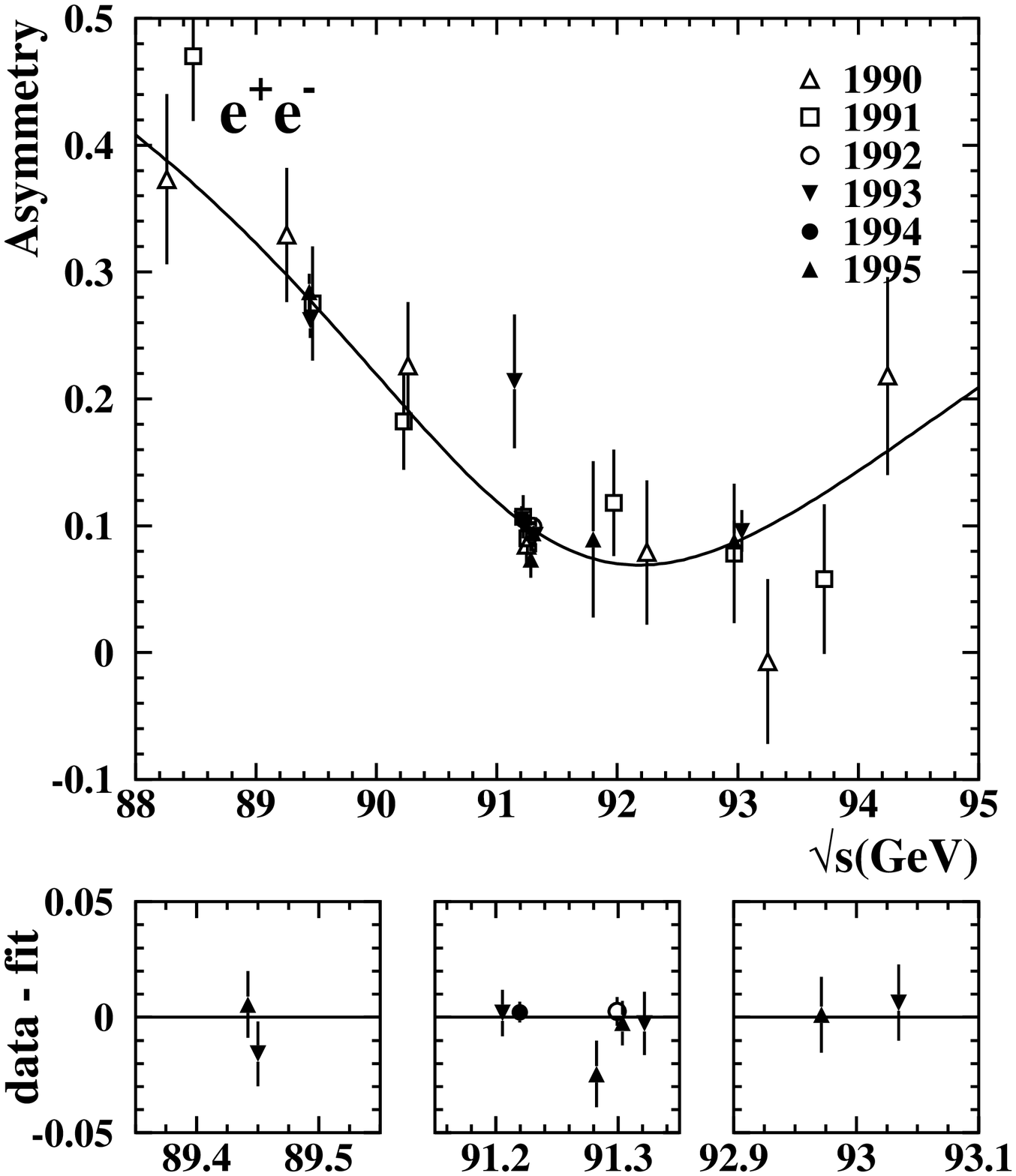,width=85mm}} 
\put(85, 110){\epsfig{file=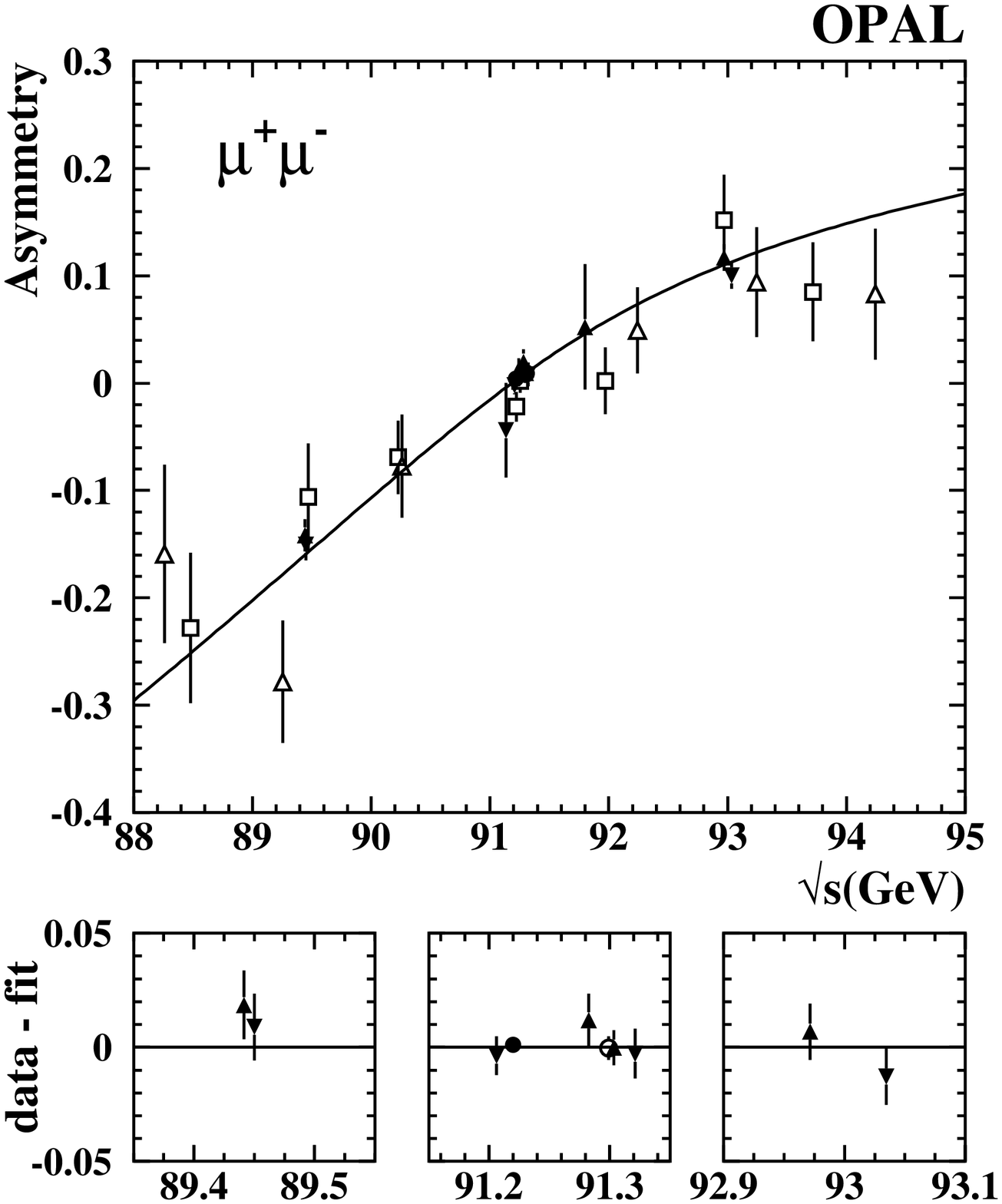,width=85mm}} 
\put(0,    0){\epsfig{file=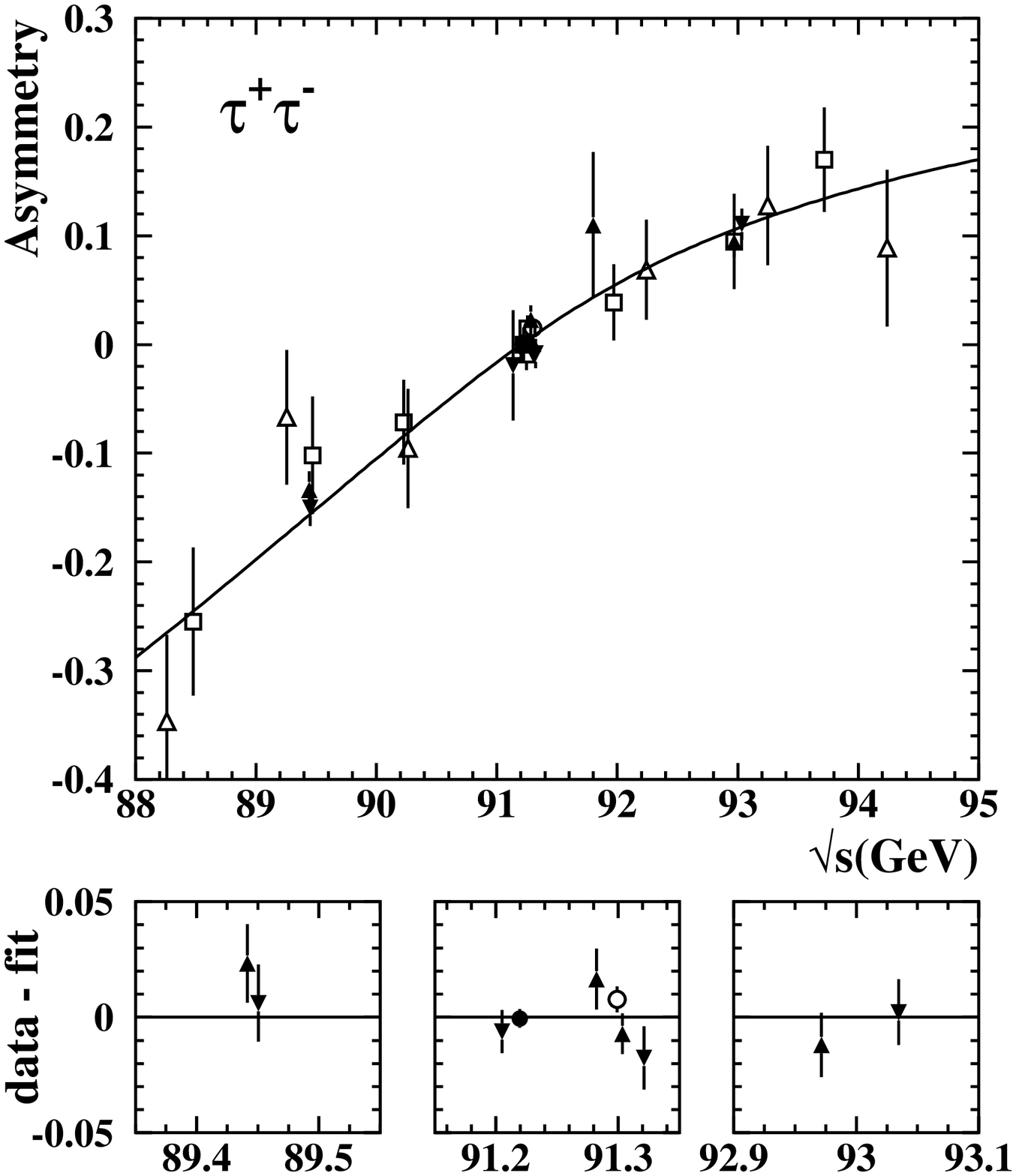,width=85mm}} 
\end{picture}
}
\caption[$\Zzero$ forward-backward asymmetries]{
The measured forward-backward asymmetry in leptonic final states as a function
of centre-of-mass energy.
The errors shown are statistical only.
The solid line is the result of the {\SLPnine} fit
to the combined leptonic and 
hadronic data (without assuming lepton universality) described
in {\SECT}~\ref{sec-ewp-leppars}.
The lower plots show the residuals to the fit.
}
\label{f-fit-afb}
\end{center}
\end{figure}


\clearpage
\newpage
\begin{figure}[bht]
\begin{center}
\mbox{\epsfxsize1.0\textwidth\epsffile{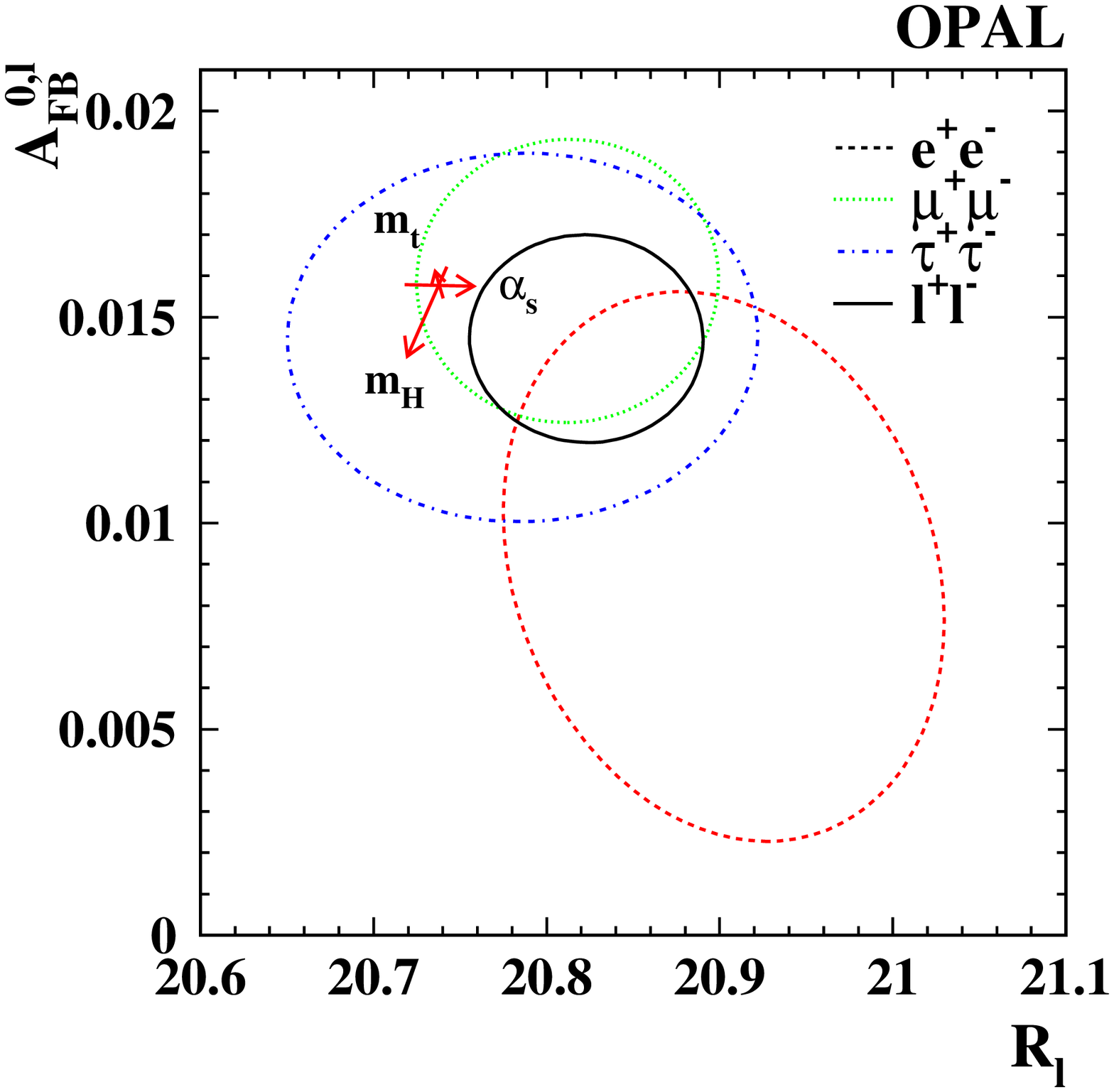}}
\caption[Probability contours in the $\Afblpol$\,--\,$\Rl$ plane]{
Contours of 68\% probability in the $\Afblpol$\,--\,$\Rl$ plane
for each of the three lepton species (dotted, dashed).
The solid contour results from a fit assuming lepton universality.
In this plot, the results for the $\tau$ are corrected for the mass
effect so that 
a direct comparison with other lepton species is possible.
The {\SM} prediction with
$\als ,\;\Mt ,\; \MH$ and $\dalh$ as specified in {\EQ}~\ref{eq-sm-pars}
is also shown as the intersection of the three arrows.
The arrows indicate the range of the variation when 
$\als ,\;\Mt$ and $\MH$ are varied within the ranges specified in
{\EQ}~\ref{eq-sm-pars}.
}
\label{f-afb0vsrl}
\end{center}
\end{figure}


\clearpage
\newpage
\begin{figure}[bht]
\begin{center}
\mbox{\epsfxsize1.0\textwidth\epsffile{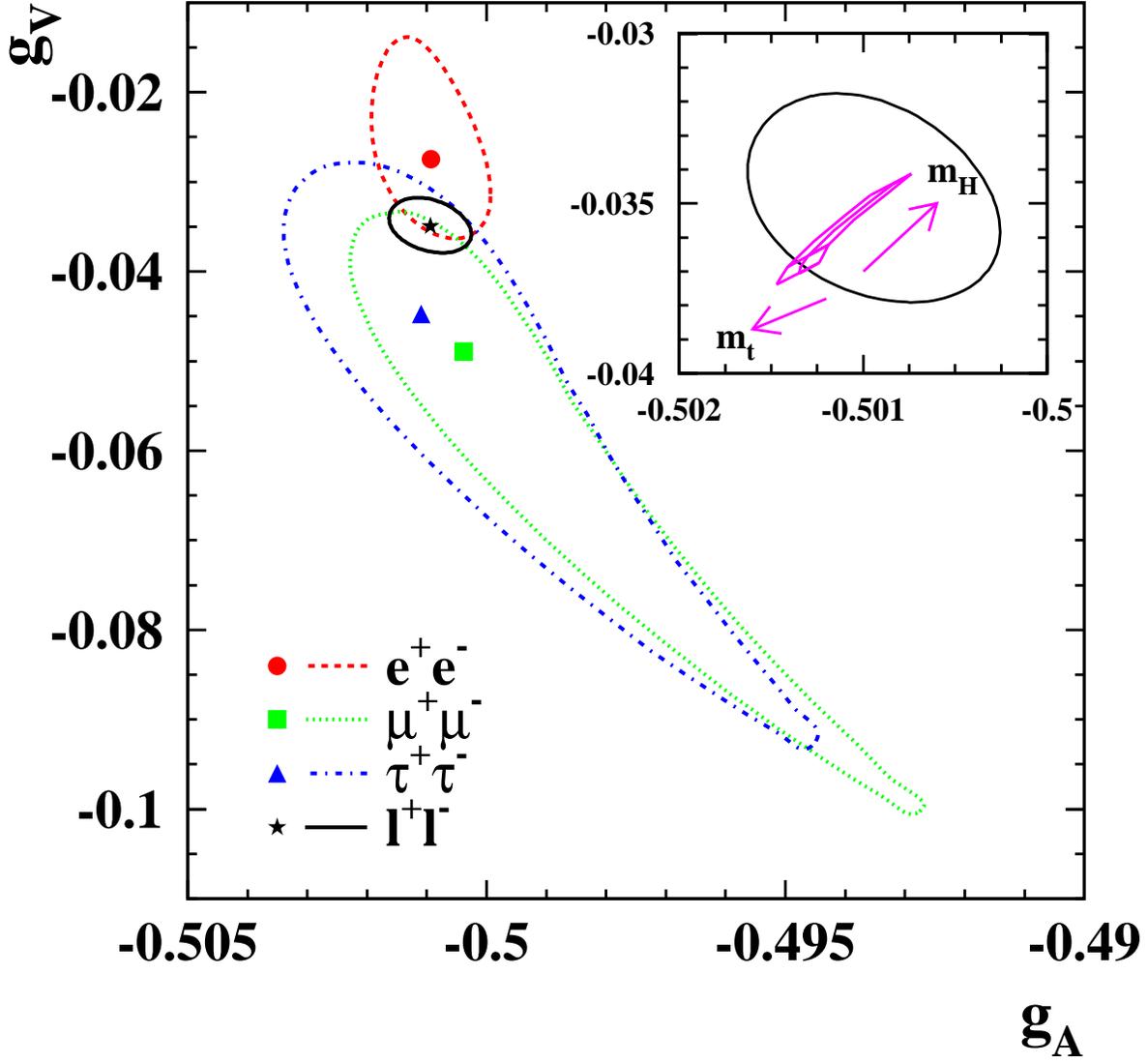}}
\caption[Probability contours in the $\gvl$\,--\,$\gal$ plane]{
Contours of 68\% probability in the $\gvl$\,--\,$\gal$ plane
for each of the three lepton species (dotted, dashed).
The solid contour results from a fit assuming lepton universality and
 is also shown enlarged in the inset figure.
Here, the band indicates the {\SM} prediction when
$\Mt$ and $\MH$ are varied as specified in equation~\ref{eq-sm-pars}.
}
\label{f-gvvsga}
\end{center}
\end{figure}

%

\clearpage
\newpage
\begin{figure}[bht]
\begin{center}
\mbox{\epsfxsize1.0\textwidth\epsffile{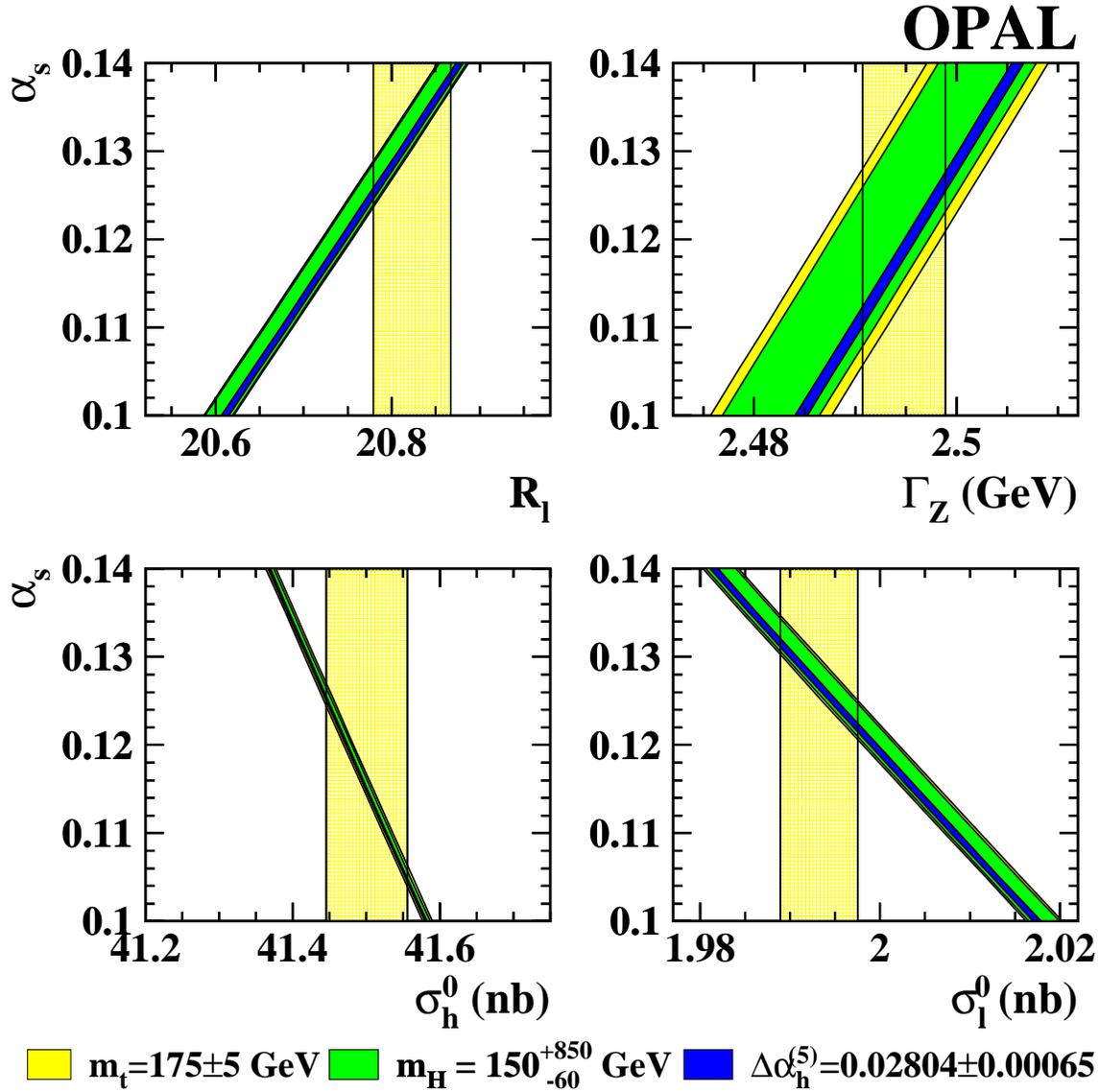}}
\caption[$\Zzero$ resonance parameters as a function of $\als$]{
Comparison of $\Zzero$ resonance parameters  with
the {\SM} prediction as a function of $\als$.
The vertical bands indicate the fit results and their horizontal widths
correspond to one standard deviation error intervals.
The hatched area shows (linearly) the variation of the {\SM}
prediction when  $\Mt$, $\MH$ and $\dalh$ are varied as specified 
in {\EQ}~\ref{eq-sm-pars} and indicated in the figure. 
}
\label{f-lsalfas}
\end{center}
\end{figure}


\clearpage
\newpage
\begin{figure}[bht]
\begin{center}
\mbox{\epsfxsize1.0\textwidth\epsffile{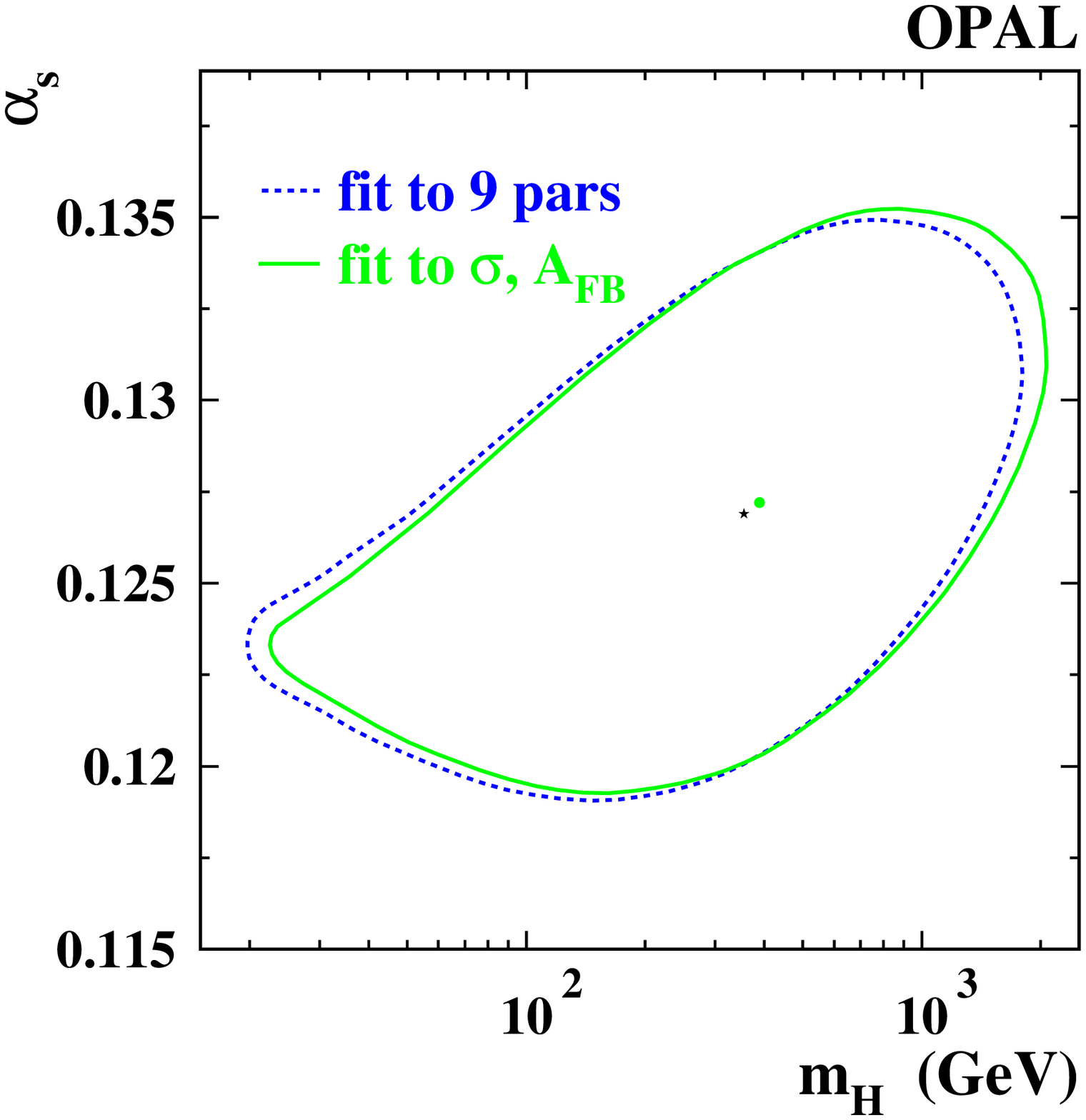}}
\caption[Probability contours in the $\als$\,--\,$\MH$ plane]{
Contours of 68\,\% probability in the $\als$\,--\,$\MH$ plane
with $\Mt$ constrained to \mbox{$174.3 \pm 5.1$ GeV}~\cite{bib-top-mass}.
%
The  undashed contour is obtained from
 the {\SM} fit to cross-sections and asymmetries. 
For the  dashed contour the fit was made on the results of the {\SLPnine}
fit. 
The  small circle and the  star give the central values
for the two fits, respectively.
In both cases a numerical evaluation of the 68\,\% C.L.\ 
contour was performed
which accounts for asymmetric or non-parabolic 
parameter  dependencies (MINUIT contour~\cite{bib-MINUIT}).
}
\label{f-mhas}
\end{center}
\end{figure}

\clearpage \newpage
%


\bibliographystyle{unsrt}
\bibliography{pr328}

\begin{thebibliography}{10}

\bibitem{bib-LEP1yr}
``Physics with Very High Energy $\ee$ Colliding Beams'', L. Camilleri \etal,
  CERN~76-18~(1976);\\ ``Physics at LEP'', CERN~86-02, eds.~J.~Ellis
  and~R.~Peccei~(1986);\\ ``Z~Physics at {\LEPI}'', CERN~89-08,
  eds.~G.~Altarelli, R.~Kleiss and~C.~Verzegnassi~(1989).

\bibitem{bib-pdg98}
Particle Data Group, C. Caso \etal , Eur.\ Phys.\ J.\ {\bf C3} (1998) 1.

\bibitem{bib-LEPEWWG}
The LEP Collaborations: ALEPH, DELPHI, L3 and OPAL, Phys. Lett. {\bf {B276}}
  (1992) 247;\\ The LEP Collaborations: ALEPH, DELPHI, L3 and OPAL, the LEP
  Electroweak Working Group and the SLD Heavy Flavour Group, ``A Combination of
  Preliminary Electroweak Measurements and Constraints on the Standard Model'',
  CERN-PPE/97-154 (December~1997); CERN-EP/2000-016 (January 2000).

\bibitem{bib-EWSM}
S.L. Glashow, Nucl. Phys. {\bf {B22}} (1961) 579;\\ S. Weinberg, Phys. Rev.
  Lett. {\bf {19}} (1967) 1264;\\ A. Salam, in ``Elementary Particle Theory'',
  ed.~N.~Svartholm, (Almqvist~and~Wiksell, Stockholm,~1968) p.~367.

\bibitem{bib-veltman}
G. 't Hooft and M. Veltman, Nucl. Phys. {\bf {B44}} (1972) 189.

\bibitem{bib-opal-ls89}
OPAL Collaboration, M.Z. Akrawy \etal, Phys. Lett. {\bf {B231}} (1989) 530;\\
  OPAL Collaboration, M.Z. Akrawy \etal, Phys. Lett. {\bf {B240}} (1990) 497;\\
  OPAL Collaboration, M.Z. Akrawy \etal, Phys. Lett. {\bf {B247}} (1990) 458.

\bibitem{bib-opal-ls90}
OPAL Collaboration, G. Alexander \etal, Z. Phys. {\bf {C52}} (1991) 175.

\bibitem{bib-opal-ls91}
OPAL Collaboration, P.D. Acton \etal, Z. Phys. {\bf {C58}} (1993) 219.

\bibitem{bib-opal-ls92}
OPAL Collaboration, R. Akers \etal, Z. Phys. {\bf {C61}} (1994) 19.

\bibitem{bib-ALEPH-final}
ALEPH Collaboration, R. Barate \etal, \EPJ {\bf{14}} (2000) 1.

\bibitem{bib-DELPHI-final}
DELPHI Collaboration, P. Abreu \etal, \EPJ {\bf{16}} (2000) 371.

\bibitem{bib-L3-final}
L3 Collaboration, M. Acciarri \etal, \EPJ {\bf{16}} (2000) 1.

\bibitem{bib-lumi-siw}
OPAL Collaboration, G. Abbiendi, \etal, \EPJ {\bf{14}} (2000) 373.

\bibitem{bib-det-opal}
OPAL Collaboration, K. Ahmet \etal, Nucl. Instrum. Meth. {\bf {A305}} (1991)
  275.

\bibitem{bib-det-SiVtx}
P.P. Allport \etal, Nucl. Instrum. Meth. {\bf {A324}} (1993) 34;\\ P.P. Allport
  \etal, Nucl. Instrum. Meth. {\bf {A346}} (1994) 476.

\bibitem{bib-det-CJ}
M. Hauschild \etal, Nucl. Instrum. Meth. {\bf {A314}} (1992) 74;\\ O. Biebel
  \etal, Nucl. Instrum. Meth. {\bf {A323}} (1992) 169.

\bibitem{bib-det-siw}
B.E. Anderson \etal, IEEE Trans. Nucl. Sci. {\bf {41}} (1994) 845.

\bibitem{bib-det-muon}
R.J. Akers \etal, Nucl. Instrum. Meth. {\bf {A357}} (1995) 253.

\bibitem{bib-det-daq}
J.T.M. Baines \etal, Nucl. Instrum. Meth. {\bf {A325}} (1993) 271.

\bibitem{bib-det-pretrig}
M. Arignon \etal, Nucl. Instrum. Meth. {\bf {A333}} (1993) 330.

\bibitem{bib-det-trig}
M. Arignon \etal, Nucl. Instrum. Meth. {\bf {A313}} (1992) 103.

\bibitem{bib-det-filter}
D.G.~Charlton, F.~Meijers, T.J.~Smith,~P.S.~Wells, Nucl. Instrum. Meth. {\bf
  {A325}} (1993) 129.

\bibitem{bib-det-offline}
J. Baudot \etal, ``A~compilation and analysis of the data management and
  computing in the OPAL experiment, 1989 to~1996'', proceedings of Computing in
  High Energy Physics (CHEP97), Berlin, Germany, 7-11~April~1997.

\bibitem{bib-mc-GOPAL}
J. Allison \etal, Nucl. Instrum. Meth. {\bf {A317}} (1992) 47.

\bibitem{bib-opal-hadew}
OPAL Collaboration, P.D. Acton \etal, Phys. Lett. {\bf {B294}} (1992) 436.\\
  OPAL Collaboration, G. Alexander \etal, Z. Phys. {\bf {C70}} (1996) 357;\\
  OPAL Collaboration, G. Alexander \etal, Z. Phys. {\bf {C72}} (1996) 1;\\ OPAL
  Collaboration, G. Alexander \etal, Z. Phys. {\bf {C73}} (1997) 379;\\ OPAL
  Collaboration, K. Ackerstaff \etal, Z. Phys. {\bf {C75}} (1997) 385;\\ OPAL
  Collaboration, K. Ackerstaff \etal, Z. Phys. {\bf {C76}} (1997) 387;\\ OPAL
  Collaboration, K. Ackerstaff \etal, \EPJ {\bf {1}} (1998) 439;\\ OPAL
  Collaboration, K. Ackerstaff \etal, \EPJ {\bf {8}} (1999) 217.

\bibitem{bib-mc-JETSET73}
T. Sj{\"o}strand, CERN-TH/6488/92.

\bibitem{bib-mc-HERWIG}
G. Marchesini \etal, hep-ph/9607393 (July 1996);\\ G. Marchesini and B.R.
  Webber, Nucl. Phys. {\bf {B310}} (1988) 461.

\bibitem{bib-mc-OPALtune-j73}
OPAL Collaboration, P.D. Acton \etal, Z. Phys. {\bf {C58}} (1993) 387.

\bibitem{bib-mc-JETSET74}
T. Sj{\"o}strand, Comput. Phys. Commun. {\bf {82}} (1994) 74.

\bibitem{bib-mc-OPALtune-j74}
OPAL Collaboration, G. Alexander \etal, Z. Phys. {\bf {C69}} (1996) 543.

\bibitem{bib-mc-KORALZ}
S. Jadach, B.F.L. Ward, Z. W{\c a}s, Comput. Phys. Commun. {\bf {79}} (1994)
  503.

\bibitem{bib-mc-BHWIDE}
S. Jadach, W. P{\l}aczek, B.F.L. Ward, Phys. Lett. {\bf {B390}} (1997) 298.

\bibitem{bib-mc-PHOJET}
R. Engel and J. Ranft, Phys. Rev. {\bf {D54}} (1996) 4244.

\bibitem{bib-mc-Vermaseren}
J.A.M. Vermaseren, Nucl. Phys. {\bf {B229}} (1983) 347;\\ R. Bhattacharya, J.
  Smith, G. Grammer, Jr., Phys. Rev. {\bf {D15}} (1977) 3267;\\ J. Smith,
  J.A.M. Vermaseren, G. Grammer, Jr., Phys. Rev. {\bf {D15}} (1977) 3280.

\bibitem{bib-mc-RADCOR}
F.A. Berends and R. Kleiss, Nucl. Phys. {\bf {B186}} (1981) 22.

\bibitem{bib-mc-FERMISV}
J. Hilgart, R. Kleiss, F. Le Diberder, Comput. Phys. Commun. {\bf {75}} (1993)
  191.

\bibitem{bib-mc-grc4f}
J. Fujimoto \etal, Comput. Phys. Commun. {\bf {100}} (1997) 128.

\bibitem{bib-LEPENE90}
R. Bailey \etal, ``LEP Energy Calibration'', CERN SL/90-95, paper presented at
  the $2^{\rm nd}$ European Particle Accelerator Conference,
  Nice,~12-16~June~1990.

\bibitem{bib-ResDepol}
L. Arnaudon \etal, Z. Phys. {\bf {C66}} (1995) 45.

\bibitem{bib-LEPENE91}
The Working Group on LEP Energy, L. Arnaudon \etal,\\ ``The Energy Calibration
  of LEP in 1991'', CERN-PPE/92-125 and CERN-SL/92-37(DI);\\ The Working Group
  on LEP Energy and the LEP Collaborations, L. Arnaudon \etal,\\ Phys. Lett.
  {\bf {B307}} (1993) 187.

\bibitem{bib-LEPENE92}
The Working Group on LEP Energy, L. Arnaudon \etal,\\ ``The Energy Calibration
  of LEP in 1992'', CERN SL/93-21~(DI).

\bibitem{bib-LEPENE93}
The Working Group on LEP Energy, R. Assmann \etal, Z. Phys. {\bf {C66}} (1995)
  567.

\bibitem{bib-LEPecal-98}
The LEP Energy Working Group, R. Assmann \etal, \EPJ {\bf 6} (1999) 187.

\bibitem{bib-mc-BHLUMItwo}
S. Jadach \etal, Comput. Phys. Commun. {\bf {70}} (1992) 305.

\bibitem{bib-mc-BHLUMI}
S. Jadach \etal, Comput. Phys. Commun. {\bf {102}} (1997) 229.

\bibitem{bib-bward}
B.F.L Ward \etal, Phys. Lett. {\bf B450} (1999) 262.

\bibitem{bib-pavia}
G. Montagna \etal, Nucl. Phys. {\bf B547} (1999) 39; \\ G. Montagna \etal,
  Phys. Lett. {\bf B459} (1999) 649.

\bibitem{bib-opal-taupol}
OPAL Collaboration, G. Alexander \etal , Z. Phys. {\bf {C72}} (1996) 365.

\bibitem{bib-fit-ZFITTER}
Dubna-Zeuthen radiative correction group, D. Bardin \etal, ``ZFITTER v.6.21 A
  Semi-Analytical Program for Fermion Pair Production in $\ee$ Annihilation'',
  DESY 99-070, hep-ph/9908433.

\bibitem{bib-alem-jeg}
S. Eidelman, F. Jegerlehner, \ZPC {\bf 67} (1995) 585.

\bibitem{bib-s-dep-width}
G.~Burgers, F.A.~Berends, W. Hollik, W.L.~van Neerven, \PLB {\bf 203} (1988)
  177;\\ D.Yu.\ Bardin, A. Leike, T. Riemann, M. Sachwitz, \PLB {\bf 206}
  (1988) 539.

\bibitem{bib-fit-TOPAZ0}
G. Montagna, O. Nicrosini, G. Passarino, F. Piccinini, Comput. Phys. Commun.
  {\bf {93}} (1996) 120.

\bibitem{bib-fit-alpha2}
F.\ Berends, W.\ van Neerven, G.\ Burgers, Nucl.\ Phys.\ {\bf {B297}} (1988)
  429.

\bibitem{bib-fit-alpha3}
G. Montagna, O. Nicrosini, F. Piccinini, Phys. Lett. {\bf {B406}} (1997) 243.

\bibitem{bib-arbuzov}
A.B. Arbuzov, ``Light pair Corrections to Electron-Positron Annihilation at
  LEP/SLC'', hep-ph/9907500.

\bibitem{bib-pcp}
D. Bardin, M. Gr{\"u}newald, G. Passarino, ``Precision Calculation Project
  Report'', hep-ph/9902452.

\bibitem{bib-degrassi}
G. Degrassi, P. Gambino, A. Vicini, \PLB {\bf 383} (1996) 219; \\ G.~Degrassi,
  P.~Gambino, A.~Sirlin, \PLB {\bf 394} (1997) 188; \\ G.~Degrassi, P.~Gambino,
  M.~Passera, A.~Sirlin, \PLB {\bf 418} (1998) 209.

\bibitem{bib-alphas-2}
K.\ G.\ Chetyrkin, J.\ H.\ K{\"u}hn, A.\ Kwiatkowski, Phys.~Rept.~{\bf 277}
  (1996) 189.

\bibitem{bib-kniehl}
B.~Kniehl, Nucl. Phys. {\bf {B347}} (1990) 86.

\bibitem{bib-czak}
A.~Czarnecki, J.H.~K{\"u}hn, \PRL {\bf 77} (1996) 3955;\\ Erratum: \PRL {\bf
  80} (1998) 893.

\bibitem{bib-fit-ALIBABA}
W. Beenakker, F.A. Berends, S.C. van der Marck, Nucl. Phys. {\bf {B349}} (1991)
  323.

\bibitem{bib-MINUIT}
F.~James, ``MINUIT'', CERN Program Library entry {\bf D506}, CERN, 1994.

\bibitem{bib-top-mass}
Particle Data Group, D.E.Groom \etal , Eur.\ Phys.\ J.\ {\bf C15} (2000) 1.

\bibitem{bib-alphas-siggi}
S.~Bethke, J.\ Phys.\ {\bf G\,26} (2000) R27.

\bibitem{bib-MH-lower}
The LEP Working Group for Higgs Boson Searches ALEPH, DELPHI, L3, and OPAL,
  ``Limits on Higgs Boson Masses from Combining the Data of the four LEP
  Experiments at Energies up to 183 GeV'', CERN-EP-2000-055.

\bibitem{bib-MH-upper}
M.~Veltman, Acta Phys.~Polon.~{\bf B8}, (1977) 475;\\ B.W.~Lee, C.~Quigg,
  H.~Thacker, Phys. Rev. {\bf D16} (1977) 1519;\\ D.~Discus, V.~Mathur, Phys.
  Rev. {\bf D7} (1973) 3111.

\bibitem{bib-stuartvanritbergen}
T.\ van Ritbergen, R.\ Stuart, \PLB~{\bf 437} (1998) 201;\\ T.\ van Ritbergen,
  R.\ Stuart, \PRL~{\bf 82} (1999) 488.

\bibitem{bib-smatrix}
A.~Leike, T.~Riemann, J.~Rose, Phys. Lett. {\bf B 273} (1991) 513; \\
  T.~Riemann, Phys. Lett. {\bf B 293} (1992) 451; \\ S.~Kirsch, T.~Riemann,
  Comp. Phys. Comm. {\bf 88} (1995) 89.

\bibitem{bib-isr-study}
S.~Jadach, B.~Pietrzyk, M.~Skrzypek \PLB {\bf 456} (1999) 77.

\bibitem{bib-fit-yfspairs}
S. Jadach, M. Skrzypek, M. Martinez, Phys. Lett. {\bf {B280}} (1992) 129;\\ M.
  Martinez and B. Pietrzyk, Phys. Lett. {\bf {B423}} (1994) 492; \\ S. Jadach,
  M. Skrzypek, B. Pietrzyk, Phys.Lett. {\bf{B456}}( 1999) 77.

\bibitem{bib-all-ls89}
Mark II Collaboration, G. S. Abrams \etal, Phys. Rev. Lett. {\bf {63}} (1989)
  724;\\ ALEPH Collaboration, D. Decamp \etal, Phys. Lett. {\bf {B231}} (1989)
  519;\\ DELPHI Collaboration, P. Aarnio \etal, Phys. Lett. {\bf {B231}} (1989)
  539;\\ L3 Collaboration, B. Adeva \etal, Phys. Lett. {\bf {B231}} (1989)
  509;\\ OPAL Collaboration, M.Z. Akrawy \etal, Phys. Lett. {\bf {B231}} (1989)
  530.

\bibitem{bib-tau-pol}
ALEPH Collaboration, D.~Buskulic \etal, \ZPC {\bf 69} (1996) 183;\\ DELPHI
  Collaboration, P.~Abreu \etal, \ZPC {\bf 67} (1995) 183;\\ L3 Collaboration,
  M.~Acciarri \etal, \PLB {\bf 429} (1998) 387;\\ OPAL Collaboration,
  G.~Alexander \etal, \ZPC {\bf 75} (1996) 365.

\bibitem{bib-sld-alr}
SLD Collaboration, K.~Abe \etal, \PRL {\bf 73} (1994) 25;\\ SLD Collaboration,
  T.~Abe \etal, ``The Final SLD Results for $\Alr$ and $\AAl$'', SLAC-PUB-8646
  (October 2000).

\bibitem{bib-alphas-soper}
L. Surguladze and D.E. Soper, Phys.~Rev.~{\bf {D54}} (1996) 4566.

\bibitem{bib-alphas-kuehn2}
K.~G.~Chetyrkin, J.H.\ K{\"u}hn and M.\ Steinhauser, \PLB {\bf 351} (1995) 33.

\bibitem{bib-4fcorr}
OPAL Collaboration, K. Ackerstaff \etal, \EPJ {\bf {2}} (1998) 441.

\bibitem{bib-mc-LEP2yr-mcws}
``LEP2 Monte Carlo Workshop : Report of the Working Groups on Precision
  Ca1culations for LEP2 Physics'', CERN-2000-009, eds. S. Jadach, G. Passarino,
  R. Pittau.

\bibitem{bib-fit-ALIERROR}
W. Beenakker and G. Passarino, Phys. Lett. {\bf B425} (1998) 199.

\bibitem{bib-fit-ALIERROR-fb}
W. Beenakker and G. Passarino, private communications.

\end{thebibliography}

\end{document}